\DocumentMetadata 
{
	lang		= en-US, 
	pdfversion  = 1.7,
	pdfstandard = a-2b,
}

\documentclass[twoside]{mitthesis}
\usepackage{listings}

\usepackage[version=4]{mhchem}

\usepackage{lipsum}
\IfPackageAtLeastTF{lipsum}{2021/09/20}{\setlipsum{auto-lang=false}}{} 
\usepackage{subcaption}

\usepackage{silence}
\usepackage{booktabs}

\usepackage{array}

\usepackage{dcolumn}
  \newcolumntype{d}[1]{D{.}{.}{#1}}

\usepackage{longtable}

\usepackage[nopatch=footnote]{microtype}

\usepackage{multicol}

\usepackage[style=ext-numeric-comp,giveninits=true,maxbibnames=10,sorting=none,language=american]{biblatex}
\usepackage{pdfpages}

    \AtEveryBibitem{%
      \ifentrytype{article}{%
        \renewbibmacro{in:}{}
		\renewbibmacro*{issue+date}{
		\iffieldundef{issue}%
        	{}%
        	{\printfield{issue}%
         	\setunit*{\addcomma\space}
        	}%
			\printdate
		}
      }{}
    }
	\renewbibmacro*{volume+number+eid}{%
        \printtext[bold]{\printfield{volume}}
        \iffieldundef{number}
          {} 
          {\printtext[parens]{\printfield{number}}} 
        \setunit{\bibeidpunct}%
        \printfield{eid}
    }

\hypersetup{%
	pdfsubject={Template for writing MIT theses with the mitthesis class},
	pdfkeywords={Massachusetts Institute of Technology, MIT},
	pdfurl={},
	pdfcontactemail={},
	pdfauthortitle={},
}

\DeclareMathOperator{\Erf}{erf}
\DeclareMathOperator{\Erfc}{erfc}
\DeclarePairedDelimiterXPP\erf[1]{\Erf\mkern1mu}(){}{#1}
\DeclarePairedDelimiterXPP\erfc[1]{\Erfc\mkern1mu}(){}{#1}

\usepackage{tikz}
\usepackage{tikz-feynman}
\tikzfeynmanset{compat=1.1.0}
\begin{document}

\title{First Demonstration of a Hybrid Cherenkov and Scintillation Detector in a Proof-of-Principle Axion Search at a Beam Dump}
\Author{Darcy A. Newmark}{Department of Physics}[B.A. Physics, University of Chicago, 2021]
\Degree{Doctor of Philosophy}{Department of Physics}
\Supervisor{Janet M. Conrad}{Professor of Physics}[Department of Physics]
\Acceptor{Scott A. Hughes}{Interim Associate Department Head}{}
\DegreeDate{May}{2026}

\ThesisDate{May 14, 2026}

%
\Reader{Janet M. Conrad}{Professor of Physics}{Department of Physics}
\Reader{Iain W. Stewart}{Otto and Jane Morningstar Professor of Science}{Department of Physics}
\Reader{Lindley A. Winslow}{Associate Department Head}{Department of Physics}

%
%
%


%
%
%
%
%
%
%
%
%

%
%


\maketitle





\begin{abstract}
%
%

This thesis presents the development and first experimental demonstration of a hybrid Cherenkov and scintillation optical detector at a beam dump facility, utilized for a proof-of-principle search for high mass ($m_a \sim \mathcal{O}(1~\text{MeV})$) axion-like particles (ALPs). This research leverages the Coherent CAPTAIN-Mills (CCM) experiment, a 10-ton liquid argon light collection detector located at Los Alamos National Laboratory. The CCM detector is instrumented with 200 photomultiplier tubes (PMTs) to achieve 50\% photo-cathode coverage. A key aspect of the detector design is that 80\% of the PMTs are coated in wavelength shifting material while the remaining 20\% of the PMTs are left uncoated. This wavelength discrimination, along with 2~ns timing resolution, allows for enhanced sensitivity to Cherenkov radiation. 

This work includes detailed modeling of $^{22}$Na calibration source data, resulting in the first detailed description of both scintillation and Cherenkov light production and propagation parameters, with systematic uncertainties, in a liquid argon light collection detector. Using this same low-energy $^{22}$Na source, this work demonstrates the first event-by-event separation of Cherenkov radiation from sub-MeV electrons in a high light-yield scintillation detector. 

Additionally, this thesis includes the development of a machine learning-based position reconstruction algorithm with $\mathcal{O}(5~\text{cm})$ spatial resolution and an energy reconstruction framework achieving $\sim 10\%$ resolution at 1~MeV.

Using both the features of Cherenkov radiation and event reconstruction algorithms developed for this thesis, this work performs a proof-of-principle ALP search at a beam dump. Four discriminating observables, leveraging the enhanced sensitivity of the uncoated PMTs to visible Cherenkov radiation in the early portion of events, directionality of Cherenkov emission, pulse shape differences between $\mathcal{O}(1~\text{MeV})$ electromagnetic final state particles and high energy fast neutrons, and spatial extent of event topologies, are constructed. These are then combined into a likelihood ratio test statistic for powerful rejection of steady state backgrounds in this analysis focusing on $\leq10~\text{MeV}$ kinetic energy events. 

Fitting for this ALP signal hypothesis in a frequentist framework yielded no significant excess of observed events over the expected background distribution. The enhanced background rejection capabilities, however, leveraging both Cherenkov and scintillation signals, enabled exclusion of new regions of the mass–coupling parameter space at the $90\%$ confidence level compared to a previous CCM analysis with greater exposure time. 

Finally, the thesis explores the broader utility of liquid argon detectors, investigating neutral current neutrino-argon interactions for supernovae neutrino flux constraints in the DUNE far detector. Additionally, ongoing detector developments targeting an ultra large liquid argon hybrid optical Cherenkov and scintillation detector are discussed as well. 

The techniques established in this thesis demonstrate the powerful background rejection capabilities of hybrid optical detectors, providing a road map for next-generation dark sector and neutrino experiments.
\end{abstract}



\chapter*{Acknowledgments}
\pdfbookmark[0]{Acknowledgments}{acknowledgments}

This work would not have been possible without the mentorship of my advisor Professor Janet Conrad. In addition to her vast knowledge of physics, I admire her dedication to teaching. Working in her group at MIT is so enjoyable and productive because of her example to always be curious and enthusiastic to teach others. 

I owe most of my technical skills to Professor Austin Schneider. This work was made possible by your vast knowledge and guidance in all things programming. Additionally, your hospitality made my yearly trips to Los Alamos both possible and enjoyable. 

Thank you Richard Van De Water for making CCM into a reality. In addition, thank you for facilitating computing resources and visits to Los Alamos; this work would not have been possible without either. 

Finally, thank you, Erik, for always supporting me. I love you. 


\tableofcontents
\listoffigures
\listoftables



\chapter{Motivations for this Thesis}\label{chap:intro}
The ultimate goal of this thesis is to study neutrinos and potential dark sector mediators. This is accomplished in two ways--- the first being developments in detector instrumentation enabling the first observation of Cherenkov light from sub-MeV electrons on an event-by-event basis in a high light-yield scintillation detector~\cite{CCM:2025kal,CCM:2025dbq}. The second technique is through Monte Carlo modeling and statistical methods to both search for new physics and model potential capabilities of future experimental searches~\cite{Newmark:2023vup}. 

The work in this thesis is carried out on the Coherent CAPTAIN-Mills experiment, a liquid argon light collection detector located at Los Alamos National Laboratory. This chapter will introduce neutrinos, describe the physics motivations for the CCM experiment, discuss the motivations for hybrid Cherenkov radiation and scintillation detection, and conclude by outlining the organization of this thesis. 

\section{The Standard Model and Neutrinos}
The Standard Model (SM) describes how all particles and fields interact. It is composed of quarks, leptons, and bosons, which differ in their electric charge and quantum spin. Neutrinos are neutral leptons and come in three ``flavors'' corresponding to the three charged leptons--- $\nu_e$, $\nu_{\mu}$, and $\nu_{\tau}$. 

\begin{figure}[t]
\centering
\begin{tikzpicture}

\begin{feynman}
  \vertex (i1) at (0, 1.5) {\(\nu_x\)};
  \vertex (i2) at (0,-1.5) {\(l\)};

  \vertex (v1) at (2, 0.7);
  \vertex (v2) at (2,-0.7);

  \vertex (f1) at (4, 1.5) {\(\nu_x\)};
  \vertex (f2) at (4,-1.5) {\(l\)};

  \diagram*{
    (i1) -- [fermion] (v1) -- [fermion] (f1),
    (i2) -- [fermion] (v2) -- [fermion] (f2),
    (v1) -- [boson, edge label=\(Z^0\)] (v2),
  };
\end{feynman}

\begin{feynman}[xshift=6cm]
  \vertex (i1) at (0, 1.5) {\(\nu_l\)};
  \vertex (i2) at (0,-1.5) {\(d\)};

  \vertex (v1) at (2, 0.7);
  \vertex (v2) at (2,-0.7);

  \vertex (f1) at (4, 1.5) {\(l\)};
  \vertex (f2) at (4,-1.5) {\(u\)};

  \diagram*{
    (i1) -- [fermion] (v1) -- [fermion] (f1),
    (i2) -- [fermion] (v2) -- [fermion] (f2),
    (v1) -- [photon, edge label=\(W^{+}\)] (v2),
  };
\end{feynman}

\end{tikzpicture}
\caption{Neutrino interactions via \(Z^0\) and \(W^{\pm}\) bosons. The neutral current channel, left, is the scattering of neutrinos off of generic targets. The charged current channel, right, is the exchange of a charged \(W^{\pm}\) boson that results in the initial neutrino being converted into its corresponding charged lepton.}
\label{fig:nu_feynman}
\end{figure}
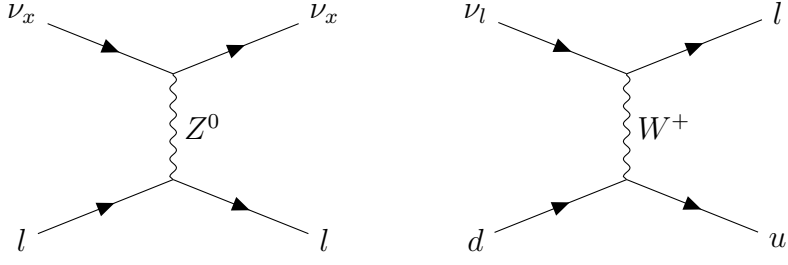

Neutrinos are only charged under SU(2) symmetry and therefore only interact via the weak force. Neutrinos can interact through both neutral current, mediated by the $Z^0$ boson, and charged current, mediated by the $W^{\pm}$ boson, exchange. Fig.~\ref{fig:nu_feynman} shows one possible Feynman diagram for each of these two interaction channels. In the diagram on the left, any flavor neutrino can scatter off of leptons, nucleons, or nuclei through exchange of the $Z^0$ boson, leaving the initial and final state particle types unchanged. Charged current exchange, demonstrated in the diagram on the right-hand side, is different in that the neutrino is converted into its corresponding lepton to conserve charge. These charged leptons typically deposit visible energy in detectors, forming the basis of many neutrino experiments.

\section{Physics Motivations for the Coherent CAPTAIN Mills Experiment}

The Coherent CAPTAIN Mills (CCM) collaboration is a small-scale experimental program comprising of approximately 10 institutions since its founding in 2018. CCM is designed to study neutrino and dark sector physics using the pulsed proton beamline and tungsten target at Los Alamos Neutron Science Center. This section will discuss the physics that motivated the design of the CCM detector, which includes neutrino oscillation studies, cross section measurements, searches for light dark matter~\cite{CCM:2021leg,CCM:2021yzc}, and searches for general portals to the dark sector like axion-like particles~\cite{CCM:2021jmk,CCM:2023itc}. 

\subsection{Neutrino Oscillations}
One of the most important phenomena that influences the field of particle physics today is neutrino oscillations. These oscillations arise when neutrino mass states propagate at different rates resulting in interference in the observed flavor basis. Neutrino oscillations were first proposed in 1957~\cite{Pontecorvo1957} but not conclusively confirmed through experimental observations until the turn of the century. The three mass eigenstates, denoted as $\nu_1$, $\nu_2$, and $\nu_3$, can be related to the flavor eigenstates through the unitary Pontecorvo–Maki–Nakagawa–Sakata (PMNS) matrix~\cite{pmns}, given in Eq.~\ref{eq:pmns}

\begin{equation}
    \begin{pmatrix}
    \nu_{e} \\
    \nu_{\mu} \\
    \nu_{\tau}
    \end{pmatrix}
    = 
    \begin{pmatrix}
    U_{e1} & U_{e2} & U_{e3} \\ 
    U_{\mu 1} & U_{\mu 2} & U_{\mu 3} \\
    U_{\tau 1} & U_{\tau 2} & U_{\tau 3}
    \end{pmatrix}
    \begin{pmatrix}
    \nu_1 \\
    \nu_2 \\
    \nu_3
    \end{pmatrix}
    \label{eq:pmns}
\end{equation}

The PMNS matrix can be fully described by four free parameters--- $\theta_{12}$, $\theta_{13}$, $\theta_{23}$, and $\delta_{CP}$, shown in Eq.~\ref{eq:pmns2}. The three mixing angles govern oscillations between the denoted mass eigenstates and the fourth parameter, $\delta_{CP}$, is a charge-parity (CP) symmetry violating phase.

\begin{equation}
    \begin{pmatrix}
    1 & 0 & 0 \\ 
    0 & \cos{\theta_{23}} & \sin{\theta_{23}} \\
    0 & -\sin{\theta_{23}} & \cos{\theta_{23}}
    \end{pmatrix}
    \begin{pmatrix}
    \cos{\theta_{13}} & 0 & \sin{\theta_{13}} e^{-i\delta_{CP}} \\ 
    0 & 1 & 0 \\
    -\sin{\theta_{13}} e^{-i\delta_{CP}} & 0 & \cos{\theta_{13}}
    \end{pmatrix}
    \begin{pmatrix}
    \cos{\theta_{12}} & \sin{\theta_{12}} & 0 \\ 
    -\sin{\theta_{12}} & \cos{\theta_{12}} & 0 \\
    0 & 0 &1
    \end{pmatrix}
    \label{eq:pmns2}
\end{equation}

For a neutrino created with a given flavor $\alpha$, the probability of observing that neutrino as flavor $\beta$ can be described using the PMNS matrix, demonstrated in Eq.~\ref{eq:prob}.

\begin{equation}
    P(\nu_{\alpha} \rightarrow \nu_{\beta}) = | \langle \nu_{\beta} | \nu(t) \rangle|^2 = |\Sigma_{i} U_{\beta i} U_{\alpha i}^* e^{-i E_i t}|^2
    \label{eq:prob}
\end{equation}

For the simple case of two neutrino oscillations, Eq.~\ref{eq:prob} can be approximated as Eq.~\ref{eq:prob_2nu}. The appearance probability depends on the mixing angle $\theta$ between the two mass states, the square of the mass splitting between the two states $\Delta m^2$, the distance traveled $L$, and the energy of the initial neutrino $E$.

\begin{equation}
    P(\nu_{\alpha} \rightarrow \nu_{\beta}) = \sin^2 2 \theta \sin^2 \left( \frac{1.27 \Delta m^2 L}{E} \right)
    \label{eq:prob_2nu}
\end{equation}

At this point in time, neutrino oscillations involving the three flavors in the Standard Model are well established through the results from many experiments~\cite{10.1093/ptep/ptac097}. The parameters that govern neutrino oscillations are $\sin^2(2\theta_{13}) = 0.086 \pm 0.002$, $\sin^2(2\theta_{12}) = 0.845 \pm 0.018$, $\sin^2\theta_{23} = 0.56^{+0.03}_{-0.05}$, $\Delta m^2_{21} = 7.41^{+0.021}_{-0.020} \times 10^{-5}~\text{eV}^2$, and $|\Delta m^2_{31}| = 2.437^{+0.028}_{-0.027} \times 10^{-3}~\text{eV}^2$. While there have been some experimental measurements of $\delta_{CP}$, it remains weakly constrained~\cite{T2K:2025wet}. 

In the past 30 years, there have also been multiple reported anomalies, with the most significant from the Liquid Scintillator Neutrino Detector (LSND)~\cite{LSND:2001aii}, Mini Booster Neutrino Experiment (MiniBooNE)~\cite{miniboone_1,miniboone_2,miniboone_3,miniboone_4} and the Baksan Experiment on Sterile Transitions (BEST)~\cite{PhysRevLett.128.232501}. Each of these experiments, when taken on their own, is consist with neutrino oscillations with a mass splitting of around $\mathcal{O}(0.5-10~\text{eV}^2)$, substantially higher than the observed $\Delta m^2_{21}$ and $|\Delta m^2_{31}|$ parameters.

The simplest explanation is the introduction of a fourth neutrino mass state which has a corresponding ``sterile" flavor state that does not interact with the weak force, referred to as ``3+1" model for the three active neutrinos and one sterile state. This would alter the observed oscillation physics while avoiding constraints on the number of active neutrinos from measurements of the width of the $Z^0$ boson~\cite{zboson_width}. Global fits, however, to a broad range of experimental oscillation data show internal tension of $3.7\sigma$ when allowing for this simple extension of the three neutrino oscillation paradigm~\cite{Hardin:2022muu}. The source of the oscillation-like signatures of the anomalies is still not well understood.

The anomaly observed by the LSND experiment is particularly relevant to this thesis because the design of CCM was motivated to probe this result. The LSND experiment measured $\overline{\nu}_e$ particles produced at the Los Alamos Meson Physics Facility (LAMPF) in the 1990s. LSND was located at an intense pion decay-at-rest source which produces monogenetic $\nu_{\mu}$ through the prompt pion decay and delayed fluxes of $\nu_e$ and $\overline{\nu}_{\mu}$ through the subsequent decay of the muon, detailed in Eq.~\ref{eq:pidar}. LAMPF utilized an 800~MeV proton beam impinging on two different target configurations, one primarily composed of high-Z material and the other composed of water, to produce intense fluxes of these neutrinos.

\begin{equation}
\begin{aligned}
    \pi^+ &\rightarrow \mu^+ + \nu_{\mu},\\
    &\quad \mu^+ \rightarrow e^+ + \nu_e + \overline{\nu}_{\mu}
\end{aligned}
\label{eq:pidar}
\end{equation}

The detector consisted of approximately 170~tons of lightly doped liquid scintillator and was instrumented with around 1200 photomultiplier tubes to measure optical photons. This detector configuration was optimized to measure $\overline{\nu}_e$ inverse beta decay interactions, where the $\overline{\nu}_e$ exchanges a $W^-$ boson with protons in the bulk material to produce a final state positron and neutron. The positron produces an initial flash of both Cherenkov and scintillation light. The neutron then captures on free protons in the material and produces a delayed gamma-ray emission at a characteristic energy of 2.2~MeV with a capture lifetime of 186~$\mu$s. This process is detailed in Eq.~\ref{eq:ibd}.

\begin{equation}
\begin{aligned}
    \overline{\nu}_e + p \rightarrow e^+ + n,\\
    n+p \rightarrow \gamma + p
\end{aligned}
\label{eq:ibd}
\end{equation}

\begin{figure}[h]
  \centering
  \includegraphics[width=0.6\linewidth]{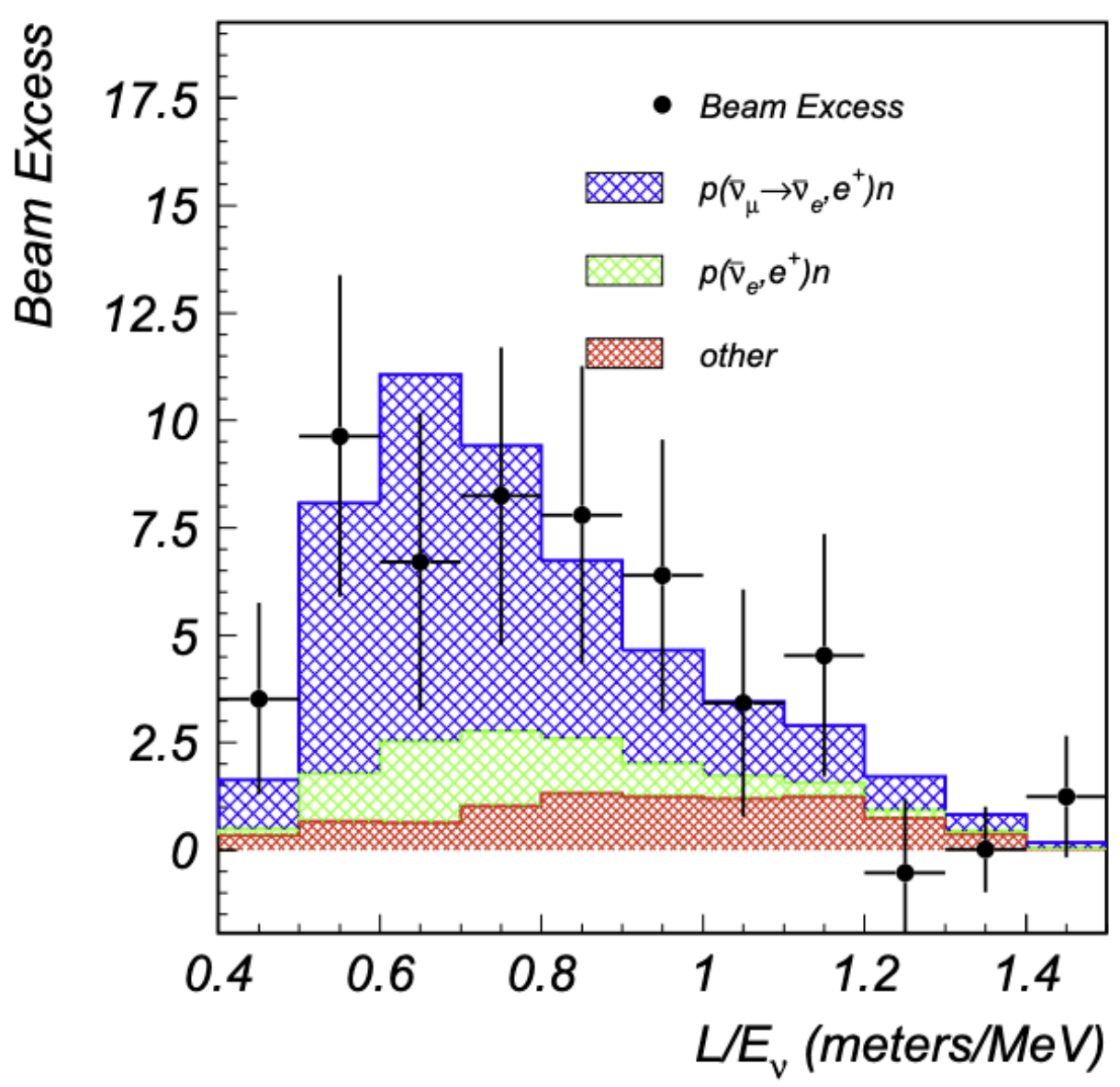}
  \caption{Measurement of $\overline{\nu}_e$ events above the expected background rate from LSND. The x-axis is parameterized as distance $L$ divided by energy $E$ for easy interpretation in the context of neutrino oscillations. The expected backgrounds to this process are displayed in the red and green histograms. The best fit to the $\overline{\nu}_{\mu} \rightarrow \overline{\nu}_e$ oscillations is the blue histogram, which requires introduction of a fourth sterile neutrino state. Figure from Ref.~\cite{LSND:2001aii}.}
  \label{fig:lsnd_excess}
\end{figure}

Given the three detection signals, prompt Cherenkov and scintillation light from the positron and delayed scintillation light from the gamma-ray, this channel has very low backgrounds. This makes it ideal for measuring $\overline{\nu}_{\mu} \rightarrow \overline{\nu}_e$ appearance. LSND measured a 3.8$\sigma$ excess of events over the expected background rate, shown in Fig.~\ref{fig:lsnd_excess}~\cite{LSND:2001aii}.

While the 3+1 explanation for the LSND anomaly is generally not favored due to the tension in the global fits, there is still a significant excess of events in a well constrained and low background interaction channel. This leads to exploration of more exotic solutions like dark sector particles as a potential answer. The CCM experiment, located at a similar pion decay-at-rest source and similar $L/E$ parameter, was designed to probe both new neutrino physics and general dark sector models that could explain the LSND anomaly.

\subsection{Neutrino Cross Sections on Argon}
CCM was originally conceptualized to measure low-energy coherent elastic neutrino nucleus scattering (CE$\nu$NS) events in liquid argon~\cite{Shoemaker:2021hvm}. This would have had two major physics results--- it would have been the first observation of this process on argon and it would have been a robust test of the three neutrino oscillation model.

CE$\nu$NS was not observed until 2017 by the COHERENT collaboration using a cesium iodide detector~\cite{COHERENT:2017ipa}. The coherent scattering process of <~50~MeV neutrinos off of target nuclei produces nuclear recoils that deposit $\mathcal{O}(10~\rm{keV})$ of energy. This faint signal is difficult for most detectors to resolve. If CE$\nu$NS events, however, can be measured, they are a very compelling channel to study neutrino interactions. At relevant neutrino energies, the CE$\nu$NS cross section is around an order of magnitude higher than the next leading process, benefiting any search for new physics. Additionally, CE$\nu$NS is still a largely unexplored interaction channel, allowing for its study to uniquely probe potential sources of new physics. 

Following up on the first observation of CE$\nu$NS events on cesium iodide, CCM was envisioned to measure CE$\nu$NS events in liquid argon, thereby constraining the total flux of active neutrinos. This would have both been validation of the SM interpretation of CE$\nu$NS and another test of the LSND anomaly. 

\begin{equation}
\begin{aligned}
    \frac{d\sigma}{dT} &\approx \frac{G_F^2 M}{2 \pi}\frac{Q_W^2}{4} F^2(Q) \left ( 2 - \frac{MT}{E_{\nu}^2} \right ),\\
    Q_W &= N - (1 - 4\sin^2\theta_W)Z
\end{aligned}
\label{eq:cevns}
\end{equation}

The CE$\nu$NS cross section as a function of recoil energy, Eq.~\ref{eq:cevns}, depends on the Fermi coupling constant $G_F$, mass of the target nucleus $M$, weak nuclear charge $Q_W$, nuclear form factor as a function of the momentum transfer $F(Q)$, recoil energy of the nucleus $T$, and energy of the incident neutrino $E_{\nu}$. Since the weak mixing angle $\theta_W$ is close to 0.231, $Q_W \approx N$, so measurements of CE$\nu$NS across different targets both test this expected event rate scaling and can probe $\theta_W$~\cite{10.1093/ptep/ptac097}. Through measurements of CE$\nu$NS in the same $L/E$ parameter space as the LSND experiment, CCM could have contributed a robust test of the sterile neutrino hypothesis through a high statistics and unique interaction channel. 

In addition to CE$\nu$NS, CCM has the potential to measure the $\nu_{e}$ charged current cross section in liquid argon using the delayed flux of $\nu_{e}$ from muon decay. This channel, while at a lower interaction rate than CE$\nu$NS, has never been observed in liquid argon at the $\mathcal{O}(10)$~MeV scale. In addition to being the first experimental study of this channel, this measurement is also very relevant to the liquid argon Deep Underground Neutrino Experiment (DUNE) physics program. One of the physics goals of the DUNE experiment is to measure neutrinos from the core collapse of supernovae, which produces large fluxes of $< 50~\text{MeV}$ neutrinos. At those energies, the DUNE detector is most sensitive to the charged current channel. Constraint of the expected cross section will greatly improve flux measurements of supernovae neutrinos. For more discussion of supernovae neutrinos in the DUNE detector see Chapter~\ref{chap:pheno}. 

\subsection{Dark Matter}
CCM is additionally motivated by the study of dark matter. Dark matter is very well motivated by experimental observations that around 85\% of the total matter density of the universe is not composed of fundamental particles in the SM--- referred to as dark matter. Based on astrophysical and cosmological observations, dark matter candidates must be must be close to neutrally charged under the electromagnetic force, non-baryonic in nature, negligible velocity compared to the speed of light (often referred to as ``cold dark matter"), and long lifetimes compared to cosmological timescales. CCM is sensitive to light dark matter models that couple to SM particles produced in the target then interact through coherent scattering off of argon nuclei in the detector~\cite{CCM:2021leg,CCM:2021yzc}.

The first evidence for matter beyond the SM dates back to 1933. Fritz Zwicky observed the Coma Cluster, containing over 1,000 galaxies around 100~Mpc away, and found that the radial velocity of the galaxies did not align with the visible mass~\cite{Zwicky:1933gu}. To account for these observations, dark matter must compose around 90\% of the mass in the Coma Cluster. While this result provided the first quantitative evidence for a substantial non-luminous matter component, it was not widely accepted until around 50 years later.

In the 1970s, Vera Rubin's study of galactic rotation curves irrefutably proved that invisible matter must account for much of the mass in galaxies~\cite{rubin}. Through measurements of the Doppler shift across the disks of spiral galaxies, Rubin and colleagues measured the velocities of stars on the edges of the galaxies. The expectation was that stars on the edges should move slower with respect to stars at the center of these galaxies, where the majority of the mass and therefore gravitation potential is concentrated. Instead, they observed that distance from the nuclei of the galaxy did not correlate to slower rotational velocities. The predominant theory is that dark matter must be contributing significantly, almost 90\%, to the mass of these galaxies to explain the observed rotation curves in context of the luminous mass.

As observational astrophysics and cosmology has continued to develop, there are even more clues that the majority of the universe is composed of dark matter. Highly sensitive radio telescopes can measure faint signals that appear almost isotropic in the microwave wavelength region, referred to as the Cosmic Microwave Background (CMB).

\begin{figure}[h]
  \centering
  \includegraphics[width=0.7\linewidth]{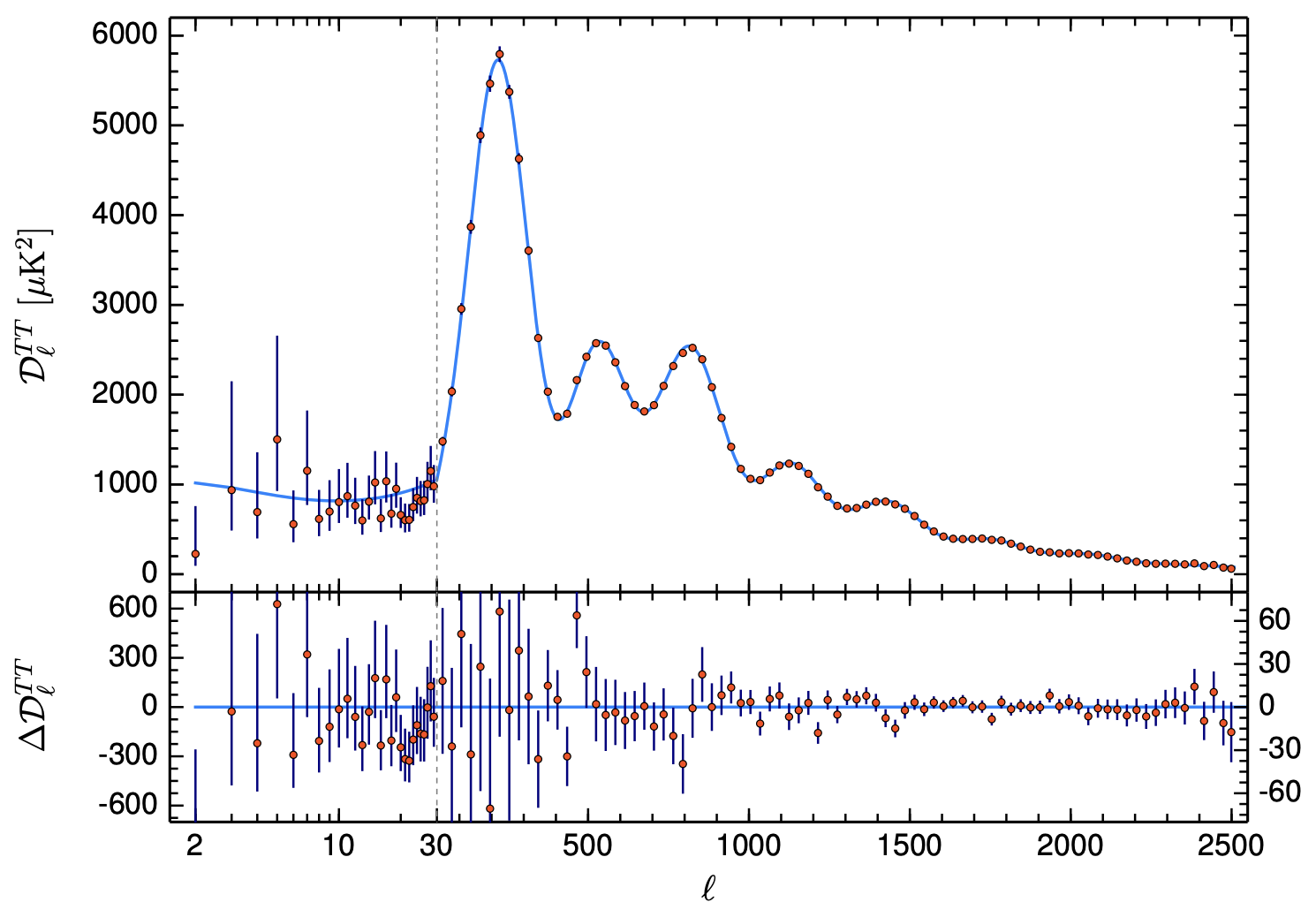}
  \caption{CMB temperature power spectrum as a function of multipole moment $l$, which represents the angular scale on the sky. The data is fit assuming the $\Lambda$CDM theoretical model, allowing for extraction of the matter density $\Omega_m$ and the baryon density $\Omega_b$. The cold dark matter density can then be extracted as $\Omega_c h^2 = 0.120 \pm 0.001$. Figure from Ref.~\cite{Planck:2018vyg}.}
  \label{fig:cmb_power_spectrum}
\end{figure}

Measurements of the CMB are directly probing the universe after recombination, the period in which electrons and protons combined to form neutral hydrogen, allowing photons to decouple and travel freely through the universe. Anisotropies in the CMB are a probe of the dark matter density during recombination~\cite{cosmicmicbackground}. Fits to the CMB temperature power spectrum, displayed in Fig.~\ref{fig:cmb_power_spectrum}, constrain the total dark matter density~\cite{Planck:2018vyg}. 

Measurements of the CMB are decomposed using spherical harmonics, allowing for extraction of the multipole moment $l$, x-axis in Fig.~\ref{fig:cmb_power_spectrum}, as a function of the temperature-temperature power spectrum, y-axis in Fig.~\ref{fig:cmb_power_spectrum}. The total matter density $\Omega_m$ is constrained using the first trough in this spectrum, $\Omega_m = 0.315 \pm 0.007$. The baryon density $\Omega_b$ is additionally parameterized using the second peak in this spectrum, $\Omega_b h^2 = 0.0224 \pm 0.0001$. The difference between the total matter and the baryon density then must be composed of dark matter, parametrized as the cold dark matter density $\Omega_c h^2 = 0.120 \pm 0.001$~\cite{Planck:2018vyg}. 

While there is even more evidence for dark matter, such as formation of large scale structure~\cite{Young:2016ala} and observations of the Bullet Cluster~\cite{Clowe:2006eq}, no such particles have been discovered. This has motivated a wide range of theoretical candidates for beyond SM particles that explain the dark matter density, spanning many orders of magnitude in mass and interaction strength. Among these, the axion is particularly well-motivated, arising from a solution to the strong CP problem while also providing a viable dark matter candidate.

\subsection{Axion Particles}
Named after the laundry detergent, the axion is a proposed pseudo-scalar boson. It both accounts for the observed dark matter in the universe and solves the so-called ``strong CP problem," thereby ``cleaning up" these anomalies. 

While CP symmetry is well known to be violated in the weak sector~\cite{cronin_fitch,BaBar:2001pki,Belle:2001zzw}, it has never been observed to be violated by the strong force. Without a conservation law, CP symmetry in the strong force is suggestive of new physics. 

CP-violation in quantum chromodynamics (QCD) can be described by an effective $\bar{\theta}$ term, included in the QCD Lagrangian, Eq.~\ref{eq:qcd_lag}, where $\alpha_s$ is the strong coupling constant and $G^{\mu \nu a}$ is the color field strength tensor (with $\tilde{G^{a}_{\mu \nu}}$ denoting its dual). While $\bar{\theta}$ can take on any value between $-\pi$ and $\pi$, neutron electric dipole moment experiments have constrained $|\bar{\theta}| < 10^{-10}$~\cite{Baker:2006ts,Abel:2020pzs}.

\begin{equation}
    \mathcal{L_{\bar{\theta}}} = -\bar{\theta} \frac{\alpha_s}{8\pi} G^{\mu \nu a} \tilde{G^{a}_{\mu \nu}}
    \label{eq:qcd_lag}
\end{equation}

In order to explain why $\bar{\theta}$ has been observed to be consistent with null while it is theoretically expected to be $\mathcal{O}(1)$, Pecci and Quinn introduced a new U(1) symmetry, U(1)$_{PQ}$~\cite{Peccei:1977hh}. Under spontaneous symmetry breaking, Nambu-Goldstone bosons can arise--- the axion in the case of U(1)$_{PQ}$. The Lagrangian can then be described as Eq.~\ref{eq:ssb_lag}, introducing the axion field $a$ and the axion decay constant $f_a$. 

\begin{equation}
    \mathcal{L_{\bar{\theta}}} = \left(\frac{a}{f_a} -\bar{\theta} \right)  \frac{\alpha_s}{8\pi} G^{\mu \nu a} \tilde{G^{a}_{\mu \nu}}
    \label{eq:ssb_lag}
\end{equation}

\begin{figure}[h]
  \centering
  \includegraphics[width=\linewidth]{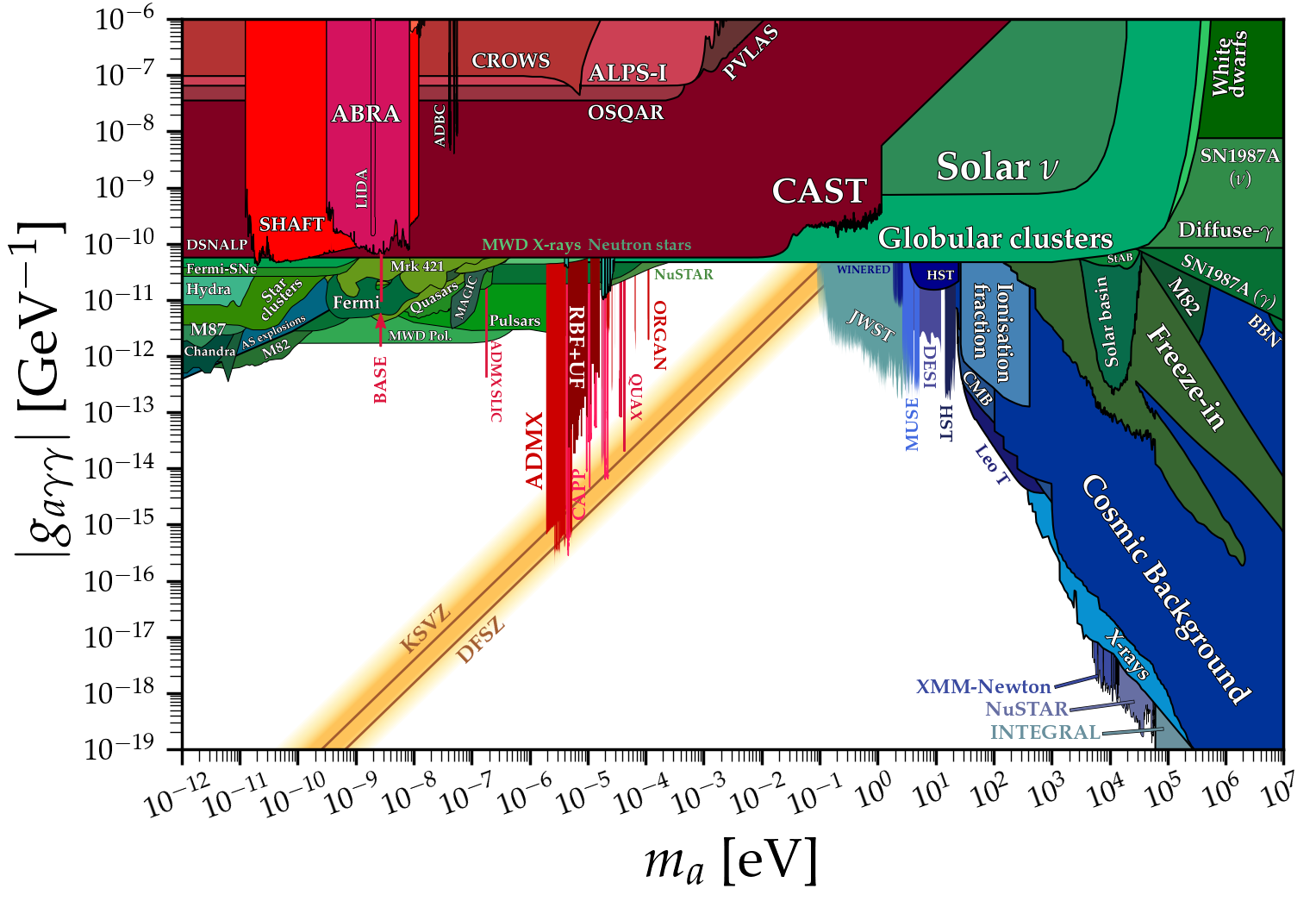}
  \caption{Existing experimental constraints on axion-like particles with coupling to photons. The QCD axion parameter space is highlighted in the gold band, corresponding to $m_a\propto 1/f_a$. Figure from Ref.~\cite{AxionLimits}.}
  \label{fig:axion_photon_limits}
\end{figure}In order to solve the strong CP problem, $a = \bar{\theta} f_a$ satisfactorily sets this term in the QCD Lagrangian to null. There are two primary models for the QCD axion, one from Kim, Shifman, Vainshtein, and Zakharov~\cite{KSVZ_1,KSVZ_2} (KSVZ), and another from Dine, Fischler, Srednicki, and Zhitnishy~\cite{DFSZ_1,DFSZ_2} (DFSZ). The KSVZ model introduces heavy vector-like quarks that couple to the axion, while the DFSZ model instead extends the Higgs sector, allowing the axion to couple directly to SM fermions.

While the axion solves the strong CP problem, there are additional axion-like particles (ALPs) with similar characteristics but do not solve the strong CP problem. Since ALPs interact similarly to axions, experiments can often search for both classes of particles, both of which are viable candidates for dark matter. 

Most searches for axions exploit their coupling to the electromagnetic force, and in particular to photons. The relevant interaction term in the Lagrangian for this interaction, Eq.~\ref{eq:alp_lagrangian_1}, depends on the axion-photon coupling $g_{a\gamma}$, the axion field $a$, and the electromagnetic field strength tensor $F_{\mu\nu}$ and its dual $\tilde{F}^{\mu\nu}$~\cite{10.1093/ptep/ptac097}. 

\begin{equation}
    \mathcal{L_{a\gamma\gamma}} = -\frac{1}{4} g_{a\gamma} a F_{\mu\nu} \tilde{F}^{\mu\nu} 
\label{eq:alp_lagrangian_1}
\end{equation}

The QCD axion parameter space is constrained by the PQ symmetry breaking scale, $m_a \propto 1/f_a$. As a result, a wide range of masses and couplings remains experimentally viable, motivating a diverse set of search strategies. Fig.~\ref{fig:axion_photon_limits} shows current constraints on the axion-photon coupling parameter space~\cite{AxionLimits}. The horizontal axis spans nearly 20 orders of magnitude in mass, while the vertical axis spans more than 25 orders of magnitude in coupling strength. The shaded QCD axion bands, corresponding to the KSVZ and DFSZ models, represent theoretically motivated targets, while the colored regions show existing laboratory and astrophysical exclusions.

While CCM is not sensitive to low mass ($m_a <  1~\text{eV}$) axions that are well motivated candidates for dark matter, CCM is sensitive to higher mass ($m_a \approx \mathcal{O}(1~\text{MeV})$) axions that could be portals to the dark sector~\cite{CCM:2021jmk}. The specific axion model explored in this thesis is discussed in detail in Chapter~\ref{chap:alp}.

\section{Motivations for Hybrid Cherenkov and Scintillation Detection}

While the physics program described above is compelling, its success ultimately hinges on achieving substantial background reduction. Searches for rare processes require impressive suppression of backgrounds, which is particularly challenging in CCM due to its location at a beam dump facility optimized for neutron production.

The physics signals of interest can be broadly divided into two classes--- electromagnetic or nuclear recoil final states. For CE$\nu$NS and light dark matter searches, the expected signal is a low-energy nuclear recoil. These events produce only scintillation light in liquid argon. The dominant background in this regime arises from low-energy electrons, primarily from $^{39}$Ar beta decays, which generate both scintillation and Cherenkov light.

For the higher energy final states expected from charged current cross section measurement and search for general dark sector models, including the axion-like particle model considered in this thesis, the final states consist of $\mathcal{O}(1-50~\text{MeV})$ electromagnetic particles that produce both Cherenkov and scintillation light. In contrast, the dominant background in this regime comes from fast neutrons, which interact hadronically in the liquid argon and produce predominantly scintillation light.

This distinction in signatures between signal and background events motivates the use of Cherenkov light identification as a powerful handle for background rejection. Traditional techniques such as pulse shape discrimination are ineffective in CCM due to the high interaction rate, which leads to significant pile-up and prevents integration of the pulse shape over the microsecond timescales required to resolve scintillation time profiles.

Hybrid Cherenkov and scintillation detection therefore provides a promising alternative approach for powerful background rejection. By exploiting differences in timing and wavelength between the two light production mechanisms, it becomes possible to distinguish signal from background on an event-by-event basis.

This strategy is also of broad interest to the wider neutrino and rare-event search community~\cite{Aberle:2013jba,Escobar:2022jau,Klein:2022lrf}. Several ongoing efforts are exploring techniques to separate Cherenkov and scintillation signals for improved signal identification and background rejection~\cite{Anderson:2022lbb,Callaghan:2023oyu,Klein:2022tqr,NuDoubt:2024jax}. 

\section{Organization of this Thesis}

This thesis will focus on the first proof-of-principle result in the search for axion-like particles at CCM, making use of a new technique to separate Cherenkov from scintillation light in liquid argon.

The thesis is organized as follows.
\begin{itemize}
  \item Chapter~\ref{chap:ccm}--- Introduction to the Lujan Spallation Facility at Los Alamos Neutron Science Center and the CCM200 detector.
  \item Chapter~\ref{chap:low_level}--- Description of the muon tagging system and the low level data processing that were instrumentation contributions by the author.
  \item Chapter~\ref{chap:om}--- Explanation of the detector calibration performed by the author using an approximately 1~MeV gamma-ray source, resulting in Ref.~\cite{CCM:2025dbq}.
  \item Chapter~\ref{chap:cherenkov}--- First demonstration of Cherenkov radiation separation from scintillation signals for sub-MeV electrons by the author, resulting in Ref.~\cite{CCM:2025kal}.
  \item Chapter~\ref{chap:ml}--- Description of event reconstruction techniques, focusing on $\leq10~\text{MeV}$ electrons, leveraging graph neural network machine learning techniques performed by the author.
  \item Chapter~\ref{chap:alp}--- Search for axion-like particles for $10^{-3}~\text{MeV} < m_a < 10~\text{MeV}$ by the author.
  \item Chapter~\ref{chap:pheno}--- Phenomenological study by the author of DUNE's ability to constrain the total flux of supernovae neutrinos through the neutral current interaction channel, resulting in Ref.~\cite{Newmark:2023vup}.
  \item Chapter~\ref{chap:conclusion}--- Plans for future results from the CCM experiment.
  \item Appendix~\ref{chap:appendix}--- Ongoing detector studies for an ultra large liquid argon hybrid Cherenkov and scintillation detector by the author in collaboration with scientists from the University of California, Berkeley.
\end{itemize}


\chapter{The Coherent CAPTAIN-Mills Detector and Experimental Facilities}\label{chap:ccm}
The Coherent CAPTAIN-Mills (CCM) Experiment is a 10-ton liquid argon light collection detector located at Los Alamos National Laboratory. The work in this thesis utilizes data collected between approximately October 15 and December 15 2022, representing a total of $1.23 \times 10^{21}$ protons on target (POT). There is an additional dataset, collected from September 2023 to March 2024, with roughly twice the POT exposure that is not considered in this work. Techniques developed in this thesis and ongoing detector calibration studies will be applied to analyses utilizing this higher statistics dataset. This chapter will discuss details of the CCM experiment, including the accelerator facilities, CCM200 detector, and detector operations. 

\section{Accelerator Facilities}
The CCM detector is a beam dump experiment located at the Los Alamos Neutron Science Center (LANSCE). Although the primary mission of LANSCE is neutron production for materials and nuclear science research, the high-intensity proton beam and tungsten target system also provide an intense source of secondary particles, including neutrinos. It was Dr. Geoffrey Mills who first recognized the potential of this facility as a neutrino source, motivating the development of a dedicated beam dump experiment in this environment. Coherent CAPTAIN-Mills is named in his honor.

\subsection{Proton Beamline}
The linear accelerator at LANSCE produces a proton beam of approximately 800~MeV kinetic energy. The proton beam is bunched in the proton storage ring (PSR) using radio frequency (RF) bunchers and quadrupole focusing magnets. This results in beam spills of approximately 20~Hz, 80~$\mu$A current, and 290~ns maximum time per beam spill.

\begin{figure}[h]
  \centering
  \includegraphics[width=0.7\linewidth]{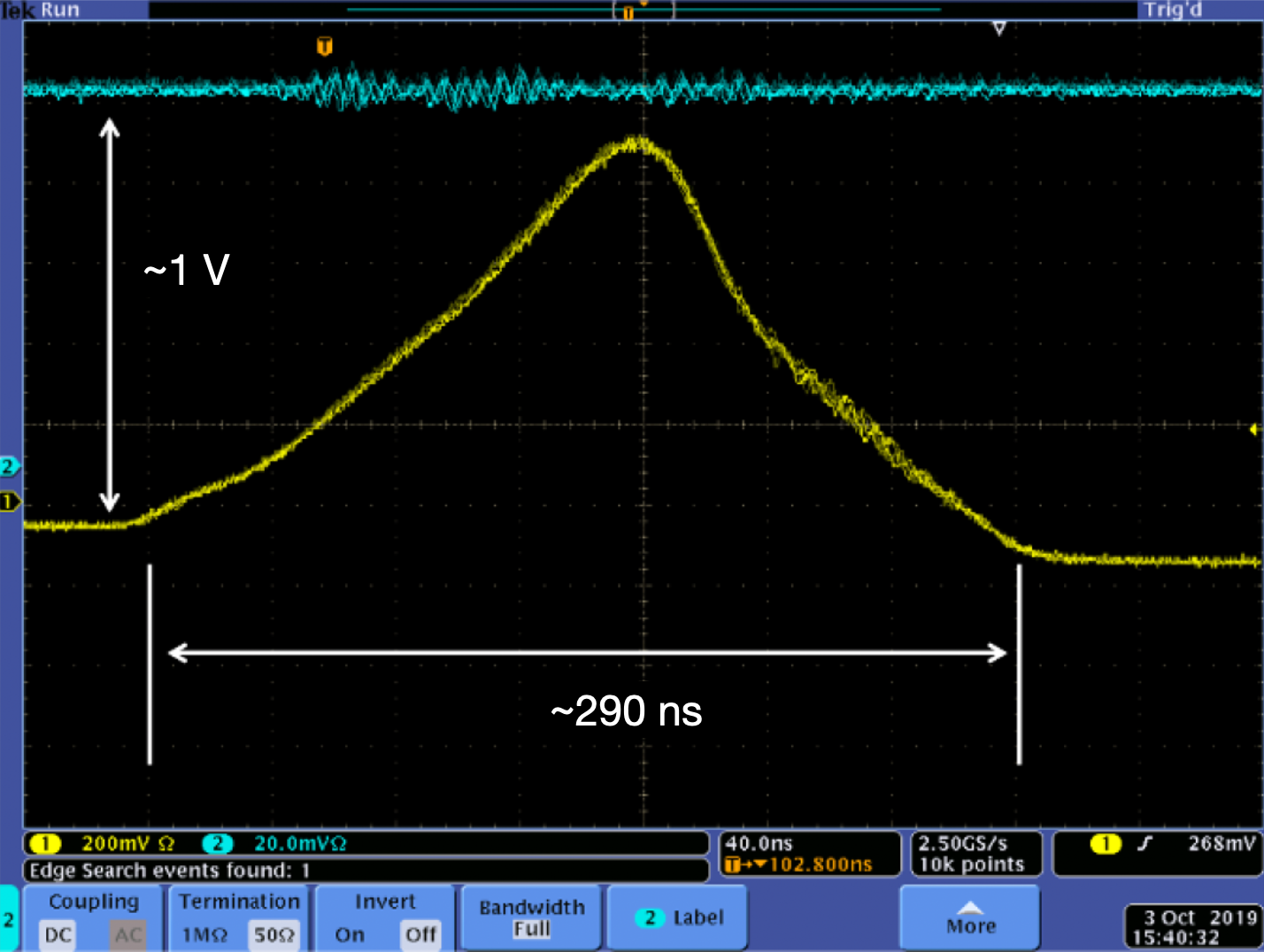}
  \caption{Example of the beam pulse time distribution. The image is captured from the Beam Current Monitor read out on an oscilloscope. The triangular beam pulse shape is evident, with a maximum time per beam spill measured as 290~ns.}
  \label{fig:beam_spill}
\end{figure}

Beam pulses are measured through the Beam Current Monitor (BCM), providing a NIM trigger signal at the CCM detector. This allows the data acquisition system at CCM to trigger on the beam spills and measure the properties of each beam spill. Fig.~\ref{fig:beam_spill} demonstrates this beam pulse on an oscilloscope. As a result of the PSR operations, the beam pulse has a triangular shape. At the widest point, the entire beam spill is approximately 290~ns long. This short beam spill allows for powerful rejection of steady state backgrounds. 

The triangular time structure arises from the way protons are accumulated in the storage ring. Small bunches of protons from the linear accelerator are injected in the PSR over a finite time window and are compressed as they circulate using RF bunching. This injection procedure is repeated multiple times, resulting in ``stacking” of the proton bunches. Protons injected earliest complete the largest number of turns in the PSR and therefore are compressed the most. Bunches injected later spend less time circulating and therefore are compressed the least. This variation of turns in the PSR leads to the observed triangular beam spill time distribution. 

\subsection{Tungsten Target}
\begin{figure}[h]
  \centering
  \includegraphics[width=0.7\linewidth]{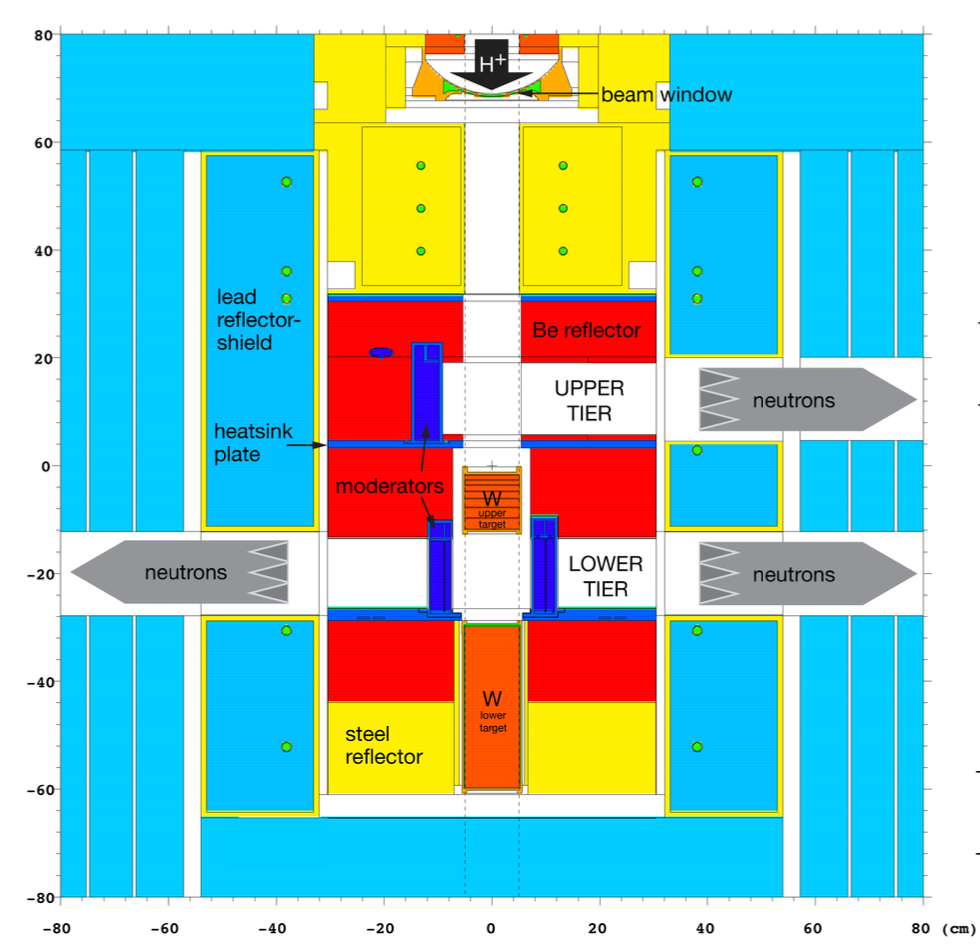}
  \caption{Mark-IV target design that was employed during the 2022 physics data collection period. The tungsten target is separated into two cylinders, an upper and lower target. The upper target cylinder is approximately 7.8~cm in height while the lower target cylinder is approximately 29.8~cm in height, both with a radius of 5~cm. Figure from Ref.~\cite{markiv_target}.}
  \label{fig:target}
\end{figure}

The CCM detector is located at the Lujan Spallation Facility, approximately 23~m away from the target and 90$^{\circ}$ off-axis from the beamline. Fig.~\ref{fig:target} shows the Mark-IV target design used during CCM200’s 2022 physics run~\cite{markiv_target}. At the core of the assembly are two cylindrical tungsten volumes, an upper and a lower target, designed to enhance thermal neutron production. These tungsten elements are surrounded by approximately 4~m of steel, 0.5~m of lead, and 4~m of concrete shielding to attenuate secondary radiation.

The proton beam is bent by 90$^\circ$ using magnetic fields before striking the tungsten target from above. Interactions of the protons with tungsten nuclei produce both hadronic and electromagnetic particle showers. Because the CCM detector is positioned 90$^\circ$ off-axis, it is primarily sensitive to the isotropic flux of particles originating from decays-at-rest, with minimal contamination from forward-peaked decays-in-flight particles.

\begin{figure}[h]
  \centering
  \includegraphics[width=\linewidth]{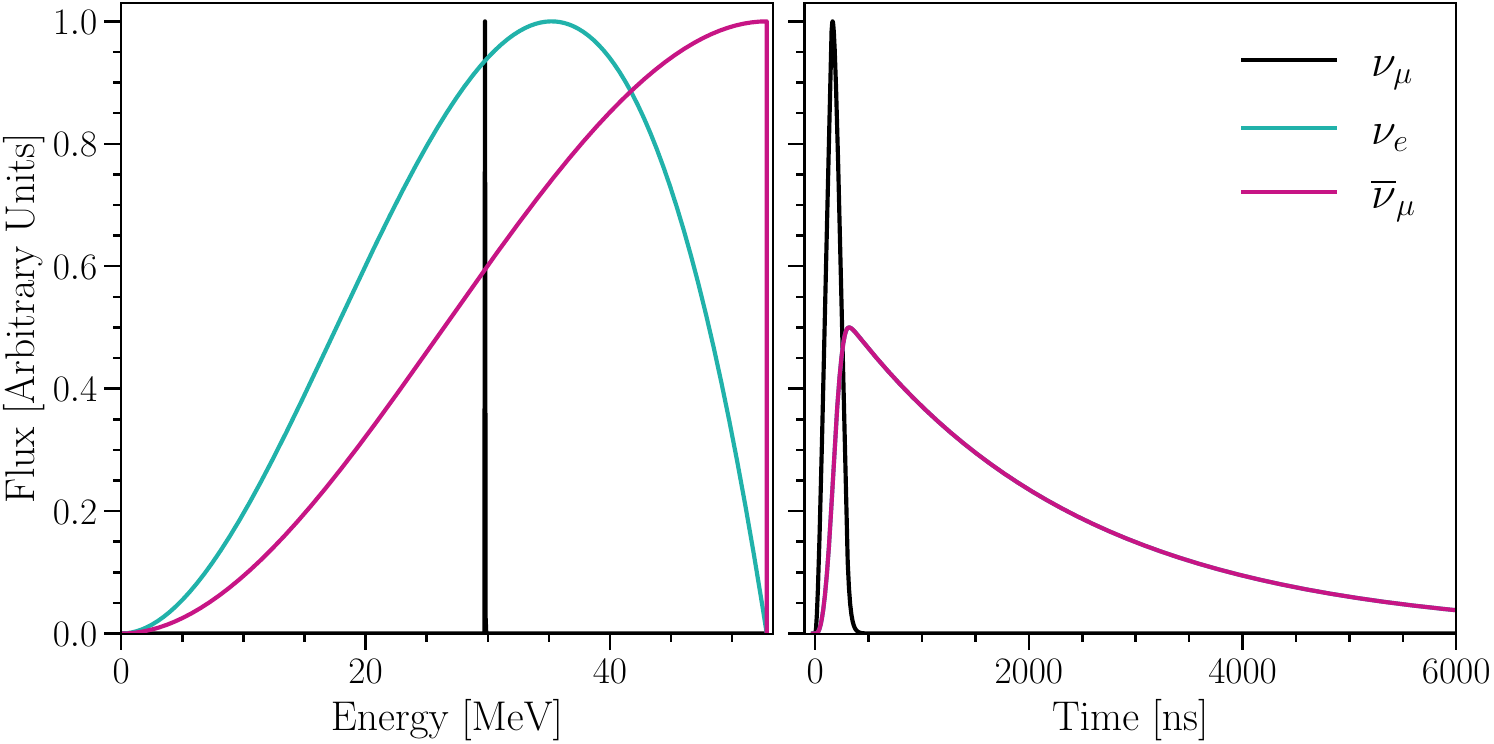}
  \caption{The left-hand plot displays the energy distribution of $\nu_{\mu}$, $\nu_e$, and $\overline{\nu}_{\mu}$ produced from $\pi^+$ decay at rest in the tungsten target. The right-hand plot displays the time distribution of the $\nu_{\mu}$, $\nu_e$, and $\overline{\nu}_{\mu}$ fluxes, which are the result of the convolution between the triangular beam pulse and the physical exponential decay lifetimes.}
  \label{fig:pidar_neutrino_energytime_spectrum}
\end{figure}

In the target, $\pi^-$ particles are predominantly captured on nuclei, while $\pi^+$ particles decay-at-rest with a lifetime of $\sim$26~ns. The resulting neutrino energy and time distributions from this decay are shown in Fig.~\ref{fig:pidar_neutrino_energytime_spectrum}. $\pi^+$ decay produces a prompt, monoenergetic flux of $\nu_{\mu}$ at approximately 30~MeV (black line). The $\nu_{\mu}$ time profile is given by the convolution of the $\sim$290~ns triangular beam spill with the $\sim$26~ns pion lifetime. The subsequent decay of the daughter muons generates delayed fluxes of $\nu_e$ and $\overline{\nu}_{\mu}$ with energies extending up to $\sim$50~MeV (cyan and magenta lines, respectively). Their time distributions are significantly broader, reflecting the additional convolution of the beam pulse and pion decay with the muon decay lifetime of $\sim$2.2~$\mu$s.

In addition to neutrinos, the target produces intense fluxes of neutrons, gamma-rays, electrons, and positrons. This enables CCM to probe dark sector physics models that couple to secondary particles generated in the target. In particular, this thesis focuses on axion-like particles produced via their coupling to gamma-rays. To accurately model this production mechanism, a detailed simulation of the target is performed to obtain a realistic distribution of primary gamma-ray kinematics, discussed in Chapter~\ref{chap:alp}.

\subsection{Beam Start Time Characterization}
CCM exploits the short beam spill window to suppress steady-state backgrounds using precise timing cuts. A key piece in this strategy is determining the start of the beam spill, denoted as $T_0$. This $T_0$ represents the earliest possible arrival time at the detector for particles traveling at the speed of light from the target. Establishing $T_0$ therefore sets the beginning of the region of interest (ROI) for prompt, beam-related signals. The end of the ROI is determined separately, based on the rapid rise in detector activity caused by beam-related neutron backgrounds.

The start time $T_0$ is measured using an auxiliary detector located in Flight Path 3 in the experimental hall. This detector is an EJ-301 liquid scintillator coupled to a photomultiplier tube, which records activity for every beam spill. Rather than determining $T_0$ on a spill-by-spill basis, it is extracted statistically over the full dataset. Specifically, $T_0$ is identified as the time corresponding to the first significant increase in activity, which marks the arrival of speed of light particles originating from the target.

The end of the ROI is determined using the pulse distribution measured directly in the CCM detector. At later times, the detector rate increases sharply due to neutron-induced backgrounds, which ultimately dominate the signal. The ROI is therefore truncated before this rise to maintain sensitivity to prompt beam-related events with minimal neutron backgrounds.

\begin{figure}[h]
  \centering
  \includegraphics[width=0.7\linewidth]{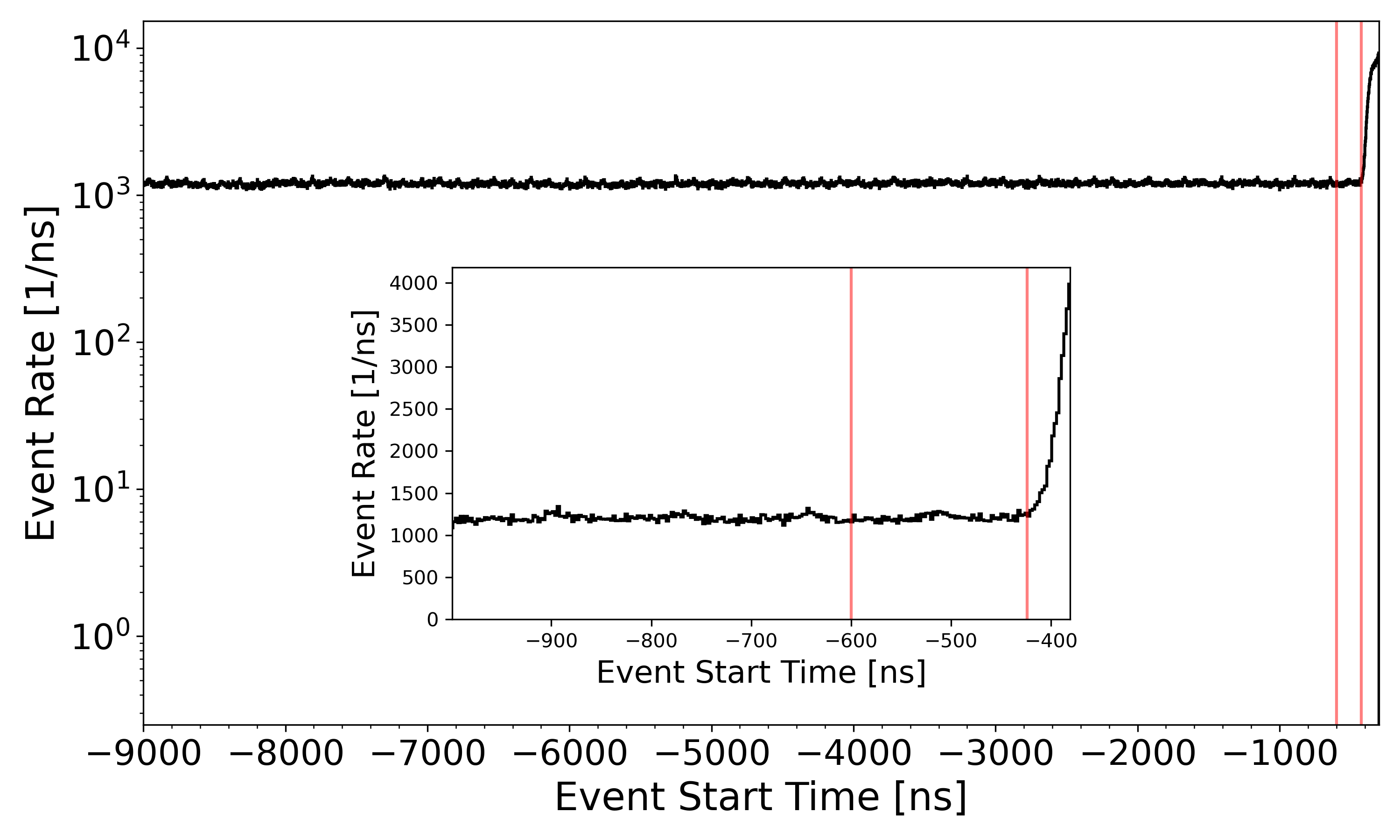}
  \caption{Distribution of event start times across the entire 2022 dataset. The red vertical lines correspond to the beam start time $T_0 = -600$~ns and the ROI end time of $-424$~ns. The exponential increase in activity due to neutron backgrounds is evident after the ROI end time. Note that for the main plot, the y-axis is in logarithmic scale while for the inset plot the y-axis is in linear scale to better demonstrate the relevant features.}
  \label{fig:ccm_timing_plot}
\end{figure}

For the 2022 physics dataset considered in this thesis, this procedure yields $T_0 = -600$~ns and an end time of $-424$~ns relative to the Beam Current Monitor. This ROI is illustrated in Fig.~\ref{fig:ccm_timing_plot}, which shows the distribution of reconstructed event start times from $-9~\mu\text{s}$ to $-300~\text{ns}$ relative to the Beam Current Monitor. The red vertical lines correspond to the ROI, with the inset plot highlighting the most relevant time region for this analysis. After the ROI end time, the sharp increase in the event rate is indicative of neutron related backgrounds. The event rate is integrated over the entire 2022 dataset and the events are selected using the procedure outlined in Chapter~\ref{chap:alp}.

There is approximately 30~ns of uncertainty at the $1\sigma$ level on this extracted $T_0$ start time. This defines a 176~ns search window for prompt beam-related signals prior to the onset of significant neutron backgrounds.

\subsection{Protons on Target}
An important quantity for characterizing the beam exposure is the total number of protons on target (POT). The accelerator division at LANSCE provides a beam current readout as a moving average sampled every minute, reported in units of $\mu$A. In order to convert this measured current into an estimate of the total POT recorded by CCM, we correlate it with independent beam measurements from the Beam Current Monitor (BCM).

\begin{figure}[h]
  \centering
  \includegraphics[width=\linewidth]{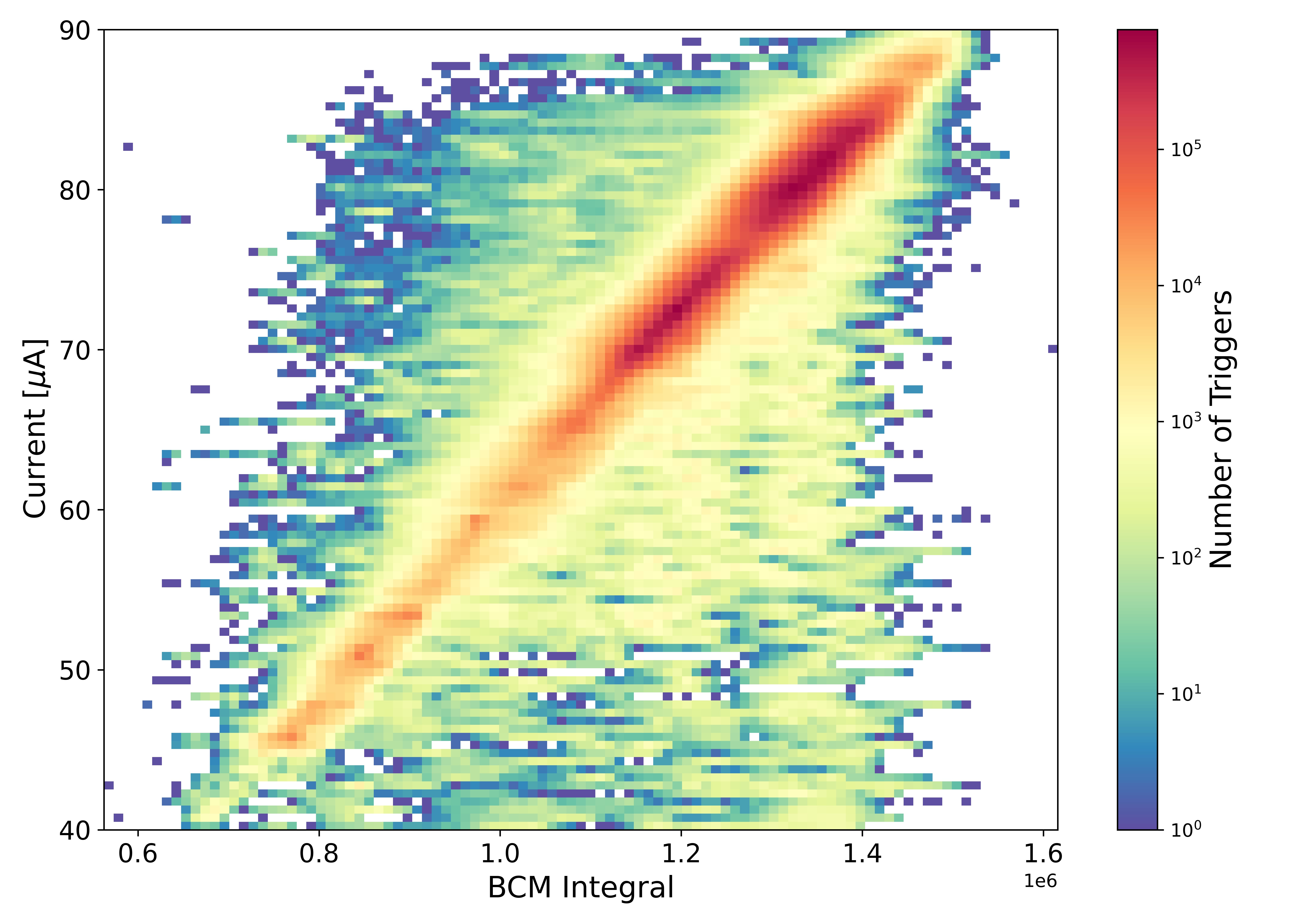}
  \caption{Two dimensional projection of the measured Beam Current Monitor (BCM) integral as a function of the beam current as reported by the accelerator division. The color represents the number of recorded beam triggers (in logarithmic scale). A clear linear relationship between the BCM integral and the current is observed, allowing for a linear fit to extract a conversion between BCM integral and current.}
  \label{fig:bcm_int_vs_current}
\end{figure}

The BCM measures the charge per beam pulse, but its raw output is in arbitrary units and is not directly convertible to physical current. Instead, we use the integral of the BCM waveform as a proxy for delivered beam intensity and correlate it with the sampled beam current provided by LANSCE. This comparison is shown in Fig.~\ref{fig:bcm_int_vs_current}, where a clear linear relationship is observed between the BCM integral (arbitrary units) and the measured beam current.

We model this relationship with a linear fit of the form $I = m \cdot \int \mathrm{BCM} + b$, where $I$ is the beam current. The best-fit parameters are $m = (6.02 \pm 0.09) \times 10^{-5}$ and $b = 0.19 \pm 1.17$.

Once the beam current is determined from the BCM integral, the corresponding number of protons is obtained using Eq.~\ref{eq:pot}, where $e$ is the elementary charge and $f_{\text{rep}} = 20~\text{Hz}$ is the beam repetition rate.

\begin{equation}
    \text{POT} = \frac{I}{e} \times f_{\text{rep}} 
\label{eq:pot}
\end{equation}

Using this calibration, we compute the POT for each beam trigger from the BCM integral. For the 2022 dataset used in this analysis, CCM recorded a total exposure of $1.23 \times 10^{21}$ POT.

\section{The CCM Detector}
Previous publications~\cite{CCM:2021leg,CCM:2021yzc,CCM:2021jmk,CCM:2023itc} and theses~\cite{Dunton:2022dez,Tripathi:2024jnq} on the CCM experiment have used the CCM120 detector. In contrast, this thesis presents the first results based on data from the upgraded CCM200 detector~\cite{CCM:2025dbq,CCM:2025kal}. 

The primary difference between the two configurations is the photon detection system. CCM120 was instrumented with 120 photomultiplier tubes (PMTs) as a testbed for the Short-Baseline Near Detector (SBND) program~\cite{SBND:2025lha}. In CCM200, this system has been replaced and expanded to a 200-PMT array to improve light collection and overall detector performance.

This section describes the CCM200 detector in detail, including its operation and expected background sources.

\subsection{The CCM200 Detector}
\begin{figure}[h]
  \centering
  \includegraphics[width=0.7\linewidth]{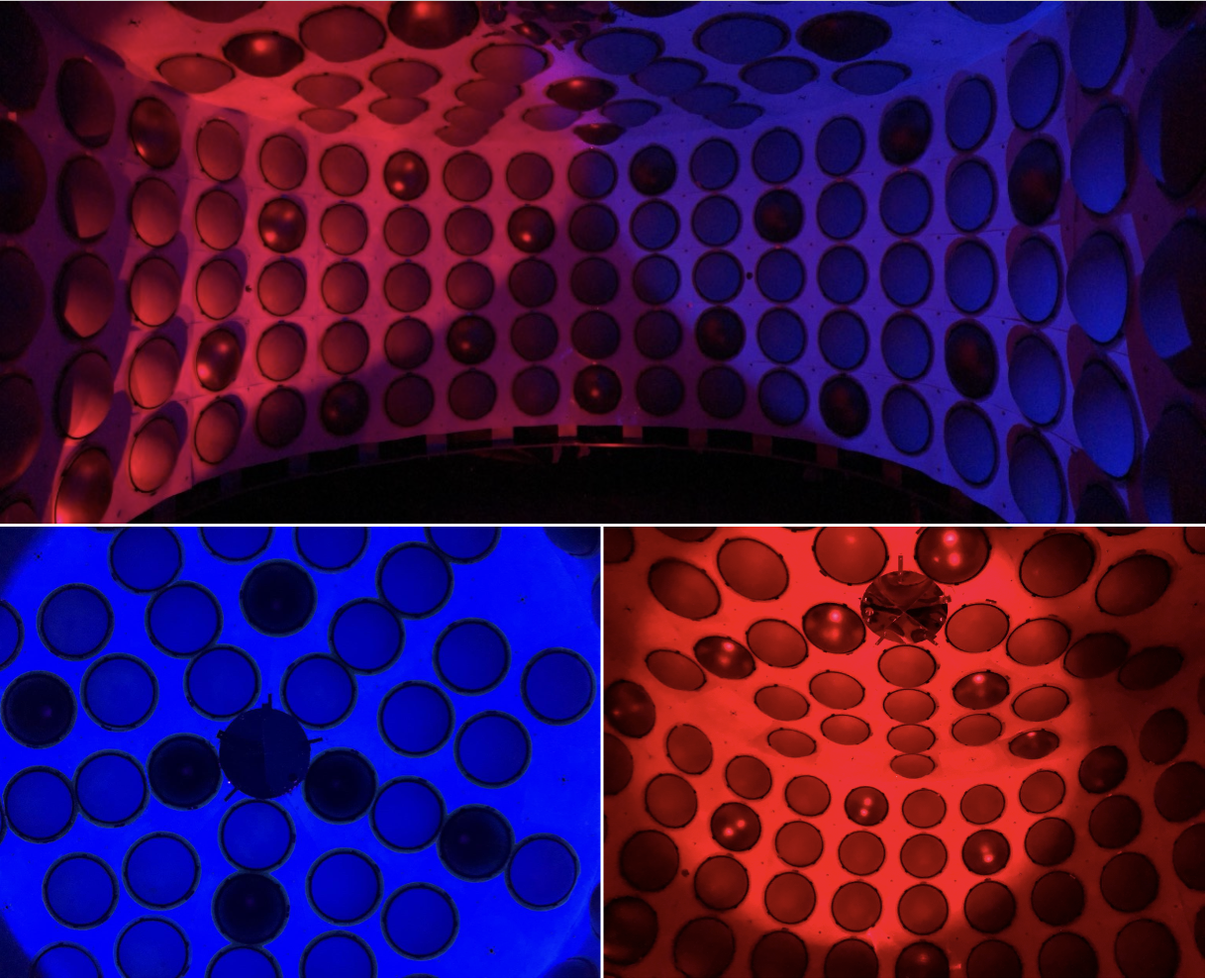}
  \caption{Interior views of the CCM200 detector. 80\% of the PMTs are coated in the wavelength shifter tetraphenyl butadiene (TPB) while 20\% of the PMTs remain bare. The coated PMTs are evident as the TPB scatters visible light, resulting in a matte appearance. The uncoated PMTs appear more reflective, from the glass surface in front of the photocathode. The uncoated PMTs are distributed roughly uniformly across the PMT array in the detector.}
  \label{fig:ccm_interior}
\end{figure}

CCM200 is the largest liquid argon light collection detector with the highest photocathode coverage. The cryostat, 2.58 m diameter $\times$ 2.25 m tall upright cylinder, is reused from the CAPTAIN experiment and can hold 10~tons of liquid argon~\cite{CAPTAIN:2013irr}. The fiducial region of the detector, enclosing approximately 7~tons of liquid argon, is instrumented with an array of 200 8'' R5912-Y002 10 stage cryogenic Hamamatsu Photonics PMTs for light collection, shown in Fig.~\ref{fig:ccm_interior}. 

There are five rows of PMTs around the barrel of the detector, each with 24 PMTs. The top and bottom end-caps are additionally instrumented with 40 PMTs each for a total 200 PMTs. The interior PMTs provide very high photocathode coverage, approximately 50\%. Additionally, there is a 3~ton optically isolated hermetic veto region instrumented with 40 1'' Hamamatsu Photonics PMTs.

Liquid argon produces scintillation light from the decay of excited dimer states, which have both a fast time constant (from the singlet spin configuration) and a long-lived time constant (from the triplet spin configuration). This scintillation light, which is discussed in depth in Chapter~\ref{chap:om}, is produced in the vacuum ultra-violet (VUV) range around 128~nm. This necessitates wavelength shifting (WLS) materials to shift the light into the visible spectrum for optimal detection efficiency by the PMTs. The CCM200 detector uses 1,1,4,4-Tetraphenyl-1,3-butadiene (TPB) to WLS the scintillation light in to the 400~nm to 550~nm wavelength region~\cite{Benson:2017vbw,doi:10.1021/j100052a011}.

The TPB is evaporatively coated on Mylar reflective foils that cover the walls of the fiducial region in the detector. The TPB layer on the foils is approximately 2.8~$\mu$m thick. Additionally, 80\% of the PMTs are evaporatively coated in TPB, forming an approximately 2.0~$\mu$m thick layer. The remaining PMTs, 40 uncoated PMTs that can be identified for their reflective surface in Fig.~\ref{fig:ccm_interior}, are optimally sensitive to visible photons produced from prompt Cherenkov radiation, discussed in Chapter~\ref{chap:cherenkov}. 

\subsection{Data Acquisition}
The PMT signals are digitized using 15 CAEN V1730/V1730S 500 MHz boards, providing 2~ns time resolution. All data acquisition (DAQ) is determined using external triggers. During normal physics data collection operations, the DAQ software triggers on the 20~Hz beam, 1~Hz random strobe, 1~Hz LED calibration, and approximately 0.5~Hz cosmic-ray triggers.

Each DAQ window is 16~$\mu$s long. For physics data collection, the beam trigger is approximately 70\% of the way through the DAQ window. This allows for almost 10~$\mu$s of data collection before the beam pulse on every trigger, referred to as ``prebeam" region. This prebeam region is used to measure and characterize the steady state backgrounds that are expected during the prompt physics region of interest.

In addition to digitizing the PMT voltage signals, every board has a copy of the trigger to allow for board-to-board timing calibrations on each trigger. This ensures proper timing of PMTs with respect to each other as well as with respect to the trigger. Additionally, the Beam Current Monitor is digitized for each beam spill allowing for data quality cuts on beam operations and calculation of the total POT as discussed previously.

All of the raw waveform data is stored on data servers for processing offline. For the 2022 dataset considered in this thesis, the raw data consists of approximately 140~terabytes. 

\subsection{Detector Operations}
The CCM200 detector is instrumented with atmospheric argon provided by Matheson Tri Gas, Albuquerque, New Mexico. The detector does not have a recirculation or filtration system. Instead, the argon continuously boils off through vents during operations at a rate of approximately 380~liters per day. The argon is replenished approximately every 3 days to ensure the liquid level never drops below the veto region. 

The manufacturer specifies maximum impurity levels of 1.95~ppm for oxygen, 2.50~ppm for nitrogen, and 0.01~ppm for water in the argon. During data taking, liquid argon conditions are continuously recorded using commercial oxygen and nitrogen analyzers. These instruments report impurity concentrations of $2.2\pm0.5$~ppm oxygen and $0.1\pm0.1$~ppm nitrogen. The observed stability of these levels throughout the data-taking period supports a reliable characterization of the optical properties, presented in Chapter~\ref{chap:om}.

The impurities in the liquid argon reduce the overall light yield and alter the structure of the scintillation time profile, discussed in Chapter~\ref{chap:om}. While this limits very low-energy measurements, it is not prohibitive to higher energy analyses. The main analysis in this thesis, search for $\mathcal{O}$(MeV) scale axion-like particles, is not significantly affected by the reduction in light yield. 

\subsection{Sources of Backgrounds}
\begin{figure}[h]
  \centering
  \includegraphics[width=0.7\linewidth]{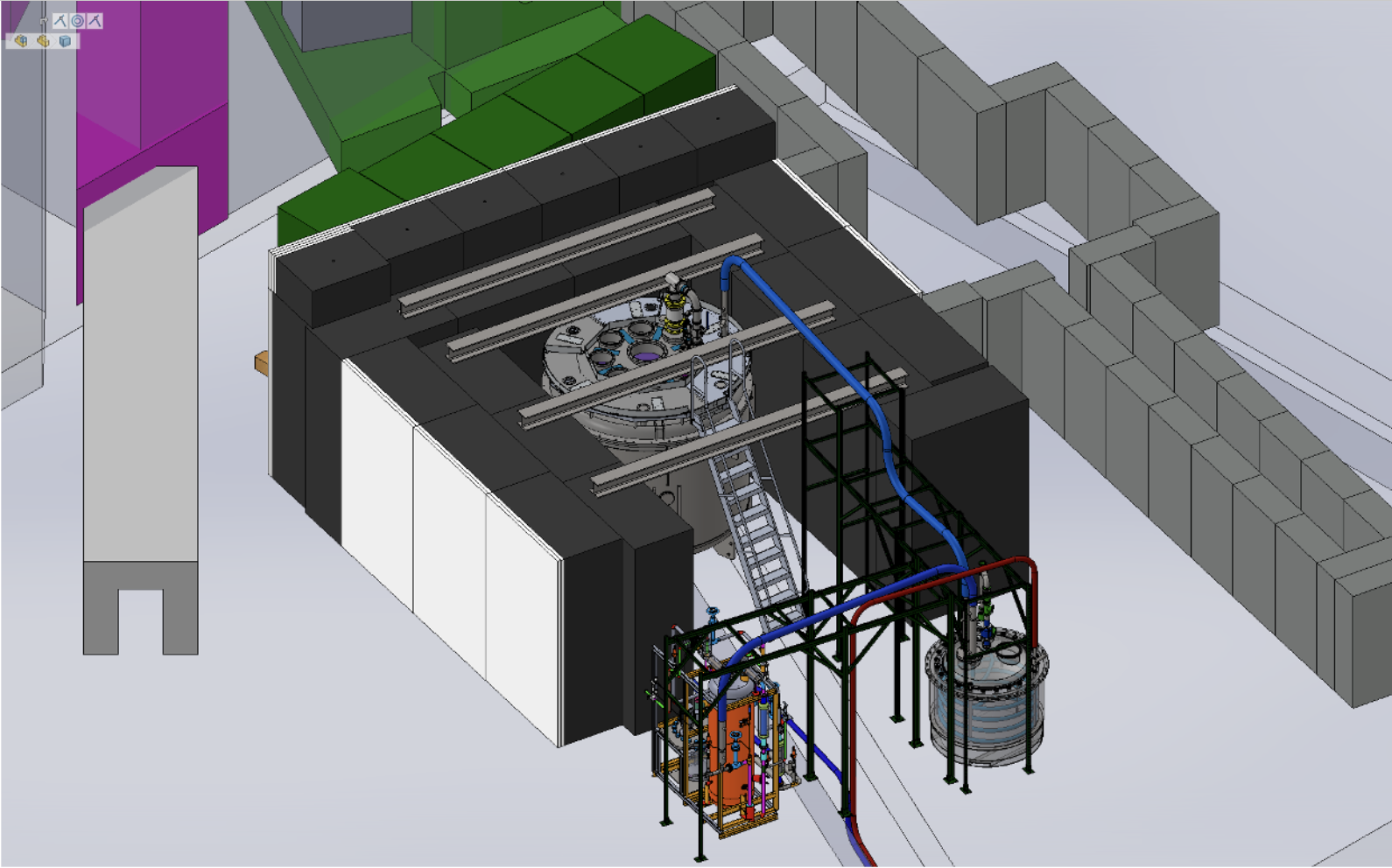}
  \caption{Rendering of the shielding configuration around the CCM200 detector. The shielding is concentrated in the direction of the target, towards the upper left-hand side of the image. There was additional shielding on the roof of the detector during the 2022 data collection period. The back face of the detector (furthest from the target) is unshielded.}
  \label{fig:ccm_shielding}
\end{figure}

The CCM detector is subject to several distinct sources of background. The first is intrinsic radioactivity from the liquid argon itself, dominated by low-energy $^{39}$Ar beta decays. These decays produce electrons with endpoint energies of approximately 0.5~MeV and contribute a steady, time-independent background rate.

A second class of backgrounds arises from cosmic-rays. Because the detector is located in an above-ground experimental hall, it is continuously exposed to a significant flux of cosmic-ray particles. In particular, cosmic-ray muons, and the subsequent Michel electrons from their decay, constitute a prominent background. However, these events are also useful for calibration purposes, as discussed in Chapter~\ref{chap:low_level}.

The dominant background contribution is beam related. This includes activation of materials in the experimental hall as well as neutrons produced at the target that propagate to the detector. The neutron backgrounds include steady-state contributions from thermal neutrons in the experimental hall, as well as fast neutrons that can fall within the region of interest.

Several mitigation strategies are employed to reduce these backgrounds:
\begin{enumerate}
    \item \textbf{Shielding}--- There is approximately 6~m of steel, 0.5~m of lead, 6~m of concrete, and interleaved layers of borated polyethylene between the target and the CCM200 detector. A schematic of the shielding configuration is shown in Fig.~\ref{fig:ccm_shielding}, where the target direction enters from the upper left-hand side. Shielding is concentrated on the sides and top of the detector facing the target, while the back remains unshielded. This configuration significantly attenuates the neutron flux and helps reduce the impacts of ambient radioactivity in the experimental hall.

    \item \textbf{Timing}--- A precise $T_0$ for near speed of light physics signals is defined using the external EJ-301 detector, as described previously. Timing cuts relative to this reference suppress the majority of steady-state and prompt beam-induced backgrounds. For the 2022 dataset used in this thesis, the final region of interest is defined within a $\sim$175~ns time window.

    \item \textbf{Event Reconstruction}--- The final and most powerful background rejection technique exploits event reconstruction based on Cherenkov light. This method is a key development of this work and is described in detail in Chapter~\ref{chap:cherenkov}, with its application to the axion-like particle search presented in Chapter~\ref{chap:alp}.
\end{enumerate}

Together, these strategies reduce the overall background rate to a level suitable for rare event searches, enabling the sensitivity required for the work presented in this thesis.

\chapter{Muon Trigger System and Low Level Data Processing}\label{chap:low_level}
This section details the hardware developments and low-level data processing carried out in the course of this thesis. To enhance the detector’s calibration capabilities, a set of cosmic-ray muon counters were constructed and deployed on top of the CCM200 cryostat. These counters provide an independent, high-energy data stream that complements the low-energy calibration channels and enables improved detector response characterization.

In parallel with the hardware efforts, this work also encompasses the full chain of low-level data processing applied after PMT waveform digitization. This includes the transformation of raw digitized signals into physically meaningful quantities through procedures such as baseline subtraction, electronics undershoot calibration, pulse reconstruction, and individual PMT charge and timing calibrations. Careful implementation and validation of these processing steps are essential for ensuring accurate event reconstruction and, ultimately, separation of Cherenkov radiation from the scintillation signals.

Together, the development of the muon counter system and the associated data processing framework enable ongoing higher energy calibration studies and form a critical foundation for the analyses presented in this thesis. 

\section{Muon Trigger System}
The CCM200 detector is located at an above ground facility at around 2,200~m in elevation, resulting in a large flux of cosmic-ray particles that are steady-state backgrounds to the beam-related physics searches. Cosmic-ray muons, produced when protons interact with particles in that atmosphere--- leading to hadronic cascades which create pions, decaying into muons, electrons, and neutrinos, can serve as both a background and a calibration source. The muons themselves deposit energy in the detector but they can also decay, producing final state electrons. These electrons from the three-body decay, referred to as Michel electrons, are an ideal high energy calibration data point as they have a well known energy spectrum with a sharp cutoff at around 50~MeV. 

While at Los Alamos National Lab working on the CCM experiment, the author created and led summer undergraduate students in building over a dozen table top muon counters to trigger on cosmic-ray muons entering the detector. The author also led integration of these external detectors into the data acquisition (DAQ) system to trigger on muons entering the detector. The author additionally performed some preliminary data analysis of the muon and subsequent Michel electrons. A full energy resolution and calibration analysis utilizing the Michel electron data is ongoing within the collaboration. 

\subsection{CosmicWatch Detectors}
The muon trigger system is based on CosmicWatch detectors~\cite{Axani:2018qzs}. These detectors are comprised of 5 x 5 x 1~cm$^3$ plastic scintillators optically coupled to a silicon photomultiplier (SiPM) for light collection. Charged particles traversing the scintillator produce scintillation light, which is converted into an electrical signal by the SiPM. This signal is then amplified and shaped by onboard analog electronics. 

\begin{figure}[h]
  \centering
  \includegraphics[width=0.7\linewidth]{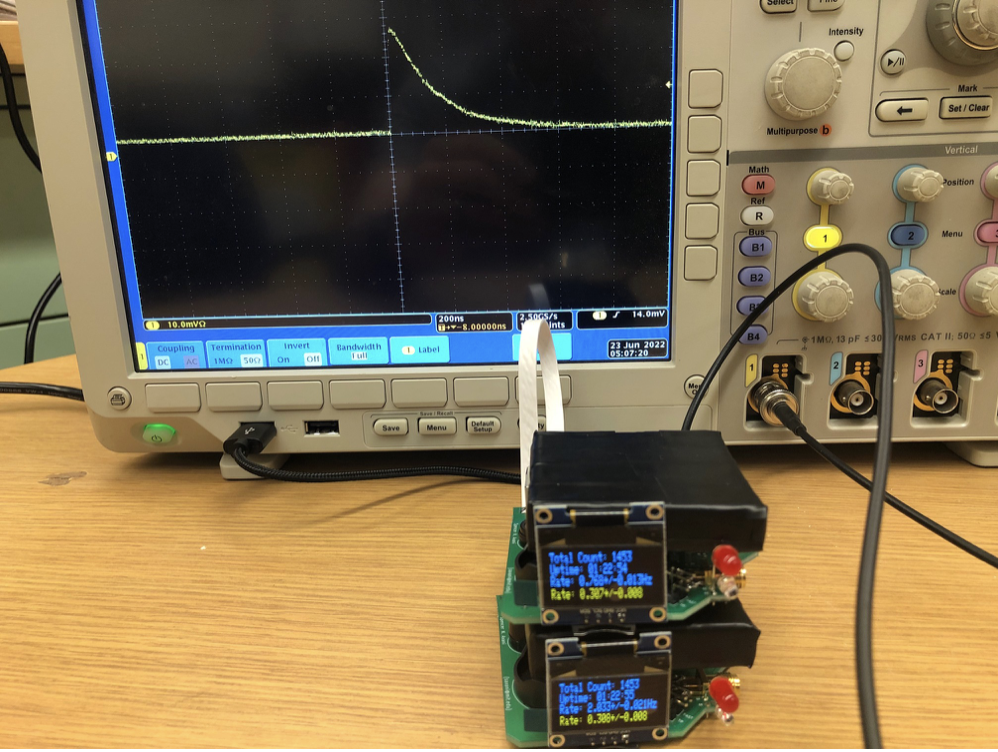}
  \caption{CosmicWatch detector characterization at Los Alamos National Laboratory. Two CosmicWatch detectors triggering in coincidence are in the front of the image. The plastic scintillator panels, covered in electrical tape to avoid light leakage, are the black rectangles. The LED readout screens are in the front of the detectors and the printed circuit boards are visible below the scintillator. The oscilloscope shows the signal readout of a coincident hit between both pairs of detectors.}
  \label{fig:cw_scope}
\end{figure}

The analog signal is digitized and processed using a microcontroller-based readout system, which enables extraction of pulse height and timing information. In this work, the detectors were primarily used as fast trigger devices, and emphasis was placed on the identification of coincident signals rather than detailed waveform reconstruction.

During laboratory characterization, the CosmicWatch detectors were connected in parallel and the digitized coincident signal was readout using an oscilloscope, displayed in Fig.~\ref{fig:cw_scope}. While low-energy radiation backgrounds can deposit charge in a single detector, requiring temporal coincidence between two CosmicWatch detectors selects for higher energy minimum ionizing particles like cosmic-ray muons. 

\begin{figure}[h]
  \centering
  \includegraphics[width=0.7\linewidth]{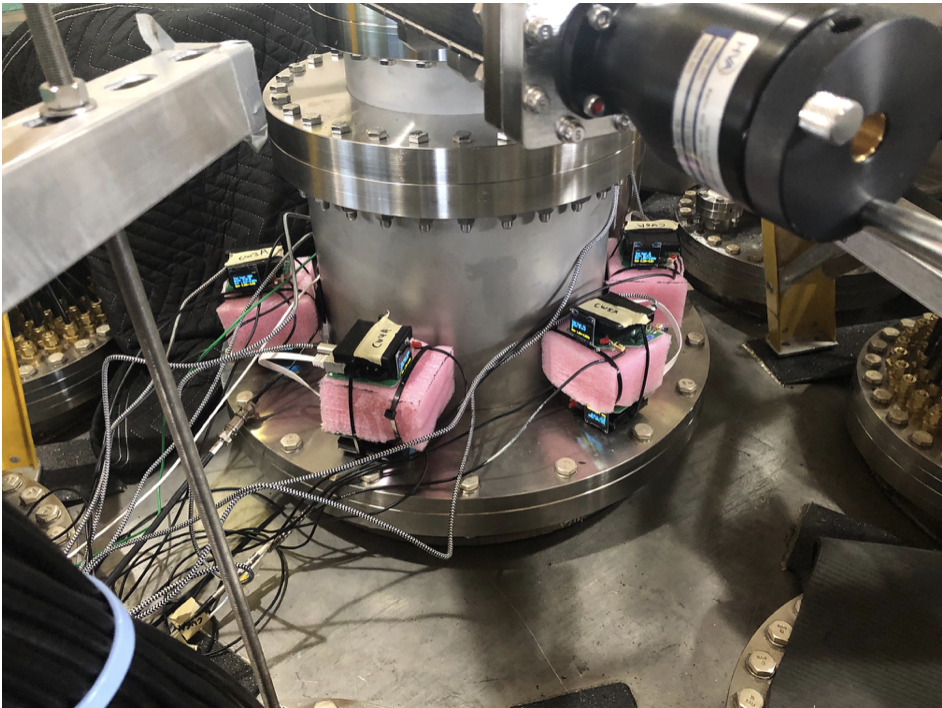}
  \caption{CosmicWatch detectors deployed on top of the CCM200 cryostat. The six pairs, total of 12 CosmicWatch detectors, are clustered around the central flange of the CCM200 detector. Each pair is separated by approximately 7~cm of material to select down-going cosmic-ray muons. The amplified signal from each board is connected to the trigger system to trigger on coincident hits in any pair of CosmicWatch detectors.}
  \label{fig:cw_insitu}
\end{figure}

There were 12 CosmicWatch detectors deployed on top of the CCM200 detector during the 2022 data collection period. The CosmicWatch detectors were assembled in vertically stacked pairs around the top central flange, illustrated in Fig~\ref{fig:cw_insitu}. The vertical pairs allows for data collection based on coincident signals, which reduces many random backgrounds to collect a very pure sample of downward-going cosmic-ray muons entering the detector.  

\subsection{Integration with the DAQ System}
Following signals from the CosmicWatch detectors, the trigger is generated using a modular NIM-based DAQ system built from LeCroy logic units. The analog output from each CosmicWatch detector is first duplicated using fan-in/fan-out modules, allowing the signal to be distributed to multiple processing paths without degradation.

Each signal is then passed through a quad discriminator, which converts the analog pulse into a standardized NIM logic pulse when the signal exceeds a fixed threshold of 100~mV. This step ensures uniform timing and suppresses low-amplitude noise contributions.

The resulting logic pulses are processed using quad logic units to form coincidences between vertically stacked detector pairs. For each pair, an \texttt{AND} operation is applied within a fixed coincidence time window to identify through-going particles that traverse both detectors. These pairwise coincidence signals define six independent muon-tagging channels.

Finally, the outputs from all detector pairs are combined using a series of \texttt{OR} operations to produce a global trigger. This logic configuration ensures that a valid trigger is issued whenever any detector pair registers a coincident event, thereby maximizing acceptance for cosmic-ray muons while maintaining strong rejection of random backgrounds. This CosmicWatch trigger was operational for all of the 2022 data collection period. 

\subsection{Michel Electron Data Sample}
Across the approximately two-month data collection period in 2022, the DAQ triggered on roughly 2.3 million coincident signals from the CosmicWatch detector pairs. This provides a high-statistics sample of both cosmic-ray muon events and Michel electrons, which are being used in ongoing high energy calibration efforts.

\begin{figure}[h]
  \centering
  \includegraphics[width=\linewidth]{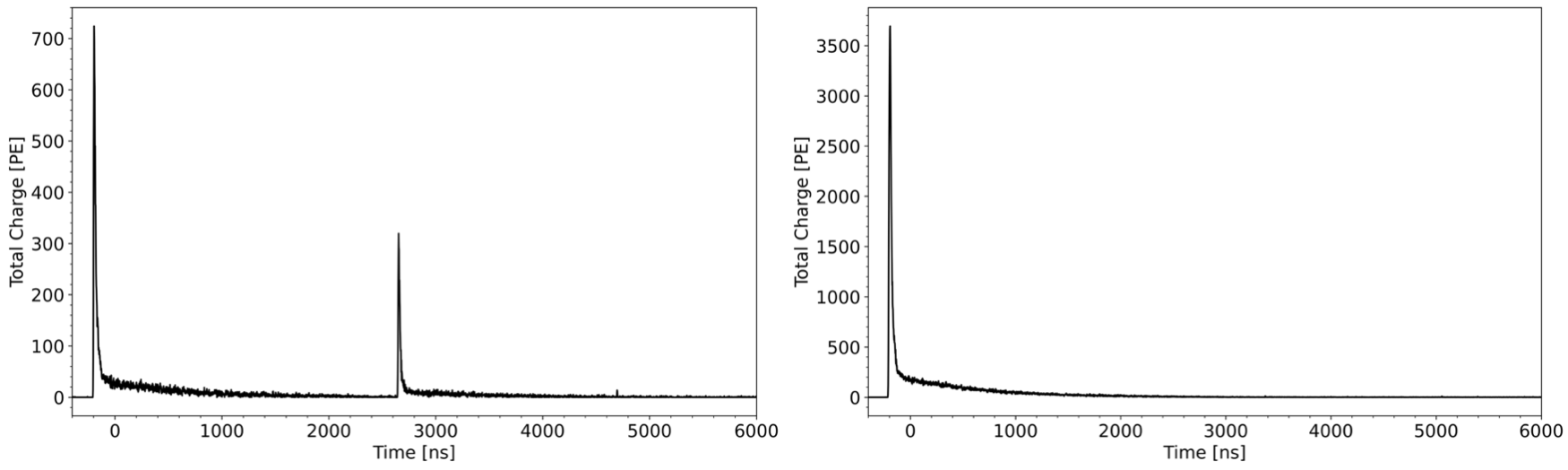}
  \caption{Example of reconstructed pulse series for cosmic trigger data. The plot on the left shows a typical charge distribution for muons that decay in the detector. The cosmic-ray muon deposits charge around the trigger at $t=0$ then, around $3~\mu\text{s}$ after the muon, there is a secondary large deposition of charge, indicative of a Michel electron. The plot on the right is a typical through-going muon distribution where there is only one distinct charge deposition from the cosmic-ray muon traversing the detector.}
  \label{fig:cw_data_example}
\end{figure}

Fig.~\ref{fig:cw_data_example} shows two example cosmic-trigger events. The x-axis represents time relative to the trigger, and the y-axis shows the total detected charge. In the left panel, a clear Michel electron candidate is visible at approximately $2.75~\mu$s after the trigger. In contrast, the right panel shows a through-going cosmic-ray muon event with no evident Michel electron.

While this thesis primarily focuses on low-energy event characterization and Cherenkov light separation, using an $\mathcal{O}(1~\text{MeV})$ calibration source, the author additionally developed a selection procedure to identify Michel electrons. Standard event-building techniques typically rely on applying a charge threshold to identify distinct groups pulses. However, this approach is not effective for cosmic-ray muon events, which deposit a large amount of charge and produce long-lived scintillation tails dominated by triplet-state emission.

Instead, a template-based method was developed using cosmic-trigger data. An average time profile of charge deposition from through-going cosmic-ray muons is constructed from a large sample of events. This template is then compared to individual cosmic triggers, and bins with excess charge relative to the expected muon-induced profile are identified. This method provides a robust way to isolate Michel electrons on top of the extended triplet light background.

\begin{figure}[h]
  \centering
  \includegraphics[width=0.7\linewidth]{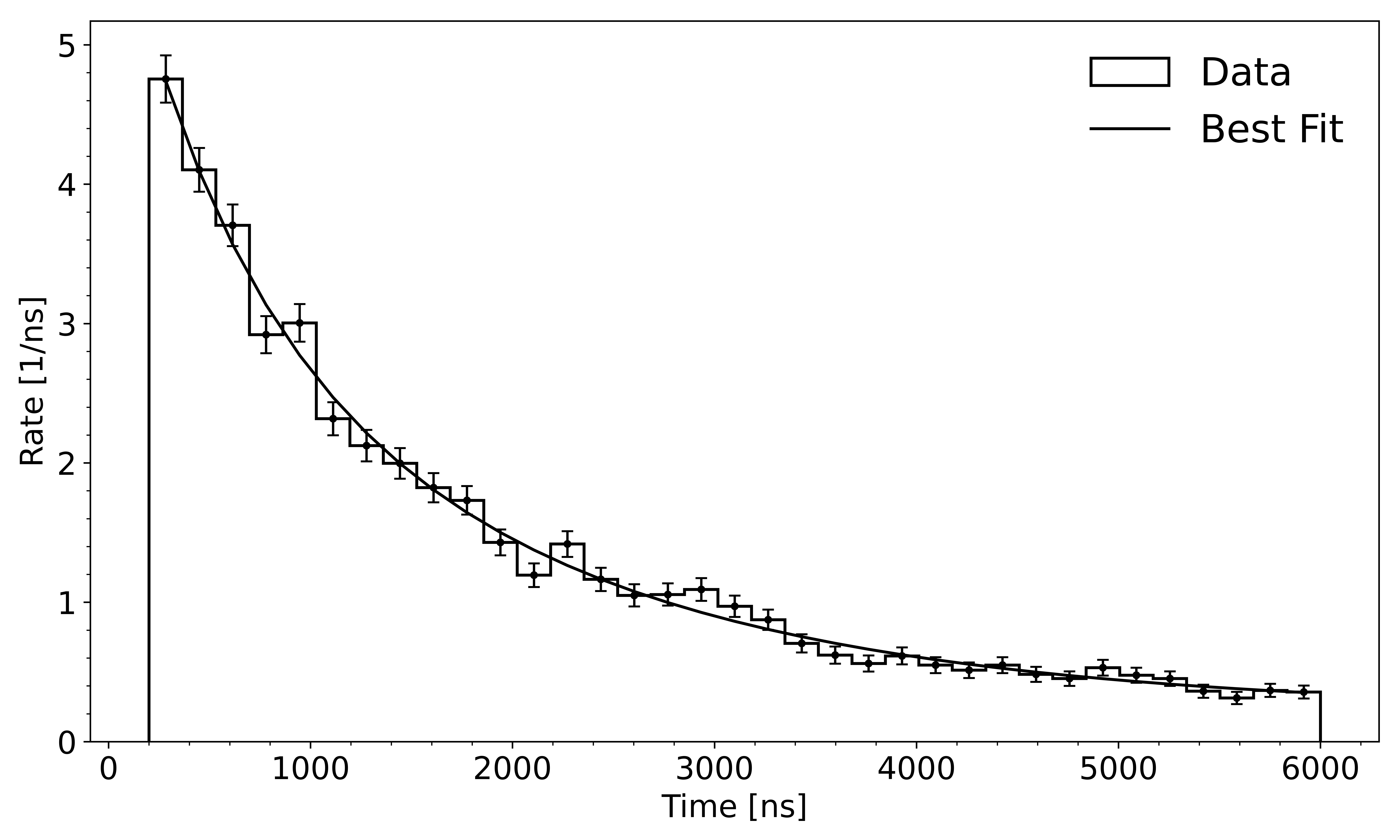}
  \caption{Distribution of decay times for identified Michel electron candidates from a subset of the data. This distribution is fit using the expected time profile given in Eq.~\ref{eq:muon_lifetime_eq} that incorporates both the muon decay and the capture effects relevant in liquid argon.}
  \label{fig:muon_lifetime}
\end{figure}

This Michel selection procedure is validated by fitting the measured decay time distribution to extract the muon lifetime. Fig.~\ref{fig:muon_lifetime} shows the decay time distribution overlaid with the best-fit lifetime model. While the free muon lifetime is well known to be approximately 2.2~$\mu$s, $\mu^-$ can undergo nuclear capture in liquid argon with a characteristic time constant of approximately 500~ns, modifying the observed lifetime~\cite{PhysRevLett.134.131801}.

\begin{equation}
    \mathcal{P}_{\mu^{\pm}}(t) = \left( \frac{1}{1+r} \right) \left( \frac{1}{\tau_d} + \frac{1}{\tau_c} \right) e^{-t \left( \frac{1}{\tau_d} + \frac{1}{\tau_c} \right)} + \left( \frac{r}{1+r} \right) \frac{1}{\tau_d} e^{-t / \tau_d}
\label{eq:muon_lifetime_eq}
\end{equation}

The $\mu^{\pm}$ time distribution, including both the decay and capture processes, is modeled as Eq.~\ref{eq:muon_lifetime_eq}, where $r$ is the relative abundance of positive to negative muons, $\tau_d$ is the free muon decay lifetime, and $\tau_c$ is the effective capture time constant in argon. The parameter $r$ is fixed to $1.32$ based on literature values~\cite{Super-Kamiokande:2024rwz}, while $\tau_d$, $\tau_c$, an overall normalization, and a constant background offset are left as free parameters in the fit.

This procedure yields best fit values of $\tau_d = 1842.65 \pm 362.82~\text{ns}$, $\tau_c = 852.70 \pm 358.11~\text{ns}$, $N = 11075.65 \pm 76.82$ for the normalization, and $c = 0.21 \pm 0.19$ for the constant offset. The $\chi^2$ between the observed data and the expectation is 38.08 for 31 degrees of freedom. These results are consistent with expectations from literature values, and ongoing Michel electron calibration studies within the collaboration are expected to further improve the precision of this measurement.

\section{Low Level Data Processing}
The low-level data processing chain converts digitized PMT waveforms into reconstructed photoelectron (PE) signals, which serve as the fundamental unit of detector data. For the data collected during the 2022 period and onward, the entire data processing framework was overhauled. Previous PE pulse reconstruction of CCM data introduced $\mathcal{O}$(10~ns) time smearing to the digitized signals. Since a major goal of this thesis is to advance searches for weakly interacting particles by leveraging Cherenkov radiation in a high light-yield scintillation detector, this level of time smearing would wash out any timing based differences between the prompt Cherenkov radiation and relatively slower scintillation emission. Part of this thesis includes developing these low-level data reconstruction tools detailed below.

\subsection{Baseline Correction}
The first step in PE reconstruction is accounting for the constant offsets in the digitized waveforms through baseline corrections. This requires a robust estimator rather than a simple average since there can be long-timescale effects like capacitor induced undershoot and exponential behavior in the digitized voltages. 

To determine the baselines for each channel, each waveform is first smoothed using an exponential moving average, followed by a short-range local averaging step over neighboring samples to suppress fluctuations. Then, an iterative outlier suppression procedure is applied, in which deviations from a locally evolving baseline estimate are exponentially down-weighted. This reduces the influence of large voltage differences and noise spikes. 

The smoothed waveform is then evaluated to determine whether it is consistent with a constant baseline or exhibits slow time variation. This is done using a $\chi^2$ goodness-of-fit metric. If the waveform is not well described by a constant baseline, further exponential smoothing is applied to mitigate long-timescale baseline relaxation effects observed in a subset of waveforms.

Finally, the variances are calculated across the smoothed waveform and regions of low variance are used to determine the baseline. The mode of these low variance samples and the median absolute deviation provide the baseline estimate and width, respectively. This approach provides a stable estimate of the baseline level even in the presence of non-linear electronics responses and noise fluctuations. 

\subsection{Electronics Undershoot Correction}
\begin{figure}[h]
  \centering
  \includegraphics[width=0.7\linewidth]{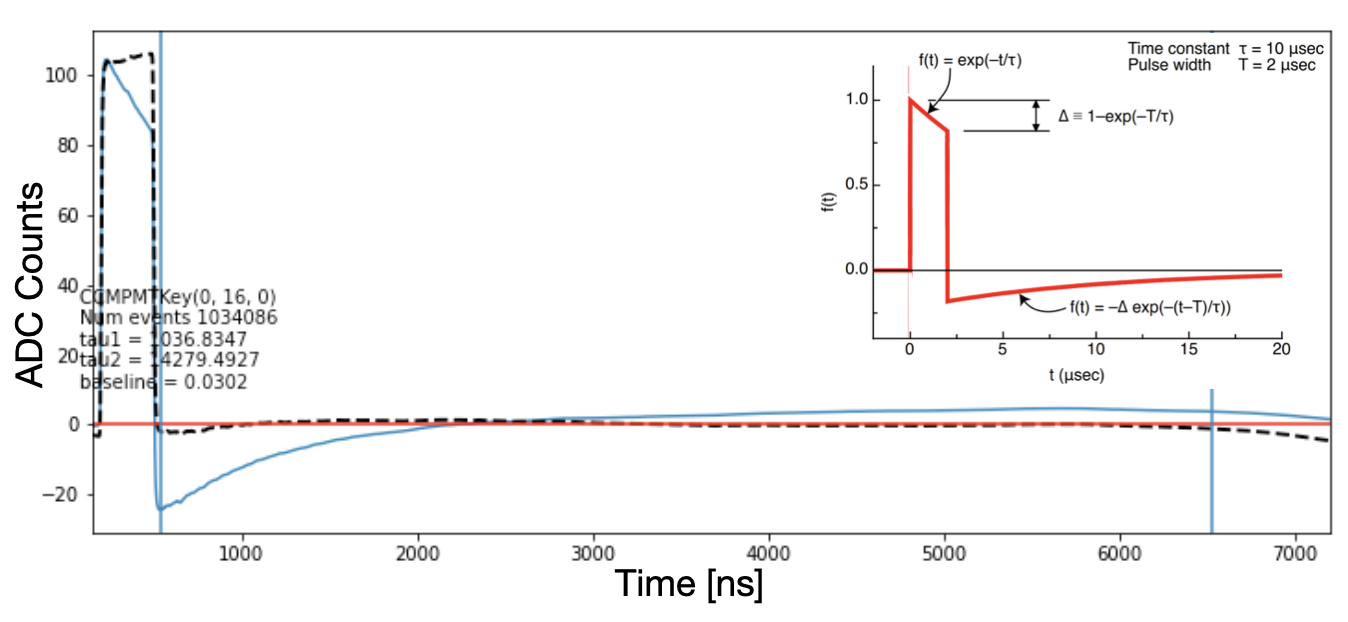}
  \caption{Illustration of the electronics induced undershoot in a single PMT. Many single PE pulses are identified and stacked over a 300~ns time window to highlight the undershoot after charge deposition (blue line). This is corrected using the procedure outlined in Eq.~\ref{eq:e_correction} and illustrated in the upper right hand side. The dashed black line is the resulting corrected waveform.}
  \label{fig:e_correction}
\end{figure}

The PMT front-end electronics are AC-coupled, resulting in long-timescale undershoot following pulses. We correct this behavior by applying a recursive inverse filter constructed from the measured time constants for each PMT. To determine these parameters, single-photoelectron waveforms are stacked over a 300~ns window for each PMT, demonstrated in Fig.~\ref{fig:e_correction}. The light blue distribution is the stacked pulses for a characteristic PMT. This stacking procedure enhances the electronics-induced undershoot, most prominent around 1~$\mu$s.

Every PMT is then fit for two time constants that describe this behavior, using Eq.~\ref{eq:e_correction}, which is demonstrated in the inset plot of Fig.~\ref{fig:e_correction} as well. Given the observed voltage $V_i$ at time step $i$, the corrected voltage $X_i$ is recovered by accounting for the integrated charge lost to the electronics' high-pass response. In Eq.~\ref{eq:e_correction}, $\Delta t$ is the sampling period and $\tau$ is the characteristic droop time constant. Since the readout chain contains multiple stages of AC-coupling, this correction is applied twice in series using distinct time constants $\tau_1$ and $\tau_2$ specific to each PMT.

\begin{equation}
X_i = \frac{V_i}{A} + (1 - e^{-\Delta t / \tau}) \sum_{j=0}^{i-1} X_j e^{-(i-j-1)\Delta t / \tau}
\label{eq:e_correction}
\end{equation}

After applying this recursive filter, the final corrected waveform is the black dashed line in Fig.~\ref{fig:e_correction}. The portion of the waveform around 1~$\mu$s, that strongly exhibited undershoot behavior, now agrees very well with null, demonstrating very little electronics effects are still present after correction.

\subsection{Pulse Unfolding}
After applying corrections to the raw waveforms to account for the baselines and electronics induced undershoots, the next crucial step in low-level data processing is to reconstruct PE pulse series. For this task, we utilize the Lawson-Hanson non-negative least squares (NNLS) algorithm to unfold PE pulses following the methods developed by the IceCube Collaboration~\cite{lawson1974least,IceCube:2013dkx}.

\begin{figure}[h]
  \centering
  \includegraphics[width=0.7\linewidth]{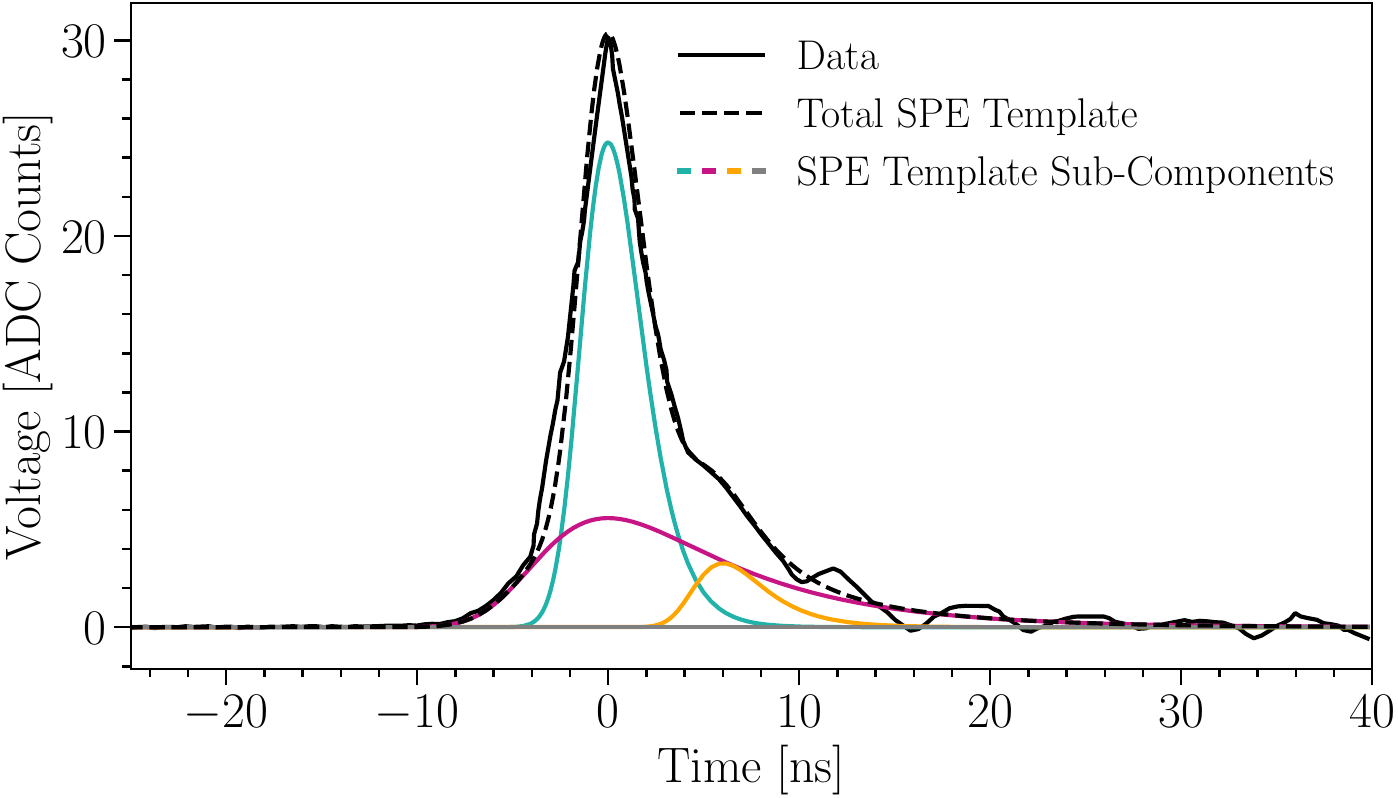}
  \caption{Average SPE distribution on a typical PMT. The data (solid black line) is fit using a combination of four templates, Eq.~\ref{eq:spe_template}, resulting in the total SPE template (dashed black line). The procedure is performed on every PMT to individually calibrate the SPE distributions for pulse unfolding.}
  \label{fig:spe_template}
\end{figure}

This template-based method requires single PE (SPE) distributions, which are characterized for every PMT. Fig.~\ref{fig:spe_template} is an example of this SPE template for a typical PMT. The data is the accumulation of around 8.75 million single pulses obtained across the entire 2022 data collection period. The pulses were selected from the digitized waveforms using a derivative based algorithm. This method requires a sequence of positive, negative, and then positive derivatives--- representing the rising, falling, then second rising edge of pulses. The last condition of a positive derivative is based on the experimentally observed overshoot after single pulses~\cite{Kaptanoglu:2017jxo}.

After accumulating single pulses on every channel, the averaged data is fit using an SPE template originally developed by the IceCube Collaboration~\cite{I3DOMCalibration,IceCube:2008qbc,IceCube:2016zyt}. This template, described in Eq.~\ref{eq:spe_template}, depends on $h$ to determine the height of the pulse and parameters $b_1$ and $b_2$ to determine the shape of the pulse. In order to accurately fit the data on every PMT, we allow four individual SPE templates (cyan, magenta, orange, and gray lines in Fig.~\ref{fig:spe_template}) to fully describe the overall SPE shape (black dashed line in Fig.~\ref{fig:spe_template}).

\begin{equation}
\begin{aligned}
w(t) &= \frac{c}{\left(e^{-(t - t_0)/b_1} + e^{(t - t_0)/b_2}\right)^8}; \\
c &= \frac{h}{
  b_1^{(8b_1)/(b_1 + b_2)} \cdot
  b_2^{(8b_2)(b_1 + b_2)} /
  (b_1 + b_2)^8
}
\end{aligned}
\label{eq:spe_template}
\end{equation}

One interesting feature of this characteristic pulse shape is the second peak contribution around time $t=6$~ns relative to the peak time, which has been observed previously for similar models of PMTs~\cite{Caldwell:2013oea}. Two of the SPE templates account for the main peak at $t=0$ and secondary peak at $t=6$~ns and the other two templates help describe the width of the characteristic SPE shape. 

After calibrating the SPE shape on every PMT, the pulse unfolding procedure extracts the time and amplitudes of pulses based on the electronics corrected waveforms using the NNLS approach. 

\begin{figure}[h]
  \centering
  \includegraphics[width=\linewidth]{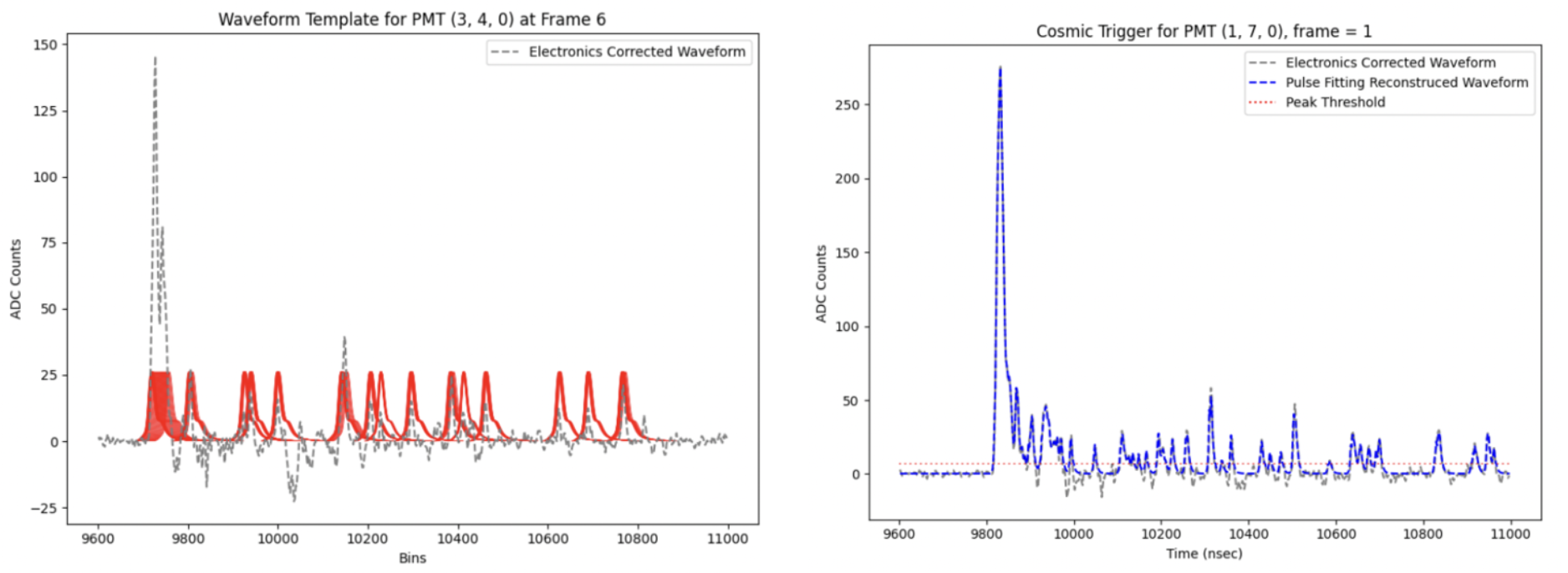}
  \caption{Illustration of pulse reconstruction procedure using the Lawson-Hanson NNLS algorithm. The left-hand side shows the electronics corrected waveform in gray and the SPE template placement in red before solving for the pulse amplitudes. The right-hand side shows the electronics corrected waveform (gray) and the unfolded pulse reconstruction waveform (blue) indicating the validity of this procedure.}
  \label{fig:pulse_unfolding}
\end{figure}

Fig.~\ref{fig:pulse_unfolding} demonstrates this process. On the left-hand side, the gray distribution is the electronics corrected waveform. The red pulses are illustrations of the SPE template for this PMT. One template is placed in every time bin over the threshold of 16~ADC counts above the baseline. Then, the regression process solves for the heights of each pulse to best match the data. The right-hand side shows this entire process, from the electronics corrected waveform in gray to the reconstructed waveform using the best fit pulse series in blue, demonstrating the agreement of this procedure with data. 

\subsection{Electron Transit Time and Single PE Calibration}
The final low-level calibration step accounts for the electron transit time and establishes a calibration of the single photoelectron (SPE) response.

The PMTs used in CCM exhibit an intrinsic transit time spread of approximately 1.7~ns, and require per-channel electron transit time calibration~\cite{hamamatsu_pmt_handbook}. This calibration was performed in two stages. The first was data-driven, using the arrival times of LED calibration pulses. Given the known positions of the LED flashers on the top and bottom of the detector, the time difference between the LED trigger and the first detected hit on each PMT was measured. By accumulating a large number of events, the rising edge of the first-hit time for each PMT was extracted very precisely, allowing relative transit time offsets to be determined and corrected so that all PMTs were aligned in time after accounting for expected physical delays.

While this procedure provided a good first-order calibration, it was further refined using simulation. In the data to Monte Carlo time-series comparisons discussed in Chapter~\ref{chap:om}, residual mis-calibrations in PMT transit times manifested as systematic timing offsets between data and simulation accumulated pulse series on each PMT. Since the electron transit time is the only PMT-specific timing offset, these differences can be directly attributed to imperfect transit time calibration.

An iterative procedure was therefore performed in which the PMT transit times were adjusted to minimize the per-PMT timing differences between data and simulation. Then the data accumulation procedure was performed using global event start times and the per-PMT timing differences were computed once again between data and simulation. This procedure was repeated until the residual differences were reduced to below $0.25~\text{ns}$, indicating a well-calibrated timing response.

\begin{figure}[h]
  \centering
  \includegraphics[width=0.7\linewidth]{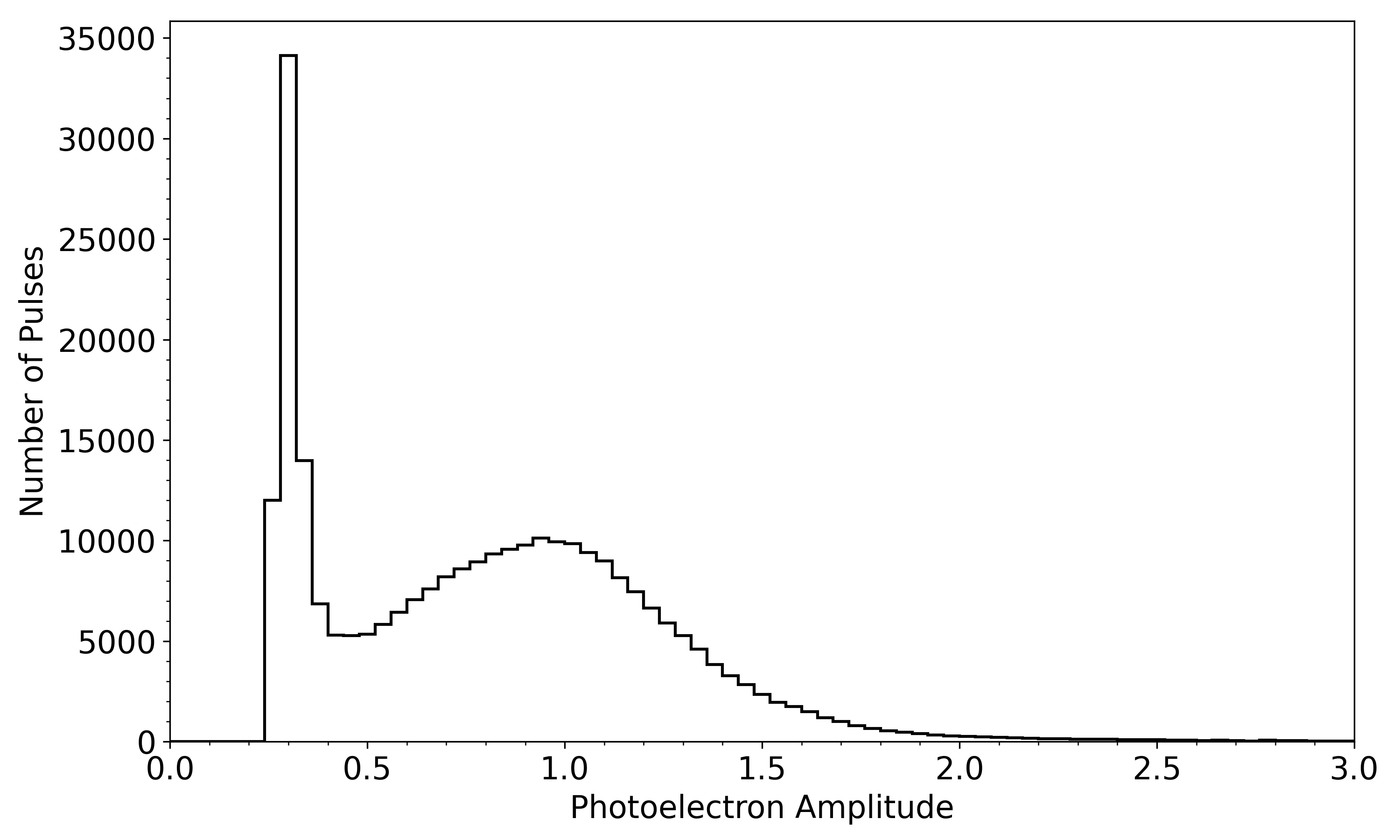}
  \caption{Distribution of single photoelectron (SPE) amplitudes for a typical PMT. SPEs are identified by requiring no other activity within 100~ns of the candidate pulse. The amplitudes of these pulses are then characterized using Eq.~\ref{eq:spe_fitfunc} which is the sum of two exponential distributions and a Gaussian distribution. The Gaussian mean parameter $\mu$ is used for calibration to ensure that the SPE distributions on every PMT are centered at an amplitude of 1.}
  \label{fig:spe_calibration}
\end{figure}

The SPE calibration proceeds separately. Single photoelectron pulses are identified on each PMT by selecting isolated pulses with no reconstructed activity within a 100~ns window before or after the candidate hit. The resulting distribution of pulse amplitudes is shown in Fig.~\ref{fig:spe_calibration} for a typical PMT.

\begin{equation}
    f(x; p_1, p_2, w_1, w_2, \mu, \sigma) = 
    \frac{p_1}{w_1} e^{-\frac{x}{w_1}} + 
    \frac{p_2}{w_2} e^{-\frac{x}{w_2}} + 
    \frac{1 - p_1 - p_2}{\sigma \sqrt{\frac{\pi}{2}} \operatorname{erf}\left(-\frac{\mu}{\sigma \sqrt{2}}\right)} e^{-\frac{(x - \mu)^2}{2\sigma^2}}
\label{eq:spe_fitfunc}
\end{equation}

The observed SPE charge distribution is modeled as Eq.~\ref{eq:spe_fitfunc}. This model consists of two exponential components to describe low-amplitude noise contributions and a Gaussian component corresponding to the true single photoelectron response. The functional form follows a parameterization used in IceCube calibration studies~\cite{I3DOMCalibration}.

The SPE distributions for each PMT are accumulated and fit according to this function. The best fit Gaussian mean parameter $\mu$ is then used to calibrate each PMT such that the SPE response is centered at 1~PE.

\chapter{Measurement of the Light Production and Propagation Parameters in Liquid Argon}\label{chap:om}

The content of this chapter summarizes and expands on Ref.~\cite{CCM:2025dbq}, on which the thesis author was the principal author.

The ability to distinguish Cherenkov radiation from scintillation light requires a detailed understanding of the light production and propagation mechanisms. This section describes the measurement of liquid argon scintillation pulse shape and light propagation parameters in the CCM200 detector. This result entailed optimization of a combined binned likelihood across 145 time bins for 191 individual PMTs for over 20 free parameters. For efficiency and accuracy in modeling the prediction across such a high dimensional parameter space, a major part of this work was developing a differentiable simulation of \texttt{GEANT4} photon propagation physics models. This is the first result of its kind--- including both simultaneous characterization of scintillation and Cherenkov light in liquid argon and the first use of differentiable simulation for experimental optical model calibration. This section will describe the theory behind liquid argon light production, calibration data collection, and fitting procedure along with best fit results. 

\section{Light Production in Liquid Argon}
While liquid argon is often utilized in particle physics experiments for charge collection, the goal of this work is to understand both the scintillation and Cherenkov radiation components in the light signals to more fully utilize the reconstructed data. This section describes both the scintillation process as well as the Cherenkov radiation in liquid argon to better understand how to characterize the two signals. 

\subsection{Scintillation Light}
Liquid argon is a prolific scintillator, producing around 40,000 scintillation photons per MeV of deposited energy. When charged particles pass through liquid argon, there are two pathways to scintillation. The first luminescence process is through self-trapped excitons~\cite{Doke:1990rza}. Analogous to electron-hole pairs in crystal scintillators, charged particles passing through the medium can excite the argon atoms which then recombine with nearby atoms to create excited dimers. As the excited dimers relax back to the ground state, they emit scintillation photons.

\begin{figure}[h]
  \centering
  \includegraphics[width=0.7\linewidth]{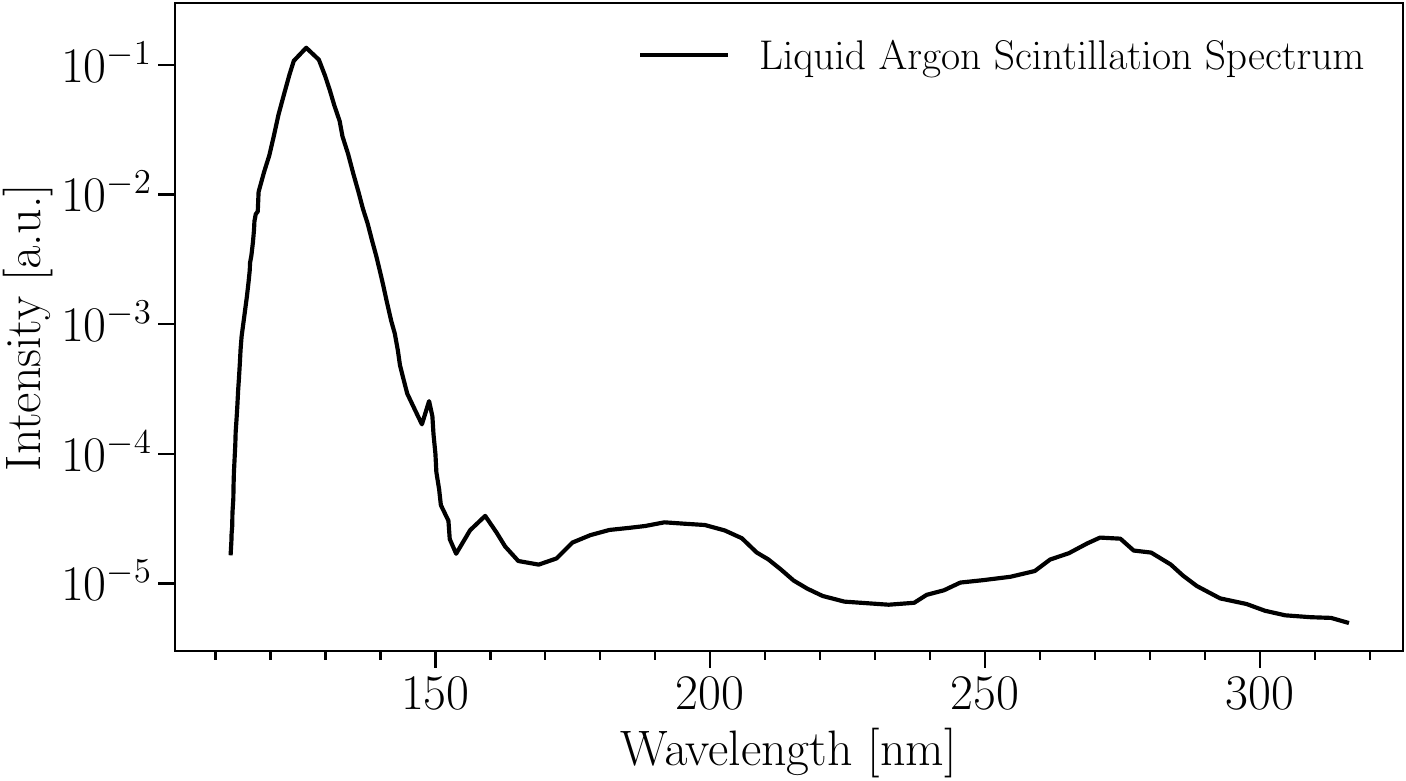}
  \caption{Measurement of liquid argon scintillation emission spectrum. The dominant emission feature is at 128~nm with a full width at half maximum of 8~nm. Subdominant emission features are theorized to be the result small levels of xenon impurities, the ``classical Left Turning Point", and the third excimer continuum~\cite{Heindl:2010zz, PhysRevA.43.6089, WIESER2000233}. Data from Ref.~\cite{Heindl:2010zz}.}
  \label{fig:lar_scint_spectrum}
\end{figure}

The second scintillation pathway occurs when charged particles ionize the argon atoms, producing free electrons and positively charged argon ions~\cite{Segreto:2020qks}. These free electrons and argon ions can then recombine, producing an argon atom in an excited energy state. This excited argon atom then once again can form an excited dimer with ground state argon atoms. This produces scintillation light when the excited dimer relaxes back to the ground state. 

Liquid argon scintillation light has two characteristic time constants from the spin configurations of the excited argon dimer~\cite{Whittington:2014aha,Segreto:2020qks,DEAP:2020hms}. The singlet spin configuration decays with a fast time constant of $\mathcal{O}$(5~ns) while the triplet spin configuration decays with a longer time constant of $\mathcal{O}$(1~$\mu$s). Additionally, experimental observations motivate a third time constant theorized to describe the electron recombination microphysics~\cite{Hofmann:2013vva,DEAP:2020hms}. Referred to as the ``recombination" time constant, this process modifies the light profile between approximately 25~ns and 175~ns. The normalized scintillation time distribution is described by Eq.~\ref{eq:lar_pulse_shape}, as given in Ref.~\cite{DEAP:2020hms}. 

\begin{equation}
I(t) = \frac{R_s}{\tau_s}e^{-t / \tau_s} + \frac{R_t}{\tau_t}e^{-t / \tau_t} + \frac{1 - R_s - R_t}{(1 + t / \tau_{rec})^2~\tau_{rec}}
\label{eq:lar_pulse_shape}
\end{equation}

In Eq.~\ref{eq:lar_pulse_shape}, the $R_s$ term describes the ratio of light emitted by the singlet state with characteristic time constant $\tau_s$. The corresponding intensity of the triplet state is described by $R_t$ with a characteristic time constant of $\tau_t$. The ad hoc recombination process is governed by the time constant, $\tau_{rec}$, and described by the third term.

For both the singlet and triplet states, the scintillation light is emitted primarily around 128~nm in the vacuum ultraviolet (VUV) range with a full width at half maximum of approximately 8~nm~\cite{Heindl:2010zz}. The measured scintillation wavelength spectrum is demonstrated in Fig.~\ref{fig:lar_scint_spectrum}, data from Ref.~\cite{Heindl:2010zz}. While the vast majority of scintillation photons are emitted very close to 128~nm, note the y-scale is logarithmic, there are subdominant emission lines as well. Xenon impurity in the experimental apparatus is believed to cause the weak emission around 149.1~nm~\cite{Heindl:2010zz}. The feature at 155~nm is the so-called ``classical Left Turning Point", which is typically a stronger emission feature for gaseous argon than liquid~\cite{PhysRevA.43.6089}. The emission between 175~nm and 250~nm is theorized to be due to the third excimer continuum~\cite{WIESER2000233}. 

As discussed in Chapter~\ref{chap:ccm}, the CCM200 detector does not have a purification system for the liquid argon. This is a unique aspect of the detector, making this the first detailed study of liquid argon not purified beyond the level at delivery. While the major impact of impurities on liquid argon scintillation is on the light yield through absorption, impurities will additionally alter the scintillation pulse shape. Nitrogen impurities are known to reduce the triplet time constant thorough quenching of the excited argon dimers~\cite{WArP:2008rgv,MicroBooNE:2022pcx}. This then additionally affects the relative intensity between the singlet and triplet states~\cite{Mavrokoridis:2011wv}. 

\subsection{Cherenkov Light}
The other piece of light production in liquid argon is through Cherenkov radiation~\cite{Cherenkov:1937mnd}. When charged particles travel faster than the speed of light in a polarizable medium, they induce dipole moments in the atoms. As those atoms relax back to the ground state, they emit Cherenkov radiation. These light signals form a coherent wavefront at a characteristic angle $\theta_C$. This angle is related to the relativistic velocity of the particle and the index of refraction in the medium through $\cos\theta_c = 1 / \beta n$.

The yield of Cherenkov radiation is governed by the Frank-Tamm formula, Eq.~\ref{eq:frank_tamm}~\cite{10.1093/ptep/ptac097}. In this relationship, the number of Cherenkov photons produced as a function of distance traveled and wavelength depends on the fine structure constant $\alpha$, charge of the incident particle $z$, wavelength of the emitted photon $\lambda$, relativistic velocity of the incident particle $\beta$, and the index of refraction in the medium $n(\lambda)$.

\begin{equation}
    \frac{d^2 N}{dx d\lambda} = \frac{2 \pi \alpha z^2}{\lambda^2} \left( 1 - \frac{1}{\beta^2 n^2(\lambda)} \right)
\label{eq:frank_tamm}
\end{equation}

\begin{figure}[h]
  \centering
  \includegraphics[width=0.7\linewidth]{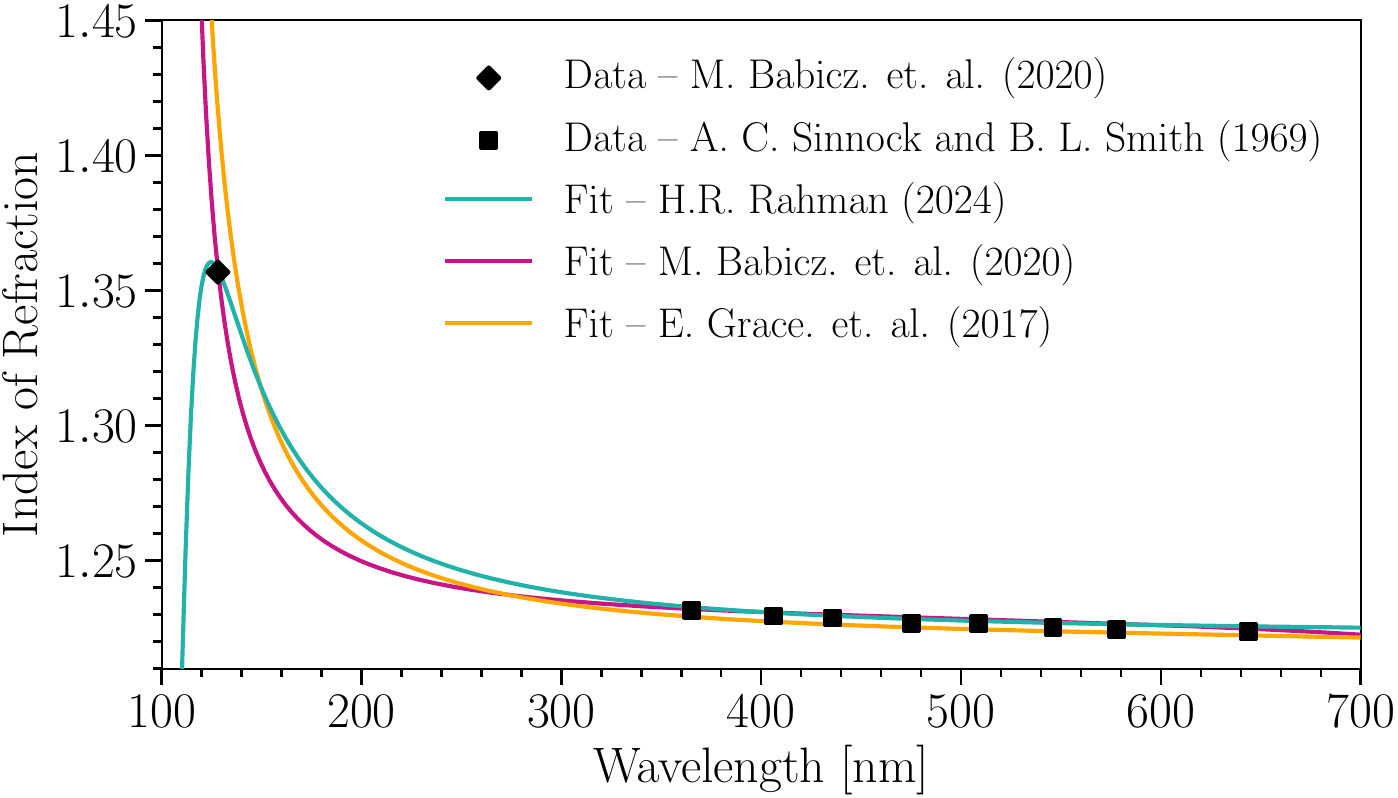}
  \caption{Existing literature values for the wavelength resolved index of refraction in liquid argon. Measurements in the visible spectrum are displayed by the square markers~\cite{Sinnock:1969zz} while the measurement in the VUV region is displayed in the diamond marker~\cite{Babicz:2020den}. Fits to the data using the Sellmeier dispersion relationship~\cite{born1999principles} are displayed in the magenta~\cite{Babicz:2020den} and orange~\cite{Grace:2015yta} lines. Fit to the data using a damped harmonic oscillator model, which accounts for absorption near the VUV resonance, is displayed in the cyan line~\cite{Rahman:2024zhp}.}
  \label{fig:rindex}
\end{figure}

The index of refraction is very important for governing both the optical properties of light in the medium as well as the yield of Cherenkov radiation. In liquid argon, there are only two experimental measurements of the wavelength resolved index of refraction. In 1969, Sinnock and Smith measured the index of refraction of visible light in liquid argon~\cite{Sinnock:1969zz}. These data points, ranging from 350~nm to 650~nm, are displayed using square markers in Fig.~\ref{fig:rindex}. In the VUV range, there is one measurement of the index of refraction from liquid argon's own scintillation light, narrowly peaked around 128~nm~\cite{Babicz:2020den}. This data point is displayed using the diamond marker in Fig.~\ref{fig:rindex}.

Since the index of refraction strongly depends on the wavelength, the Sellmeier dispersion relationship is generally utilized to fit for this relationship~\cite{born1999principles}. Eq.~\ref{eq:sellmeier} describes the expansion of the index of refraction as a function of wavelength. The Sellmeier coefficients $a_0$ and $a_i$ normalize each term and the $\lambda_i$ represent the absorption resonances in the medium. As the wavelength approaches the resonance, the denominator in Eq.~\ref{eq:sellmeier} goes to zero, resulting in a diverging index of refraction as $\lambda \rightarrow \lambda_i$. 

\begin{equation}
n^2(\lambda) = a_0 + \sum_i \frac{a_i \lambda^2}{\lambda^2 - \lambda_i^2}
\label{eq:sellmeier}
\end{equation}

Using the Sellmeier dispersion relationship and measured experimental data, Ref.~\cite{Babicz:2020den} and Ref.~\cite{Grace:2015yta} fit the index of refraction as a function of wavelength, shown in the magenta and orange lines, respectively, in Fig.~\ref{fig:rindex}. These functional forms for the index of refraction diverge to infinity at approximately 106.6~nm where there is an absorption resonance in liquid argon~\cite{Lane1968}. This VUV range is of particular interest because of both its proximity to the 128~nm scintillation emission and because of the implications for Cherenkov radiation.

As demonstrated in Eq.~\ref{eq:frank_tamm}, the number of emitted Cherenkov photons is proportional to $1 / n^2(\lambda)$. So in practice, if using the Sellmeier dispersion relationship to describe the index of refraction as a function of wavelength, one needs to implement an arbitrary lower bound in wavelength. This arbitrary bound then limits the Cherenkov production and sets an energy threshold for Cherenkov radiation.

\begin{equation}
    n = a_0 + a_{UV} \left ( \frac{\lambda_{UV}^{-2} - \lambda^{-2}}{ (\lambda_{UV}^{-2} - \lambda^{-2})^2 + \gamma_{UV}^2 \lambda^{-2}} \right )
\label{eq:ho_rindx}
\end{equation}

For a more thorough treatment of the index of refraction around the VUV resonance, this work utilizes a damped harmonic oscillator model for the index of refraction as a function of wavelength~\cite{Rahman:2024zhp}. This model includes both a real component for the index of refraction and an additional imaginary component that describes the absorption near the VUV resonance. This parameterization no longer diverges to infinity at the VUV resonance, a clear advantage over the Sellmeier dispersion relationship. 

Eq.~\ref{eq:ho_rindx} describes the damped harmonic oscillator model for the wavelength resolved index of refraction. The $a_0$ and $a_{UV}$ are similar to the Sellmeier coefficient terms while there is an additional $\gamma_{UV}$ parameter which describes the absorption at the VUV resonance. This relationship is fit to the experimental data in liquid argon, best fit parameter are detailed in Table~\ref{table:ho_rindex_paramters}, and the resulting fit is displayed as the cyan line in Fig.~\ref{fig:rindex}.

While all of the fits for the wavelength resolved index of refraction in Fig.~\ref{fig:rindex} behave similarly above approximately 300~nm, the fit using the harmonic oscillator model turns over at the VUV resonance due to the absorption while the fits using the Sellmeier dispersion relationship diverge to infinity around 106~nm. 

\begin{table}[h]
    \centering
    \begin{tabular}{lcc}
     \toprule
      \textbf{Parameter} & \textbf{Fit Value}  \\
      \midrule
      $a_0$ & $1.10232$  \\
      $a_{UV}$ & $0.00001058~\rm{nm}^{-2}$  \\
      $\gamma_{UV}$ & $0.002524~\rm{nm}^{-1}$ \\
      \bottomrule
    \end{tabular}
  \caption{Parameters in the damped harmonic oscillation model for the wavelength resolved index of refraction. The fit values are obtained from constraining the model using experimental data~\cite{Sinnock:1969zz,Babicz:2020den} for the index of refraction as a function of wavelength in liquid argon. The model and data are from Ref.~\cite{Rahman:2024zhp}.}
  \label{table:ho_rindex_paramters} 
\end{table}

As part of this work characterizing light production and propagation in liquid argon, we implement the damped harmonic oscillator model for the index of refraction. We additionally fit for the $\gamma_{UV}$ parameter, while fixing $a_0$ and $a_{UV}$ to the literature values displayed in Table~\ref{table:ho_rindex_paramters}~\cite{Rahman:2024zhp}.

\section{\texorpdfstring{$^{22}$Na}{22Na} Calibration Source}
This work and the following Chapter~\ref{chap:cherenkov} both utilize data from a $^{22}$Na calibration source. $^{22}$Na has two decay pathways~\cite{BASUNIA201569}. The primary one, with around a 90\% branching ratio, is $\beta^+$ decay. This produces a low-energy positron, with an endpoint energy of approximately 0.546~MeV, and an excited state of $^{22}$Ne. When the neon atom decays back to the ground state, it emits a 1.275~MeV gamma-ray. The secondary decay channel, with around 10\% branching ratio, is through electron capture, producing only the 1.275~MeV gamma-ray from the neon de-excitation. 

\subsection{Data Selection}
Data was collected with a 3~$\mu$Ci $^{22}$Na calibration source inserted at the origin of the detector while the accelerator was not running. The source is encapsulated in approximately 1~mm of stainless steel and inserted into the detector using a thin steel rod through the central flange on the top of the detector. Since the source is encapsulated in stainless steel, the positron emitted in the primary $\beta^+$ decay pathway promptly annihilates before entering the liquid argon. This produces two 0.511~MeV gamma-rays from the annihilation process.

Therefore, the decay of $^{22}$Na produces either a single 1.275~MeV gamma-ray from the electron capture reaction or a 1.275~MeV and two 0.511~MeV gamma-rays from the $\beta^+$ decay and prompt positron annihilation. These low-energy gamma-rays then typically Compton scatter in the liquid argon to produce around 1~MeV and sub-MeV electrons which scintillate and can emit Cherenkov radiation.

Data was collected using a 20~Hz fixed trigger with a 16~$\mu$s data acquisition (DAQ) window. Multiple events can then be identified over each DAQ window. The events are identified using a charge over a certain threshold algorithm. We required at least 20~PE of recorded charge in the entire detector across a 20~ns time window to identify an event. After identifying events, a finer grained event finding procedure is run requiring only 3~PE in the detector in a single 2~ns time bin to define the start time for each reconstructed event. After event finding, the following data quality cuts were applied. 

\begin{enumerate}
    \item \textbf{Cosmic-ray muon}--- This cut removes entire DAQ windows with identified cosmic-ray muons to avoid contamination from the long-lived triplet scintillation emission. Without significant overburden on top of the CCM200 detector, cosmic ray muons are relatively common. As cosmic-ray muons have $\mathcal{O}(\text{GeV})$ of kinetic energies, this cut removes DAQ windows with more than 200~PE reconstructed in the entire detector in a 2~ns window. This reduces background contamination from triplet light as well as the possibility of Michel electron emission from the muon decay. 
    
    \item \textbf{Surrounding event}--- The next cut removes events without adequate time separation between both the previous and the subsequent reconstructed event. The previous event requirement is no other reconstructed events in the 2.2~$\mu$s time window preceding the event, aimed to reduce the possibility of stray light from previous triplet emission. Additionally, in order to accumulate a highly pure sample of events and fit for long-time scale behavior, we require that the next reconstructed event start time is at least 2~$\mu$s after the current event's reconstructed start time. This avoids any overlapping events that would complicate fitting the time structure of the data.
    
    \item \textbf{Radius}--- The final cut isolates events with roughly isotropic spatial distribution of light emission. Using the first 20~ns of each event, the positions are roughly reconstructed as the charge weighted average of the PMT locations. This cut then requires the reconstructed radius to be $\leq25$~cm relative to the origin. This removes a large proportion of measured background events which are more likely to occur in the outer volume of the detector. 

\end{enumerate}

\begin{figure}[h]
  \centering
  \includegraphics[width=0.7\linewidth]{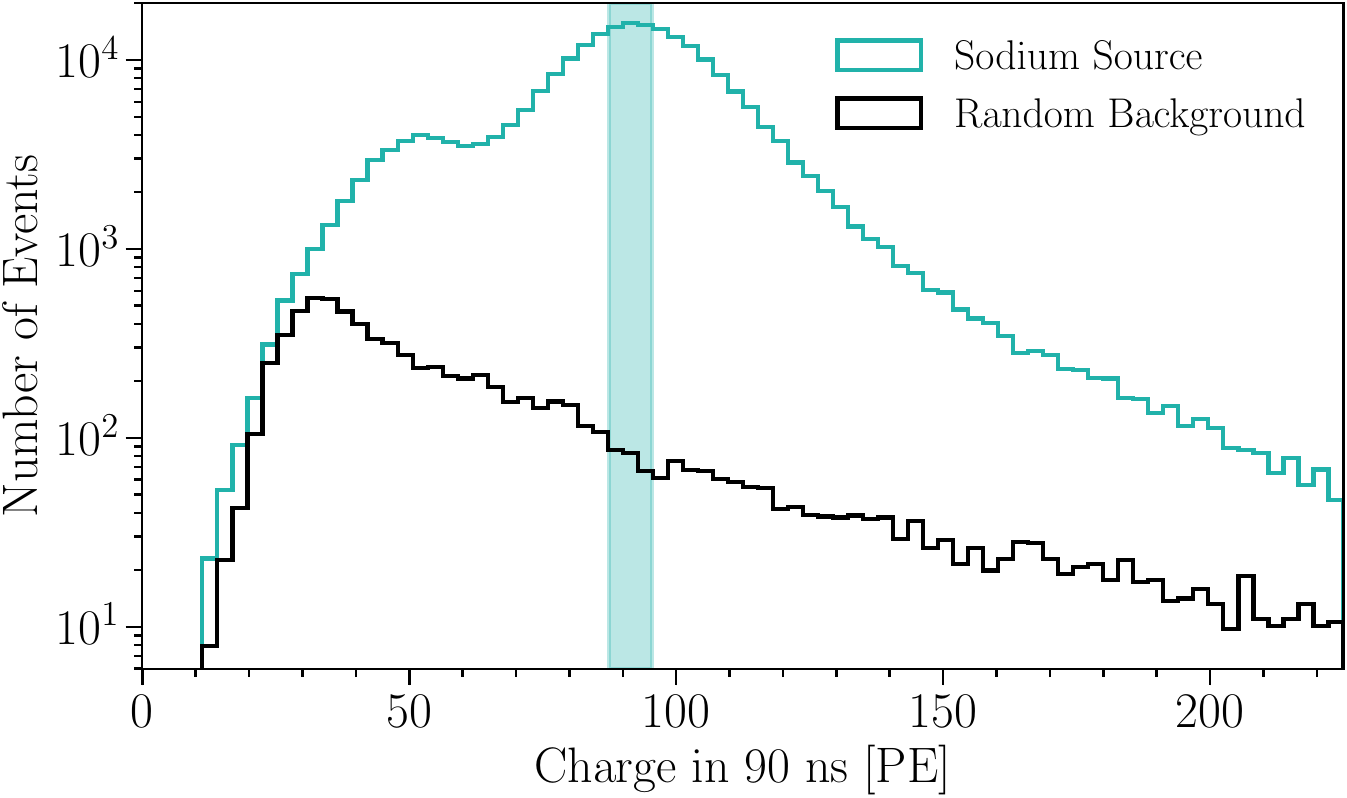}
  \caption{Charge distribution of reconstructed events in the detector. The black line represents reconstructed events observed without any source inserted in the detector as a measure of the random backgrounds. The cyan line represents events observed when the $^{22}$Na radioactive source was inserted. The two decay pathways are clearly visible at around 50~PE and 100~PE, corresponding to electron capture and $\beta^+$ decay, respectively.}
  \label{fig:sodium_charge}
\end{figure}

After applying the cuts listed above, Fig.~\ref{fig:sodium_charge} demonstrates the distribution of the charge in the first 90~ns of selected events as a proxy for energy. The cyan line represents data collected while the sodium source was inserted in the detector and the black line is data collected without the source as a measure of the random backgrounds.

When the source was inserted, there are two clear peaks in the charge distribution. The first, at approximately 50~PE, corresponds to the lower energy electron capture decay pathway. The second, at approximately 90~PE, corresponds to the higher energy $\beta^+$ decay channel. Events within $\pm$4~PE of the high energy peak, demonstrated in the shaded cyan band, were selected and accumulated across each PMT for fitting against Monte Carlo simulations.

The measured random background distribution and data collected while the sodium source was inserted are scaled to the same exposure time, indicating the expected rate of background contamination. For the selected charge region, the random background contamination rate is $\leq0.52\%$ of the events, indicating a highly pure sample of sodium data. 

This procedure selected approximately 50,000 events in the sodium data sample. After selection, events were accumulated in time for each PMT relative to the global event start times. This accumulated data per PMT was then compared to Monte Carlo simulations for optimization. 

\section{Fitting Strategy}
A major component of this work was creating a differentiable simulation model. Differentiable simulation, a common technique in machine learning applications, allows for analytic computation of an expectation based on the gradients with respect to physical parameters~\cite{degrave2016differentiable,6386109,newbury2024reviewdifferentiablesimulators}. In this case, where the simulation quantity is photons incident on the PMT photocathode surface, constructing these gradients involves extensive tracking of optical photons in the detector. By tracking emission wavelength, distance traveled, wavelength shift, and characteristics of the photon at detection, we developed an analytic framework to re-weight the probability of photon observation under various physical scenarios. 

Tracking the distance traveled, for example, allows for analytic re-weighting of the expectation as a function of absorption length. The probability that a photon is observed for a given absorption length is the ratio of the exponential probability distribution functions for the simulated absorption length and the desired absorption length, which only requires the distance traveled in liquid argon as input. In aggregate, this produces accurate predictions across the parameter space without needing to re-simulate.

This procedure has two benefits. The first is the reduction in the computation time. Optical simulations are very costly, especially in such a prolific scintillator like liquid argon. Full Monte Carlo simulations across a parameter space with over 20 dimensions is computationally prohibitive. Being able to compute the expectation analytically, as opposed to the full optical simulations, vastly reduces the computation time, making this work possible. The second benefit of differentiable simulation is the computation of the gradient information. This enables use of efficient gradient-based minimizers, improving fit convergence and reducing optimization time~\cite{minimizer1,minimizer2}. 

\subsection{\texttt{GEANT4} Simulation}
\begin{figure}[h]
  \centering
  \includegraphics[width=0.7\linewidth]{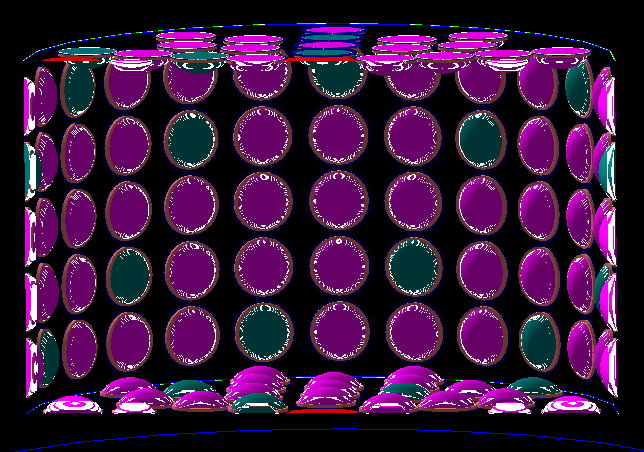}
  \caption{Rendering of the CCM200 interior geometry. The 8'' PMTs and the cryostat are fully modeled in \texttt{GEANT4}. The magenta PMTs are coated in TPB while the green ones are uncoated. The sodium source and its insertion mechanism are additionally fully described in \texttt{GEANT4} for this work.}
  \label{fig:geant4_visualization}
\end{figure}

This work utilized the Monte Carlo simulation framework \texttt{GEANT4} to simulate the $^{22}$Na decays for comparison to data~\cite{GEANT4:2002zbu}. The detector is fully described in \texttt{GEANT4}, visualization of the active region is demonstrated in Fig.~\ref{fig:geant4_visualization}. The \texttt{GEANT4} description of the CCM200 detector includes the cryostat, supporting structures, calibration source and deployment rod, fully modeled 8'' PMTs, and the bridle and frill structures supporting the PMTs.

In the \texttt{GEANT4} simulation, the majority of the physics models are from the \texttt{FTFP\_BERT\_HP} physics list, with some exceptions. Electromagnetic interactions are simulated using the \texttt{G4EmStandardPhysics\_option4} physics list. To improve the accuracy of photon interactions in the relevant energy range, the default Compton scattering implementation is replaced with the \texttt{G4PenelopeComptonModel}~\cite{GEANT4:2002zbu}.

Optical processes are included via \texttt{G4OpticalPhysics}; however, the standard \texttt{G4Cerenkov} class is modified to accommodate the damped harmonic oscillation model for the index of refraction. The standard \texttt{G4Cerenkov} class integrates the Frank-Tamm formula over the index of refraction assuming normal dispersion (index of refraction decreasing with increasing wavelength). For normal dispersion, only one boundary where the Cherenkov threshold is met ($\beta n(\lambda) > 1$) needs to be determined for the integration procedure. While the damped harmonic oscillation model also exhibits normal dispersion for wavelengths above the VUV resonance, below that resonance the index of refraction exhibits anomalous dispersion (index of refraction increasing with increasing wavelength). In this case, two boundaries need to be determined where the Cherenkov threshold is valid. Therefore the integration procedure in the \texttt{G4Cerenkov} class was adapted to account for this difference when determining the bounds for integration of the Frank-Tamm formula.

In order to allow for differentiation, the \texttt{SensitiveDetector} class implementation included extensive photon tracking. The entire history of every photon was cataloged until detection. This includes the creation process, time, location, and wavelength at creation, all wavelength shifts and subsequent daughter particles, total path length, time, location, and wavelength at detection. This allowed us to create a nominal $^{22}$Na decay simulation set that could be used for re-weighting when fitting. This simulation set was then processed through the same event identification, selection cuts, and accumulation procedure as data. 

\subsection{Differentiation Techniques}
To evaluate the expected signal at alternative points in parameter space, each simulated photoelectron (PE) is reweighted on an event-by-event basis. The weight is constructed from ratios of probability density functions (PDFs) evaluated at the nominal simulation point and at the target parameter values, making use of forward-mode automatic differentiation~\cite{icecube_phystools}. Because detailed photon-level information is retained in the simulation, this approach naturally incorporates variations in both the temporal and spatial distributions arising from changes in photon production, transport, and detection. The primary parameters included in the fit, along with their treatment in the reweighting framework, are summarized below.

\begin{enumerate}
\item \textbf{Scintillation Pulse Shape}--- The time structure of observed photons is primarily dominated by the liquid argon scintillation physics. In order to fit for the various components of the pulse shape, described in Eq~\ref{eq:lar_pulse_shape}, photons are re-weighted analytically by the ratio of the desired to the nominal pulse shape probability. Using the creation process that was tracked for every photon, this re-weighting procedure was only applied to photons created through the scintillation mechanism and not photons created through Cherenkov radiation. 

\item \textbf{Photon Absorption}--- The absorption length of photons in liquid argon depends strongly on the purity of the argon. Since the choice to not filter the liquid argon for the CCM200 detector is unique in liquid argon experiments, this parameter is least constrained by existing experimental results. This work fits for the wavelength resolved photon absorption length. Through information on the wavelength and distance traveled for a photon, the probability of observation can be analytically computed for different absorption lengths using the ratio of exponential decay PDFs; $\mathcal{P}(d, x(\lambda)) = e^{-d / x(\lambda)}$ for the distance traveled $d$ and the wavelength resolved absorption length $x(\lambda)$.   

\item \textbf{Index of Refraction}--- As discussed previously, the index of refraction is important in both optical photon interactions and Cherenkov radiation. Since the dependence of observed photons on the index of refraction is non-intuitive, we did not treat this parameter analytically. To preserve the physics modeling, we created multiple nominal simulation sets at various points in the index of refraction parameter space. We then interpolated across the index of refraction parameter space to generate the expectation. 

\item \textbf{Scattering Lengths}--- Rayleigh scattering is the dominant scattering process for short-wavelength photons in liquid argon. This work also considers Mie scattering, which becomes relevant at longer wavelengths. In pure liquid argon, atomic argon is too small to support Mie scattering. However, in CCM200, where the argon is unfiltered, larger molecular impurities can introduce Mie scattering. Modeling the combined impact of these scattering processes analytically is non-trivial. As with the index of refraction, we instead generated multiple simulation sets spanning the scattering parameter space and interpolated between them to obtain the expectation as a function of the scattering lengths.

\item \textbf{PMT Timing Response}--- The last major component of this fit is the response of the PMTs. The PMT time distribution, including post-pulsing effects, were modeled analytically. For this work, we used Gumbel distributions to model the PMT timing response~\cite{gumbel1958statistics}. Observed photons were re-weighted analytically using the ratio of the PDFs, $P(t, \mu, \sigma) = \frac{1}{\sigma} u(t) e^{-u(t)}$ for $u(t) = e^{-(t - \mu) / \sigma}$ for the location of the pulse $\mu$ and its shape $\sigma$.  
\end{enumerate}

\subsection{Systematic Uncertainties}
In similar analyses, parameter uncertainties are often derived from a profiled likelihood ratio test statistic and interpreted using Wilks’ theorem~\cite{Wilks:1938dza}. In this case, however, residual mismatches between the data and the simulation lead to artificially small uncertainty estimates when using that approach. Similar behavior is observed in other optical model fitting procedures and uncertainties are not reported~\cite{DEAP:2020hms}. In this work, we account for this behavior by instead adopting an alternative procedure that quantifies how much the model parameters must vary to accommodate the observed data on each PMT individually. 

Concretely, the optical model fit is performed independently for each PMT, yielding a distribution of best-fit parameter values. The spread of this distribution is then used to define parameter uncertainties: we assign $1\sigma$ intervals based on the marginal highest posterior density (HPD) region for each parameter. In the same way, uncertainty bands on the predicted observables are obtained from the HPD region of the corresponding expectations evaluated at these 191 best-fit parameter sets.

\section{Fit Procedure and Results}
 For this analysis, both data and simulated events are accumulated in time for each individual PMT over the interval from $-30~\text{ns}$ to $2~\mu\text{s}$ relative to the event start times. Variable sized time bins are employed to retain fine grained time structure at the beginning of events and ensure high statistics in the low light-yield tail of events. Bins of $5~\text{ns}$ are used between $-30~\text{ns}$ and $-10~\text{ns}$, followed by finer $2~\text{ns}$ bins from $-10~\text{ns}$ to $80~\text{ns}$ to resolve the detailed structure of the pulse rise and fall. Beyond this region, from $80~\text{ns}$ to $2~\mu\text{s}$, a coarser bin width of $20~\text{ns}$ is adopted to reflect the reduced statistics in the slowly varying triplet component.

Out of the full set of 200 PMTs, 191 are included in the fit. The remaining nine channels are omitted due to persistent issues such as elevated noise, instability, or other performance concerns. This fitting procedure optimizes the combined binned likelihood across all 191 PMTs and 145 time bins per PMT. The comparison between data and simulation is performed using an effective likelihood formalism developed for finite Monte Carlo statistics~\cite{Arguelles:2019izp}. The best fit point is obtained by minimizing the negative log-likelihood using the \texttt{L-BFGS-B} gradient-based optimization algorithm~\cite{minimizer1,minimizer2}.

The rest of this section presents the best-fit model prediction and examines key components of the fit in detail, including the extracted parameter values and their associated uncertainties. Parameters constrained by external inputs are also identified.

\subsection{Overall Time Structure}
\begin{figure}[h]
  \centering
  \includegraphics[width=0.7\linewidth]{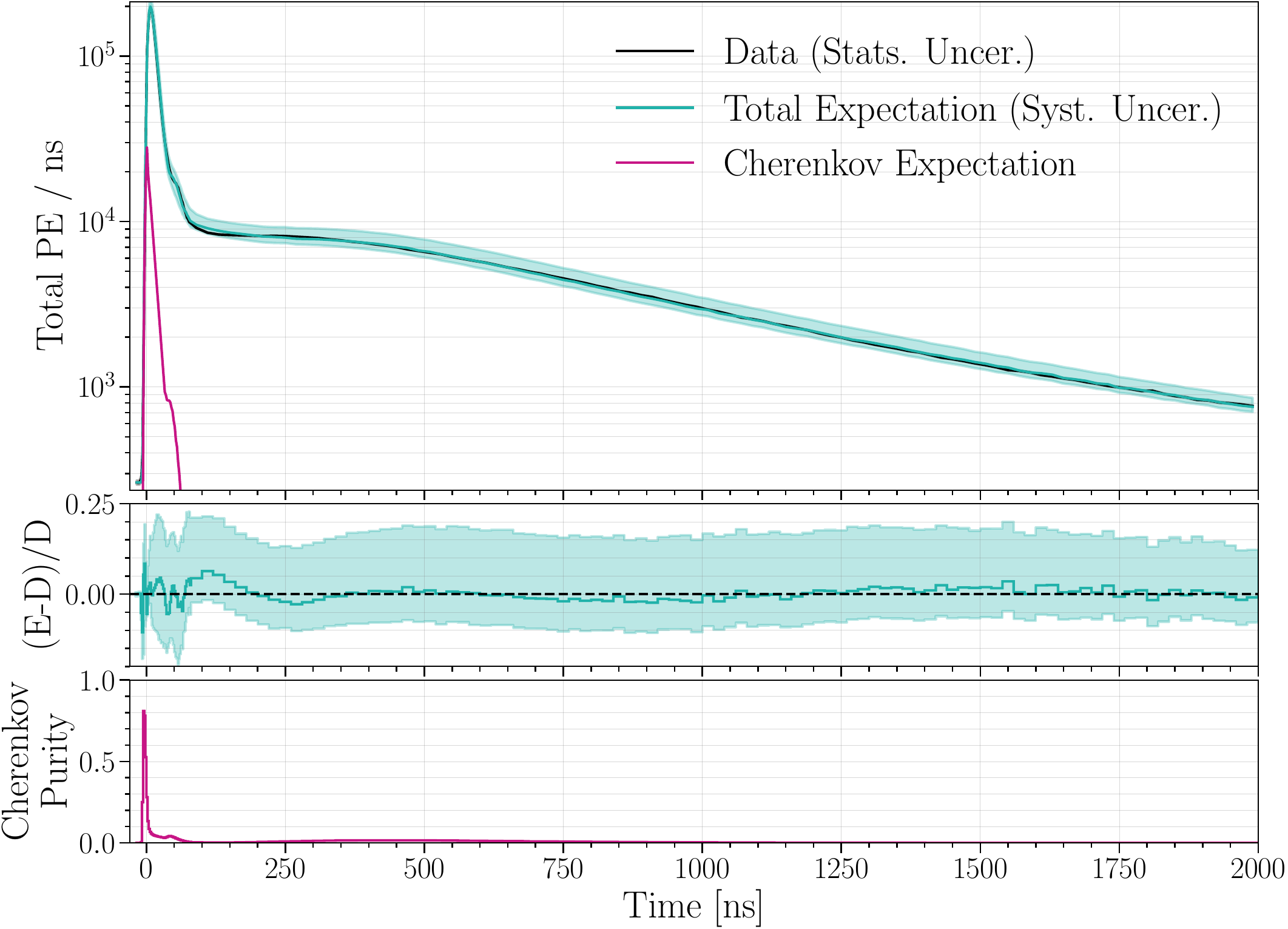}
  \caption{Data and expectation for the accumulated $^{22}$Na events across all PMTs across the entire fitting time region. The Cherenkov radiation component of the expectation is additionally displayed separate from the total expectation in the magenta line. The top panel is the total charge per ns, the middle panel is the residual between the data and expectation, and the bottom panel shows the ``Cherenkov Purity" of the expectation, defined as the ratio charge emitted from Cherenkov radiation to the total expected charge.}
  \label{fig:long time scale}
\end{figure}

Fig.~\ref{fig:long time scale} shows the data and best fit expectation accumulated across all PMTs across the entire time region used for fitting. The x-axis is the time relative to the reconstructed event start time, spanning from -30~ns to 2~$\mu$s. The top panel in this figure details the total charge accumulated across all events per time bin. The black line is the measured data, with $1\sigma$ gray shaded error bands obtained from Poisson statistics. The cyan line is the total expectation, which combines both the expected scintillation and Cherenkov PEs with measured random background per PMT. The cyan shaded bands are the $1\sigma$ systematic uncertainties obtained from the procedure outlined above. The magenta line isolates only the Cherenkov component of the expectation. This will be informative as we examine the overall time structure in different time regions. 

The middle panel of this figure shows the residual between data and the expectation with the propagated error bands. Across the entire fitting time window, the data and expectation agree within $10\%$. In the long-time scale time region, dominated by triplet scintillation light emission, the prediction agrees with data to the $<5\%$ level. 

The bottom panel in this figure is the ``Cherenkov purity.'' This is defined as the proportion of Cherenkov expectation to the total expectation. Again, this will become very interesting in the early time regions when Cherenkov light is most dominant. 

\begin{figure}[h]
  \centering
  \includegraphics[width=0.7\linewidth]{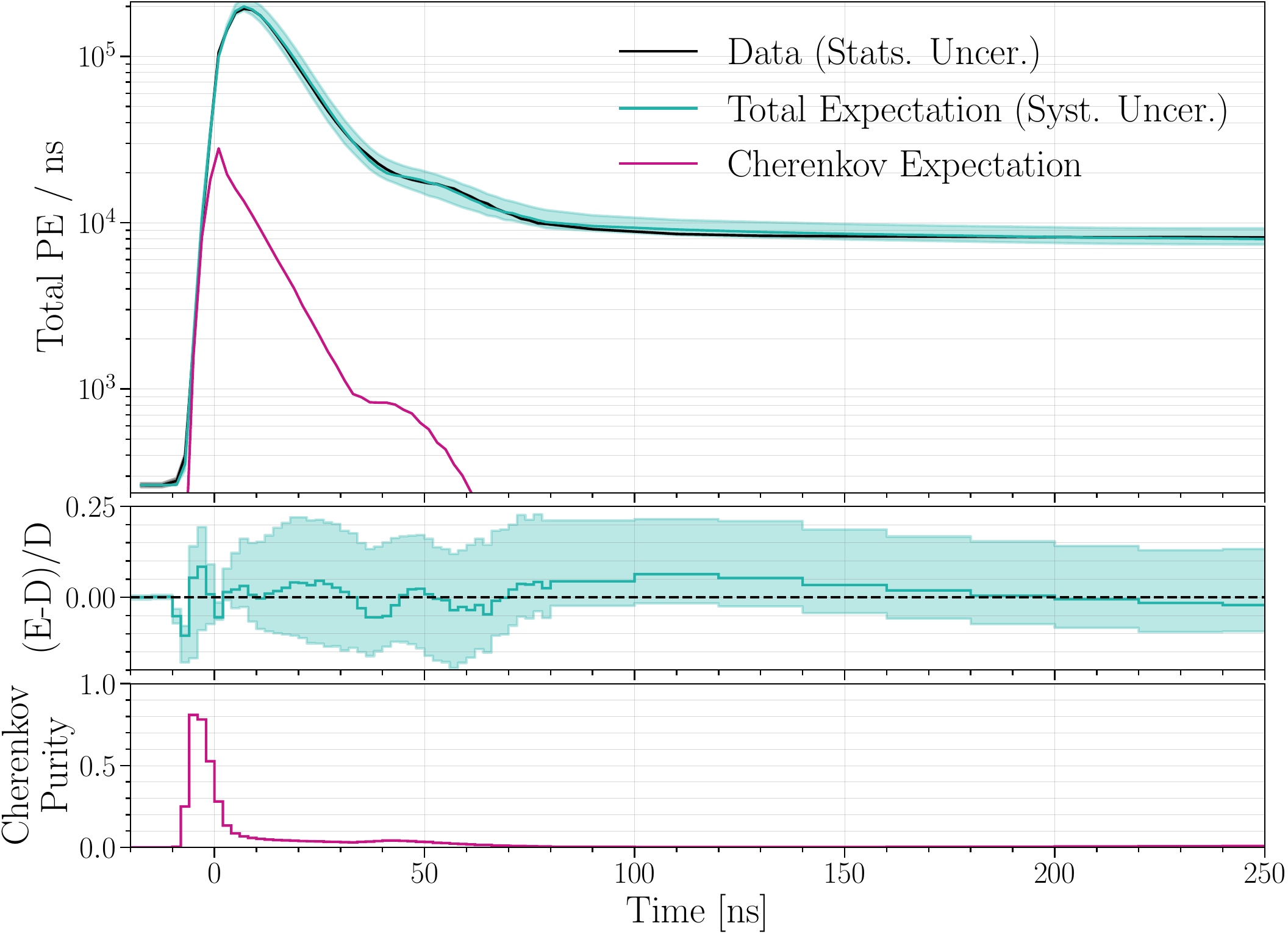}
  \caption{Data and expectation for $-30~\text{ns} < t < 250~\text{ns}$ time region. In this region, the most obvious feature is PMT post-pulsing around 50~ns. Data and expectation agree to the $\leq5\%$ level, indicating the accuracy of the PMT post-pulsing model employed in this fit.}
  \label{fig:medium time scale}
\end{figure}

Fig.~\ref{fig:medium time scale} shows the data and expectation between -30~ns and 250~ns. In this time range, PMT post-pulsing is evident around 50~ns. The $\leq5\%$ agreement between the central values of the data and Monte Carlo expectation in the time region indicate that the model of PMT post-pulsing adopted in this fitting procedure accurately represents the data.

\begin{figure}[h]
  \centering
  \includegraphics[width=0.7\linewidth]{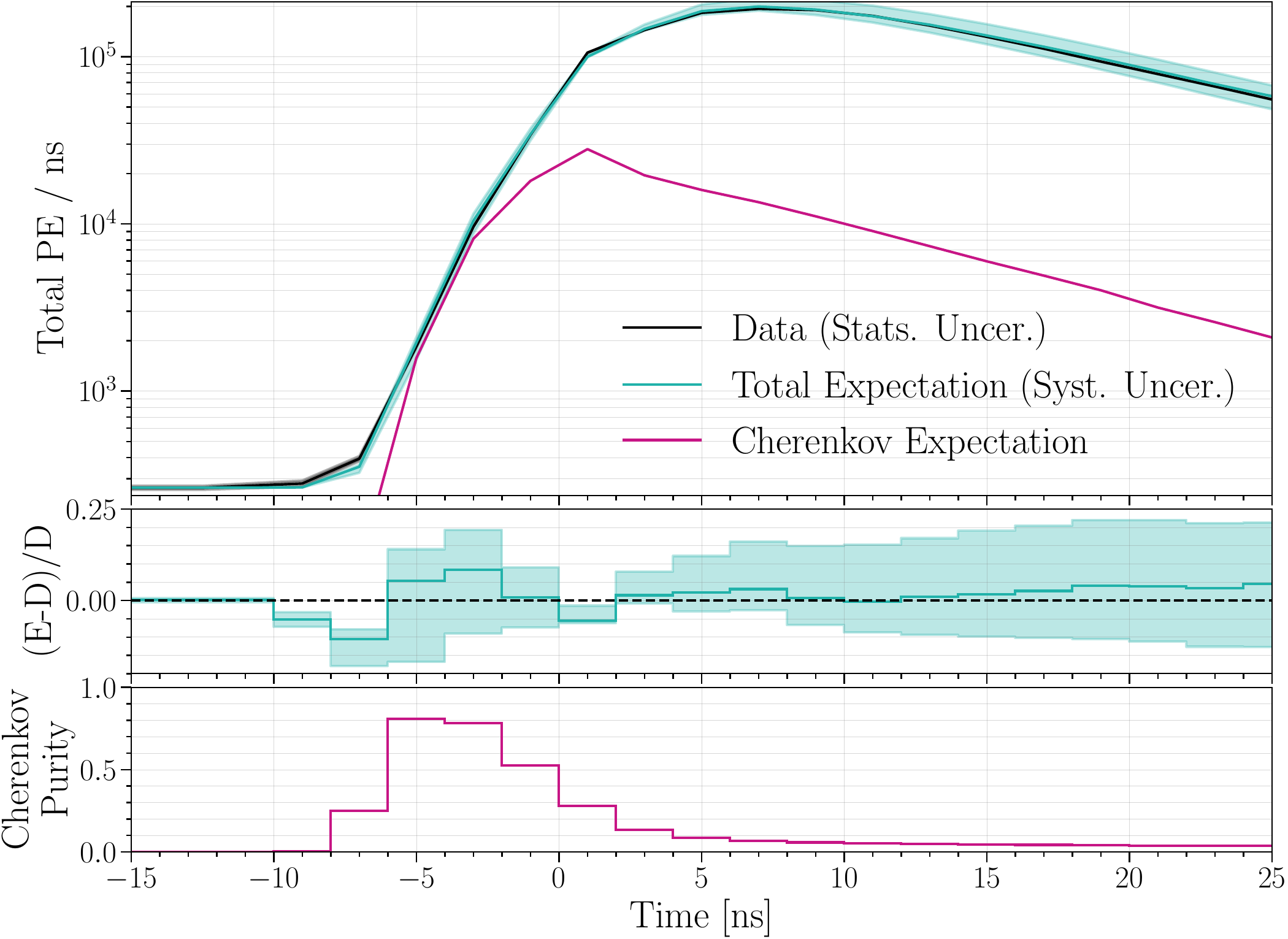}
  \caption{Data and expectation for $-15~\text{ns} < t < 25~\text{ns}$ time region. In this very early time region, the Cherenkov purity reaches a maximum of around 0.8 at approximately -5~ns relative to the event start time. Additionally, the Cherenkov radiation is a $\geq10\%$ effect until 5~ns after the event start time, necessitating modeling of both scintillation and Cherenkov light in this fitting procedure. For more discussion on the Cherenkov light in the $^{22}$Na source, see Chapter~\ref{chap:cherenkov}.}
  \label{fig:short time scale}
\end{figure}

Finally, Fig.~\ref{fig:short time scale} demonstrates the data and expectation very close to the reconstructed event start times, between -15~ns and 25~ns. Across all PMTs, the Cherenkov purity reaches a maximum of approximately 0.8 around -5~ns relative to the reconstructed event start times. The Cherenkov component of the expectation does not drop below 10\% until around 5~ns after the reconstructed event start times. This necessitates fitting for both scintillation and Cherenkov components of the expected light simultaneously, an original aspect to this work. See Chapter~\ref{chap:cherenkov} for a full discussion of Cherenkov light in this calibration source.

\subsection{Scintillation Light Parameters}
\begin{table}[h]
    \centering
    \begin{tabular}{lcc}
     \toprule
      \textbf{Parameter} & \textbf{Central Value} &  \textbf{Uncertainties} \\
      \midrule
      $R_s$ & $0.367$ & $-0.015, +0.017$ \\
      $R_t$ & $0.633$ & $-0.015, +0.017$ \\
      $\tau_s$ & $4.28~\mathrm{ns}$ & $-0.42~\mathrm{ns}, +1.20~\mathrm{ns}$ \\
      $\tau_t$ & $588.80~\mathrm{ns}$ & $-3.30~\mathrm{ns}, +3.65~\mathrm{ns}$ \\
      \bottomrule
    \end{tabular}
  \caption{Best fit scintillation pulse shape parameters, and corresponding $1\sigma$ uncertainty regions, obtained in this work. See Eq.~\ref{eq:lar_pulse_shape} for the full implementation of the pulse shape parameters.}
  \label{table:light_table} 
\end{table}

Table~\ref{table:light_table} details the best fit values and $1\sigma$ systematic uncertainty bands on the parameters that govern the scintillation light time structure. While Eq.~\ref{eq:lar_pulse_shape} has the data-driven intermediate time constant $\tau_{rec}$, we allowed for this third component in the fit but did not find any preference for an additional time component. 

This fit results in preference for $0.367^{+0.017}_{-0.015}$ ratio of singlet light, $0.633^{+0.017}_{-0.015}$ ratio of triplet light, $4.28^{+1.20}_{-0.42}$~ns singlet time constant, and $588.80^{+3.65}_{-3.30}$~ns triplet time constant. While the triplet state of liquid argon scintillation emission often found to have around 1.5~$\mu$s characteristic time constant in pure liquid argon, impurities can quench this time constant~\cite{WArP:2008dyo,WArP:2008rgv,Jones:2013bca}. Impurities, additionally, change the ratio between singlet and triplet light emission. These results reflect the measured ppm-level of oxygen, nitrogen, and water contaminants in the CCM200 detector. 

\subsection{TPB Characteristics}
The TPB response is fully implemented in \texttt{GEANT4}, using measured inputs for both the re-emission spectrum and the absorption length~\cite{Benson:2017vbw,doi:10.1021/j100052a011}. We examined how variations in the absorption length, particularly in the exponential tail overlapping the re-emission spectrum, affect the results. These variations were found to have a negligible impact on the fit, and therefore both the absorption length and re-emission spectrum were fixed to their literature values.

We also studied the impacts of a wavelength-shifting time constant. Simulations of short time-scale constants indicated that an exponential time constant of 0.3~ns provides the best agreement with the data. To reduce the number of free parameters in subsequent fits of other light-related quantities, this value was held fixed. Although longer time-scale structure in the TPB re-emission has been observed, it was neglected here because the fit is restricted to a relatively short window of 2~$\mu$s following the event start time~\cite{DEAP:2020hms}.

\subsection{Photon Absorption}
One of the ideal properties of pure liquid argon is that it does not absorb optical photons with wavelengths greater than 113~nm~\cite{Neumeier:2012cz}; this is discussed further in the context of Cherenkov light separation in Chapter~\ref{chap:cherenkov}. Photon absorption, however, is very sensitive to the presence of impurities in liquid argon. These impurities can both quench the scintillation process, evident in the reduced triplet emission time constant, and absorb emitted scintillation photons. This is evident through the range of literature values for the absorption lengths for UV photons in liquid argon, ranging from 66~cm to 50~cm~\cite{ISHIDA1997380, ArDM:2016jbw}. Additionally, there is evidence that photon absorption in liquid argon depends strongly on wavelength~\cite{Neumeier:2012cz,Neumeier:2015lka,Fields:2020wge}. 

\begin{equation}
    \lambda_{att} = \frac{d}{\ln \left(\frac{1}{1 - e^{-a (\lambda - b)}} \right)}
    \label{eq:uv_absorption}
\end{equation}

For this reason, this work fits for wavelength resolved absorption length based on the formalism in Ref.~\cite{Neumeier:2012cz}, described in Eq.~\ref{eq:uv_absorption}. The absorption length, $\lambda_{att}$, depends on a constant $d$, parameter $a$ that controls the shape of the absorption length, and $b$ which is related to the first excimer continuum.

\begin{figure}[h]
  \centering
  \includegraphics[width=0.7\linewidth]{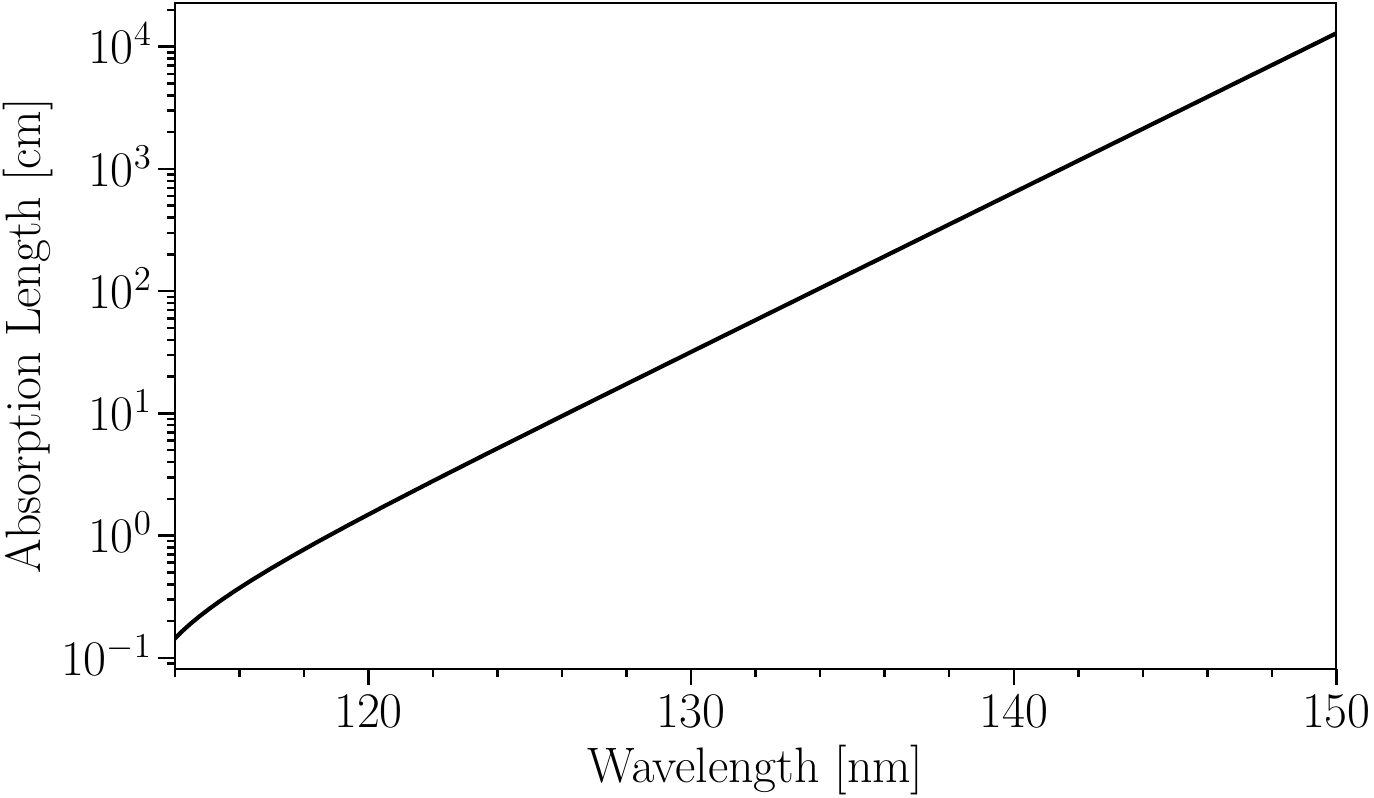}
  \caption{Best fit wavelength resolved absorption length. At 128~nm, the absorption length is 17.42~cm. This then increases exponentially with wavelength. The absorption length parameterization is based on Ref.~\cite{Neumeier:2012cz}.}
  \label{fig:uv_abs}
\end{figure}

In this work, we allowed the normalization parameter $d$ and the shape parameter $a$ to vary while fixing the parameter $b$ to the literature value of 113.2~nm. Fig.~\ref{fig:uv_abs} shows the best fit absorption length as a function of wavelength. The fit prefers $a = 0.30$ and $d = 0.194$, resulting in an absorption length of 17.42~cm at 128~nm, which increases exponentially with wavelength. In addition to this strong absorption in the VUV, we allowed for absorption in the visible spectrum due to impurities in the liquid argon. This work prefers an absorption length of 98.25~cm integrated from 300~nm to 400~nm.

While the uncertainty estimation procedure in this work is useful in addressing deviations in the measurements across all PMTs, it, however, cannot constrain the photon absorption length. The absorption length is most strongly constrained by the combination of measurements across all PMTs, as each PMT is a different distance away from the photon source. In the uncertainty estimation procedure, where each PMT is fit individually, that distance-dependent constraining information is lost, resulting in artificially very large ranges in the absorption length across the fits to individual PMTs. Therefore this work does not quote uncertainties on the measured wavelength resolved absorption length.

\subsection{Birks' Law}
This work additionally investigates the non-linearity of scintillation emission. The exciton formation in the scintillation process introduces non-linear quenching factor that can be described by Birks' Law, Eq. \ref{eq:birk}~\cite{Birks:1951boa}. The scintillation light yield per unit distance, $dL / dx$, depends on the scintillation efficiency $S$, energy deposited per unit distance $dE / dx$, and $k$, often referred to as Birks' constant.

\begin{equation}
\frac{dL}{dx} = \frac{S \cdot \frac{dE}{dx}}{1 + k \cdot \frac{dE}{dx}}
\label{eq:birk}
\end{equation}

The ICARUS collaboration characterized Birks' coefficient in liquid argon and found that this model of non-linear quenching agrees with data~\cite{AMORUSO2004275}. Since ICARUS employs an electric field to collect charge in addition to scintillation emission from the liquid argon, their measurement of Birks' constant, $k = 0.0486 \pm 0.0006$~(kV/cm)((g/cm$^2$)/MeV), is across electric drift fields $0.1 < E < 1.0$~kV/cm. This work additionally uses a modified version of Birks' law, Eq.~\ref{eq:e_field_birk}, which introduces an additional $\varepsilon$ term that is the electric field strength. In this modification, $Q$ is the initial ionization and $Q_0$ is the collected charge. 

\begin{equation}
Q = \frac{Q_0}{1 + \frac{k}{\varepsilon} \cdot \frac{dE}{dx}}
\label{eq:e_field_birk}
\end{equation}

While this measurement is valid for liquid argon with an electric field applied, it diverges to infinity as the electric field approaches 0. Therefore, this measurement is not applicable in liquid argon detectors without any electric field. As an alternative, this work utilizes a measurement of the linear energy transfer (LET) for 1~MeV electrons in liquid argon without an applied electric field~\cite{Doke:1990rza}. From the measurement of the LET, Eq.~\ref{eq:scint_eff_birk} allows for calculation of Birks' constant given the $dE / dx$. 

\begin{equation}
\text{LET} = \frac{1}{1 + k \cdot \frac{dE}{dx}}
\label{eq:scint_eff_birk}
\end{equation}

The $dE / dx$, also referred to as the stopping power, of 1~MeV electrons in cryogenic liquid argon (87$^\circ$K), is 1.386~MeV cm$^2$/g~\cite{PDG2024}. Using the measurement of the LET and Eq.~\ref{eq:scint_eff_birk} results in $k = 0.295$~(g/cm$^2$)/MeV for 1~MeV electrons in liquid argon without an applied electric field. While characterizing the liquid argon properties in this work, we investigated the effects of deviations to this literature value for Birks' constant. We found Birks' constant to be degenerate with an overall shift in the normalization in this study of $^{22}$Na decays. Since the decay produces $\mathcal{O}$(1~MeV) gamma-rays, these particles are too low-energy to provide sensitivity to Birks' constant. Ongoing calibration work studying Michel electrons may be able to make a robust measurement Birks' constant in liquid argon without any electric field applied. 

\subsection{Index of Refraction}
This works uses a damped harmonic oscillator model for the wavelength resolved index of refraction, as described in Eq.~\ref{eq:ho_rindx}. Since the index of refraction we well measured in the visible wavelength region between approximately 350~nm and 650~nm~\cite{Sinnock:1969zz}, but only one experimental constraint exists in the VUV region at 128~nm~\cite{Babicz:2020den}, this works allows the $\gamma_{UV}$ parameter, which controls the behavior around the VUV resonance, to vary.

\begin{figure}[h]
  \centering
  \includegraphics[width=0.7\linewidth]{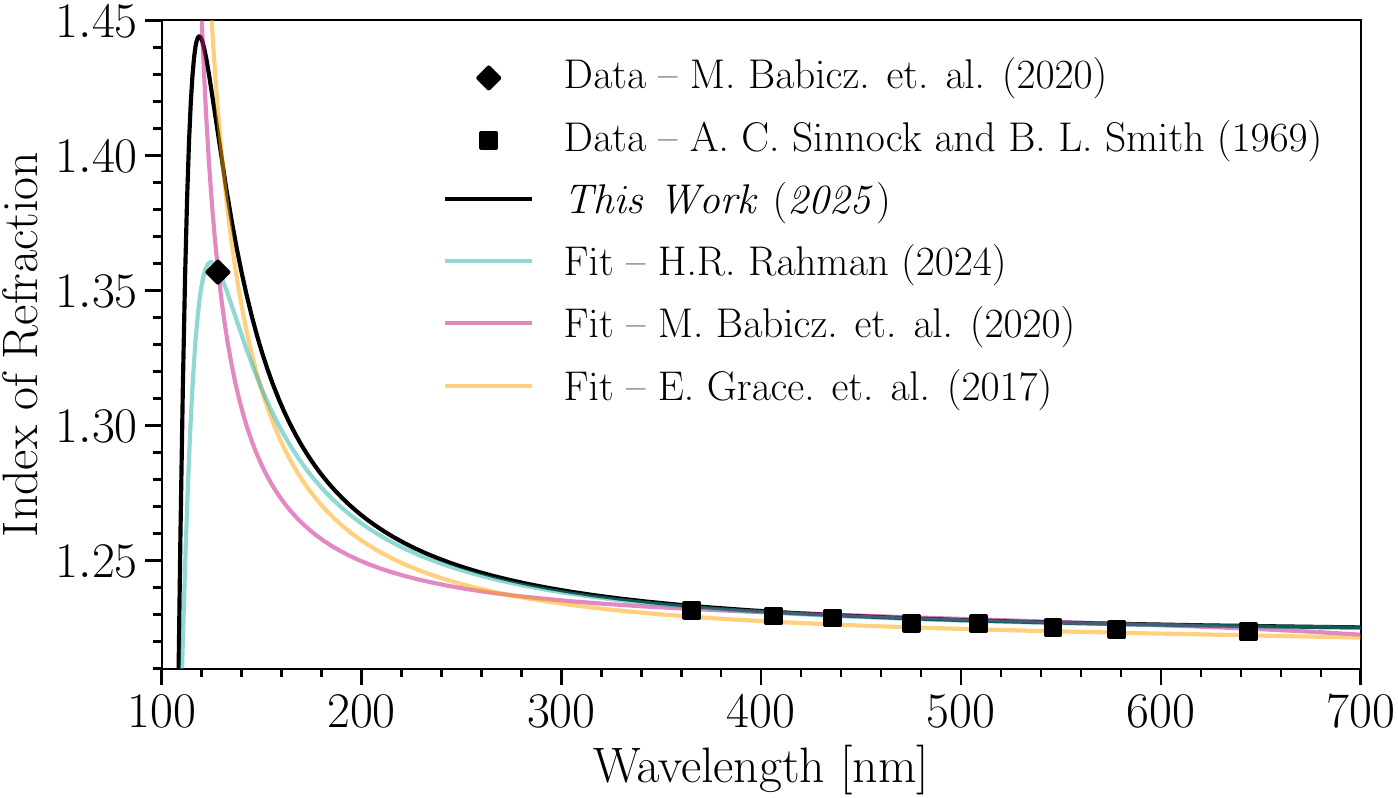}
  \caption{Existing literature values for the wavelength resolved index of refraction in liquid argon~\cite{Sinnock:1969zz,Babicz:2020den,Grace:2015yta,Rahman:2024zhp}. Additionally, the results of this work are displayed in the black line. Using the damped harmonic oscillator model for the index of refraction, Eq.~\ref{eq:ho_rindx}, we find preference for $\gamma_{UV}=0.0018$.}
  \label{fig:best_fit_rindex}
\end{figure}

We find preference for $\gamma_{UV}=0.0018$, the resulting index of refraction as a function of wavelength is displayed as the black line in Fig.~\ref{fig:best_fit_rindex} along with existing literature values. Additional experimental data, especially between 150~nm and 250~nm where the index of refraction transitions from steeply falling to relatively constant, would be useful in further constraining this parameter. 

To simplify the fitting procedure, we first fit for the index of refraction then fixed it for uncertainty estimation. Therefore, we do not report an uncertainty on the index of refraction. 

\subsection{Rayleigh and Mie Scattering}
An important property of light propagation is scattering. This work considers two different scattering processes--- Rayleigh and Mie scattering.

\begin{figure}[h]
  \centering
  \includegraphics[width=0.7\linewidth]{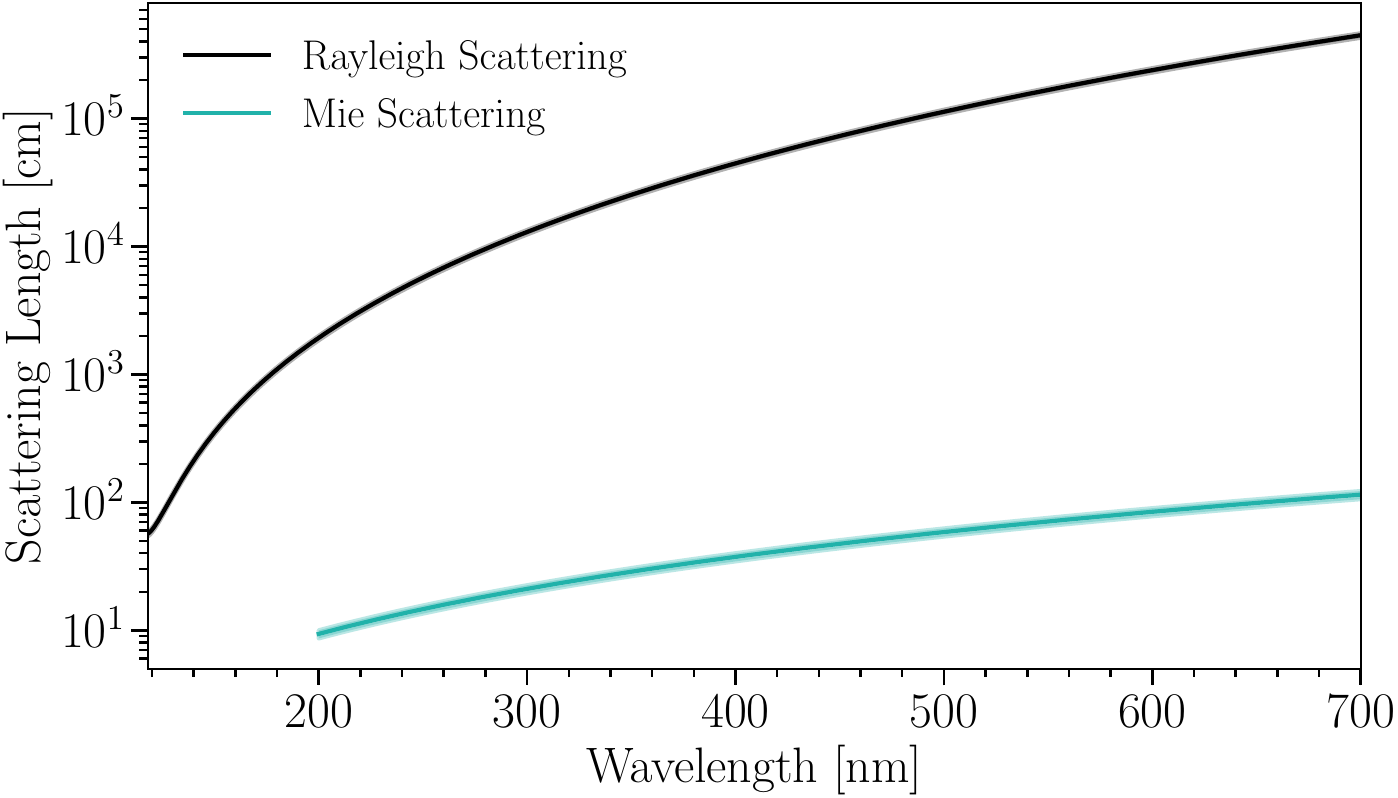}
  \caption{Best fit Rayleigh and Mie scattering lengths as a function of wavelength. At 128~nm, we find preference for $99.98^{+3.56}_{-4.52}~\rm{cm}$ Rayleigh scattering length. For Mie scattering, which primarily affects long-wavelength photons, we find preference for $9.37^{+0.63}_{-0.73}~\rm{cm}$ scattering length at 200~nm. See the text for more discussion on Mie scattering in liquid argon.}
  \label{fig:rayl_mie_scattering}
\end{figure}

In liquid argon, the argon atoms serve as targets for Rayleigh scattering due to their small atomic size. The Rayleigh scattering length depends on both the wavelength of the incident photon and the index of refraction in the medium, described in Eq.~\ref{eq:rayl}~\cite{landau1984electrodynamics}. Using this relationship, this work allows the overall normalization of the Rayleigh scattering length as a function of wavelength to vary and we find preference for a scattering length of $99.98^{+3.56}_{-4.52}~\rm{cm}$ at 128~nm. This best fit Rayleigh scattering length, with $1\sigma$ uncertainties, is shown in Fig.~\ref{fig:rayl_mie_scattering}.

\begin{equation}
l^{-1} = \frac{16 \pi^3}{6 \lambda^4} \left[ k T \rho^2 \kappa_T \left(\frac{(n^2-1)(n^2+2)}{3} \right)^2 \right]
\label{eq:rayl}
\end{equation}

This measurement is relatively consistent with other literature values, which vary between 55~cm and 100~cm~\cite{Grace:2015yta,Babicz:2020den,Seidel:2001vf}. Potentially, the Rayleigh scattering length may be effectively longer in the CCM200 detector due to the lack of filtration. Since absorption and Rayleigh scattering are competing processes for VUV photons, extinction due to absorption can increase the measurement of the effective Rayleigh scattering length. 

In addition to Rayleigh scattering, this work characterizes Mie scattering in the liquid argon. While Rayleigh scattering is mostly dominant for short wavelength photons and requires scattering targets that are much smaller than the wavelength of the incident photon, Mie scattering primarily affects long-wavelength photons~\cite{Mie1908}. Moreover, Mie scattering requires scattering targets that are on the order of the same size as the incident photons. In pure liquid argon, you would not expect any Mie scattering since the argon atoms are small. In CCM200, however, the contaminants introduce sources of Mie scattering targets. 

One potential target for Mie scattering in CCM200 is from the TPB. Both the Mylar reflective foils on the walls of the detector and 80\% of the PMTs are evaporatively coated in TPB. Other liquid argon experiments with similarly evaporatively coated TPB have found that TPB particulates can leech into the main volume of the detector~\cite{Asaadi:2018ixs,Ignarra:2014yqa}. Since TPB forms large crystal structures, this would serve as a target for Mie scattering for long-wavelength photons. Another potential source of Mie scattering targets is dust contamination. While efforts were made to reduce outside contaminants into the fiducial region of the CCM200 detector, it was not constructed in a dedicated clean room and it is possible that small level of contaminants would be present. A final potential source of contaminants that would facilitate Mie scattering is the known water impurity. The liquid argon manufacturer quotes a water impurity level of 0.01~ppm, which could form ice crystals in the fiducial region of the detector. 

Since the exact level of each contaminant is difficult to precisely know, this work fits for an effective Mie scattering length using a $1 / \lambda^2$ scaling dependence. Since this process is only relevant for longer wavelength photons, we only fit for the scattering length above 200~nm. Fig.~\ref{fig:rayl_mie_scattering} additionally demonstrates the best fit Mie scattering length, with $1\sigma$ systematic uncertainty bands. At 200~nm, this fit prefers a Mie scattering length of $9.37^{+0.63}_{-0.73}~\rm{cm}$, which then scales as $1 / \lambda^2$. Dedicated study of Mie scattering in liquid argon would greatly benefit future liquid argon detectors, especially those aiming to resolve Cherenkov radiation in addition to scintillation emission. 

\subsection{PMT Timing Characteristics}

The final component of this work is the PMT time response. While the majority of photoelectrons are amplified and digitized with a small time spread, so-called ``post-pulsing" effects are well known~\cite{MiniBooNE:2006fhd,Kaptanoglu:2017jxo,DEAP:2017fgw,Abbasi_2010,Brigatti_2005}. Short-time scale post-pulsing can be caused by electrons back-scattering off of a dynode stage, which introduces a time delay in the voltage signal. Long-time scale post-pulsing can be the result of heavy elements in the PMT ionizing and propagating through the same amplification chain. These elements will travel much slower than the typical electrons, resulting in $\mathcal{O}$($\mu$s) scale post-pulsing. 

\begin{figure}[h]
  \centering
  \includegraphics[width=0.7\linewidth]{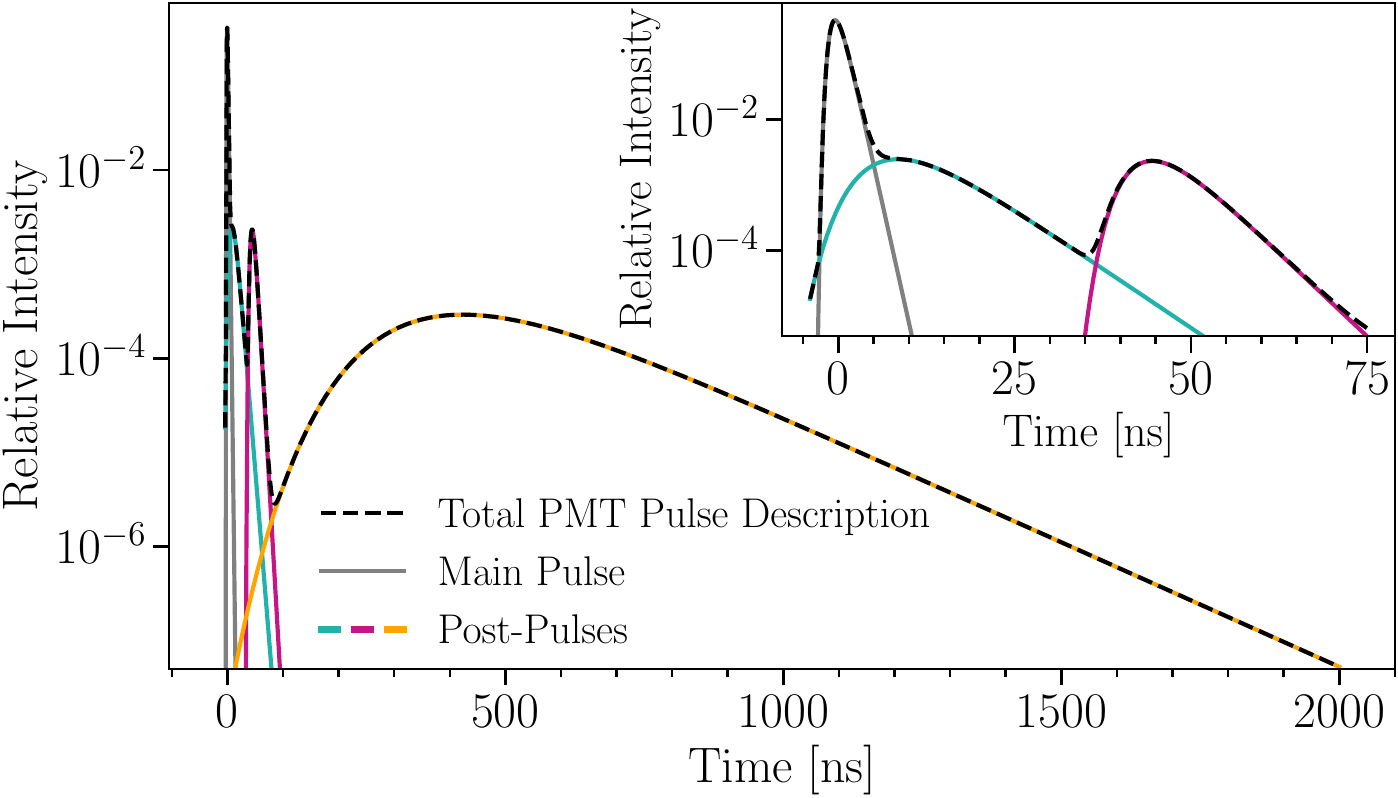}
  \caption{Best fit PMT time distribution. The main PMT pulse (gray distribution) is constrained at -0.45~ns and the fit prefers three additional post-pulses (cyan, magenta, and orange distributions). The post-pulses are centered at 8.47~ns, 44.51~ns, and 423.24~ns, see Table~\ref{table:pmt_params} for the full description of the best fit PMT time distribution.}
  \label{fig:best_fit_late_pulse}
\end{figure}

\begin{table}[h]
    \centering
    \begin{tabular}{cccc}
     \toprule
      \textbf{Pulse} & \textbf{Location~[ns]} &  \textbf{Shape~[ns]} & \textbf{Probability} \\
      \midrule
      Main Pulse & $-0.45$ & $0.9$ & $0.803$ \\
      Post-Pulse 1 & $8.47~[+2.42, -1.20]$ & $6.00~[+2.12, -1.52]$ & $0.040~[+0.038, -0.040]$ \\
      Post-Pulse 2 & $44.51~[+1.00, -7.21]$ & $4.26~[+4.77, -0.21]$ & $0.027~[+0.017, -0.008]$ \\
      Post-Pulse 3 & $423.24~[+0.63, -1.22]$ & $163.84~[+1.71, -1.49]$ & $0.13~[+0.028, -0.001]$ \\
      \bottomrule
    \end{tabular}
  \caption{Best fit PMT time distribution parameters. Each pulse is described using a Gumbel distribution~\cite{gumbel1958statistics}. The main pulse is fixed based on literature values for the spread in PMT electron transit times~\cite{hamamatsu_pmt_handbook}. The three additional post-pulses, with the corresponding $1\sigma$ systematic uncertainties, are the results of this fitting procedure. See Fig.~\ref{fig:best_fit_late_pulse} for the full PMT time distribution.}
  \label{table:pmt_params} 
\end{table}

In this work, we utilize the Gumbel distribution to describe the PMT time response. We fix the response of the main pulse based on the quoted spread of the electron transit times~\cite{hamamatsu_pmt_handbook} and find preference for three additional post-pulses that modify the time structure. The time distribution of the PMT response is demonstrated in Fig.~\ref{fig:best_fit_late_pulse} and the parameters, with $1\sigma$ systematic uncertainties, are in Table~\ref{table:pmt_params}.

In the first 75~ns of the PMT pulse, displayed in the inset plot of Fig.~\ref{fig:best_fit_late_pulse}, we find two additional post-pulses that modify the PMT response, one centered at 8.47~ns and the second at 44.51~ns. In addition, we find preference for a post-pulse at approximately 423.24~ns, primarily affecting the time structure in the triplet region. Overall, we find that the post-pulses are around a 19.7\% effect compared to the main pulse. For reference, the PMTs in CCM200 are operated around 1700~V and draw approximately 750~$\mu$A of current. 

\section{Applications}
This study provides the first detailed characterization of both liquid argon scintillation and Cherenkov radiation production and propagation in a large, light-collection–only detector, with explicit quantification of uncertainties on the scintillation parameters. It further incorporates modern techniques such as differentiable simulation, enabling efficient treatment of a high-dimensional optimization problem without sacrificing fidelity in the underlying physics model. Together, these developments establish a useful reference framework for future experimental efforts in similar detector configurations.

Beyond its immediate application in CCM200 data analysis, the model is well-suited for studies of next-generation detectors extending beyond the CCM200 design. In particular, it can be used to evaluate trade-offs between detector performance and cost, especially in scenarios where argon purification systems represent a significant design consideration.

More broadly, this work underpins a central goal of this thesis--- the separation of Cherenkov radiation in a high light-yield scintillation detector, which is the focus of Chapter~\ref{chap:cherenkov}.


\chapter{Separation of Cherenkov and Scintillation Light in $^{22}$Na Calibration Source}\label{chap:cherenkov}

The content of this chapter summarizes and expands on Ref.~\cite{CCM:2025kal}, on which the thesis author was the principal author.

One of the major outcomes of this thesis is the first observation of Cherenkov light from sub-MeV electrons in a high light-yield scintillation detector on an event-by-event basis. This chapter will discuss the motivations for hybrid optical detectors that can resolve both Cherenkov and scintillation signals, how the two signals were separated in calibration data using the CCM200 detector, and the implications of this result. 

\section{Optical Detectors}
\begin{table}[h!]
    \centering
    \begin{tabular}{lcc}
        \toprule
        \textbf{Quality} & \textbf{Scintillation Light} & \textbf{Cherenkov Light} \\
        \midrule
        Intensity & $\mathcal{O}(10^4)$ & $\mathcal{O}(10^2)$ \\
        Direction & Isotropic & Directional \\
        Timing & $\mathcal{O}$(ns) & $\mathcal{O}$(ps) \\
        Photon Wavelength & Narrow emission spectrum & Broad spectrum \\
        \bottomrule
    \end{tabular}
    \caption{Comparison of scintillation and Cherenkov light properties. Methods to separate the two signals typically rely on leveraging the differences in timing and wavelength.}
    \label{tab:cherenkov_vs_scint}
\end{table}

As discussed in Chapter~\ref{chap:intro}, neutrinos and potential dark sector candidates are very weakly interacting particles, making detection and study difficult. One proven and cost effective way to study these particles is by using large optical detectors like CCM200. These detectors instrument a bulk material with photo-detecting sensors around the edges, allowing for resolution of single photon signals. The bulk material provides both the targets for interactions to study and produces the light signals to reconstruct event properties.

Typically, the bulk material dictates if it is a scintillation or Cherenkov detector. The key differences between the two signals are the intensity, wavelength, timing, and directionality, summarized in Table.~\ref{tab:cherenkov_vs_scint}.

While scintillation light is quite prolific, high light-yield scintillators like liquid argon emit around $10^4$ photons per MeV of deposited energy, Cherenkov radiation is much dimmer, producing only around $10^2$ photons per MeV of deposited energy. This leads to resolving Cherenkov light in a scintillation detector being something like finding a needle in a haystack.

There is also a difference in the direction of light emission between the two processes. Scintillation light is emitted isotropically while Cherenkov radiation is a emitted at characteristic angle $\cos\theta_C = 1 / \beta n(\lambda)$, for the relativistic velocity $\beta$, and index of refraction $n(\lambda)$ of light in the medium. Detection of this directional signature can be very helpful for both rejecting backgrounds and selecting specific event topologies.

In terms of timing, scintillation light is emitted with a characteristic decay time constant, typically $\mathcal{O}$(ns), while Cherenkov radiation is emitted very promptly with only around $\mathcal{O}$(ps) delays.

The last difference between the two signals is the wavelength spectrum of the emitted photons. Scintillation light is emitted in a narrow wavelength region, typically then wavelength shifted into the visible spectrum to improve detection efficiency. Cherenkov light is emitted across a broad spectrum of wavelengths. The number of emitted photons is proportional to $1/\lambda^2$ for the wavelength $\lambda$. Typically, the timing and wavelength are the two strongest discriminants between scintillation and Cherenkov light.

\subsection{Scintillation Detectors}

Scintillation detectors utilize prolific scintillators as the bulk material, such as liquid argon or oil-based liquid scintillators. These materials produce copious numbers of emitted photons, allowing for low energy thresholds and excellent energy resolution~\cite{BOREXINO:2020aww,KamLAND:2013rgu,JUNO:2024jaw}. 

\subsection{Cherenkov Detectors}

Cherenkov detectors typically utilize water or mineral oil to measure Cherenkov radiation emitted when relativistic charged particles traverse the medium. These detectors can reconstruct direction of motion as well as Cherenkov ring topology to discriminate between particle types~\cite{Patterson:2009ki,Super-Kamiokande:2019gzr}. Because of the relatively low intensity of Cherenkov radiation, these detectors typically have worse energy resolution and higher energy thresholds than scintillation detectors~\cite{SNO:2024vjl,Super-Kamiokande:2024qbv,MiniBooNE:2020pnu}.

\subsection{Need for Hybrid Optical Detectors}

The ideal optical detector that can best search for rare events with excellent precision and background rejection capabilities leverages both the scintillation and Cherenkov signals, dubbed a ``hybrid" optical detector. For over a decade, hybrid optical detection has been an instrumentation goal to advance rare event searches~\cite{Aberle:2013jba}. Hybrid detectors were specifically highlighted during the 2021 Snowmass process as a major instrumentation goal because of their potential to advance new physics searches like neutrinoless double beta decay and measurements of solar neutrinos~\cite{Escobar:2022jau,Klein:2022lrf}. 

Neutrinoless double beta decay is a major field of research in neutrino physics because the observation of such a process definitively proves that neutrinos have a Majorana spinor component, potentially elucidating the origin of neutrino mass and the matter to anti-matter asymmetry in the early universe. Hybrid optical detectors are so attractive for searches for neutrinoless double beta decay because the two signals allow for excellent energy resolution and low backgrounds in addition to being more cost effective for larger target volumes--- potentially providing sensitivity into the normal hierarchy regime. The signal for neutrinoless double beta decay are two outgoing electrons at the energy of the Q-value for the isotope while the backgrounds are typically single electrons or photons from radioactive decay. The scintillation portion of hybrid detection allows for the excellent energy resolution necessary to resolve the Q-value while the Cherenkov light provides a very nice handle on two electron final states, producing two Cherenkov cones, from single electron backgrounds, with only a single Cherenkov cone. 

In the case of solar neutrinos, the direction reconstruction of hybrid detectors allows for very nice correlation with the known direction of the sun. Additionally, the low energy threshold allows for measurement of few MeV solar neutrinos, probing the transition region between vacuum and matter-dominated oscillations. As unmeasured, this is a very interesting region of parameter space that can inform both models of solar fusion processes and search for non-standard oscillations or interactions in the neutrino sector. 

Given all of these motivations, there is a lot of active research into hybrid optical detectors. In 2022, the Borexino collaboration published the first statistical observation of Cherenkov radiation from sub-MeV electrons~\cite{PhysRevLett.128.091803}. Then in 2024, the SNO+ experiment additionally observed Cherenkov radiation in a high light-yield scintillation detector on an event-by-event basis for $>5$~MeV electrons~\cite{PhysRevD.109.072002}. The results detailed in this thesis, and published in Ref.~\cite{CCM:2025kal}, is the first event-by-event observation of Cherenkov radiation produced from sub-MeV electrons in a scintillation detector providing an important advancement in the instrumentation community. This result is additionally the first demonstration of a hybrid detector utilizing liquid argon as the detection medium, which has many implications.

\section{Liquid Argon as a Hybrid Detector Medium}

While liquid argon is a common detection medium for neutrino detectors, it is typically used in the context of time projection chambers (LArTPCs). LArTPCs utilize electric fields to drift ionized charges across the detector volume for readout. This provides a very nice spatial resolution of $\mathcal{O}(1~\text{mm})$. Additionally, LArTPCs utilize photon detection systems to measure the fast scintillation signals, allowing for full reconstruction in three dimensions with the combination of two spatial readout planes and event start times from the scintillation light. Optical detectors like CCM200, however, utilize liquid argon for only the light emission. Without an electric field drifting ionized electrons, optical detectors utilizing liquid argon have much higher photon yields. In addition to the prolific scintillation light, liquid argon has multiple advantages for hybrid detection systems. 

\subsection{Photon Absorption}
The biggest advantage of liquid argon is that, in the absence of contaminants, it does not absorb optical photons above approximately 113~nm~\cite{Neumeier:2012cz}. This has two major implications for hybrid Cherenkov and scintillation detection. The first being that visible Cherenkov photons are not attenuated before detection. While oil-based liquid scintillators typically have very long absorption lengths for visible photons, $\mathcal{O}$(20~m), the absorption in ultra-large detectors becomes non-negligible~\cite{Beretta:2025rwj}. In the case that an ultra-large detector, like a DUNE far detector module with the longest detector axis of around 60~m, is necessary to attain fiducial mass for rare event searches like neutrinoless double beta decay, liquid argon has a distinct advantage over oil-based liquid scintillators since it does not intrinsically absorb optical photons.

The second implication of the lack of photon absorption is that liquid argon based hybrid detectors have the potential to capture the short wavelength portion of the Cherenkov emission spectrum. Because of the absorption in oil-based liquid scintillators, wavelength shifters (WLS) need to be doped within the bulk material to shift VUV scintillation light into the visible spectrum. This means that short-wavelength Cherenkov photons are WLS and re-emitted isotropically, losing their directionality component. In the case of pure liquid argon, however, UV Cherenkov radiation can propagate to the edges of the detector without absorption, then WLS on the photo-detector and with a high probability be re-emitted directly into the photo-detector. This greatly increases the photo-statistics of Cherenkov light in liquid argon detectors since the intensity goes as $1/\lambda^2$. 

\subsection{Timing Considerations}
Liquid argon additionally allows for easier separation between the prompt Cherenkov and the slower scintillation signals. This, again, is done in two ways. The first being that liquid argon itself is a slower scintillator compared to oil-based liquid scintillators. As discussed in Chapter~\ref{chap:om}, liquid argon has a fast scintillation time constant of $\mathcal{O}$(5~ns). For many liquid scintillators, the scintillation time constant is around $\mathcal{O}$(3~ns)~\cite{Elisei:1997tw,Borexino:2008gab}. This small difference in scintillation time constants, along with fast photo-detection systems with potential of $\mathcal{O}(100~\text{ps})$ resolution, can be exploited to better separate the very prompt Cherenkov radiation from the scintillation light. 

The second way that liquid argon allows for easier temporal separation between scintillation and Cherenkov light is again because of the WLS instrumentation. WLS on the edges of the detector, as is the case in liquid argon detectors, introduces time delays compared to WLS in the bulk of the detector, as is the case in oil-based liquid scintillator detectors. In liquid argon, photons must propagate to the edges of the detector, WLS, then propagate into a photo-sensor. This additional propagation, between WLS and detection, can introduce time delays between prompt visible Cherenkov radiation that does not WLS and VUV scintillation light. In oil-based liquid scintillators, however, because all of the WLS is happening in the bulk material, it does not introduce additional propagation time delays. 

\subsection{Cryogenic Nature}
One final advantage of liquid argon for hybrid detection is that the cryogenic nature allows for multiple types of photo-detection systems. While PMTs are a typical choice for photo-detection in optical detectors, there is growing research into silicon photomultipliers (SiPMs). SiPMs are an active area of research for improving detector instrumentation because of their low operating voltage, cost, and radioactivate backgrounds~\cite{Wang:2022zsv}.

For hybrid detectors specifically, SiPMs offer advantages in both timing and wavelength discrimination compared to traditional PMTs. Because of the condensed amplification structure, SiPMs offer very good timing resolution, $\mathcal{O}$(100~ps), which is crucial for Cherenkov separation. Furthermore, SiPMs have broader wavelength sensitivity than PMTs, offering additional avenues for separation of the two signals based on wavelength. 

SiPMs, however, have very high dark rate currents at room temperature, as would be the operating conditions for oil-based liquid scintillators. At cryogenic temperatures, as necessary for liquid argon detectors, the dark rate current drops by several orders of magnitude~\cite{Wang:2022zsv,Anfimov:2020ikk,Ozaki:2020liq}. This enables additional photo-detection systems in liquid argon based hybrid detectors that are not present in oil-based liquid scintillators. 

\subsection{Approach to Scintillation and Cherenkov Light Separation in CCM}
The CCM200 detector utilizes fast timing and wavelength-discriminating photo-detectors to separate scintillation from Cherenkov signals in liquid argon.

The R5912 cryogenic PMTs installed in CCM200 have an intrinsic spread in the electron transit time of approximately 1.7~ns~\cite{hamamatsu_pmt_handbook}, setting the absolute time resolution. Since the CAEN digitizer boards digitize the voltage signals at a rate of 500~MHz, providing digitization every 2~ns, we use the 2~ns digitization bins as our timing resolution. This timing resolution enables distinction between early Cherenkov radiation and scintillation light emitted with a characteristic time constant of around 5~ns. 

The combination of PMTs coated in wavelength shifting TPB and bare PMTs allows for discrimination based on wavelength. In the very early time region of an event, the first hits on the uncoated PMTs are more likely to be from visible Cherenkov light. Only after VUV light, from both Cherenkov radiation and scintillation, has wavelength shifted and propagated back into an uncoated PMT will the delayed signal arrive. Isolating the early time region on the uncoated PMTs preferentially selects for visible Cherenkov light. 

This work separating Cherenkov from scintillation signals on an event-by-event basis utilizes $\beta^+$ decay events from the $^{22}$Na calibration data discussed in Chapter~\ref{chap:om}. For the this decay pathway, the $^{22}$Na source produces a low-energy $e^+$ with an endpoint at approximately 0.55~MeV and a 1.275~MeV gamma-ray from de-excitation of an excited state of Ne$^{22}$. The calibration source is encapsulated in stainless steel while deployed in the detector, meaning the $e^+$ very promptly annihilates, producing two 0.511~MeV gamma-rays. So in essence, the $^{22}$Na calibration source produces three gamma-ray signals isotropically. In order to emit Cherenkov radiation, these gamma-rays must interact in the detector, typically through Compton scattering, to produce relativistic electrons. This work is focused on isolating Cherenkov radiation in the uncoated PMTs in the $^{22}$Na calibration data. 

\section{Response of Uncoated PMTs}
\begin{figure}[h]
  \centering
  \includegraphics[width=0.7\linewidth]{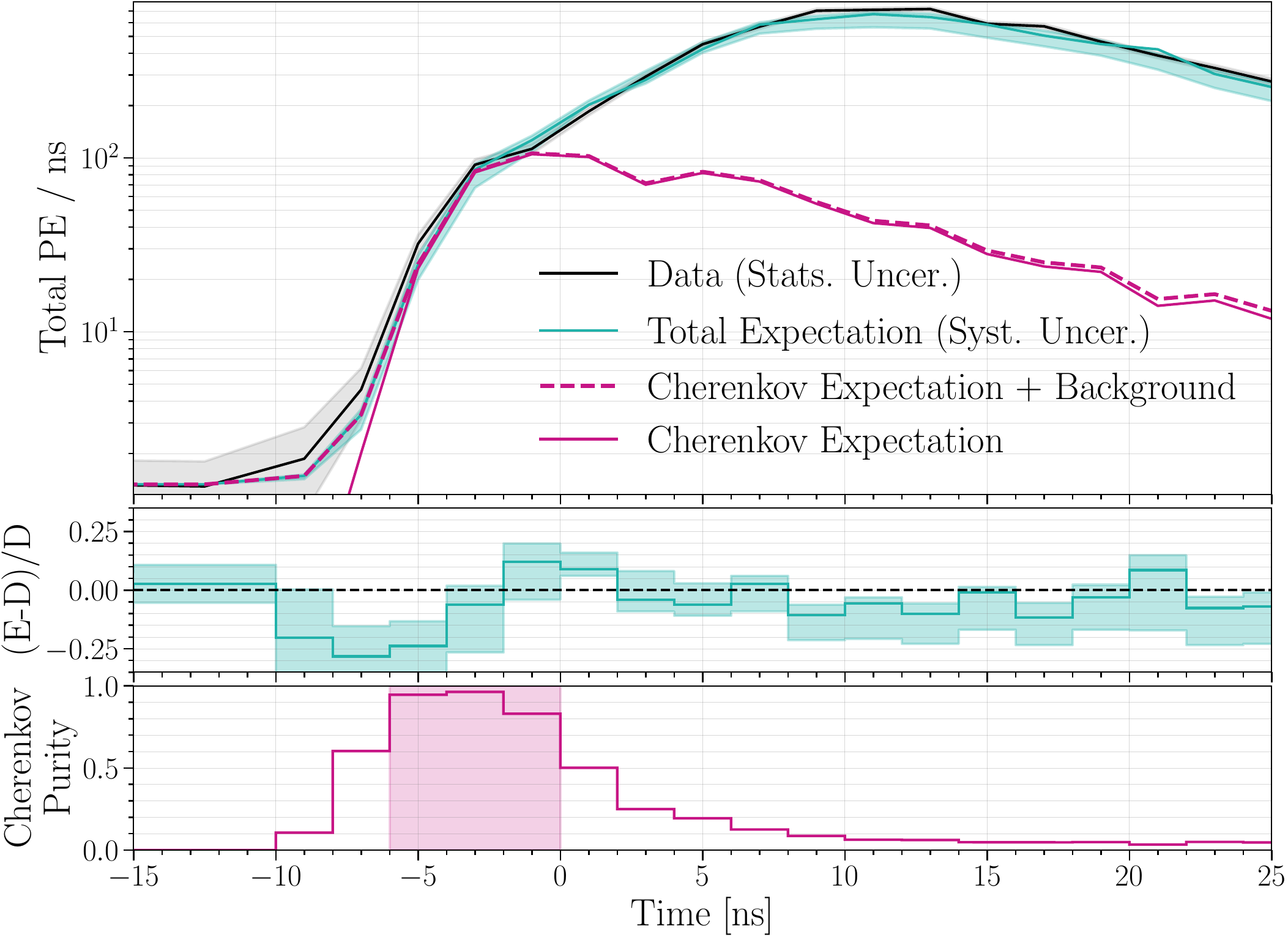}
  \caption{Observed data and Monte Carlo expectation for the $^{22}$Na calibration source on an uncoated PMT, utilizing the resulting optical model calibration from Chapter~\ref{chap:om}. For the purposes of Cherenkov radiation separation, the prompt time region around the reconstructed event start time ($t=0$) is the focus of this comparison. Cherenkov radiation (magenta line) produces a clear signal in the early time regions. Based on the Cherenkov purity (bottom panel), $-6 \leq t < 0$~ns is identified as the ``Cherenkov enhanced" time region utilized for this analysis.}
  \label{fig:typical_uncoated}
\end{figure}

Based on the optical model calibration, Chapter~\ref{chap:om}, Fig.~\ref{fig:typical_uncoated} demonstrates the response of a typical uncoated PMT to the $^{22}$Na calibration data.

Data was accumulated across approximately 50,000 sodium events, aligned on the event start times. The gray band demonstrates the measured detected photons with statistical uncertainty. The total expectation from the Monte Carlo simulation is demonstrated in the cyan line, with $1\sigma$ systematic uncertainty bands. This total expectation is obtained by combining the simulated detected photons from scintillation and Cherenkov radiation with the measured random backgrounds for each PMT.

The magenta line displays the Cherenkov-only portion of the expectation. As expected, on a typical uncoated PMT there is a peak in the very early portion of the event, at approximately $t=-3$~ns, due to the prompt Cherenkov radiation. The Cherenkov expectation does extend to long time scales, at a much lower rate than the scintillation contribution, due to WLS and propagation processes that UV Cherenkov photons can undergo. 

The middle plot displays the residual between the data and the expectation with the $1\sigma$ uncertainty band. This demonstrates $\pm15\%$ or better agreement within the $1\sigma$ bands between the data and the expectation. 

The bottom plot is a measure of the Cherenkov purity, obtained as the ratio of the Cherenkov-only portion of the expectation to the total expectation. It is clear that the early time region in the uncoated PMTs is highly sensitive to Cherenkov radiation. The highlighted band, defined as $-6 \leq t < 0$~ns, is selected as the ``Cherenkov enhanced" time region. This time region has $>90\%$ purity of Cherenkov hits. The rest of this analysis isolates the uncoated PMTs in the Cherenkov enhanced time region to separate visible Cherenkov radiation from scintillation signals. 

Fig.~\ref{fig:typical_uncoated} is the only direct fit between data and simulation, the result of the analysis described in Chapter~\ref{chap:om}, in this section. The rest of this chapter utilizes the expectation at the best fit without any additional changes to the simulation. 

\section{Cherenkov Separation}
\begin{figure}[h]
  \centering
  \includegraphics[width=0.7\linewidth]{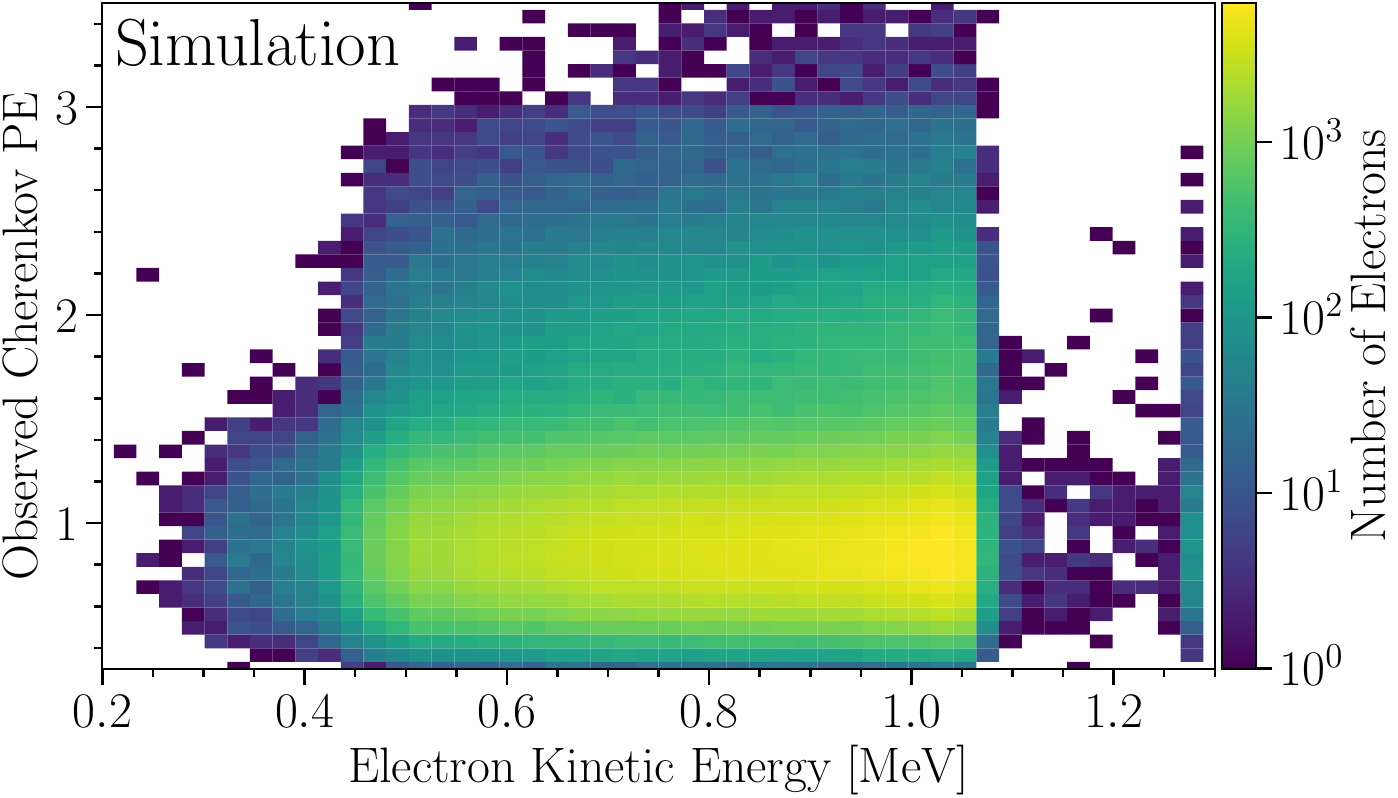}
  \caption{Two dimensional distribution between electron kinetic energy and observed Cherenkov PE obtained in the simulation of $^{22}$Na decays. This distribution demonstrates that the majority of observed Cherenkov PEs are produced from electrons with 0.5~MeV to 1~MeV kinetic energy.}
  \label{fig:electron_ke_vs_cherenkov}
\end{figure}

The $^{22}$Na calibration source produces gamma-rays, which then must interact to produce relativistic charged particles to scintillate and emit Cherenkov radiation. Fig.~\ref{fig:electron_ke_vs_cherenkov} examines the correlations between kinetic energy of electrons produced from the gamma-rays and total observed Cherenkov photons on the uncoated PMTs in the Cherenkov enhanced time region. This plot is made using only simulation.

The higher energy 1.275~MeV gamma-ray produced in the $^{22}$Na decay typically interacts through Compton scattering in the liquid argon. The Compton edge is evident in Fig.~\ref{fig:electron_ke_vs_cherenkov} between approximately 1~MeV and 1.275~MeV. The majority of electrons that produce observed Cherenkov photons have approximately 0.5~MeV to 1.0~MeV of kinetic energy. This plot also demonstrates the approximately 0.2~MeV threshold for electron kinetic energy to produce Cherenkov radiation in liquid argon. Note that since the majority of $^{22}$Na decay events do not produce detectable Cherenkov radiation on the uncoated PMTs in this early time region, the y-axis of Fig.~\ref{fig:electron_ke_vs_cherenkov} begins at 0.3~PE for legibility in the color scale. 

\begin{figure}[h]
  \centering
  \includegraphics[width=0.7\linewidth]{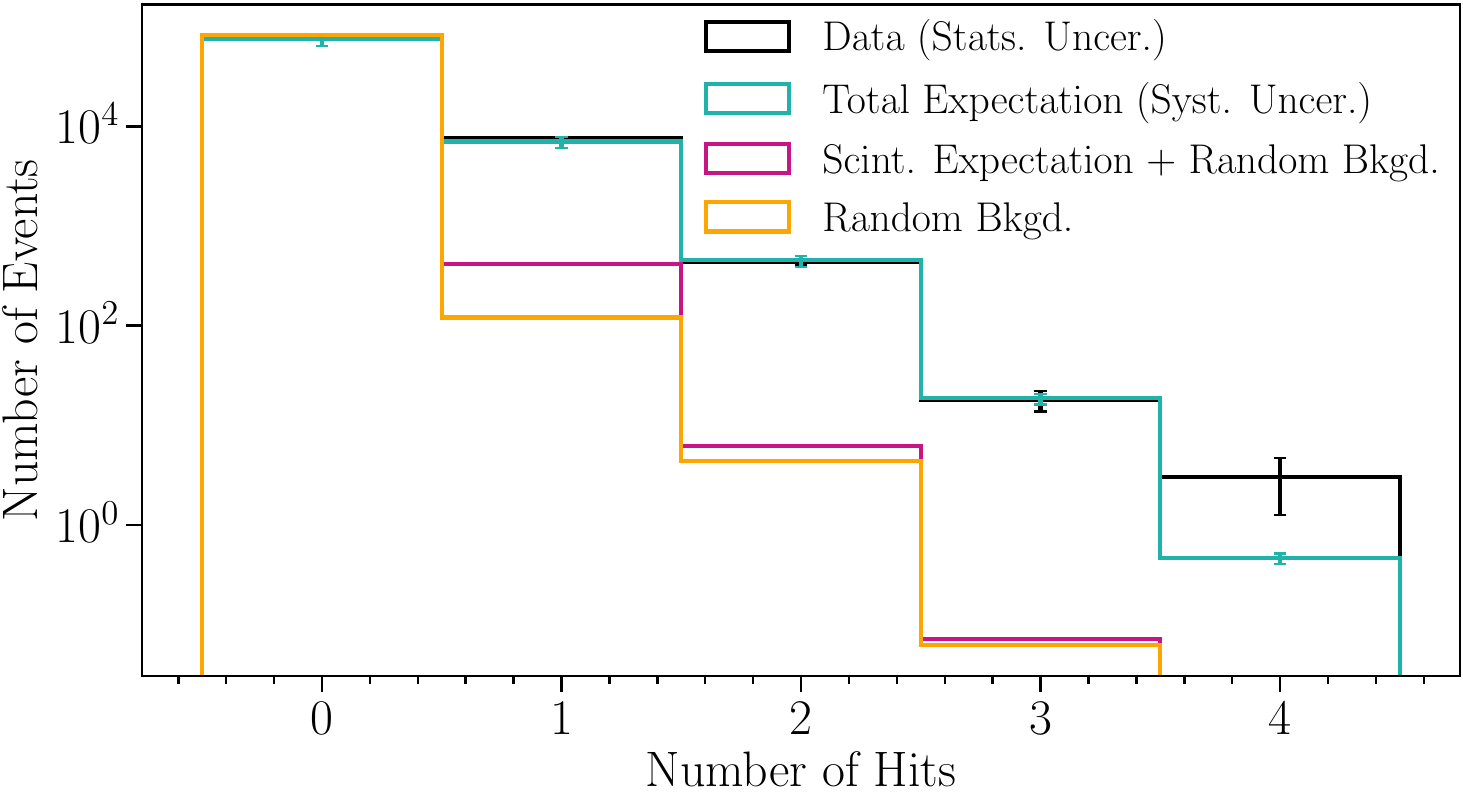}
  \caption{Number of hits on the uncoated PMTs in the Cherenkov enhanced time region. The observed $^{22}$Na data is displayed in the black distribution while the total Monte Carlo expectation is the cyan distribution. The magenta distribution represents backgrounds to Cherenkov radiation (namely scintillation photons and random backgrounds). The order of magnitude difference between the total expectation (cyan) and the background (magenta) is due to Cherenkov radiation.}
  \label{fig:nhits}
\end{figure}

On an event-by-event basis, Fig.~\ref{fig:nhits} demonstrates the number of hits between observed data and simulation on the uncoated PMTs in the early time region. The data distribution is the black line with statistical error bars. The total expectation, obtained from the simulation of scintillation and Cherenkov photons and measured random background per event, is the cyan distribution with $1\sigma$ systematic uncertainty bars. This total expectation agrees with the measured data very well. The magenta distribution shows the ``backgrounds" to Cherenkov identification--- scintillation light and random background hits while the orange distribution shows only the random backgrounds. 

The order of magnitude difference between the total expectation and the scintillation and random background components is due to Cherenkov radiation. Selection of 1 or more hits in the early time region on the uncoated PMTs has a 94.8\% purity of Cherenkov hits. This cut additionally selects 9.78\% of data events and 9.10\% of simulation events. This is very promising that using only the uncoated PMTs, which provide approximately 10\% photo-coverage of the detector, a simple requirement of hits in the early time region selects a highly pure sample of events with Cherenkov radiation. 

\begin{figure}[h]
  \centering
  \includegraphics[width=0.7\linewidth]{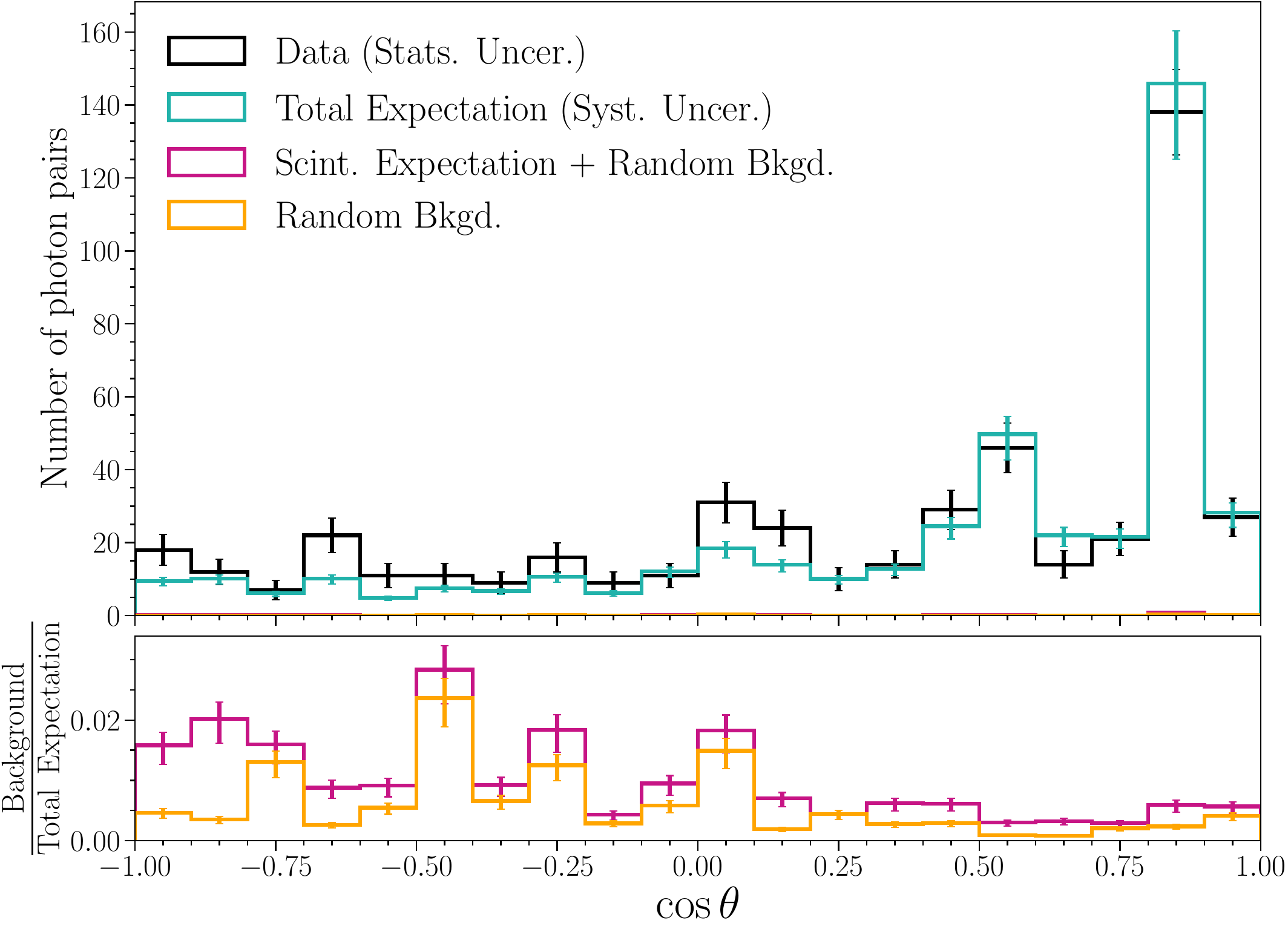}
  \caption{Distribution of opening angles for events in the $^{22}$Na data dataset with two or more hits in the early time region on the uncoated PMTs. Data and expectation have a clear preference for $0.8 \leq \cos\theta < 0.9$, which corresponds to the expected Cherenkov angle for $0.7 < KE < 1.0$~MeV kinetic energy electrons. The bottom panel shows the shape difference between the background distributions (magenta and orange) and the total expectation, indicating that the backgrounds are not as forward peaked.}
  \label{fig:costheta}
\end{figure}

One of the most important aspects of Cherenkov radiation is the directional emission. In order to study directional preference of emitted photons, events with two or more hits on the uncoated PMTs in the early time region were selected. The opening angle was then calculated using the known location of the sodium source, the origin of the detector, and the locations of the PMTs with hits, demonstrated in Fig.~\ref{fig:costheta}.

The data, black distribution, and total expectation, cyan distribution, have a clear preference for $0.8 \leq \cos\theta < 0.9$ and agree across all bins within $2\sigma$ uncertainty bars. For this analysis, selecting visible Cherenkov photons with an index of refraction $n = 1.22$ in liquid argon, Cherenkov opening angle of $0.8 \leq \cos\theta < 0.9$ corresponds to $0.7 < KE < 1.0$~MeV kinetic energy electrons. Since the expectation assuming only scintillation light or random backgrounds is an order of magnitude lower than the total expectation with Cherenkov radiation, the bottom plot in Fig.~\ref{fig:costheta} demonstrates the shape differences. This panel shows the shape differences between the backgrounds, which are emitted isotropically, and the expectation with directional Cherenkov radiation. There is an increase in preference for backwards angles in the background expectations compared to the forward angle expected for Cherenkov emission. 

The data and the total expectation have a $\chi^2$ of 30.12 for 20 degrees of freedom. The data and the scintillation and random background only expectation have a $\chi^2$ of 473.60 for 20 degrees of freedom. Using a $\Delta\chi^2$ test, we reject the scintillation and random background only hypothesis with $>5\sigma$ confidence, indicating event-by-event observation of Cherenkov light from sub-MeV electrons in LAr. 

\section{Data Driven Validation}
One final check on this analysis is to perform the same procedure on a data set without any expected Cherenkov radiation. In addition to the $^{22}$Na decay data, CCM collected calibration data from $^{57}$Co decays. The $^{57}$Co source was inserted in a stainless steel capsule at the origin of the detector during data collection. This source decays through electron capture interactions, producing low-energy gamma-rays $<136$~keV~\cite{Auble:1977pdy}. These gamma-rays then primarily interact through photo-electric effect and Compton scattering in the liquid argon, producing sub-Cherenkov threshold electrons. 

\begin{figure}[h]
  \centering
  \includegraphics[width=0.7\linewidth]{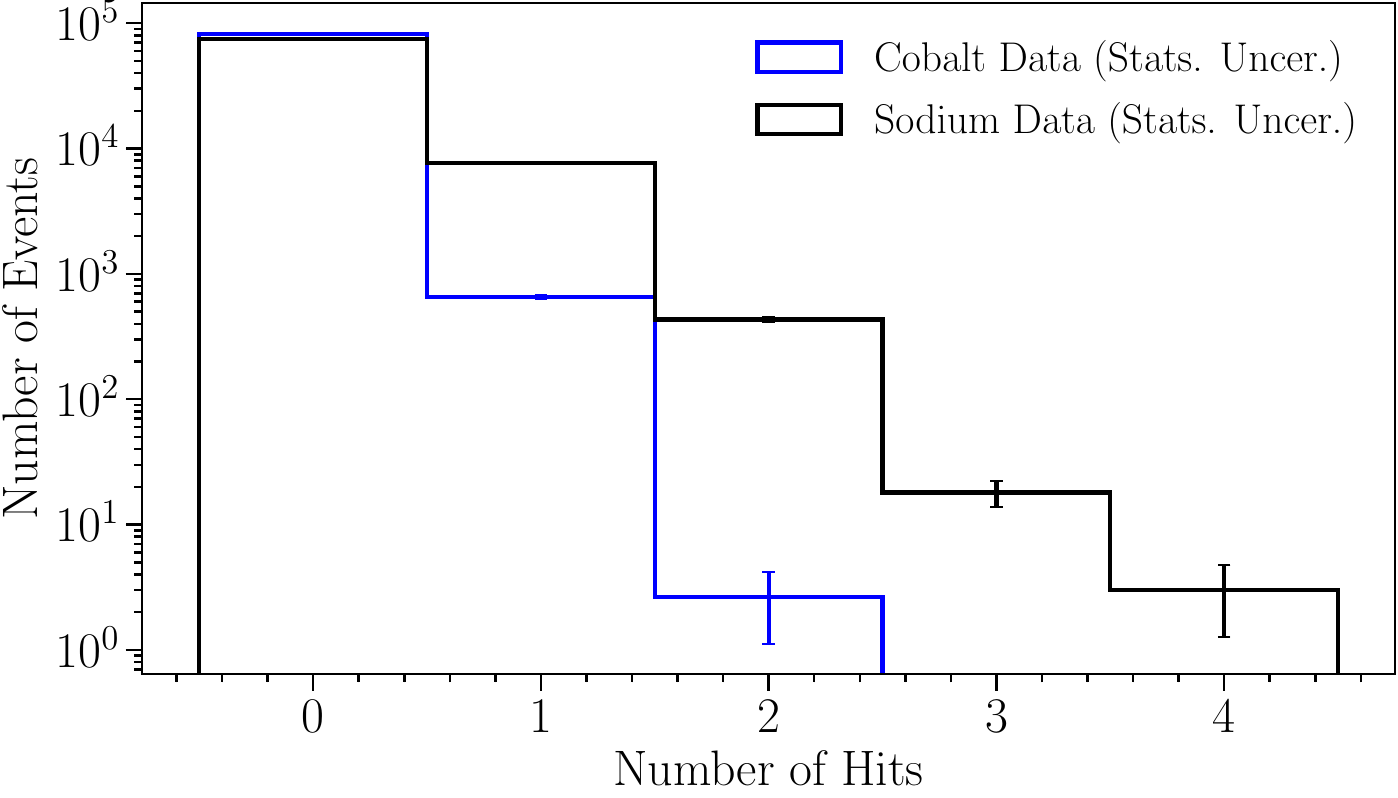}
  \caption{Number of hits on the $^{22}$Na and $^{57}$Co observed datasets in the Cherenkov enhanced time region on the uncoated PMTs. The $^{57}$Co data, which produces charged particles below Cherenkov threshold, has 0.79\% of events with $\geq1$ hit. This is in line with the expected random background activity based on the analysis of $^{22}$Na data.}
  \label{fig:cobalt_nhits}
\end{figure}

Fig.~\ref{fig:cobalt_nhits} demonstrates the number of hits in the early time region between the $^{22}$Na and $^{57}$Co data using the same procedure described previously. For the $^{57}$Co data, without any expected Cherenkov radiation, only 0.79\% of events have one more hit under this selection criteria. This is in line with the expected rate of background hits from the study of the $^{22}$Na source, which is 0.51\% of events that had one or more hits. This further supports the analysis of the $^{22}$Na data that this selection procedure isolates Cherenkov radiation. 

\section{Implications}
One of the major results of this thesis is the first event-by-event observation of Cherenkov radiation from sub-MeV electrons in the high light-yield scintillation detector. This was possible through fast timing resolution, combination of coated and uncoated PMTs, and detailed modeling and study of $^{22}$Na calibration data. This answers a major call to the instrumentation community to develop hybrid detectors for improved rare event searches. These results additionally highlight the benefits of liquid argon for this application, a counterpoint to the ongoing efforts to engineer oil or water based scintillators for this purpose~\cite{Anderson:2022lbb,Callaghan:2023oyu,Klein:2022tqr}. The rest of this thesis is concerned with utilizing these results to implement Cherenkov discrimination in an analysis searching for axion-like particles. 

\chapter{Event Reconstruction of Low-Energy Electrons}\label{chap:ml}
Chapter~\ref{chap:cherenkov} demonstrated one of the primary achievements of this thesis--- the identification of Cherenkov radiation within a high light-yield scintillation detector. While this capability provides a powerful handle for isolating electromagnetic events, it is not sufficient on its own to enable a search for axion-like particles. This physics search additionally requires reconstruction of both the event vertex position and deposited energy in order to suppress backgrounds and understand potential signals. 

To meet this need, a complete event reconstruction framework was developed as part of this work. Because the analysis of CCM200 data collected in 2022 relies on an entirely new software processing chain, all reconstruction algorithms were developed and validated using CCM200 data for this thesis. The position reconstruction leverages the \texttt{GraphNeT} machine learning library, which was adapted and extended to train on detector specific geometry, and attains around 5~cm resolution in each spatial dimension. The energy reconstruction algorithm leverages the reconstructed positions and the charge deposited to achieve around 12\% energy resolution at 1~MeV of true energy. 

In the following sections, the architecture, training procedure, and performance of the position reconstruction model are described, followed by the development of the energy reconstruction algorithm and its integration into the full analysis pipeline.

\section{Position Reconstruction}
Position reconstruction, using the time and geometric pattern of reconstructed photoelectrons (PE), is a powerful tool for discriminating background events from potential signals. Radioactive backgrounds during beam operations, and steady-state cosmic-ray backgrounds, enter the detector from the surrounding environment. As a result, these events are more likely to reconstruct near the edges of the active region. Signal events, however, are expected to be uniformly distributed throughout this region. Position reconstruction, therefore, is a strong handle for improving the overall signal-to-background ratio.

Previous CCM analyses employed a simple charge-weighted average of PMT positions to roughly estimate the interaction vertex. While efficient to implement, this method neglects information contained in the photon timing and light propagation effects, resulting in $\mathcal{O}$(20~cm) position resolution in each spatial dimension. In contrast, modern optical detectors like DEAP, IceCube, and SNO+ employ likelihood-based or machine learning reconstruction techniques that model photon propagation and detector response in detail, achieving significantly improved spatial resolution~\cite{DEAP:2025igy,IceCube:2022kff,SNO:2021wcl,}.

While traditional likelihood based methods can provide excellent position resolution, machine learning based methods can still improve upon this resolution and are often less computationally intensive to deploy~\cite{DEAP:2025igy,Abbasi:2022ypr}. Graph neural networks (GNNs), in particular, provide a natural framework to represent the PMT array as nodes and the relations between them as edges. Transformer-based architectures, furthermore, is a natural choice to encode time-series data like reconstructed pulses, allowing the model to fully utilize basic units of data for position reconstruction. 

For this thesis, the author developed a position reconstruction algorithm based on transformer graph neural networks. This approach leverages both the timing and spatial information of reconstructed pulses to significantly improve vertex resolution and background discrimination relative to previous CCM methods.

\subsection{\texttt{GraphNeT} Based Model}
The position reconstruction model is built using the open-source \texttt{GraphNeT} toolkit~\cite{Sogaard:2022qgg}, with a \texttt{DeepIce}~\cite{deepice} backbone tailored for irregular detector data. The input to the model consists of reconstructed PMT photoelectron pulse series, where each pulse is characterized by features such as charge, time, PMT spatial coordinates, and PMT coating status. These pulses form an unordered set of variable length, reflecting the sparse and asynchronous nature of detector readout.

Each pulse is first embedded into a high-dimensional latent space using a Fourier feature encoder. This encoding maps continuous spatial and temporal coordinates into a periodic representation, allowing the model to more effectively resolve fine-grained structure and long-range correlations in both space and time.

The \texttt{DeepIce} architecture processes this set of embedded pulses using a stack of Transformer blocks operating on an implicit, fully-connected attention graph~\cite{vaswani2017attention}. Rather than explicitly defining edges between PMTs, the self-attention mechanism dynamically learns pairwise relationships between all pulses. Crucially, the attention layers incorporate spacetime-relative positional encodings, meaning that attention weights depend not only on learned features but also on relative distances in space and time between pulses. This inductive bias is well-matched to the underlying physics, where causal and geometric relationships govern the propagation of light in the detector.

Within each Transformer block, multi-head attention layers aggregate information globally across the event, while feed-forward networks refine the learned representations. This enables the model to capture both local structures (e.g., clustered hits) and global event topology (e.g., light propagation patterns) without relying on handcrafted features or fixed graph connectivity.

To obtain an event-level prediction, a learned classification token (\texttt{[CLS]}) is prepended to the input sequence. Through successive attention operations, this token aggregates information from all pulses and serves as a global summary of the event. The final hidden state of the \texttt{[CLS]} token is passed to a regression head to predict the reconstructed interaction position.

The model is trained using the LogCosh loss function, Eq.~\ref{eq:logcosh}, which provides robustness to outliers by smoothly interpolating between quadratic behavior for small residuals and linear behavior for large residuals~\cite{logcoshloss}. Here, the residual is defined as $\Delta x_i = \hat{y}_i - y_i$, where $y_i$ is the true position and $\hat{y}_i$ is the model prediction. The loss is implemented in a numerically stable form to avoid overflow for large residuals. Optimization is performed using the \texttt{Adam} optimizer~\cite{adamoptimizer}.

\begin{equation}
    \text{LogCosh}(y, \hat{y}) = \sum_{i=1}^{n} \left[ \Delta x_i + \ln(1 + e^{-2\Delta x_i}) - \ln(2) \right]
\label{eq:logcosh}
\end{equation}

\subsection{Training Data Preparation}
A high statistics set of electron simulation is used for training. Approximately 6 million electrons were simulated uniformly across the detector, extending in the veto region, with isotropic directions. The energy of these events is sampled from a uniform distribution between 0.1 and 30~MeV. The truth-level position used for the training objective is the injection location of each electron event.

The effects of the DAQ and low-level timing calibrations are emulated in the simulation. Since the digitization of the 200 PMTs is spread across approximately 15 boards, board-to-board timing calibration is achieved through the alignment on the copy of the NIM trigger pulse on each board. This procedure is mimicked in the simulation through realistic sampling of board time offsets based on the measured distribution across each board and applied to the appropriate PMTs on each board. Additionally, the electron transit time calibration procedure is applied to the simulation by adding the measured electron transit times for each PMT. Then the effects of digitization are emulated by binning the simulated PEs in 2~ns bins. Finally, the added NIM pulse time and electron transit times are subtracted from the pulse series to replicate the grid of reconstructed PE times observed for each PMT. 

After that, a data-driven noise model is overlaid on the simulation to further model the observed data. For this procedure, pulse series from the 2022 physics data collection period are randomly sampled. Pulses recorded in the roughly 10~$\mu$s before the beam trigger are utilized to accurately represent the steady-state noise present in the physics data. These pulses are incorporated with the simulated pulse series to faithfully reproduce the noise on each PMT. 

The simulated events then undergo the same event finding procedure as collected data. This defines individual events as when the total charge in the detector crosses a threshold of 10~PE in a 10~ns window. For training purposes, only events with reconstructed event start times within 20~ns of the true event start time are used to remove entirely noise-based events from the training sample. 

Finally, the pulse series are then extracted into column-oriented data file format using \texttt{Apache Parquet} for training. To reduce the amount of data used in training, adjacent reconstructed pulses are re-binned from the digitizer-based 2~ns bins to 6~ns bins. Additionally, the systematic uncertainties derived in the measurement of the liquid argon scintillation pulse shape, discussed in Chapter~\ref{chap:om}, are taken into account. For each event, a scaling factor on the normalization of the charge is sampled using a normal distribution centered at 1 with a standard deviation of 0.10 to represent 10\% uncertainty on the total charge. This is then applied to the reconstructed charge, resulting in a distribution of simulated events to more accurately model the data. Furthermore, to avoid the model learning low-level correlations like binning artifacts, a small offset to the the time of the reconstructed pulse is sampled from a normal distribution centered at 0 with a standard deviation of 0.5~ns and added to the pulse time. 

\subsection{Position Reconstruction Performance}
\begin{figure}[h]
  \centering
  \includegraphics[width=\linewidth]{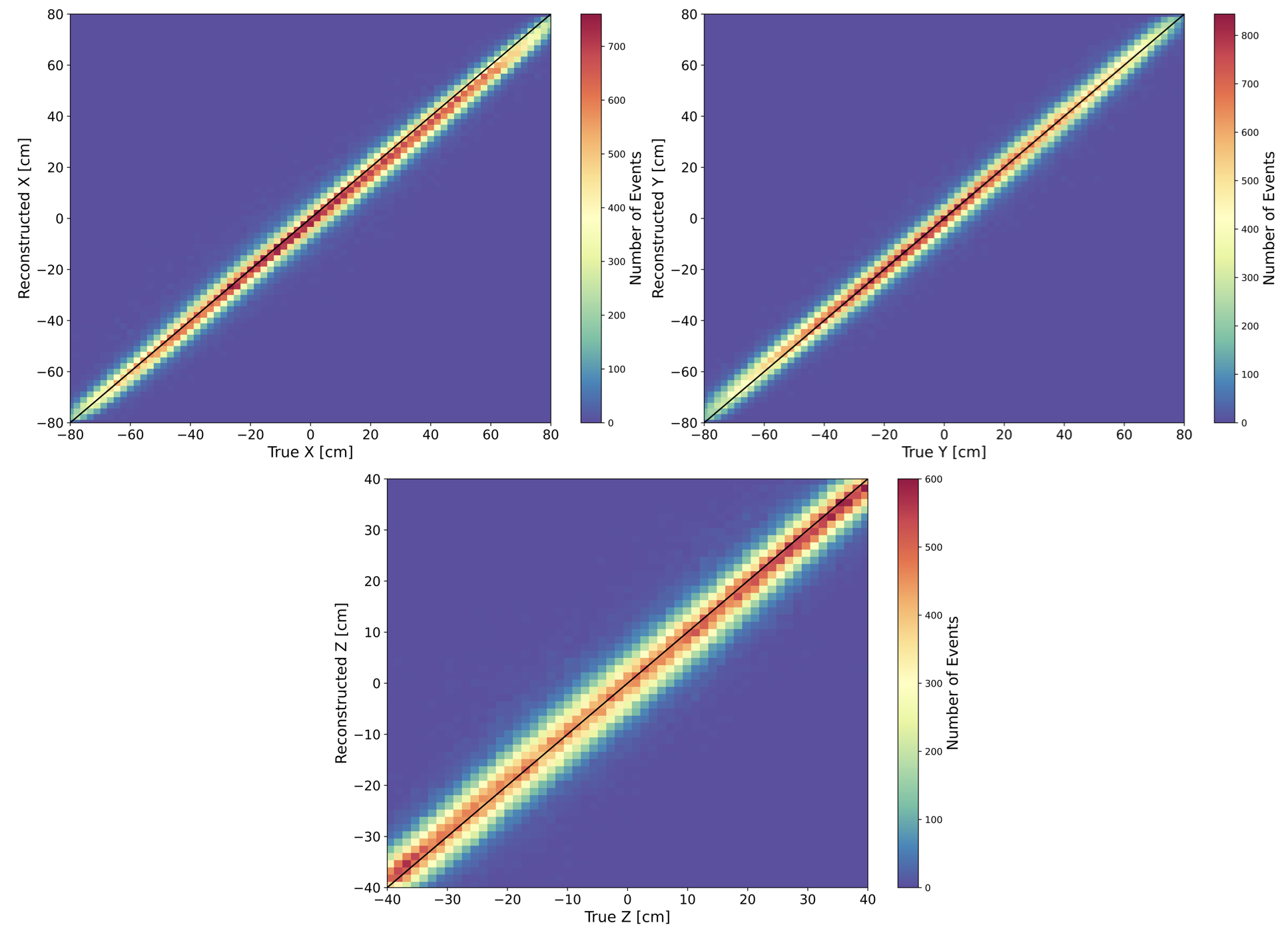}
  \caption{Two dimensional distributions of the true and reconstructed positions obtained from the \texttt{GraphNeT} based position reconstruction model. The top left plot shows the $x$-dimension, top right plot shows the $y$-dimension and the lower plot shows the $z$-dimension. There is additionally a line at $y=x$ in each plot to guide the viewer in evaluating the reconstruction performance.}
  \label{fig:true_vs_reco_xyz}
\end{figure}

After training on A100 GPUs to convergence, the reconstruction performance is evaluated using a held-out validation dataset comprising 10\% of the full simulated sample. This dataset is excluded from training and is used exclusively to assess the model's performance. 

Fig.~\ref{fig:true_vs_reco_xyz} shows the correlation between true and reconstructed vertex positions in the $x$, $y$, and $z$ dimensions for events passing fiducial selection criteria. These two-dimensional distributions provide a direct visual diagnostic of both bias and resolution: an ideal reconstruction would yield a narrow diagonal band with unit slope and minimal scatter. Deviations from this behavior indicate systematic offsets or degradation in resolution.

The distributions are shown after applying fiducial cuts on the reconstructed event position, requiring a reconstructed radial coordinate $r_\mathrm{reconstructed} \leq 80~\mathrm{cm}$ and $|z_\mathrm{reconstructed}| \leq 40~\mathrm{cm}$. These cuts remove events reconstructed near the detector boundaries, corresponding to the outer 20~cm of the active volume. This region is expected to have increased contamination from external radioactivity and detector-edge effects.

\begin{figure}[h]
  \centering
  \includegraphics[width=\linewidth]{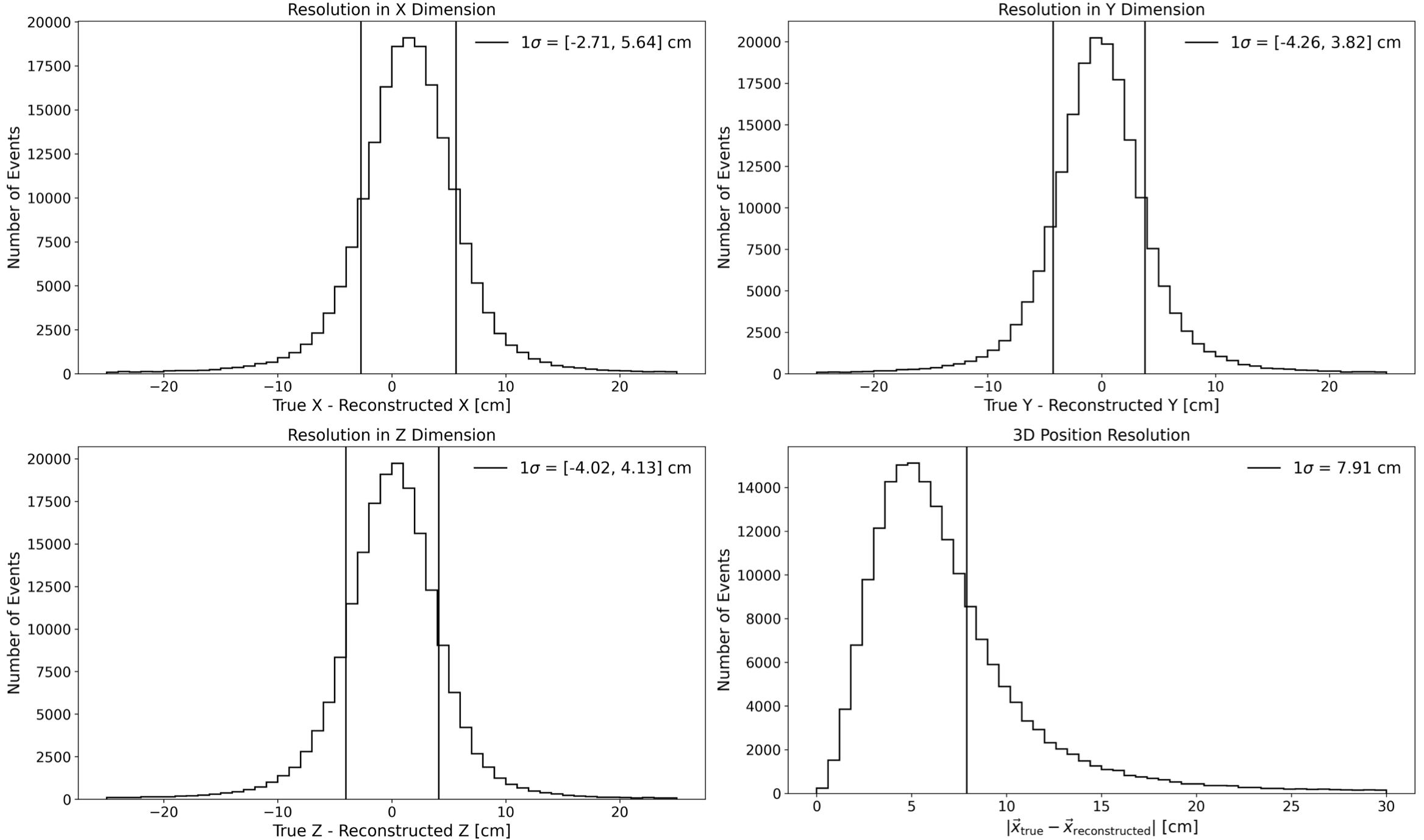}
  \caption{Residuals for position reconstruction performance. The $x$, $y$, $z$ dimensions show around 5~cm resolution at the $1\sigma$ level. The total three-dimensional position resolution is 7.91~cm at the $1\sigma$ level.}
  \label{fig:resolution_xyzalldir}
\end{figure}

Fig.~\ref{fig:resolution_xyzalldir} quantifies the corresponding reconstruction residuals, defined as the difference between true and reconstructed positions in each coordinate, $\Delta x = x_\mathrm{true} - x_\mathrm{reconstructed}$ (and analogously for $y$ and $z$). These residual distributions directly characterize the spatial resolution and any residual bias in the reconstruction. The $1\sigma$ resolution is computed using the highest posterior density interval of each distribution, providing a robust measure that is less sensitive to non-Gaussian tails.

Across all three spatial dimensions, the reconstruction achieves a consistent performance, with an approximately isotropic resolution at the level of 7.9~cm at $1\sigma$, as summarized in the lower right-hand plot in Fig.~\ref{fig:resolution_xyzalldir}. This corresponds to a significant improvement in spatial precision relative to previous CCM reconstruction approaches, which typically achieved resolutions on the order of several tens of centimeters. Quantitatively, this represents approximately a four-fold improvement, enabling improved fiducialization and enhanced background rejection near detector boundaries.

\subsection{Validation on Calibration Data}
In order to validate the position reconstruction performance on real detector data, the trained model is applied to sodium calibration data. This calibration dataset provides a controlled and well-understood data sample, allowing for a direct comparison between reconstructed event distributions in data and Monte Carlo.

The sodium calibration data was collected at the beginning of the 2022 run cycle using a deployed sodium radioactive source positioned at the detector origin, described in detail in Chapter~\ref{chap:om}. Data was acquired using an external 20~Hz trigger with a 16~$\mu$s data acquisition (DAQ) window.

Candidate events are identified by requiring at least 10~PE observed in the full detector within a 10~ns time window. Due to the relatively long DAQ window, multiple events may be reconstructed within a single trigger window. To reduce contamination in the events from delayed scintillation photon emission from the long-lived triplet excited dimer state, an additional timing cut is applied. Specifically, events occurring within 1.5~$\mu$s of a previous reconstructed event are removed.

Since the external trigger does not exclusively select sodium interactions, the dataset collected during source deployment consists of a mixture of true sodium decays and irreducible background events. These backgrounds include environmental radioactivity, $^{39}$Ar decays throughout the fiducial region, and electronics noise processes, which are expected to be approximately spatially uniform across the detector volume.

\begin{figure}[h]
  \centering
  \includegraphics[width=\linewidth]{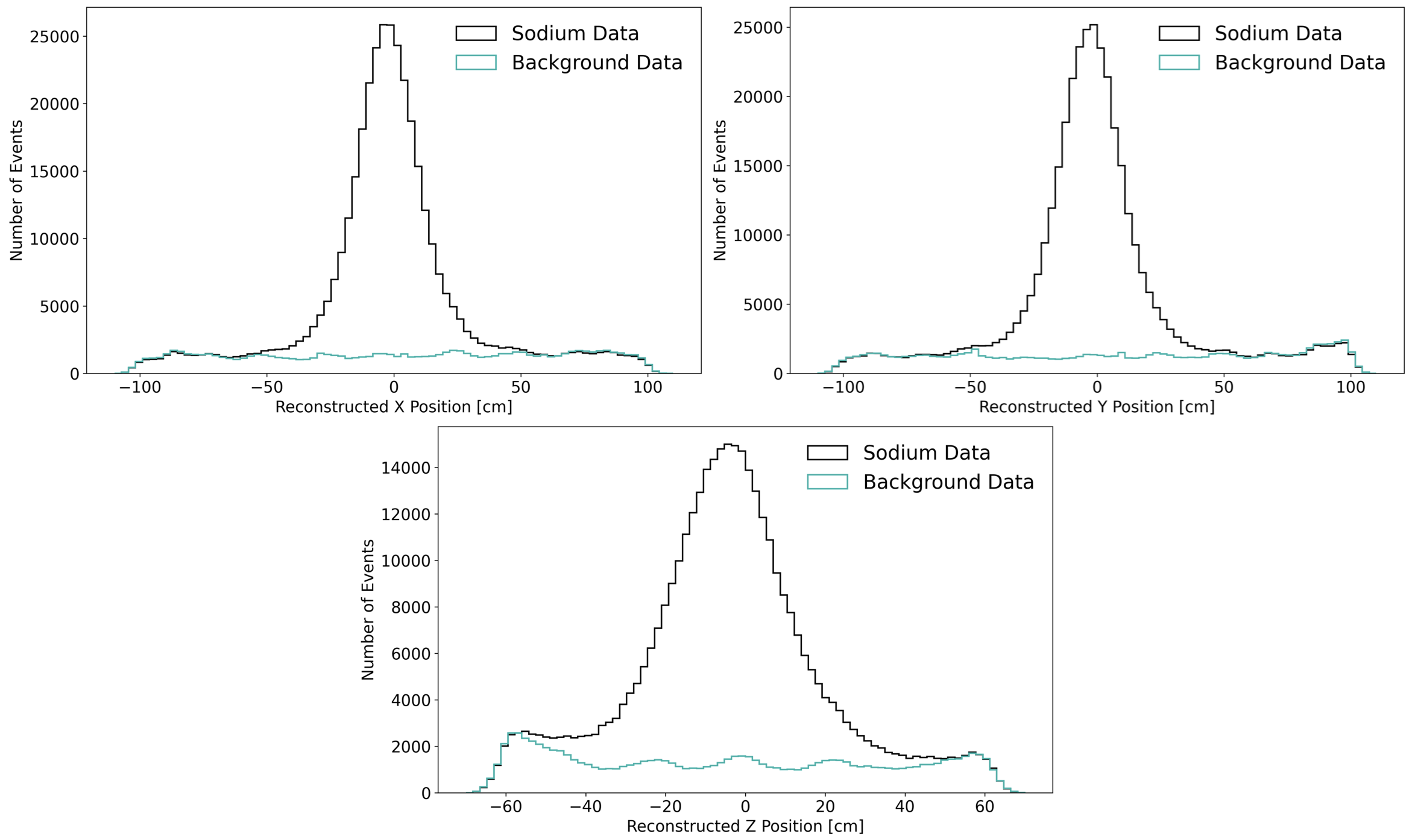}
  \caption{Reconstructed position distributions on $^{22}$Na data and a background data sample. The sodium data (black) position reconstruction shows clear preference for the origin of the detector, as expected. The background data (cyan) is roughly uniformly distributed across the active region of the detector.}
  \label{fig:sodium_data}
\end{figure}

Fig.~\ref{fig:sodium_data} shows the reconstructed position distributions for both data collected while the sodium source was inserted in the detector and background-only data (data collected with the sodium source removed) in the $x$, $y$, and $z$ dimensions. The two samples are normalized to the same exposure time to allow a direct comparison of event rates and spatial structure. As expected, the sodium dataset exhibits a pronounced peak near the detector origin, corresponding to the known deployed source position. In contrast, the background-only dataset shows a much flatter spatial distribution, consistent with an approximately uniformly distributed background component.

\begin{figure}[h]
  \centering
  \includegraphics[width=\linewidth]{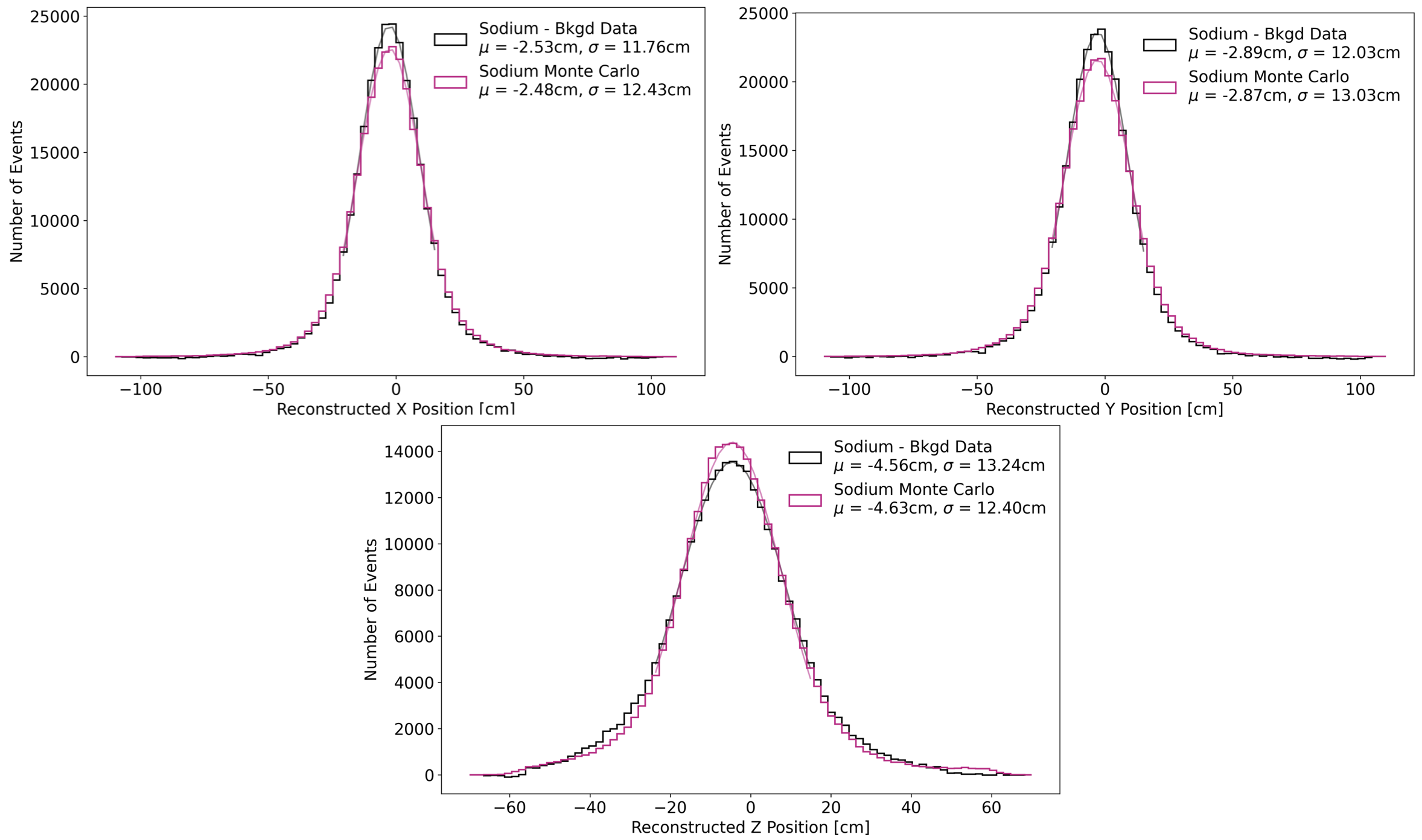}
  \caption{Background subtracted sodium data compared to Monte Carlo simulation in reconstructed position. Both distributions are fit with a Gaussian function to characterize the location and spread of the reconstructed positions. All dimensions agree at the $\mathcal{O}(1~\mathrm{cm})$ level.}
  \label{fig:sodium_datavsmc}
\end{figure}

To isolate the contribution from sodium decays, a background subtraction is performed using the background-only dataset scaled to the same exposure time as the source data. This subtraction yields an estimate of the spatial distribution of sodium-induced interactions, which can then be directly compared to Monte Carlo predictions of the sodium decays in the detector. Fig.~\ref{fig:sodium_datavsmc} shows the resulting comparison between data and simulation after background subtraction.

To quantify agreement, each subtracted distribution is fitted with a Gaussian distribution to extract the mean reconstructed position and the spatial width. The mean values are consistent between data and simulation at the $\mathcal{O}(1~\mathrm{cm})$ level, indicating negligible reconstruction bias for source-like events near the detector center. The fitted widths, $\mathcal{O}(10~\mathrm{cm})$, while agreeing quite well between data and simulation are larger than the expected position resolution from the electron validation dataset. This is due to the fact that the sodium source emits gamma-rays and not electrons directly. The gamma-rays travel finite mean free paths before depositing energy via secondary electron production. This introduces an additional spatial smearing relative to a point-like energy deposition of electrons, broadening the observed reconstructed distribution.

Overall, the reconstructed spatial distributions and their agreement with Monte Carlo demonstrate that the position reconstruction performs consistently on real calibration data, correctly reproducing both the location of the source and the spatial distribution of the reconstructed events. 

\section{Energy Reconstruction}
In addition to position, the reconstruction of the energy is a central task for event characterization. In liquid argon detectors, energy reconstruction is closely tied to the number of photoelectrons (PEs) observed. Energy deposited by a particle produces prolific scintillation photons and the total number of observed photons is a proxy for the deposited energy. In this sense, the detector behaves similarly to a calorimeter, where the observed light yield is approximately proportional to the true energy deposition.

However, this proportionality is not uniform throughout the detector volume. The observed charge depends not only on the deposited energy, but also on the event position due to several effects--- geometric light collection efficiency, absorption and scattering of scintillation photons, and PMT position dependent efficiencies. As a result, identical energy depositions at different locations can produce significantly different observed numbers of PEs.

To account for these effects, the energy reconstruction model must incorporate position-dependent corrections. In this work, we develop an energy reconstruction that leverages the reconstructed positions and the charge deposited in a fixed time region to learn this mapping directly. This provides an improved estimate of the energy reconstruction compared to previous CCM analyses that used a single charge to energy calibration factor across the entire detector. 

\subsection{Position Dependent Charge Scaling}
For the energy reconstruction, we adopt a simplified approach in which a position-dependent scaling factor is used to convert the measured charge into deposited energy. Specifically, we consider the integrated charge collected within the first 90~ns of an event and map this quantity to energy using the reconstructed event position.

The scaling relation is derived from an isotropic simulation of electrons distributed throughout the detector volume with energies sampled uniformly from 0.1 to 30~MeV. For each simulated event, the interaction vertex is first reconstructed using the methods described above, after which the detector volume is voxelized into 8~cm bins in $x$, $y$, and $z$ coordinates. Within each voxel, we compute the median ratio $Q/E$, where $Q$ is the observed charge and $E$ is the true energy of the event. This procedure yields a discrete map of the detector response, which is subsequently smoothed using a spline interpolation to provide a continuous mapping from reconstructed position to a charge-to-energy conversion factor.

\begin{figure}[h]
\centering
    \includegraphics[width=\linewidth]{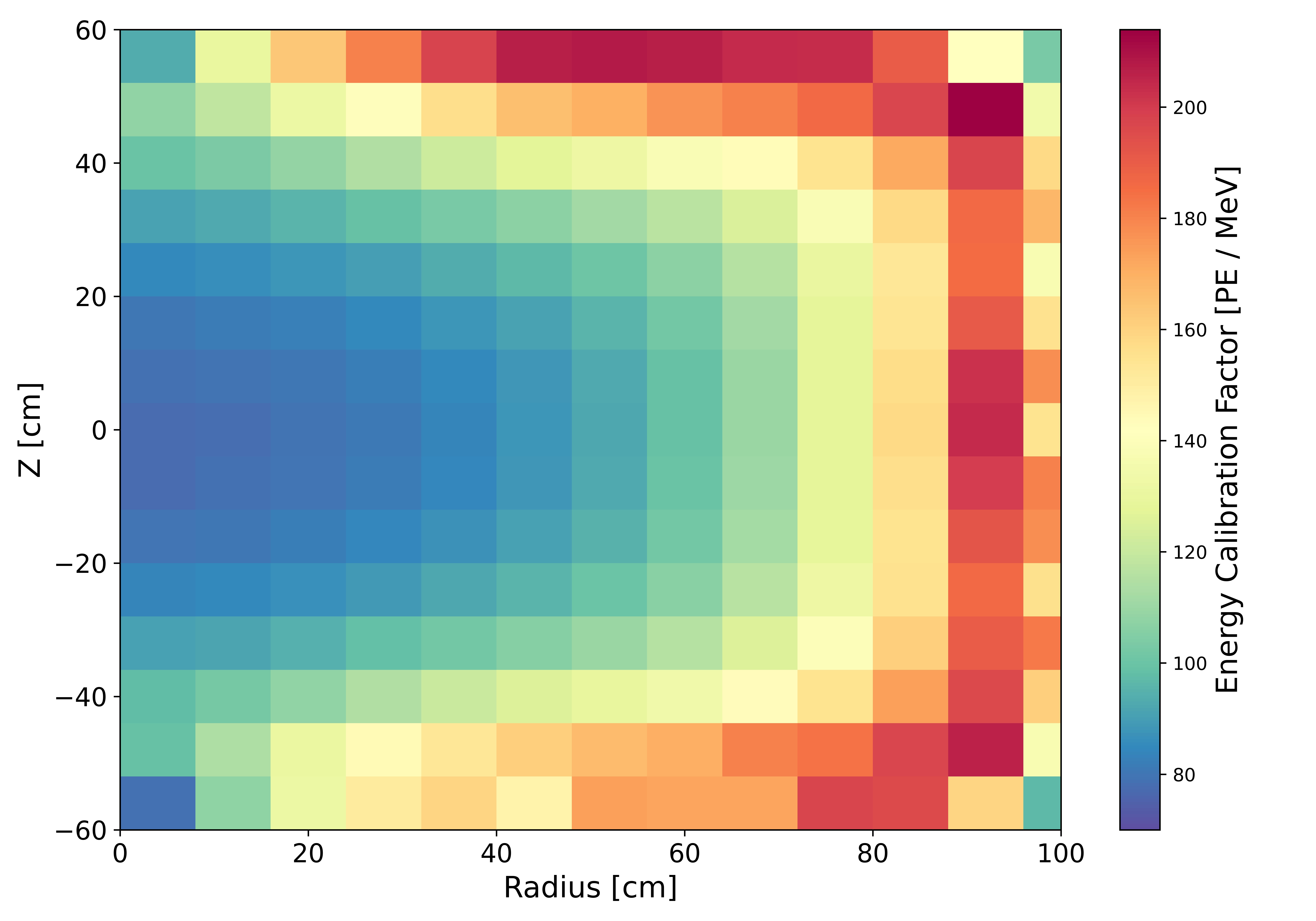}
    \caption{Two dimensional distribution of the energy calibration factor as a function of the reconstructed radial and $z$ positions. Near the origin of the detector, the energy calibration value is around 80~PE/MeV while this increases to around 200~PE/MeV near the edges of the detector, motivating the need for a position-dependent energy reconstruction.}
\label{fig:spline_scaling_vs_position}
\end{figure}

The resulting scaling factors as a function of radial and $z$ position are shown in Fig.~\ref{fig:spline_scaling_vs_position}. The response exhibits a smooth spatial dependence, with a larger $Q/E$ scaling factor towards the detector boundaries relative to the central region as photons emitted closer to the PMTs are more likely to be observed. Across the entire active region of the detector, this is more than a factor of two change in the charge-to-energy calibration factor, emphasizing the importance of a position-informed energy reconstruction. 

This approach does not explicitly account for energy losses due to particles escaping the detector volume, and therefore is most reliable for contained, low-energy ($< 10$~MeV) events. In this regime, the method provides a robust and stable energy estimate. In contrast, machine learning-based regression models were observed to exhibit pathological behavior at low charge, frequently overestimating the energy for events with few observed PE.

Finally, the use of a restricted 90~ns integration window is motivated by the timing constraints for physics searches in the prompt time region after the proton beam hits the tungsten target. This physics signal region is confined to a $\sim$175~ns window preceding a large rate of neutron-induced background events. Limiting the charge integration window mitigates contamination from pile-up events, ensuring that the reconstructed energy reflects the primary interaction.

\subsection{Energy Reconstruction Performance}
\begin{figure}[h]
\centering
\includegraphics[width=0.7\linewidth]{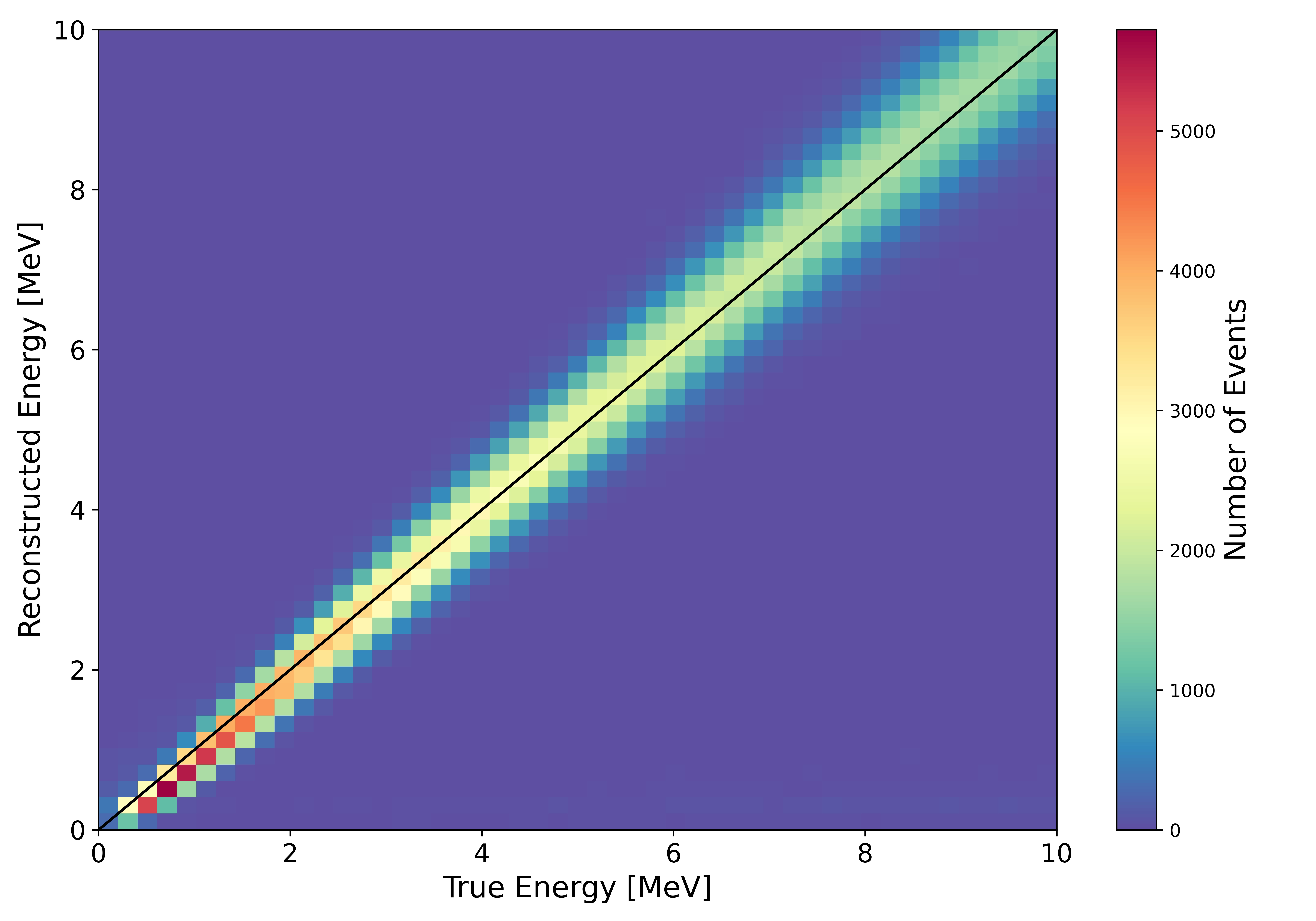}
\caption{True vs reconstructed energies in simulated events. First, the positions are determined using the \texttt{GraphNeT} position reconstruction model. Then the energy is reconstructed using the charge observed in the first 90~ns of each event and the position-dependent energy scaling factors, displayed in Fig.~\ref{fig:spline_scaling_vs_position}.}
\label{fig:true_vs_reco_energy}
\end{figure}

The performance of the energy reconstruction is evaluated by comparing the reconstructed energy to the true energy of the event in simulation, focusing on true kinetic energies $<10~\text{MeV}$. Fig.~\ref{fig:true_vs_reco_energy} shows the distribution of true versus reconstructed energies after applying the same fiducial selection used in the position reconstruction analysis ($r_{\mathrm{reconstructed}} < 80$~cm, $|z_{\mathrm{reconstructed}}| < 40$~cm). A clear linear correlation is observed across the energy range, indicating that the position-dependent scaling procedure successfully recovers an approximately unbiased estimate of the true energy.

To quantify the reconstruction performance, we compute the energy resolution as a function of true energy. The resolution is defined as the width of the $68\%$ highest posterior density interval of $(E_{\mathrm{reconstructed}} - E_{\mathrm{true}})$, normalized to the mean true energy in each bin, and expressed as a percentage. This definition provides a robust measure of the spread that is less sensitive to non-Gaussian tails than a simple standard deviation.

\begin{figure}[h]
\centering
\includegraphics[width=0.7\linewidth]{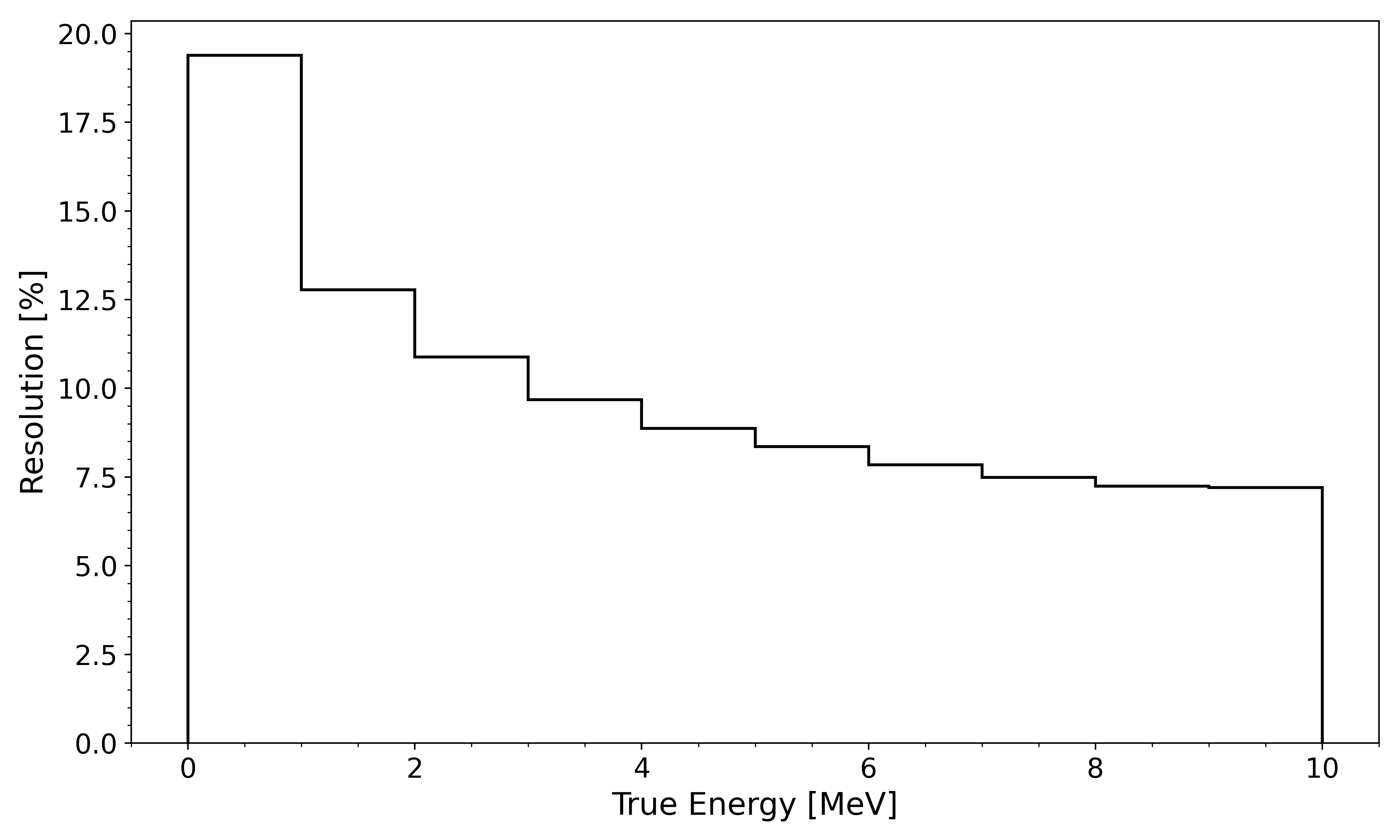}
\caption{Resolution of the energy reconstruction method as a function of true energy. As this work focuses on low-energy calibration, only particles with true kinetic energies $\leq10~\text{MeV}$ are evaluated in determining the energy resolution.}
\label{fig:resolution_energy}
\end{figure}

Fig.~\ref{fig:resolution_energy} shows the resulting energy resolution as a function of true energy up to 10~MeV. As this thesis is focused on lower energy event characterization utilizing the $^{22}$Na source, only reconstructed energies $\leq10~\text{MeV}$ are considered in the axion-like particle physics search. In this region, the energy reconstruction results in approximately 12.5\% resolution at 1~MeV which then improves to roughly 7.5\% resolution at 10~MeV, approximately following expected stochastic behavior. 

Overall, this reconstruction approach provides a stable and well-understood energy estimator in the low-energy regime most relevant for this thesis. The ongoing Michel electron calibration work, utilizing the results of Chapter~\ref{chap:low_level}, will address the energy reconstruction from approximately $10~\text{MeV} < E < 50~\text{MeV}$. 

\subsection{Validation on Calibration Data}
The energy reconstruction can be validated using the sodium calibration dataset, which provides a well-understood, localized source of low-energy electromagnetic activity near the center of the detector. This dataset serves as an important cross-check of the reconstruction chain, allowing for a direct comparison between data and Monte Carlo under controlled conditions.

\begin{figure}[h]
\centering
\includegraphics[width=\linewidth]{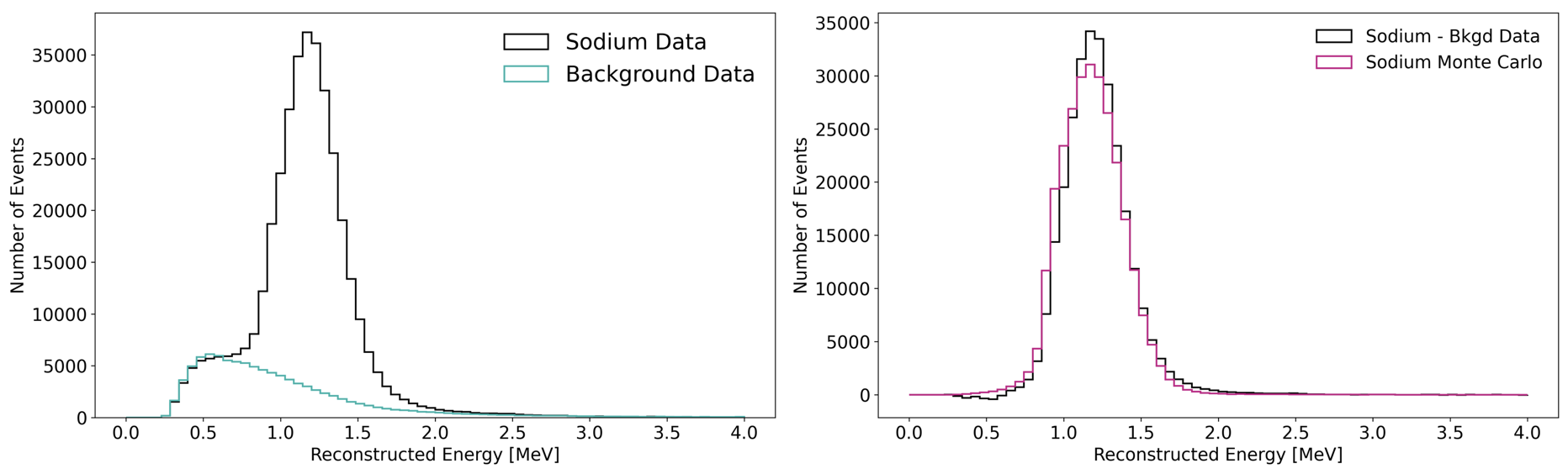}
\caption{Energy reconstruction validation using the sodium source data. The left-hand plot demonstrates the reconstructed energy distributions for both data collected when the sodium source was inserted in the detector and a background sample of data. Right-hand plot demonstrates direct comparison of background subtracted sodium distribution to the Monte Carlo simulation. The energy distributions are centered at the same value of approximately 1.2~MeV and have $\leq10\%$ differences in the shapes of the distributions.}
\label{fig:sodium_datamc_energy}
\end{figure}

Fig.~\ref{fig:sodium_datamc_energy}, left plot, shows the reconstructed energy spectra for the sodium calibration data alongside the corresponding background dataset. The sodium data exhibits a clear excess over background, forming a well-defined peak consistent with the expected energy deposition from the calibration source. The background spectrum, taken under identical conditions but without the source, is used to isolate the signal contribution.

To enable a more direct comparison with simulation, the background-subtracted data spectrum is constructed and compared to the sodium Monte Carlo prediction, as shown in Fig.~\ref{fig:sodium_datamc_energy}, right plot. The reconstructed energy distributions exhibit good agreement in the absolute energy scale. The shapes of the two distributions agree within approximately $\leq 10\%$, as expected from the systematic uncertainty observed in Chapter~\ref{chap:om}. This level of agreement indicates that the position-dependent charge scaling derived from simulation provides a reasonable description of the detector response in data.

Residual discrepancies at the $\mathcal{O}(\leq10\%)$ level may arise from a combination of factors, including imperfect modeling of the detector optical response, uncertainties in the absolute light yield, and differences between the true and reconstructed source position.

Overall, this comparison demonstrates that the energy reconstruction procedure, though relatively simple, captures the dominant detector effects and provides a reliable estimate of the deposited energy in the regime relevant for the calibration source. This validation builds confidence in applying the same reconstruction to physics data, where an accurate and stable energy scale is essential.


\chapter{Axion-Like Particle Analysis for $10^{-3}~\text{MeV} < m_a < 10~\text{MeV}$}\label{chap:alp}
The culminating physics analysis performed on the 2022 CCM200 physics dataset in this thesis is a search for axion-like particles (ALPs). ALPs are hypothetical pseudo-scalar bosons that arise in many extensions of the Standard Model, particularly in theories involving spontaneously broken global symmetries. While they are distinct from the QCD axion, which was originally proposed to solve the strong CP problem, ALPs share many phenomenological similarities, including weak couplings to photons and matter. CCM is sensitive to higher mass ALPs, $10^{-3}~\text{MeV} < m_a < 10~\text{MeV}$, which could be mediators in a larger dark sector. 

The general strategy of this analysis is to search for events in excess of the expected background within a narrow time window following the arrival of the proton beam on the target. In particular, we focus on a $\sim$175~ns region in which prompt, beam-related signals are expected, but prior to the arrival of a large flux of background neutrons produced in the beam dump. This timing-based approach provides a powerful handle for isolating potential ALP-induced signals from the dominant backgrounds.

A key feature of this analysis is the use of a data-driven background model. The CCM data acquisition system records approximately 10~$\mu$s of detector activity prior to the beam spill, referred to as the ``prebeam'' region. This sample provides a high-statistics measurement of steady-state backgrounds under identical detector conditions, allowing for a robust estimation of environmental contributions. However, this method is inherently insensitive to beam-correlated backgrounds that occur within the signal region, and therefore cannot fully constrain in-time backgrounds such as fast neutrons or other prompt processes.

The analysis of the data is performed within a frequentist framework. Signal expectations are obtained by generating Monte Carlo samples at each point in the ALP parameter space and processed through the same reconstruction and selection pipeline as the data. The resulting templates are compared to data using a likelihood-based test statistic. To account for the finite statistics of the simulated samples, an effective log-likelihood formulation is employed that incorporates the statistical uncertainty of the Monte Carlo prediction, ensuring an unbiased fit even in low-yield regimes~\cite{Arguelles:2019izp}.

The results presented in this thesis include three sources of systematic uncertainties introduced as nuisance parameters to the fit. These parameters govern the expected rate of background events, observed protons on target (POT), and expected $T_0$ that speed of light particles can reach the detector. At each point in the parameter space, the nuisance parameters are profiled over to obtain the likelihood test statistics.

Finally, this analysis builds upon the detector calibration and reconstruction techniques developed earlier in this thesis, particularly the characterization of low-energy ($<10$~MeV) events using sodium calibration data, improved position and energy reconstruction techniques, and an improved understanding of Cherenkov light characteristics in the detector. These studies inform the event selection criteria used in the ALP search, enabling efficient signal identification while maintaining strong background rejection in the relevant energy regime.

\section{ALP Theoretical Model}
This analysis targets axion-like particles (ALPs) that couple predominantly to photons. Unlike the QCD axion, the ALP mass $m_a$ and coupling strength $g_{a\gamma}$ are treated as independent parameters, allowing for a broad and phenomenologically rich parameter space.

The interaction relevant for this search is the effective two-photon coupling, Eq.~\ref{eq:alp_lagrangian}, where $g_{a\gamma}$ is the ALP-photon coupling strength, $a$ is the ALP field, and $F_{\mu\nu}$ is the field strength tensor ($\tilde{F}^{\mu\nu}$ denoting its dual). 

\begin{equation}
    \mathcal{L_{a\gamma\gamma}} = -\frac{1}{4} g_{a\gamma} a F_{\mu\nu} \tilde{F}^{\mu\nu}
\label{eq:alp_lagrangian}
\end{equation}

For photon-coupled ALPs, the dominant production mechanism in a fixed-target environment is the Primakoff process~\cite{Thompson:2023jbo,CCM:2021jmk}. In this process, an incident photon coherently scatters off the Coulomb field of a nucleus, converting into an ALP. The coherence of the interaction leads to a cross section that scales approximately as $Z^2$, where $Z$ is the atomic number of the target nucleus. This provides a significant enhancement for high-$Z$ materials such as tungsten, which is used as the primary target at the Lujan Spallation Facility.

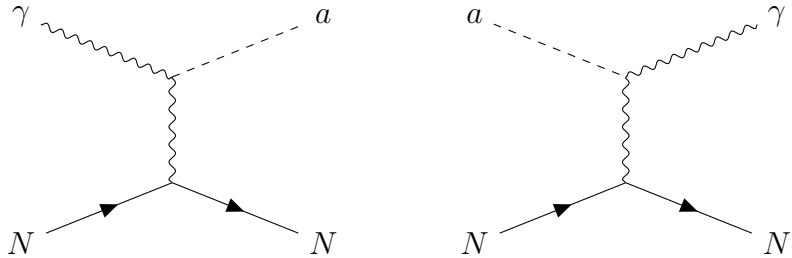
\begin{figure}[h]
\centering
\begin{tikzpicture}
\begin{feynman}[]
  \vertex (i1) at (0, 1.5) {\(\gamma\)};
  \vertex (i2) at (0,-1.5) {\(N\)};

  \vertex (v1) at (2, 0.7);
  \vertex (v2) at (2,-0.7);

  \vertex (f1) at (4, 1.5) {\(a\)};
  \vertex (f2) at (4,-1.5) {\(N\)};

  \diagram*{
    (i1) -- [photon] (v1) -- [scalar] (f1),
    (i2) -- [fermion] (v2) -- [fermion] (f2),
    (v1) -- [photon] (v2),
  };
\end{feynman}
\begin{feynman}[xshift=6cm]
  \vertex (i1) at (0, 1.5) {\(a\)};
  \vertex (i2) at (0,-1.5) {\(N\)};

  \vertex (v1) at (2, 0.7);
  \vertex (v2) at (2,-0.7);

  \vertex (f1) at (4, 1.5) {\(\gamma\)};
  \vertex (f2) at (4,-1.5) {\(N\)};

  \diagram*{
    (i1) -- [scalar] (v1) -- [photon] (f1),
    (i2) -- [fermion] (v2) -- [fermion] (f2),
    (v1) -- [photon] (v2),
  };
\end{feynman}
\end{tikzpicture}
\caption{Feynman diagrams for the ALP Primakoff production process (left) and ALP inverse Primakoff detection channel (right).}
\label{fig:primakoff_feynman}
\end{figure}

The inverse Primakoff process, in which an ALP converts back into a photon in the presence of a nuclear electric field, provides a complementary detection channel. This process is particularly important for low-mass ALPs ($m_a < \mathcal{O}(10~\mathrm{keV})$), where the ALP decay length is typically much larger than the detector dimensions, suppressing visible decays. The corresponding production and detection Feynman diagrams are shown in Fig.~\ref{fig:primakoff_feynman}.

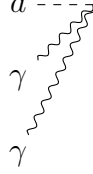
\begin{figure}[h]
\centering
\begin{tikzpicture}
\begin{feynman}
  \diagram [horizontal=i to v] {
    i [particle=\(a\)] -- [scalar] v,
    v -- [photon] f1 [particle=\(\gamma\)],
    v -- [photon] f2 [particle=\(\gamma\)],
  };
\end{feynman}
\end{tikzpicture}
\caption{Feynman diagram for ALP diphoton decay process.}
\label{fig:diphoton_feynman}
\end{figure}

For larger ALP masses, the dominant observable signature arises from the decay $a \rightarrow \gamma\gamma$, shown in Fig.~\ref{fig:diphoton_feynman}. The total decay width for this process is given in Eq.~\ref{eq:diphoton_decay_width}.

\begin{equation}
    \Gamma_{a\rightarrow\gamma\gamma} = \frac{g^2_{a\gamma\gamma} m_a^3}{64 \pi} 
\label{eq:diphoton_decay_width}
\end{equation}

This leads to a proper lifetime $\tau = 1/\Gamma$. The strong $m_a^3$ dependence implies that heavier ALPs decay much more rapidly, making the diphoton decay signature observable within the detector volume. Conversely, for small $m_a$ or weak coupling $g_{a\gamma}$, the decay length can exceed the detector size, and scattering-based detection channels dominate.

The total and differential cross sections for ALP production and detection channels are computed using implementation in the \texttt{alplib} framework~\cite{thompson_alplib_2023}. This package incorporates realistic nuclear and atomic form factors, including the parameterization of Tsai~\cite{Tsai:1986tx}, which accounts for screening effects and finite nuclear size corrections in Primakoff scattering. These effects are essential for accurately modeling the momentum transfer dependence and overall normalization of the production rate in a high-$Z$ target.

\section{Monte Carlo Simulation of the Expected Signal}
The expected ALP signal in the detector is modeled using a two-stage Monte Carlo simulation chain. This factorized approach allows the production of a realistic initial gamma-ray flux at the target to be separated from the subsequent ALP production, propagation, and detection processes, improving both computational efficiency and modularity of the simulation framework.

In the first stage, the interactions in the target are simulated to obtain the kinematic spectrum of gamma-rays. These gamma-rays serve as the primary particle for the subsequent ALP production process. Simulation of the protons on the target provides a robust estimate of this initial gamma-ray flux. 

In the second stage, ALP production, propagation, and detection are simulated using the \texttt{SIREN} injection toolkit~\cite{Schneider:2024eej}. This step models the conversion of photons into ALPs, their subsequent transport to the detector, and the detection processes which produce final state visible particles in the detector. The resulting final state gamma-rays in the detector are simulated and then used as input to the reconstruction and selection pipeline to estimate the expected signal distribution in the analysis.

\subsection{Target Simulation}
The Mark-IV target and surrounding shielding are fully modeled in \texttt{GEANT4} for the target simulation. This simulation is used to predict the energy spectrum, spatial distribution, and momentum distribution of gamma-rays produced by proton interactions in the target material, which serve as the primary production mode for ALPs in this analysis.

\begin{figure}[h]
  \centering
  \includegraphics[width=0.7\linewidth]{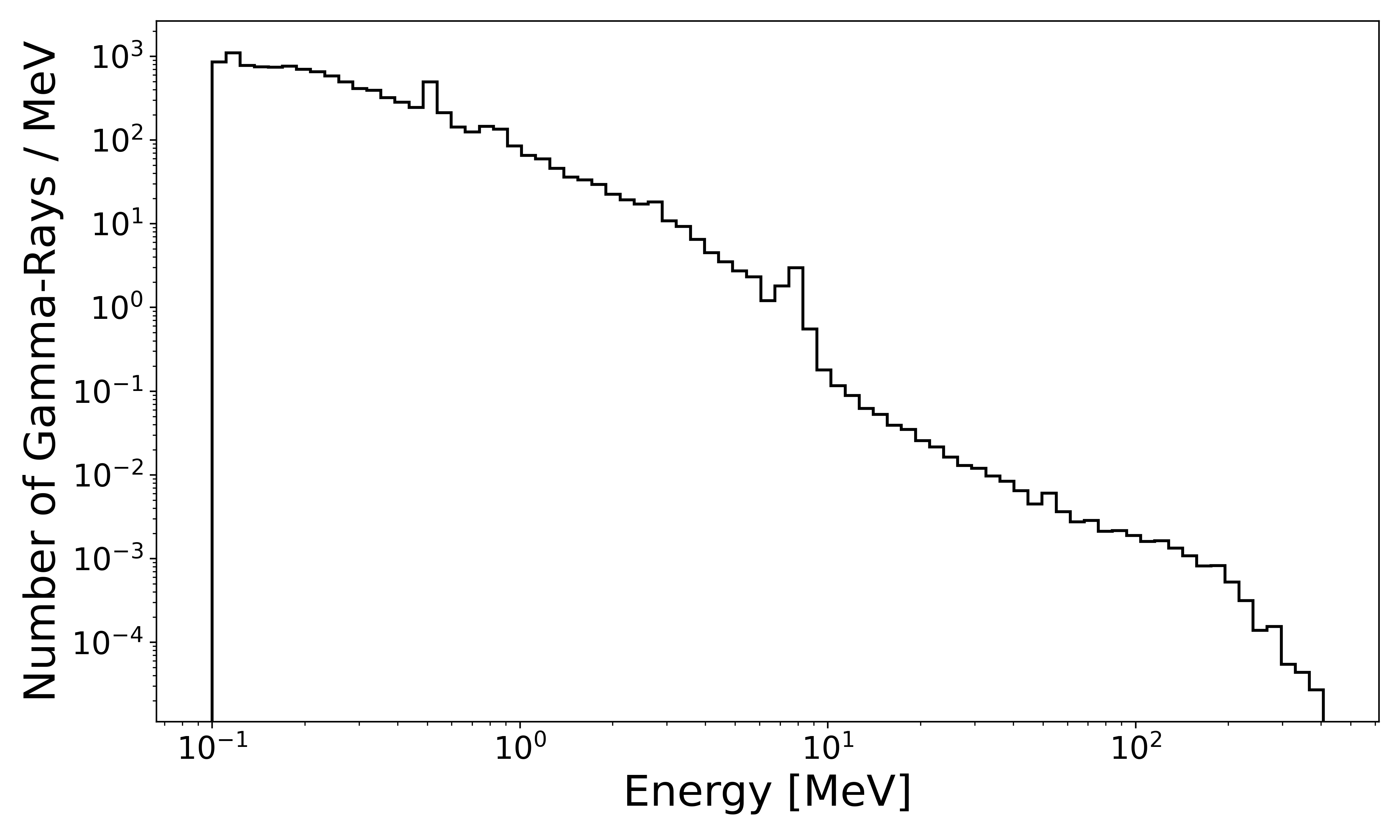}
  \caption{\texttt{GEANT4} Monte Carlo energy distribution of gamma-rays produced in the target, normalized a single proton on target. To obtain enough statistics, $10^4$ protons were injected on the target and the resulting gamma-rays are tracked through the materials to generate a realistic flux distribution for the ALP production process.}
  \label{fig:photon_flux}
\end{figure}

For the simulation, 800~MeV protons are injected incident from the top-down along the central axis of the target geometry, reproducing the experimental beam configuration. The hadronic and electromagnetic interactions within the target are modeled using the \texttt{QGSP\_BIC\_AllHP} physics list, which includes a detailed treatment of low-energy hadronic processes and nuclear de-excitation, allowing for an accurate description of nuclear excitation and subsequent gamma-ray emission.

All gamma-rays produced within the target volume are fully tracked through the geometry. For each gamma-ray, its kinematic properties (energy, position, and direction) are recorded at each step of its propagation. In order to correctly weight the contribution of gamma-rays for the ALP production, each recorded gamma-ray is assigned a weight proportional to the step length over which it propagates within the target material. This provides an effective sampling of the gamma-ray production density along the particle track, ensuring that regions of higher interaction probability contribute appropriately to the ALP injection sample.

To obtain a high-statistics estimate of the gamma-ray flux, $10^4$ protons are simulated. The resulting gamma-ray energy distribution, normalized per proton on target, is shown in Fig.~\ref{fig:photon_flux}. This distribution forms the input flux model used in the subsequent ALP production and propagation simulation.

\subsection{ALP Simulation}
To simulate the production and detection of ALPs, this work employs the \texttt{SIREN} injection framework~\cite{Schneider:2024eej}, which allows for efficient sampling of rare processes with flexible event reweighting. The simulation begins by sampling the kinematic distributions of prompt gamma-rays produced in the target, as obtained from the full \texttt{GEANT4} target simulation described in the previous section. These gamma-rays serve as the parent population for ALP production.

For each sampled gamma-ray, ALP production is modeled via the Primakoff process, in which a gamma-ray converts into an ALP in the electromagnetic field of a nucleus. The differential production rate is evaluated on an event-by-event basis, and the outgoing ALP kinematics are sampled accordingly. Since the gamma-ray flux emerging from the target is approximately isotropic, a geometric biasing scheme is applied to preferentially sample ALPs emitted in the direction of the detector. This bias is later accounted for through event weights to preserve the correct physical normalization.

\begin{figure}[h]
  \centering
  \includegraphics[width=0.7\linewidth]{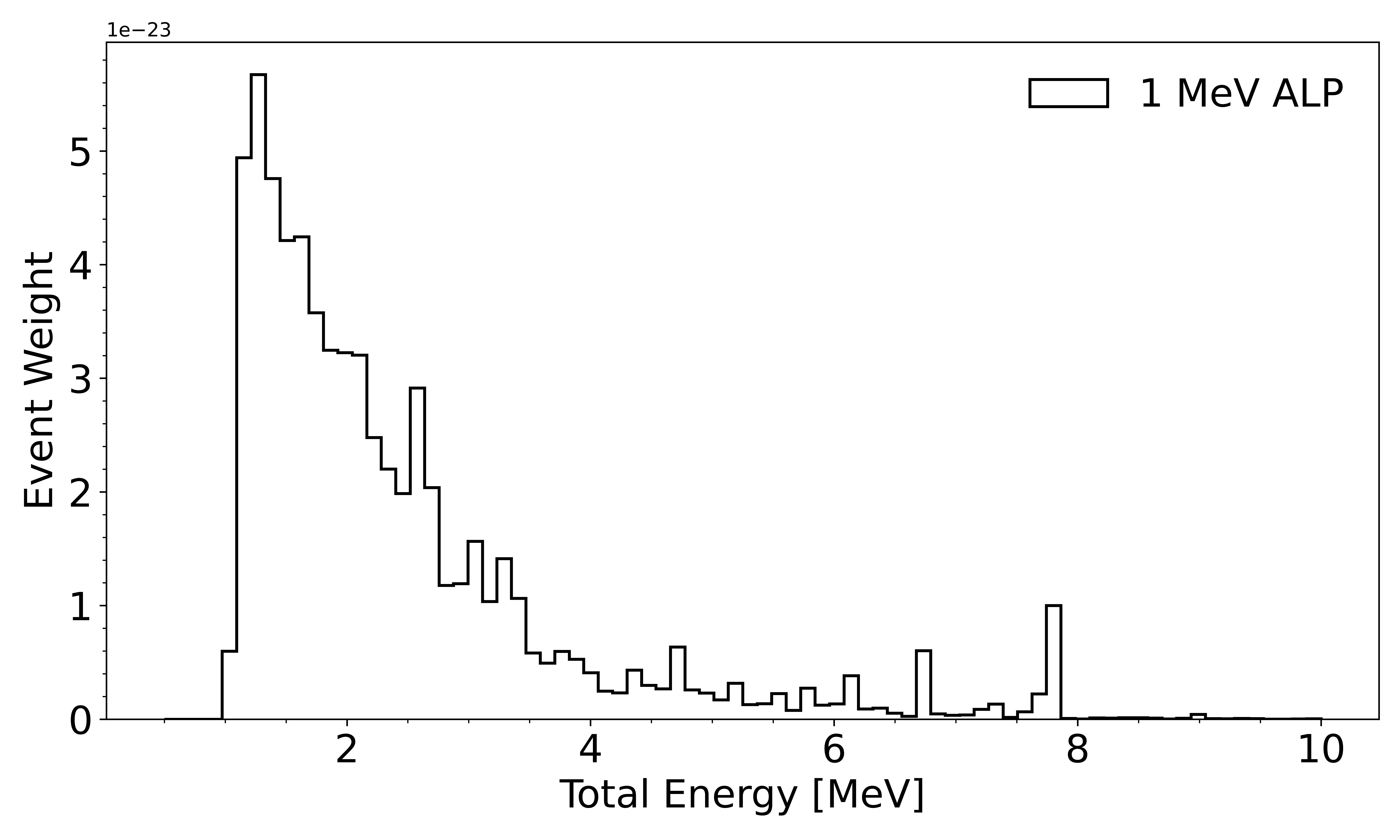}
  \caption{Example output of the full ALP simulation chain. In this case, a 1~MeV ALP was considered and simulated using the \texttt{SIREN} injection framework. This distribution shows the total energy of the one or two gamma-ray final states from the two possible ALP detection channels. This distribution peaks at the mass of the ALP and follows the underlying gamma-ray energy distribution as shown in Fig.~\ref{fig:photon_flux}.}
  \label{fig:1mev_alp_energy}
\end{figure}

Once produced, the ALP is propagated in a straight line toward the detector, neglecting interactions due to its weak coupling. Along its trajectory, two possible detection channels are considered: inverse Primakoff scattering ($a + N \rightarrow \gamma + N$) and diphoton decay ($a \rightarrow \gamma\gamma$). For each event, the probability of interaction or decay within the detector volume is computed based on the relevant cross section or decay width, and the specific secondary process is sampled using Monte Carlo methods. The final-state kinematics, either a single photon from inverse Primakoff scattering or two photons from diphoton decay, are then generated.

Each simulated event is assigned a total weight given by the product of three components: (i) the production weight from the initial gamma-ray flux and Primakoff matrix element, (ii) the geometric biasing factor used to enhance detector-pointing trajectories, and (iii) the detection probability weight, which encodes the likelihood of interaction or decay within the fiducial volume. This weighted-event approach enables efficient sampling of an otherwise extremely low-probability signal.

Fig.~\ref{fig:1mev_alp_energy} shows the total energy of the final-state gamma-rays in the detector for a 1~MeV ALP following the full injection procedure. The event weights correspond to a coupling of $g_{a\gamma} = 10^{-6}$~GeV$^{-1}$ for a single proton on target. The energy distribution turns on at the ALP mass, as expected from energy conservation, and reflects the underlying energy spectrum of the parent gamma-rays produced in the target. 

Following the injection stage, the final-state photons are passed through the full detector simulation chain. This includes optical photon simulation and propagation using the \texttt{GEANT4}-based detector model and PMT response model. Detector-specific timing effects, such as board-to-board synchronization offsets and electron transit time variations, are applied to accurately reproduce the timing structure of real data. In addition, data-driven noise is overlaid on the simulated waveforms to match the observed detector conditions. This processing is identical to that used for the Monte Carlo samples described in Chapter~\ref{chap:ml}, ensuring consistency between reconstruction training and physics analysis.

A key advantage of the \texttt{SIREN} framework is that each simulated event retains sufficient information to be reweighted to arbitrary ALP model parameters. To construct the signal model used in the statistical analysis, the ALP parameter space is discretized into a grid of $100$ mass points and $100$ coupling values, yielding a total of $10^4$ parameter points. The mass parameter space tested in this analysis ranges from $10^{-3}$~MeV to 10~MeV while the coupling strength $g_{a\gamma}$ parameter space ranges from $10^{-6}~\text{GeV}^{-1}$ to $10^{-3}~\text{GeV}^{-1}$, both of which are divided into 100 discrete points uniformly in logarithmic spacing. Rather than generating independent simulations at each point, events are generated only across the mass grid. For each mass hypothesis, both the Primakoff $\rightarrow$ inverse Primakoff and Primakoff $\rightarrow$ diphoton channels are simulated with sufficient statistics to ensure at least $\mathcal{O}(10^4)$ events remain after all selection criteria. The dependence on the ALP-photon coupling $g_{a\gamma}$ is then incorporated entirely through event reweighting, exploiting the known scaling of the production and detection processes with $g_{a\gamma}$.

This strategy dramatically reduces the computational cost of the simulation campaign, from $10^4$ independent samples to $10^2$, while maintaining full coverage of the parameter space and preserving statistical precision across the grid.

\section{Analysis Cuts}
The primary objective of this analysis is to suppress background events while maintaining high signal efficiency for ALP-induced electromagnetic (EM) final states. A first layer of event selection is constructed using the position and energy reconstruction algorithms developed in Chapter~\ref{chap:ml}. Simple cuts on total charge, reconstructed vertex position, and reconstructed energy are effective at rejecting a large fraction of instrumental and radiological backgrounds, as well as events originating outside the fiducial volume.

However, these conventional observables alone are not sufficient to best distinguish signal from background. The central goal of this thesis is to exploit the presence of Cherenkov light in a high light-yield scintillation detector to enhance sensitivity to EM final states. While scintillation light dominates the total photon yield, Cherenkov emission provides additional information through its prompt timing, directional structure, and distinct spatial distribution.

To leverage this information, four discriminating variables are constructed after applying basic charge, position, and energy cuts to target key features of EM particles. These variables probe: (i) the preferential detection of prompt Cherenkov photons on the uncoated PMTs, (ii) the directional anisotropy of early light, (iii) differences in pulse shape arising from prompt versus delayed light components, and (iv) the spatial extent of the event topology. Together, these observables provide some sensitivity to the underlying particle type and interaction kinematics.

These variables are combined into a likelihood ratio test statistic to separate signal from background in a multivariate framework. This approach enables a more powerful discrimination than any individual cut, while retaining interpretability in terms of physically motivated features. The performance of these cuts and the resulting signal efficiency and background rejection are evaluated in detail in the following section.

\subsection{Leveraging Cherenkov Radiation}
Chapter~\ref{chap:cherenkov} demonstrated that Cherenkov light can be isolated on the uncoated PMTs within an early time window using $^{22}$Na calibration data collected at the detector origin. Although these events involve sub-MeV electrons, where Cherenkov emission is relatively limited, they provide an idealized benchmark for studying the separation of Cherenkov and scintillation light. In this configuration, the source position is well known and centrally located, minimizing detector edge effects. Additionally, the low deposited energy reduces scintillation contamination in the prompt time region, and the uniform event topology enables straightforward timing alignment across events.

Extending this technique to physics data introduces several additional challenges. In the signal region of interest, events are distributed throughout the detector volume and span energies up to $10~\mathrm{MeV}$. As a result, detector edge effects become significant, altering photon propagation and PMT acceptance. Furthermore, the increased scintillation light yield at higher energies leads to greater contamination in the early time window, reducing the relative visibility of Cherenkov photons. These effects complicate the direct application of the calibration-based Cherenkov isolation strategy.

Accurate timing alignment is particularly critical for isolating prompt Cherenkov light. For the sodium calibration data, a simple event-finding metric, requiring 3~photoelectrons (PE) within a 2~ns window, provides a consistent reference time due to the uniformity of event energies and positions. However, for events with varying energies and interaction locations, such a fixed threshold does not correspond to a common physical point in the event development.

To address this, a constant fraction discrimination (CFD) approach is adopted to define the event start time. While an initial event finder (requiring 10~PE within a 10~ns window) is still used to identify candidate events, the timing reference for subsequent variable construction is taken as the point at which 20\% of the total integrated charge in the detector has been collected for each event. This definition provides a more robust and event-independent alignment of the prompt light component across the sample.

The electromagnetic-sensitive variables introduced in the following section are evaluated on three distinct datasets: steady-state background events from the prebeam region, simulated ALP signal events spanning the full mass range, and neutron-induced events from the prompt region of the neutron wall. This comparison enables a comprehensive assessment of their discriminating power between electromagnetic signal and background processes.

\subsubsection{Number of Hits on the Uncoated PMTs}
The first discriminating variable considered is the number of reconstructed photoelectron hits on the uncoated PMTs within an early time window. For the ALP analysis, this window is defined as $-6~\mathrm{ns}$ to $-2~\mathrm{ns}$ relative to the CFD-based event start time. This region is chosen to preferentially select prompt light, where Cherenkov emission is most prominent.

The uncoated PMTs provide enhanced sensitivity to Cherenkov light in this early time region. In the absence of a wavelength shifter, these PMTs are largely insensitive to the primary VUV scintillation spectrum and instead detect only wavelength-shifted scintillation light and the visible component of Cherenkov radiation. As a result, early hits on the uncoated PMTs are a useful proxy for the presence of Cherenkov light.

\begin{figure}[h]
  \centering
  \includegraphics[width=0.7\linewidth]{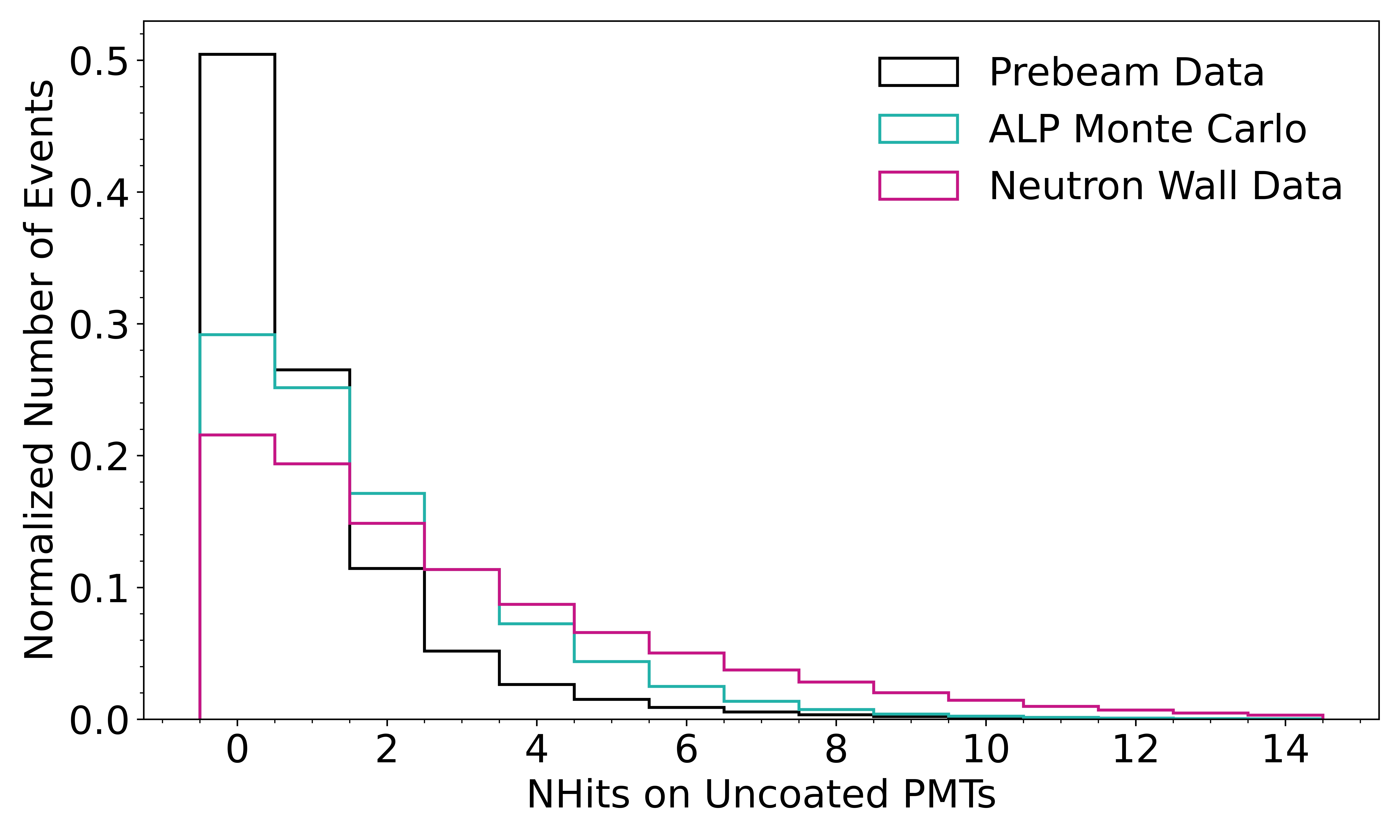}
  \caption{Number of hits on the uncoated PMTs for the prebeam data events, ALP Monte Carlo, and neutron wall data events.}
  \label{fig:llr_uncoated_hits}
\end{figure}

Fig.~\ref{fig:llr_uncoated_hits} shows the distribution of the number of hits on the uncoated PMT in this early time window for prebeam data, ALP Monte Carlo, and neutron-wall data. The prebeam and ALP distributions are both dominated by events with zero hits, reflecting the generally low light yield in this narrow time window. However, the ALP sample exhibits a higher fraction of events with non-zero hits, consistent with the presence of prompt Cherenkov emission from EM final states.

In contrast, neutron-wall events show a significantly broader distribution, with a larger fraction of events containing multiple hits on the uncoated PMTs. This behavior is consistent with increased scintillation light yield at higher energies, leading to greater contamination of the early time region.

These distributions illustrate that a simple threshold cut on the number of uncoated PMT hits, such as that used in the sodium calibration study of Chapter~\ref{chap:cherenkov}, is not sufficient in this energy regime for the ALP analysis. At higher energies and for spatially distributed events, the separation between Cherenkov-dominated and scintillation-dominated signatures using the number of hits on the uncoated PMTs becomes less powerful, motivating the need for a more sophisticated, multivariate approach.

\subsubsection{Spread in the Directionality}
The next variable used to isolate electromagnetic signal-like events exploits the intrinsic directionality of Cherenkov radiation. Unlike scintillation light, which is emitted isotropically, Cherenkov emission is strongly directional, providing a powerful handle for signal discrimination.

To quantify this effect, we consider only hits within the prompt time window, defined as $-6~\mathrm{ns}$ to $-2~\mathrm{ns}$ relative to the CFD-based event start time. Within this region, we construct the variable $C$, defined in Eq.~\ref{eq:cid}, which measures the degree of alignment of PMT hit directions. Specifically, $C$ is the squared magnitude of the charge-weighted average of unit vectors pointing from the reconstructed vertex $\vec{x}_{\text{vertex}}$ to the positions of hit PMTs $\vec{x}_i$.

\begin{equation}
    C = \left | \left | \sum_i  \left(\frac{q_i}{Q}\right) \frac{\vec{x}_i - \vec{x}_{\text{vertex}}}{|| \vec{x}_i - \vec{x}_{\text{vertex}} ||} \right | \right |^2
\label{eq:cid}
\end{equation}

Here, $q_i$ is the charge recorded in the $i$-th PMT and $Q = \sum_i q_i$ is the total charge in the prompt time window. By construction, $C$ ranges between 0 and 1: values near 0 correspond to isotropic light distributions, while values approaching 1 indicate highly collimated, co-linear hit patterns consistent with directional Cherenkov emission.

\begin{figure}[h]
  \centering
  \includegraphics[width=0.7\linewidth]{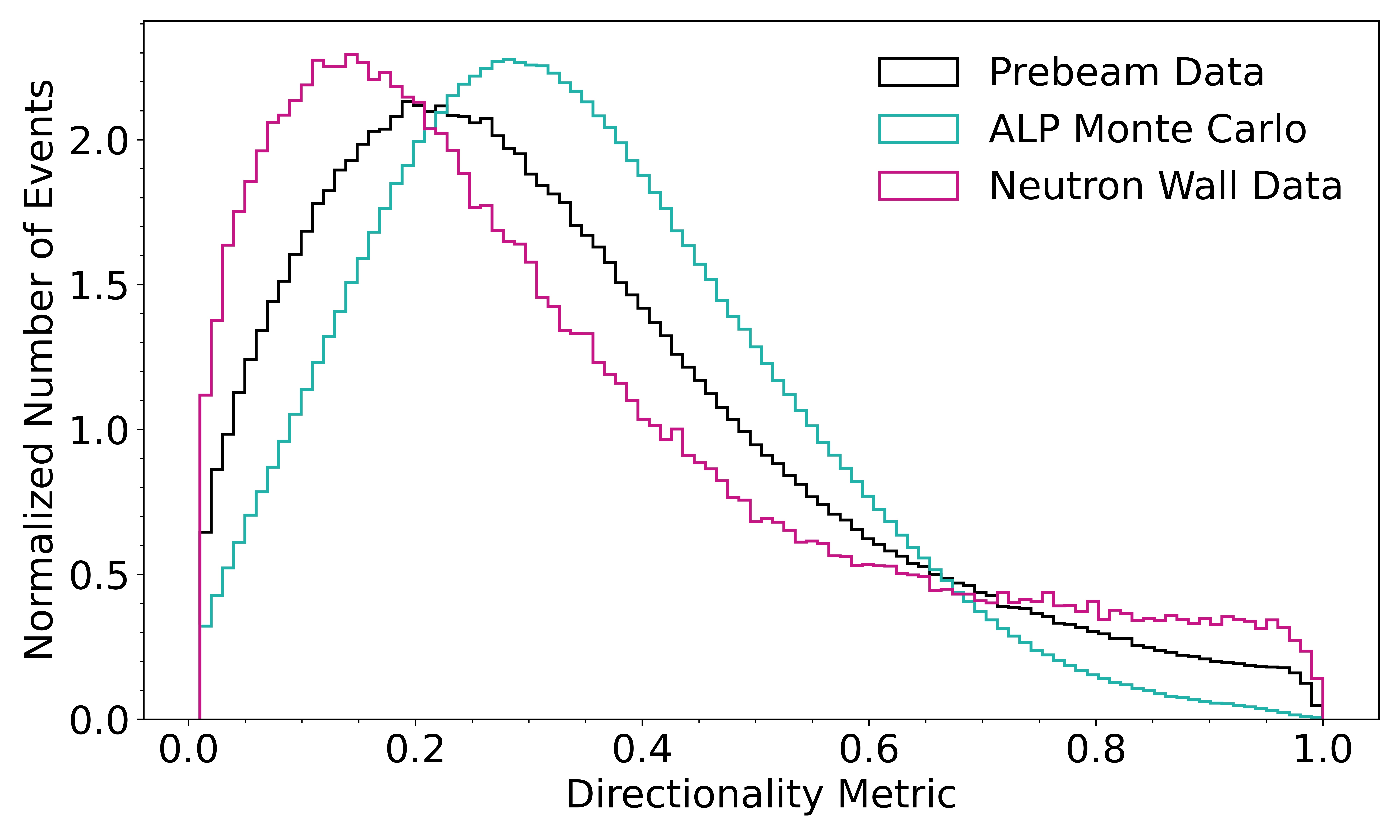}
  \caption{Distribution of directionality metric defined in Eq.~\ref{eq:cid} across the data samples considered. ALP Monte Carlo events have generally higher directionality metrics (consistent with directional light emission). Events from the neutron wall have generally lower direction metrics as expected for isotropic light emission.}
  \label{fig:llr_cid}
\end{figure}

Fig.~\ref{fig:llr_cid} shows the distribution of $C$ for prebeam data, ALP Monte Carlo, and neutron-wall data events. As expected, the ALP signal exhibits a distribution shifted toward higher values of $C$ (peaking around $\sim 0.3$), reflecting the presence of directional Cherenkov light from EM final states. In contrast, neutron-wall events, dominated by scintillation light with little intrinsic directionality, peak at lower values (around $\sim 0.15$). The prebeam data falls between these two cases, with a peak near $\sim 0.2$, consistent with a mixture of electromagnetic backgrounds, instrumental noise, and residual activity such as thermal neutrons from previous beam spills.

While this variable alone does not provide complete separation, it captures a key physical distinction between signal and background events and serves as an important input to the combined likelihood discriminant.

\subsubsection{Pulse Shape Ratio}
The third input to the likelihood ratio discriminant is a pulse shape–based observable. In conventional liquid argon scintillation detectors, pulse shape discrimination (PSD) is a powerful technique for separating nuclear recoils from electromagnetic interactions by exploiting differences in scintillation time profiles arising from differing ionization densities ($dE/dx$)~\cite{DEAP:2021axq,FIORILLO2006372,Agostini:2015boa,DarkSide-20k:2017zyg}. However, a true PSD approach is not directly applicable in this analysis because of the experimental conditions.

The CCM experiment operates in a pulsed beam environment with significant beam-related backgrounds. Extending integration windows to the microsecond scale, as is typical in PSD analyses, would introduce substantial contamination from pile-up events. As a result, the charge integration window is restricted to $\mathcal{O}(100~\mathrm{ns})$ timescales to balance photo-statistics against pile-up concerns. 

Despite this limitation, pulse timing information remains useful when adapted to the experimental constraints. In this analysis, the pulse shape ratio is defined as the fraction of charge collected in an early time window relative to the total prompt charge, Eq.~\ref{eq:psr}, where times are defined relative to the CFD-based event start time.

\begin{equation}
    \text{PSR} = \frac{Q (0~\text{ns} < t < 20~\text{ns})}{Q (0~\text{ns} < t < 90~\text{ns})}
\label{eq:psr}
\end{equation}

\begin{figure}[h]
  \centering
  \includegraphics[width=0.7\linewidth]{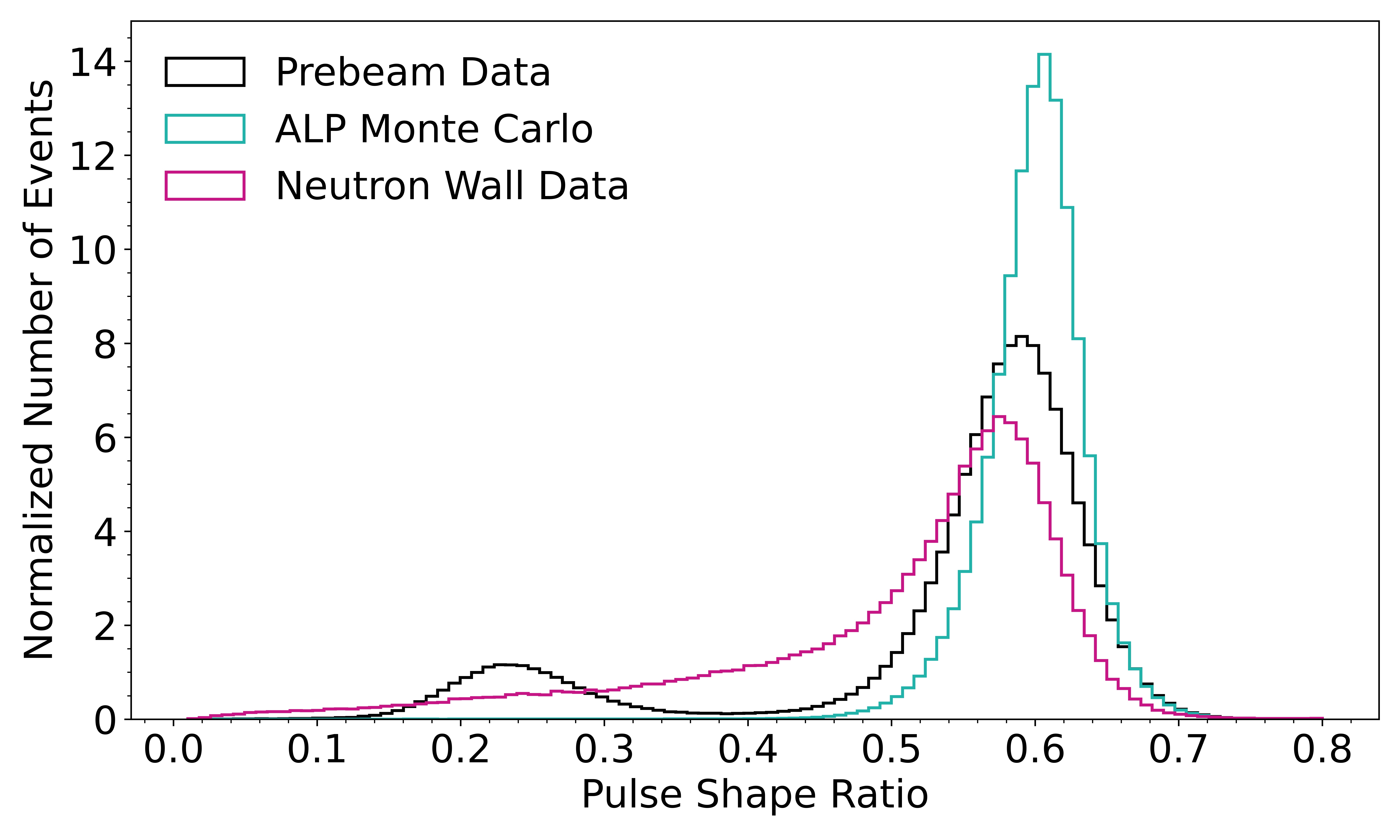}
  \caption{Pulse shape ratio distributions across the ALP, prebeam, and neutron wall datasets. While the ALP Monte Carlo events are narrowly peaked around 0.6, the prebeam and neutron wall datasets are broader and skewed towards lower values.}
  \label{fig:llr_pulse_shape}
\end{figure}

Fig.~\ref{fig:llr_pulse_shape} shows the resulting distribution for ALP Monte Carlo, prebeam data, and neutron-wall data. The ALP signal is narrowly peaked around $\sim 0.6$. In contrast, both prebeam and neutron-wall samples exhibit broader distributions shifted toward lower values, indicating a larger fraction of delayed light components. The prebeam sample also contains a subpopulation centered near $\sim 0.25$, which has been identified as consistent with non-physical noise events and overlapping physical events.

While this variable does not constitute a true PSD observable in the traditional sense, it provides complementary timing information that improves separation between signal and background within the constraints of the beam environment.

\subsubsection{Spatial Spread of Events}
The final variable used in the likelihood ratio discriminant characterizes the spatial extent of charge deposition within an event. This observable exploits the expected topological differences between signal and background interactions. ALP signal events, consisting of one or two final-state gamma-rays, tend to produce relatively localized electromagnetic interactions. In contrast, fast neutron backgrounds can undergo multiple scatters and secondary interactions throughout the detector volume, leading to more spatially extended charge distributions.

\begin{equation}
    \text{RMS} =  \sqrt{ \frac{\sum_i q_i ||\vec{x}_i - \vec{x}_{\text{vertex}}||^2}{\sum_i q_i} }
\label{eq:rms}
\end{equation}

To quantify this behavior, we define a charge-weighted root-mean-square (RMS) distance of PMT hits from the reconstructed vertex, Eq.~\ref{eq:rms}, where $\vec{x}_i$ is the position of the $i$-th PMT, $\vec{x}_{\text{vertex}}$ is the reconstructed event vertex, and $q_i$ is the recorded charge.

This quantity is computed using only hits in the early time window ($-6~\mathrm{ns}$ to $-2~\mathrm{ns}$ relative to the CFD start time) in order to reduce contamination from delayed and re-emitted scintillation light, which would otherwise artificially broaden the spatial distribution. By construction, the RMS represents the charge-weighted spatial scale of the event: smaller values correspond to more localized energy depositions, while larger values indicate more diffuse or multi-site interactions.

\begin{figure}[h]
  \centering
  \includegraphics[width=0.7\linewidth]{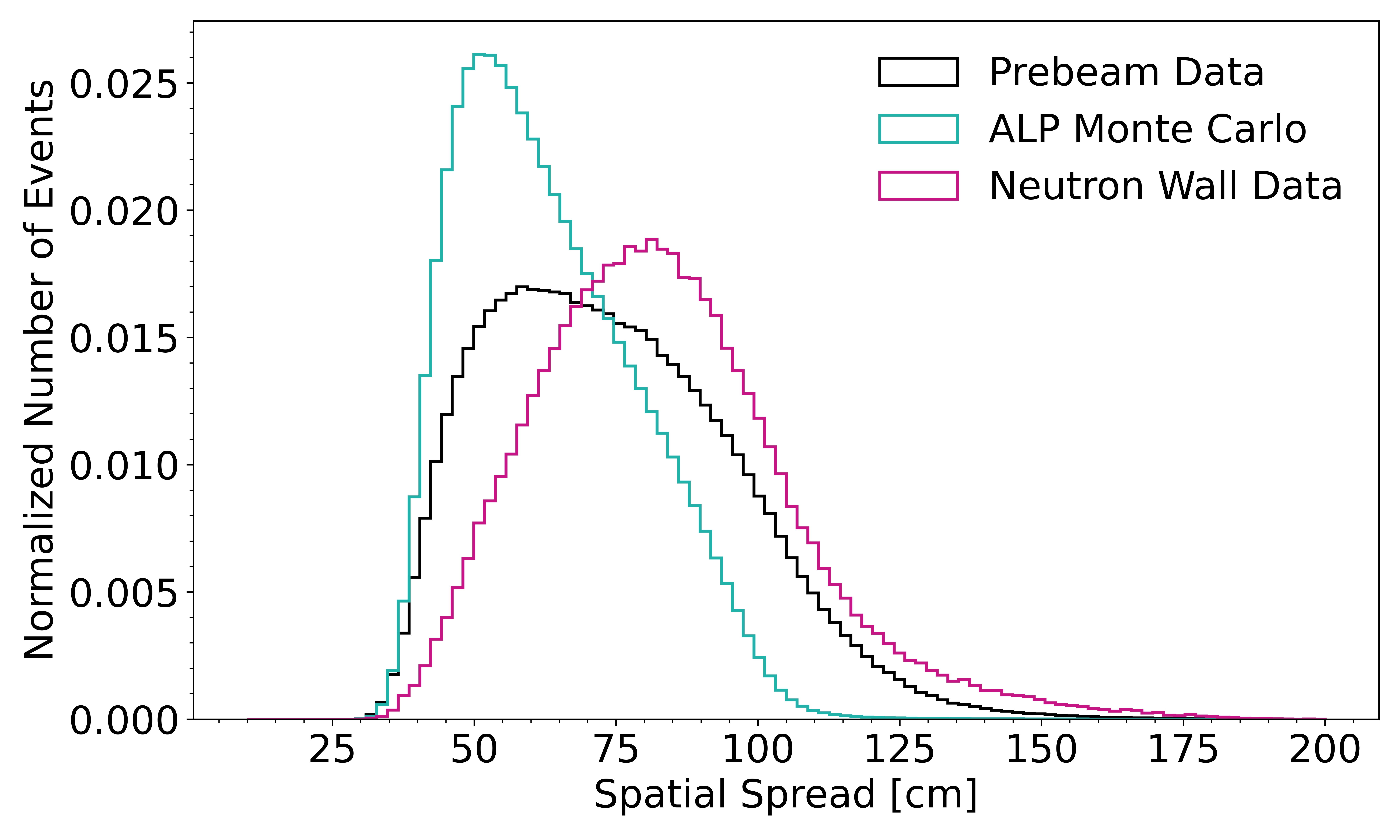}
  \caption{Spatial spread of reconstructed charge in the early time region. This metric, defined in Eq.~\ref{eq:rms}, can help separate point-like energy depositions from EM particles from extended energy depositions from neutrons that interact multiple times across the detector.}
  \label{fig:llr_rms}
\end{figure}

Fig.~\ref{fig:llr_rms} shows the resulting RMS distributions for ALP Monte Carlo, prebeam data, and neutron-wall data. The ALP signal is shifted toward smaller values, peaking around $\sim 50~\mathrm{cm}$, consistent with relatively localized electromagnetic activity. In contrast, neutron-wall events exhibit a broader distribution with a peak near $\sim 80~\mathrm{cm}$, reflecting their more spatially extended interaction topology. The prebeam sample lies between these two cases, peaking near $\sim 60~\mathrm{cm}$, with a noticeable tail toward larger RMS values, consistent with a mixture of background processes and non-physical events.

While this variable alone does not provide complete separation, it captures an important topological distinction between signal and background and contributes complementary information to the overall likelihood discriminant.

\subsubsection{Likelihood Ratio Test}
The four discriminating variables described above are combined into a single test statistic using a likelihood ratio framework. For each variable $x_k$, probability density functions (PDFs) are constructed for both the signal and background hypotheses. The signal PDFs are derived from ALP Monte Carlo, while the background PDFs are obtained from the prebeam data sample.

\begin{equation}
    \mathrm{LLR} = \sum_{k} \log(p_{\mathrm{sig}}(x_k)) - \log(p_{\mathrm{bkg}}(x_k))
\label{eq:llr}
\end{equation}

For uncorrelated variables, the joint likelihood for an event is taken as the product of the individual PDFs. The log-likelihood ratio (LLR) is then defined as Eq.~\ref{eq:llr}, where $p_{\mathrm{sig}}(x_k)$ and $p_{\mathrm{bkg}}(x_k)$ are the signal and background PDFs evaluated at the observed value of the $k$-th variable. This formulation is equivalent to the logarithm of the ratio of the joint likelihoods under the signal and background hypotheses.

In practice, the PDFs are constructed from normalized histograms of each variable and interpolated to provide continuous probability estimates. A small regularization term is added to avoid numerical instabilities in regions with low statistics.

\begin{figure}[h]
  \centering
  \includegraphics[width=0.7\linewidth]{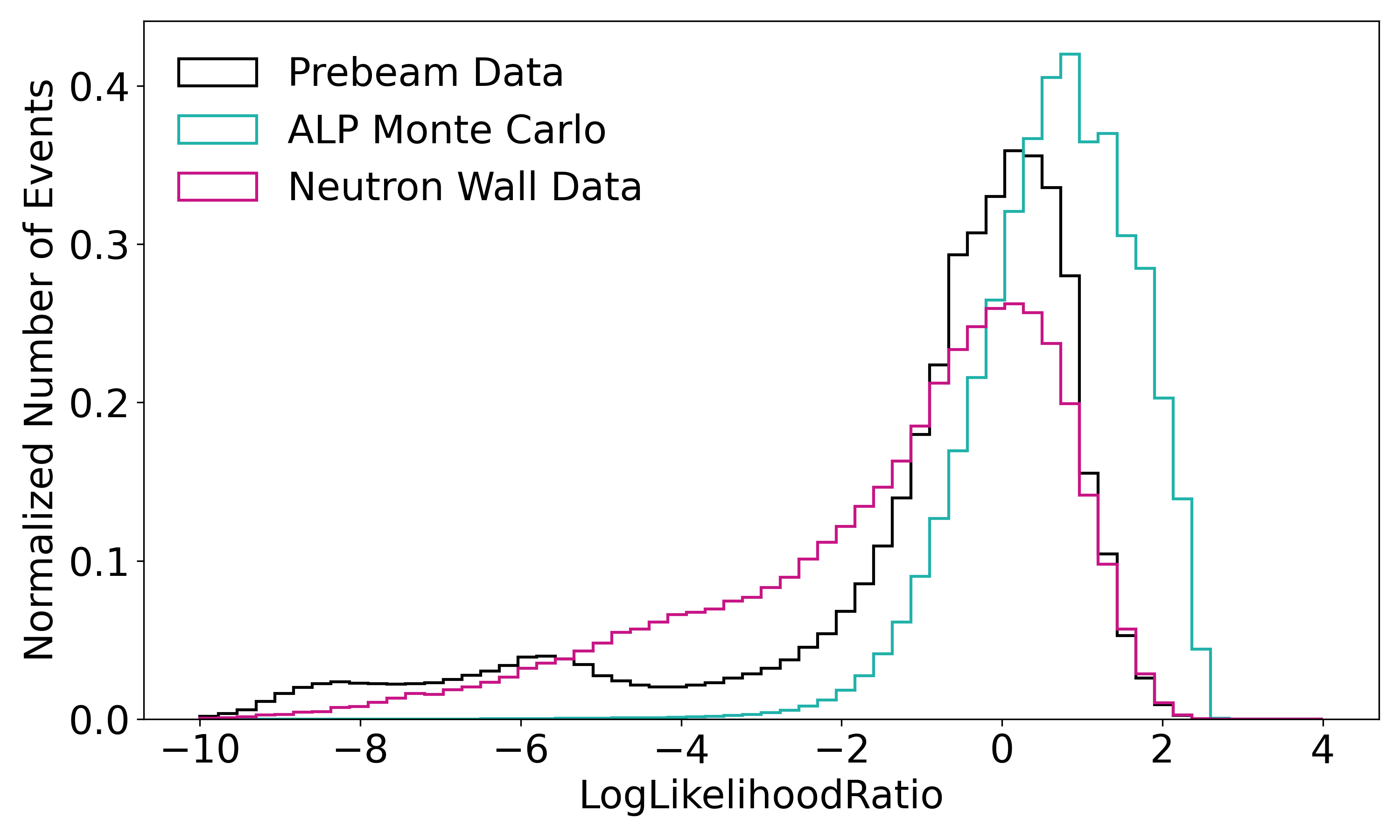}
  \caption{Log of the likelihood ratio (LLR) test statistic. This variable is computed by combining the probability distribution functions for the four previously defined event variables. A requirement of $\text{LLR} > 1$ is chosen to maintain adequate signal selection efficiency while removing many of the sources of backgrounds.}
  \label{fig:final_llr}
\end{figure}

Fig.~\ref{fig:final_llr} shows the resulting LLR distributions for ALP Monte Carlo, prebeam data, and neutron-wall data. The ALP signal is shifted toward higher values of the test statistic, peaking near $\sim 1$, while both prebeam and neutron-wall events populate lower values and exhibit broader tails. A selection requirement of $\mathrm{LLR} > 1$ is applied to define the signal region. This threshold is chosen to retain high signal efficiency while still providing significant rejection of background events.

\subsection{Final Event Selection}
The final event selection is designed to suppress prebeam background while retaining sensitivity to low-energy electromagnetic signatures expected from ALP decays. The full set of cuts and their impact on prebeam data are summarized in Table~\ref{table:alp_cuts}.

\begin{table}[h]
    \centering
    \begin{tabular}{l|c}
        \toprule
        \textbf{Cut} & \textbf{Prebeam Selection Efficiency [\%]}  \\
        \midrule
        Charge & 49.75 \\
        Position  & 18.29 \\
        Energy  & 80.93 \\
        Charge + Position + Energy & 7.19 \\
        Charge + Position + Energy + LLR & 0.49 \\
        \bottomrule
    \end{tabular}
    \caption{Selection efficiencies of cuts utilized in the ALP analysis for the prebeam data sample.}
    \label{table:alp_cuts}
\end{table}

An initial charge requirement selects events with total detected charge between 30~PE and 3000~PE within the first 90~ns of the event window. This removes both low-charge noise-like events and high-charge activity inconsistent with the signal hypothesis. 

Geometric containment of events to the active region is enforced through position cuts, requiring reconstructed vertices to satisfy $R < 80~\text{cm}$ and $-40~\text{cm} < Z < 40~\text{cm}$. These cuts reduce backgrounds originating near detector boundaries and poorly reconstructed events.

An energy selection of $0.2~\text{MeV} < E < 10~\text{MeV}$ is then applied. The upper bound reflects the current calibration regime, which is optimized using the $^{22}$Na source at $\mathcal{O}(1~\text{MeV})$. While this analysis focuses on low-energy reconstruction, ongoing work using Michel electrons is expected to extend reliable calibration up to $\sim 50~\text{MeV}$, enabling future analyses to probe higher-energy signals.

Finally, a likelihood ratio cut is applied using the log-likelihood ratio (LLR) test statistic described in the previous section. Events are required to satisfy $\mathrm{LLR} > 1$, providing additional discrimination between signal-like and background-like event topologies.

The cumulative effect of these selections is a strong suppression of prebeam backgrounds, with a final selection efficiency of 0.49\%.

\begin{figure}[h]
  \centering
  \includegraphics[width=0.7\linewidth]{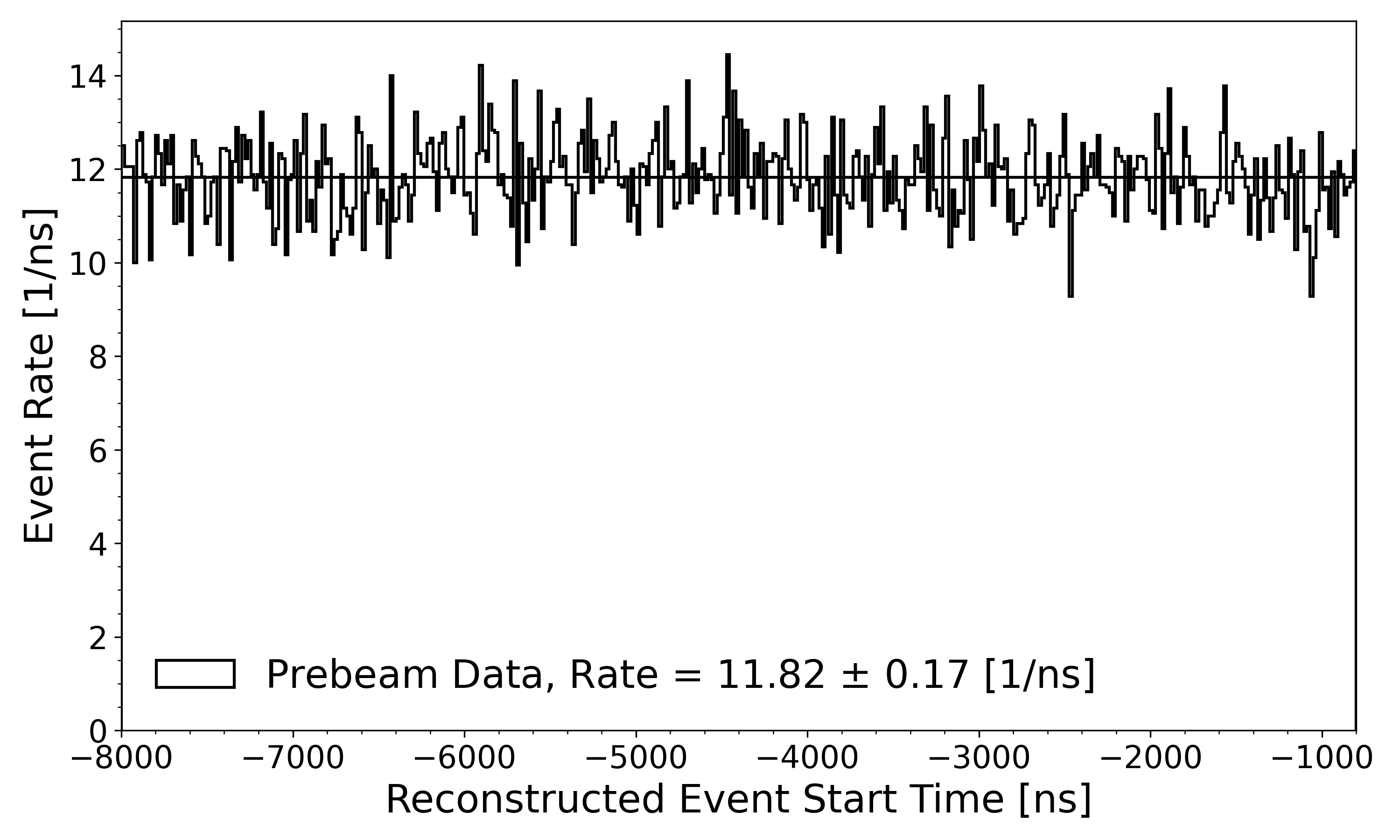}
  \caption{Time distribution of prebeam events selected after applying all analysis cuts. This data is fit with a uniform distribution to obtain an expected background event rate of $11.82 \pm 0.17$ events per ns.}
  \label{fig:prebeam_time_after_cuts}
\end{figure}

After applying the full selection criteria, the time distribution of the surviving prebeam events is shown in Fig.~\ref{fig:prebeam_time_after_cuts}. The event rate is given in units of events per ns and integrated over the full $1.23 \times 10^{21}$ POT dataset. The distribution is consistent with a steady-state background, as expected for the prebeam data.

To quantify this, the time spectrum is fit with a constant function, yielding a best-fit rate of $11.82 \pm 0.17$ events per ns. The goodness-of-fit is evaluated using a Pearson $\chi^2$ test, resulting in $\chi^2 = 427.76$ for 399 degrees of freedom, corresponding to a reduced $\chi^2$ of 1.07. This indicates good agreement with the hypothesis of a time-independent background rate, with no significant evidence for residual time structure after the selection cuts.

\begin{figure}[h]
  \centering
  \includegraphics[width=0.7\linewidth]{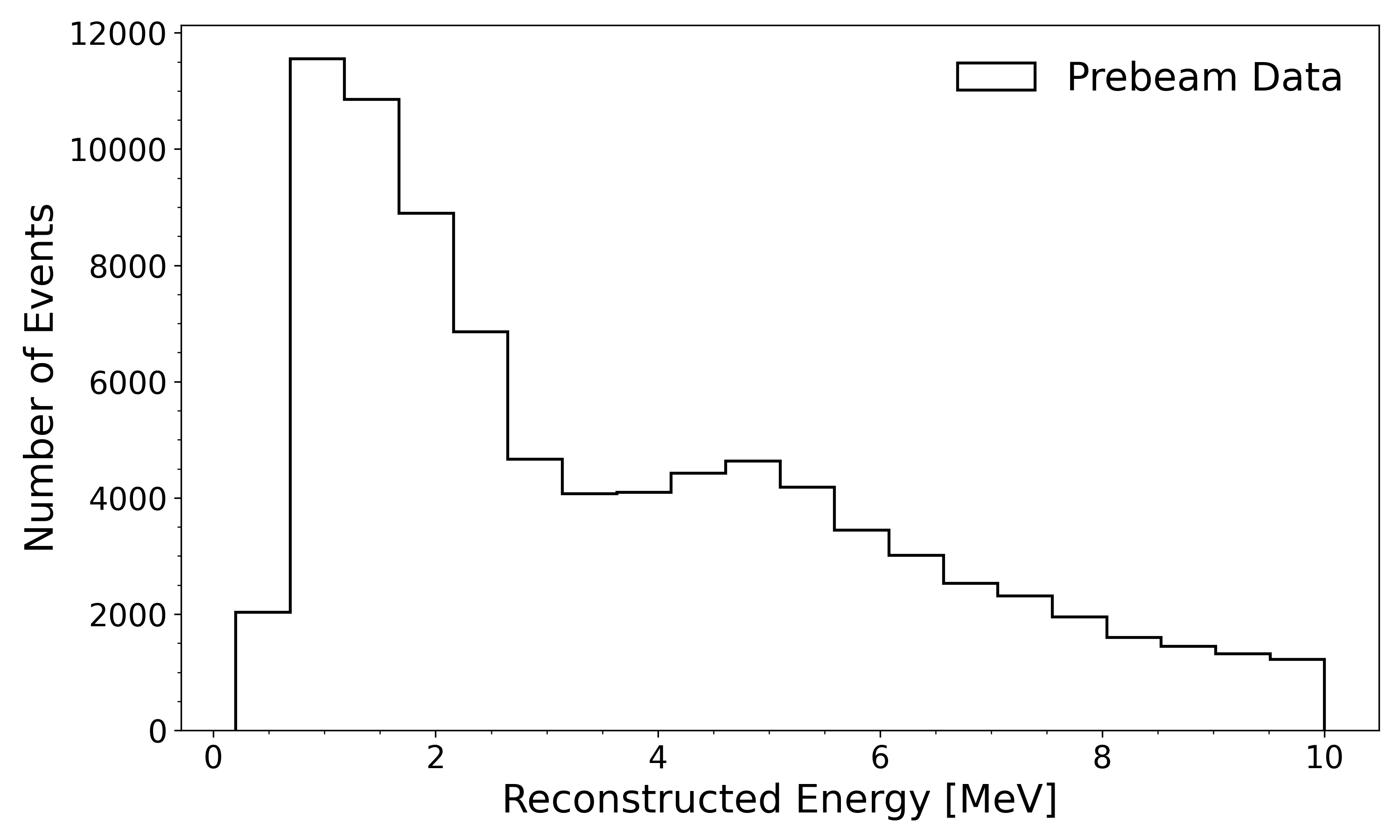}
  \caption{Reconstructed energy distribution of prebeam events selected after applying all analysis cuts. The feature around 5~MeV could be consistent with gamma-rays produced from neutron capture on argon atoms, but more investigation is necessary.}
  \label{fig:prebeam_energy_after_cuts}
\end{figure}

\begin{figure}[h]
  \centering
  \includegraphics[width=0.7\linewidth]{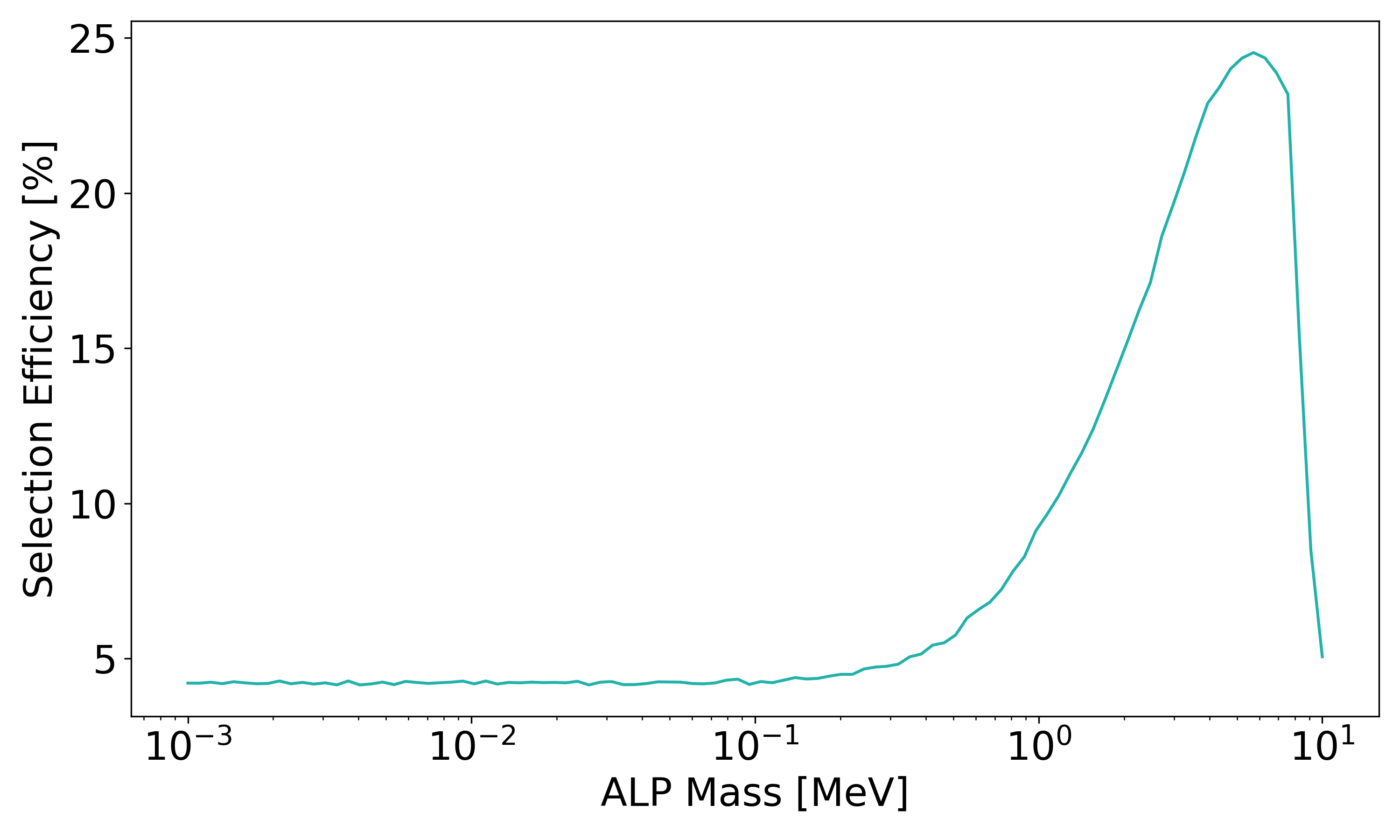}
  \caption{Signal efficiency as a function of ALP mass. The selection efficiency increases with ALP mass, reaching a maximum value of around 25\% for 6~MeV ALPs. Above that mass, the efficiency drops steeply due to the maximum reconstructed energy cut of 10~MeV.}
  \label{fig:alp_mass_eff}
\end{figure}

The reconstructed energy spectrum of selected prebeam events is shown in Fig.~\ref{fig:prebeam_energy_after_cuts}. The distribution exhibits a primary peak near $\sim 1$~MeV, consistent with low-energy electromagnetic background events that survive the likelihood ratio cut to select events characteristic of Cherenkov radiation. 

In addition, a secondary feature is observed around $\sim 5$~MeV. One possible explanation is gamma-ray emission from neutron capture on argon nuclei in the detector. Thermal neutrons can undergo capture via $^{40}\text{Ar}(n,\gamma)^{41}\text{Ar}$, producing de-excitation gamma rays with energies up to approximately 6.1~MeV. Such neutrons may originate from prior beam activity and persist in the experimental hall, contributing to the prebeam background. While this interpretation is consistent with the observed energy scale of the feature, further studies are required to confirm its origin.

Now let us consider the final event selection for the ALP physics signal. The selection efficiency for signal events depends strongly on the ALP mass. Fig.~\ref{fig:alp_mass_eff} shows the total efficiency after all cuts as a function of the ALP mass over the range $10^{-3}$~MeV to 10~MeV.

At low masses ($< 0.3~\text{MeV}$), the efficiency is approximately constant at $\sim 2\%$, reflecting the limited visible energy deposited by low-mass ALPs. As the mass increases, the efficiency rises, reaching $\sim 10\%$ near 1~MeV. The efficiency continues to increase with mass, peaking at approximately 25\% around 6~MeV. For higher masses, the efficiency sharply drops due to the imposed upper energy cut at 10~MeV, which truncates higher-energy signal events.

\section{Axion-Like Particle Fit}
With the event selection criteria established, the final step of the analysis is to perform a fit for the ALP signal. This procedure is based on a template likelihood approach, in which the prebeam data defines the null (background-only) hypothesis, while the simulated ALP samples in addition to the background expectation define the signal hypothesis. The observed data in the physics window are selected using the same reconstruction and selection criteria as both templates, with the additional requirement that event start times lie within the time window $-600~\mathrm{ns}$ to $-424~\mathrm{ns}$ relative to the Beam Current Monitor. This timing region is motivated by the arrival time of the first speed of light gamma-rays observed in the external EJ-301 detector, as discussed in Chapter~\ref{chap:ccm}.

The test statistic is constructed from a binned likelihood evaluated over a two-dimensional space in reconstructed energy and event start time. Incorporating timing information provides additional discriminating power, as beam-related signal events are expected to follow the characteristic time structure of the proton pulse, in contrast to the approximately uniform distribution of steady-state background events.

The fit is performed in a frequentist framework by evaluating the likelihood across a grid of $10^4$ points in the ALP parameter space, in addition to the null hypothesis. At each point, signal templates are generated by performing the full ALP simulation procedure described previously, followed by reconstruction and selection. The resulting templates are compared to the observed data by minimizing an effective log-likelihood function that accounts for limited Monte Carlo statistics~\cite{Arguelles:2019izp}. Confidence intervals are derived under the assumption of Wilks' theorem~\cite{Wilks:1938dza}.

\subsection{Fitting Procedure}
To construct the templates used in the fit, events are binned in reconstructed energy and time. The time axis is divided into 20 uniform bins spanning $-600~\mathrm{ns}$ to $-424~\mathrm{ns}$, corresponding to the interval between the earliest possible arrival of speed-of-light particles at the detector and the onset of neutron-induced activity. The energy axis is divided into 12 uniform bins between $0.2~\mathrm{MeV}$ and $10~\mathrm{MeV}$.

The observed data are filled directly into this two-dimensional $(E, t)$ histogram. For the background and signal hypotheses, however, the templates are constructed in a hybrid manner to reduce statistical fluctuations. The energy distributions are obtained by binning the simulated ALP or prebeam events directly, while the time distributions are modeled using analytic or semi-analytic probability density functions (PDFs), normalized to the expected number of events in the energy distribution. This approach avoids introducing additional variance from finite Monte Carlo statistics in the time dimension.

For the null hypothesis, the prebeam time distribution is consistent with a uniform distribution. The probability for an event to fall within a time bin $[t_j, t_{j+1}]$ is therefore given by integrating a constant rate given in Eq.~\ref{eq:prebeam_time_distribution_pdf} where $b$ is the background event rate per unit time. 

\begin{equation}
    P_j^{\text{bkg}} = \int_{t_j}^{t_{j+1}} b~dt = b (t_{j+1} - t_j),
\label{eq:prebeam_time_distribution_pdf}
\end{equation}

The signal timing distribution is more complex and is constructed by convolving three effects: the beam time profile, the physical propagation delay of the ALP from production to detection, and reconstruction-level timing offsets. Each simulated ALP event is assigned a total time offset $\Delta t_i$ that accounts for both the propagation time and reconstruction effects. The intrinsic beam structure is modeled as an equilateral triangular of width $\Delta t_{\text{beam}} = 290~\mathrm{ns}$, starting at $t_0 = -600~\mathrm{ns}$.

The probability for a signal event with offset $\Delta t_i$ to fall within a time bin $[t_j, t_{j+1}]$ is then given by Eq.~\ref{eq:signal_time_distribution_pdf} where $F_{\text{beam}}(t)$ is the cumulative distribution function (CDF) of the triangular beam profile.

\begin{equation}
    P_{ij}^{\text{sig}} = F_{\text{beam}}(t_{j+1} - t_0 - \Delta t_i) - F_{\text{beam}}(t_j - t_0 - \Delta t_i),
\label{eq:signal_time_distribution_pdf}
\end{equation}

The expected number of signal events in each $(E_i, t_j)$ bin is obtained by summing over all simulated events with weights $w_i$, described in Eq.~\ref{eq:signal_weights}.

\begin{equation}
    N_{ij}^{\text{sig}} = \sum_{k \in E_i} w_k , P_{kj}^{\text{sig}}.
\label{eq:signal_weights}
\end{equation}

Finally, the fit is performed by minimizing a binned effective likelihood over the two-dimensional histogram~\cite{Arguelles:2019izp}. The likelihood function is given in Eq.~\ref{eq:leff} and depends on the observed counts $k$, $\alpha = \frac{(\sum_i w_i)^2}{\sum_i w_i^2} + 1$, and $\beta = \frac{\sum_i w_i}{\sum_i w_i^2}$ for the weights for each event $w_i$.

\begin{equation}
    \mathcal{L_{\text{eff}}}(k, \alpha, \beta) = \frac{\Gamma(k+\alpha)}{\Gamma(k+1) \Gamma(\alpha)} \frac{\beta^{\alpha}}{(1+\beta)^{k+\alpha}}
\label{eq:leff}
\end{equation}

\subsection{Sources of Systematic Uncertainties}
Several sources of systematic uncertainty are considered in this analysis and incorporated directly into the fitting framework through the introduction of nuisance parameters. These nuisance parameters are profiled at each point in the parameter space during the likelihood minimization, ensuring that their impact on the signal sensitivity is consistently propagated to the final results. The dominant systematics arise from uncertainties in the background normalization, the total number of protons on target, and the absolute timing of the beam.

\subsubsection{Rate of Background Events}
One leading source of systematic uncertainty in this analysis is associated with the expected rate of steady-state background events. As shown in Fig.~\ref{fig:prebeam_time_after_cuts}, the time distribution of prebeam data after all analysis cuts is well described by a uniform distribution, indicating no significant time-dependent structure of the background within the analysis window. A fit to this distribution yields a best-fit background rate of 11.82 events per ns, with a corresponding $1\sigma$ uncertainty of 0.17 events per ns.

To account for this uncertainty in the fit, the background normalization is treated as a free nuisance parameter constrained by a Gaussian prior. The prior is centered at the measured value of 11.82 events per ns, with a standard deviation of 0.17, reflecting the statistical precision of the prebeam measurement. This treatment allows the fit to adjust the overall background level within its uncertainty while penalizing deviations from the nominal expectation through the prior.

\subsubsection{Total Observed POT}
An additional systematic uncertainty arises from the determination of the total number of protons on target during the 2022 physics data-taking period. As described in Chapter~\ref{chap:ccm}, the POT is calculated from a calibration relating the recorded Beam Current Monitor integral to the beam current reported by the LANSCE accelerator division. This relationship is obtained through a linear fit, which introduces an uncertainty of approximately 5\% on the overall normalization of total POT.

This uncertainty directly affects the predicted signal yield, as the expected number of ALP-induced events scales linearly with the total POT. To incorporate this effect, a multiplicative nuisance parameter is introduced that rescales the effective POT in the signal prediction. This parameter is assigned a Gaussian prior centered at unity with a width of 0.05, corresponding to the estimated fractional uncertainty. During the fit, this allows for variations in the signal normalization while maintaining consistency with the external POT calibration.

\subsubsection{Start of the Beam Time}
The third systematic uncertainty considered in this analysis is associated with the absolute timing of the beam relative to the detector. In particular, there is an estimated uncertainty of approximately 30~ns in the earliest time at which speed-of-light particles originating from the target can arrive at the detector, discussed in Chapter~\ref{chap:ccm}.

This uncertainty is especially relevant because the signal model depends explicitly on the convolution of the ALP time-of-flight distribution with the beam time profile. Any shift in the assumed beam start time therefore translates directly into a distortion of the predicted signal timing distribution.

To account for this effect, the beam start time is treated as a nuisance parameter in the fit. A Gaussian prior is imposed with a central value of $-600$~ns and a standard deviation of 30~ns, reflecting the uncertainty in the timing calibration. Profiling over this parameter allows the fit to accommodate small shifts in the signal timing structure while constraining them within the experimentally determined bounds.

\subsubsection{Ongoing Investigations of Systematic Uncertainty}
Additional sources of systematic uncertainty are currently under investigation within the collaboration but are not included in the present analysis due to their sub-dominant impact relative to statistical uncertainties.

Ongoing efforts within the collaboration aim to extend the calibration of the detector response to higher energies, up to approximately 50~MeV, using samples of Michel electrons. This work will provide a more robust validation of the energy reconstruction and enable future analyses to extend the fit range beyond 10~MeV with improved confidence. 

In addition, there are ongoing studies utilizing events observed in the neutron wall to better characterize potential in-time backgrounds by extrapolating their contribution into the physics time window. While preliminary results suggest that this effect is small for the current dataset and selection criteria, it may become relevant for higher-statistics analyses and will be incorporated in future iterations of this work.

Analyses utilizing higher energy events and the full 2022 and 2023/2024 datasets, the latter of which has approximately twice the POT exposure of the former, will include both of these sources of systematic uncertainties as necessary. 

\subsection{Expected Sensitivity}
\begin{figure}[h]
  \centering
  \includegraphics[width=0.7\linewidth]{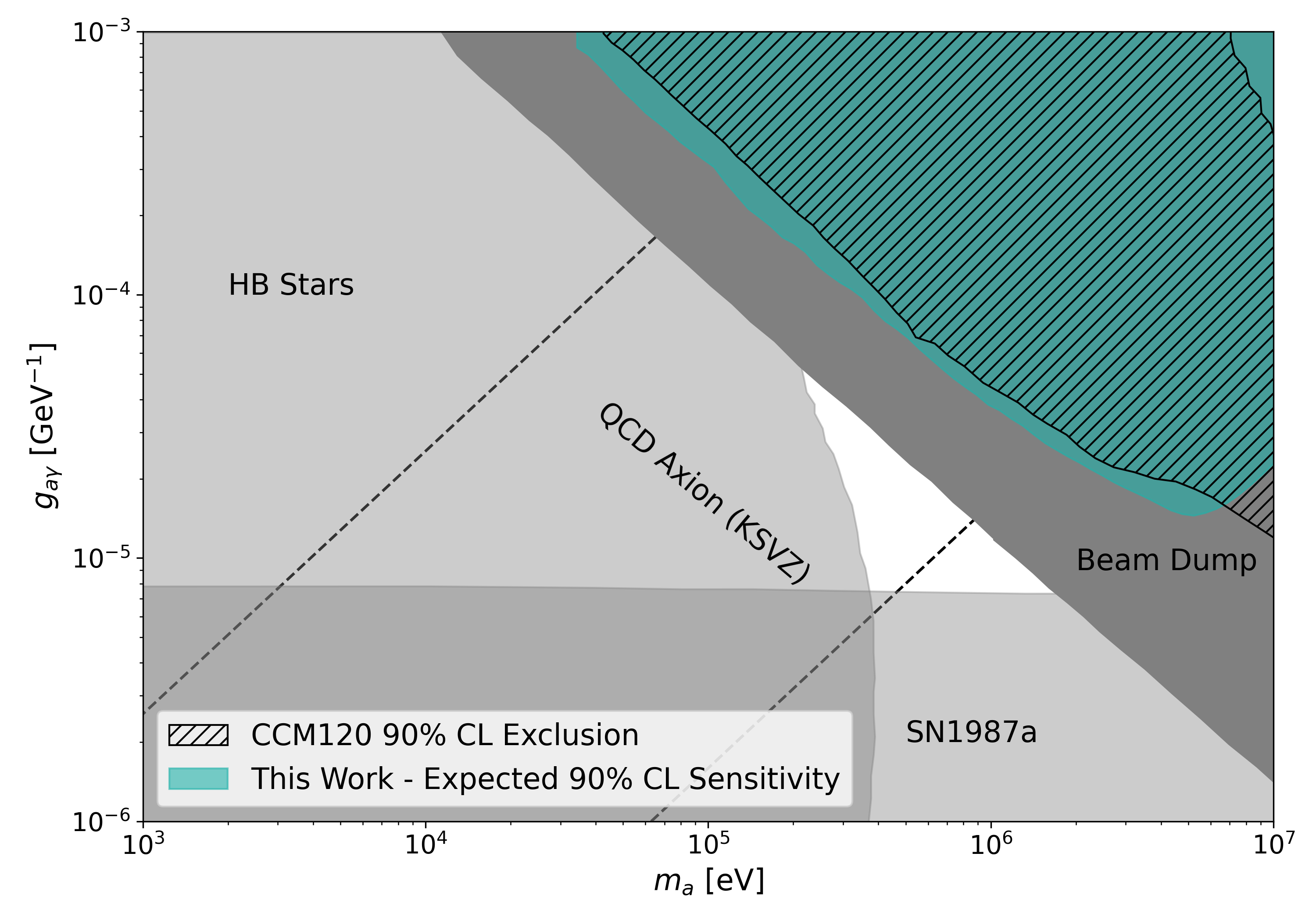}
  \caption{Expected sensitivity at 90\% confidence level of this work (cyan) and relevant existing experimental limits (gray).}
  \label{fig:expected_sensitivity}
\end{figure}

The expected sensitivity of this analysis is evaluated under the background-only hypothesis. In this approach, the prebeam data, after the full set of selection cuts, is used as a proxy for the observed dataset.

For each point in the ALP parameter space, defined by the mass and photon coupling $(m_a, g_{a\gamma})$, the full fitting procedure described previously is applied. This includes the construction of the likelihood function, incorporation of signal and background probability density functions in time, and profiling over all nuisance parameters associated with systematic uncertainties. The resulting likelihood is then compared to that of the null (background-only) hypothesis through a likelihood ratio test statistic.

The confidence intervals are derived assuming the validity of Wilks' theorem, which allows the test statistic to be interpreted in terms of a $\chi^2$ distribution in the asymptotic limit. Under this assumption, the 90\% confidence level exclusion contour is obtained by identifying the set of parameter points for which the test statistic exceeds the corresponding critical value.

The resulting expected sensitivity is shown in Fig.~\ref{fig:expected_sensitivity}. This curve represents the median exclusion that would be obtained in the absence of a signal, given the observed background rates and statistical fluctuations consistent with the prebeam dataset. The dashed lines represent the allowed parameter space for the KSVZ QCD axion model~\cite{KSVZ_1,KSVZ_2}.

Also shown on the figure are existing constraints in this region of parameter space. These are dominated by previous beam dump experiments, which probe similar production and decay mechanisms for ALPs and set strong bounds at comparable masses and couplings~\cite{JAECKEL2016482,doi:10.1142/S0217751X9200171X}. In addition, constraints from astrophysical observations are overlaid, including limits derived from energy loss arguments in horizontal branch (HB) stars and from the neutrino signal observed in the supernovae core collapse in 1987 (SN1987a)~\cite{PhysRevD.75.013004,PhysRevLett.93.171104,PhysRevLett.97.151802,PhysRevLett.98.131802,PhysRevLett.98.050402,PhysRevLett.99.121103,DeRocco:2020xdt,Masso_2005}. These astrophysical bounds provide complementary sensitivity, particularly at low masses and small couplings, though they rely on model-dependent assumptions about stellar environments and ALP interactions.

Despite utilizing approximately 70\% of the total POT compared to the previous CCM120 result~\cite{CCM:2021jmk}, this analysis is expected to achieve improved sensitivity across a significant portion of the parameter space. This gain is primarily driven by enhanced background rejection, made possible through the utilization of Cherenkov radiation in the detector. The improved particle identification techniques reduce the steady-state background contamination, thereby increasing the signal-to-background ratio and strengthening the overall exclusion reach.

\subsection{Fit Results}
The full likelihood fit is performed on the observed data including three nuisance parameters corresponding to the dominant sources of systematic uncertainty described above. For the null hypothesis, the fit yields a negative log-likelihood of 578.28 with a profiled background event rate of 11.74 events per ns, which is within the $1\sigma$ standard deviation bands.

For the signal hypothesis, the global best-fit point is found at $m_a = 0.18$~MeV and $g_{a\gamma} = 1.63 \times 10^{-4}~\text{GeV}^{-1}$, with a negative log-likelihood of 577.34. The corresponding profiled nuisance parameters are consistent with expectations: a total POT scaling of 0.99, a beam start time shift of $-584.52$~ns, and a background event rate of 11.67 events per ns. The small shifts in these nuisance parameters, all within their corresponding $1\sigma$ standard deviation bands, indicate that the fit does not require significant deformation of the nominal model to accommodate the data.

\begin{figure}[h]
  \centering
  \includegraphics[width=0.7\linewidth]{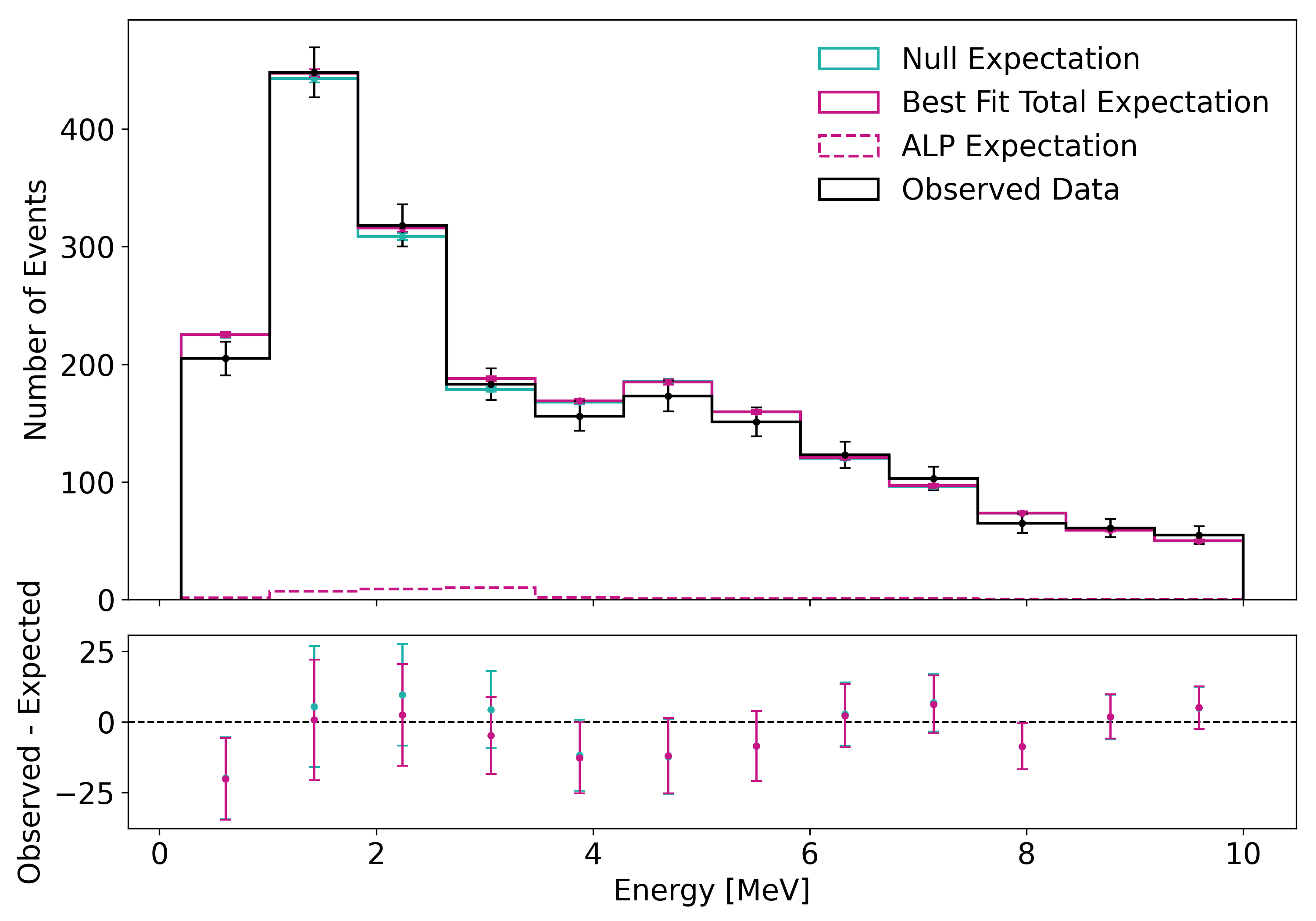}
  \caption{Energy distribution showing the observed data (black line), null hypothesis expectation (cyan line) and the best fit signal hypothesis expectation (magenta line). The bottom panel shows residual between the null and signal hypotheses with the observed data.}
  \label{fig:fitting_energy}
\end{figure}

\begin{figure}[h]
  \centering
  \includegraphics[width=0.7\linewidth]{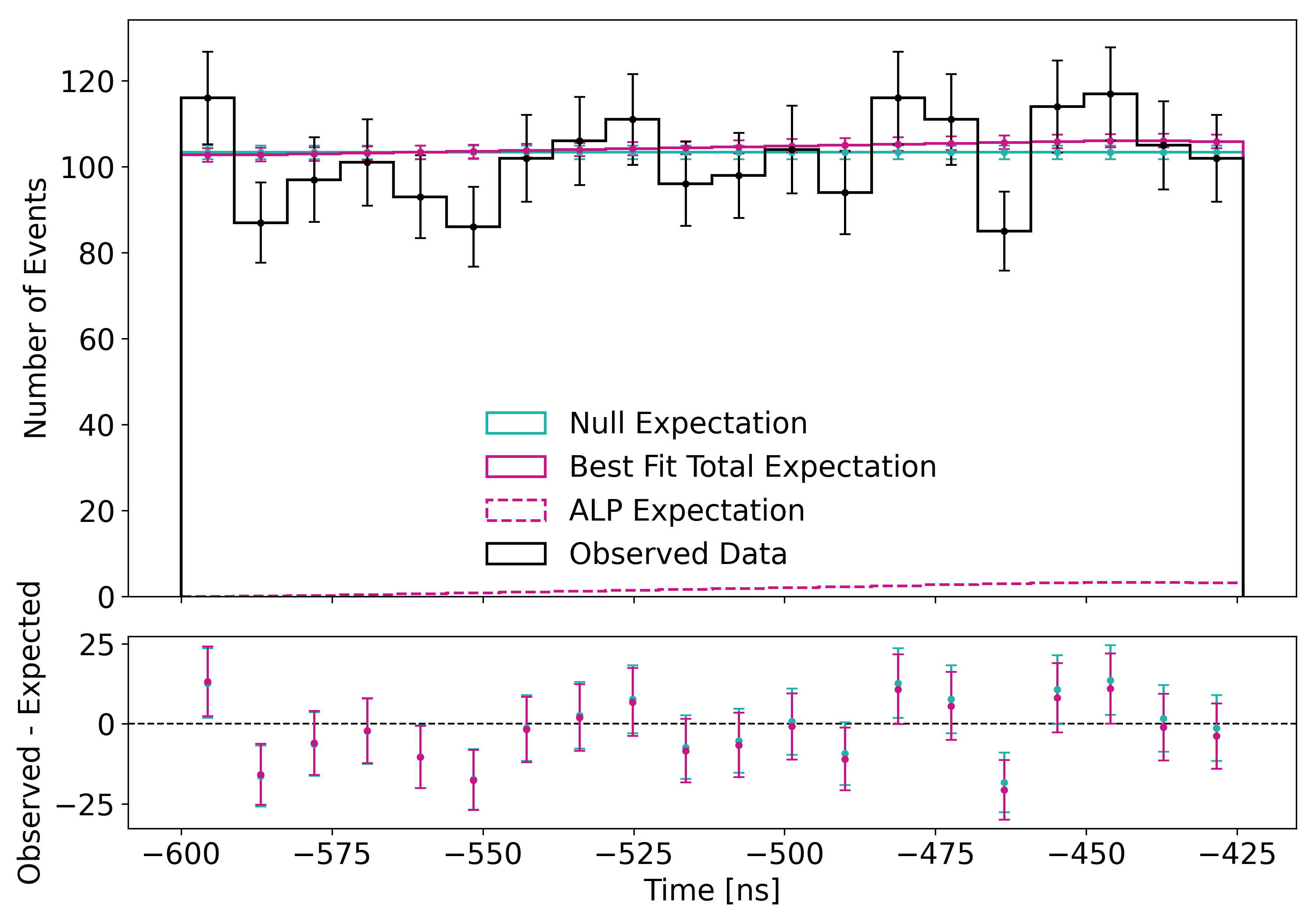}
  \caption{Time distribution of observed data, null expectation, and signal expectation. The bottom panel shows the residual between the observed data and the expectation.}
  \label{fig:fitting_time}
\end{figure}

The difference in negative log-likelihood between the signal and null hypotheses is $\Delta \text{NLL} = 0.94$. Interpreting this using Wilks' theorem with two degrees of freedom (corresponding to the signal mass and coupling) leads to a local significance of $0.86\sigma$. This level of significance is consistent with a statistical fluctuation, and therefore no evidence for an ALP signal is observed in the data.

Fig.~\ref{fig:fitting_energy} shows the one-dimensional projection of the energy distribution from this fitting procedure. The observed data are shown in black, while the cyan curve represents the background-only expectation and the magenta curve shows the best-fit signal-plus-background model. The dashed magenta line indicates the signal-only contribution. The preferred signal introduces a localized excess in the energy range between 1 and 3~MeV, although the statistical uncertainty in this region remains large. The residual panel illustrates the difference between the observed data and both hypotheses, demonstrating that the inclusion of the signal provides only a marginal improvement to the fit quality.

Fig.~\ref{fig:fitting_time} shows the corresponding one-dimensional time distribution. The background-only hypothesis follows a uniform distribution, as expected for the steady-state backgrounds. In contrast, the ALP signal hypothesis produces a characteristic triangular distribution, beginning at approximately $-584$~ns with a width of about 290~ns, reflecting the timing structure of the signal model. The observed data (black) does not exhibit a statistically significant preference for this structure, although small fluctuations are present.

\begin{figure}[h]
  \centering
  \includegraphics[width=\linewidth]{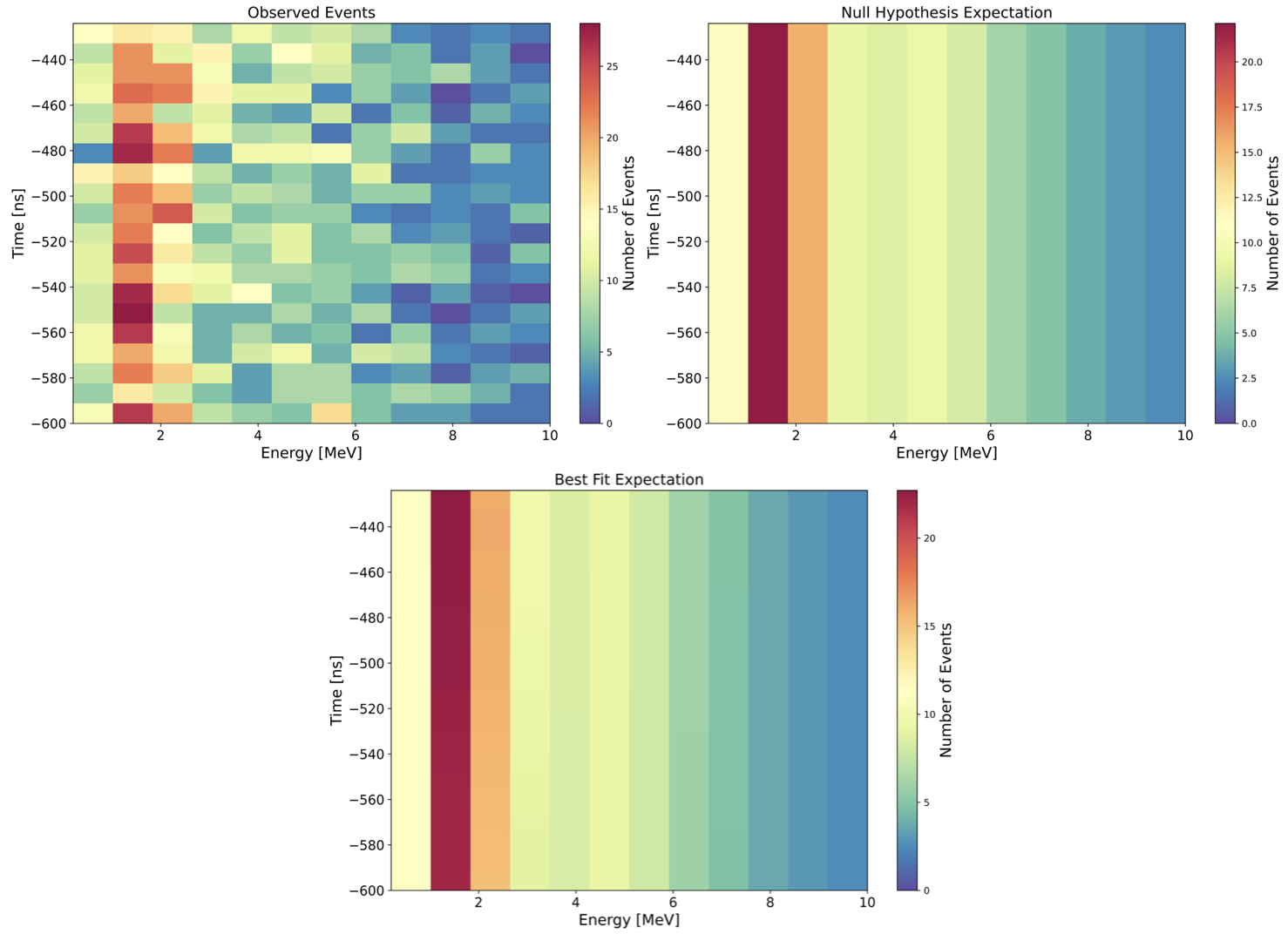}
  \caption{Two-dimensional energy and time distributions utilized in this fitting procedure. Upper left-hand side shows the observed data, upper right hand side is the null expectation, and lower plot is the best fit signal hypothesis.}
  \label{fig:fitting_energy_time}
\end{figure}

Fig.~\ref{fig:fitting_energy_time} presents the two-dimensional energy versus time distributions that form the basis of the likelihood fit. The observed data are shown in the upper left panel, the background-only expectation in the upper right, and the best-fit signal-plus-background model in the lower panel. The signal contribution introduces a correlated structure in energy and time that is not present in the background model; however, this feature is not strongly preferred by the data.

\begin{figure}[h]
  \centering
  \includegraphics[width=\linewidth]{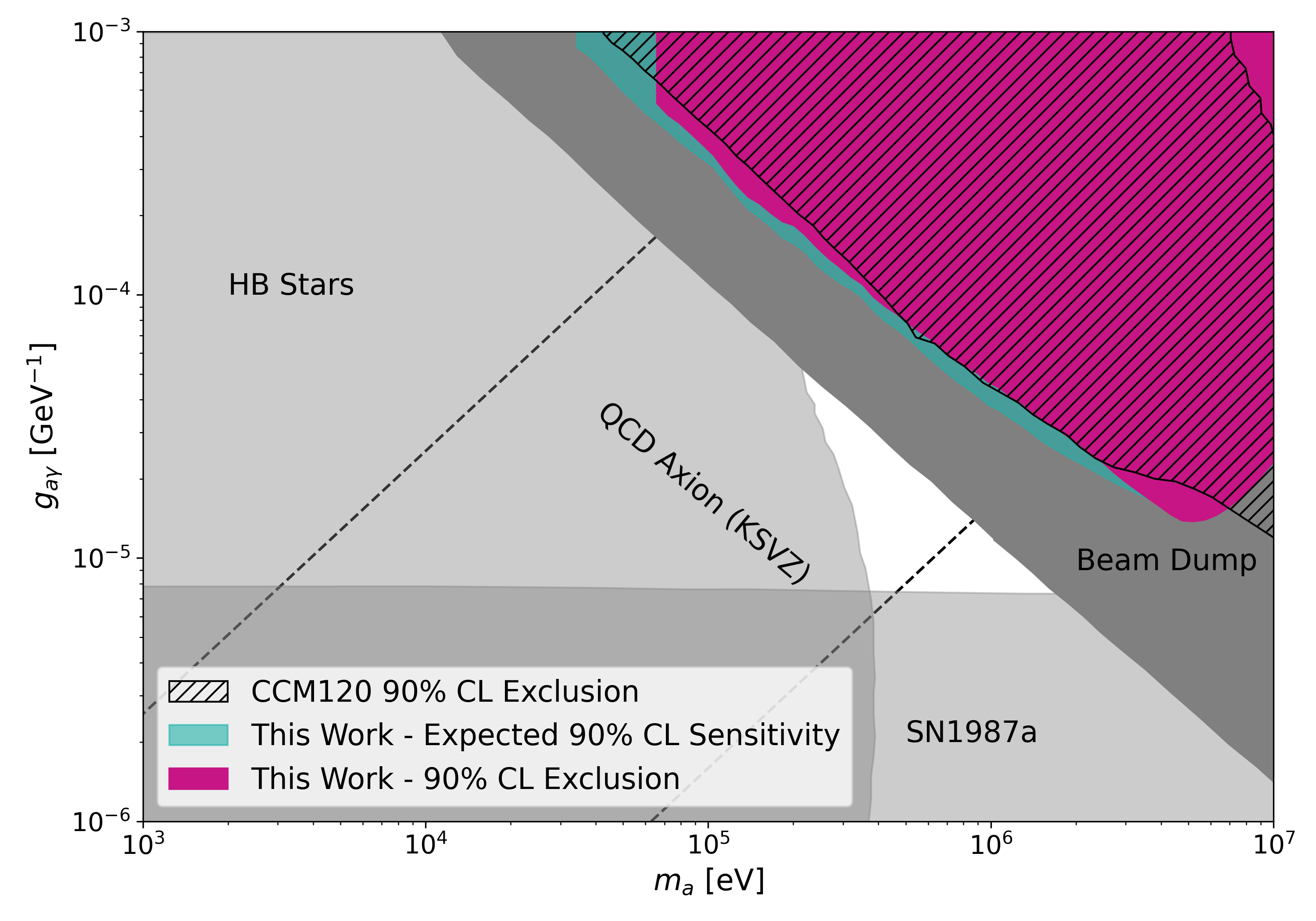}
  \caption{90\% confidence level excluded region in the ALP parameter space as a result of this analysis (magenta). This work does not find significant preference for ALP, allowing for determination of the excluded regions of parameter space at 90\% confidence level. This analysis is able to improve on the CCM120 excluded bounds despite 70\% the total POT because of the improved background rejection techniques leveraging Cherenkov radiation.}
  \label{fig:actual_sensitivity}
\end{figure}

Fig.~\ref{fig:actual_sensitivity} shows the resulting exclusion limits from this analysis, along with the expected sensitivity. The observed limits (magenta) are broadly consistent with the expected sensitivity, indicating that the observed data behaved as anticipated under the background-only hypothesis. Despite less total POT exposure, this result extends the exclusion of parameter space beyond previous CCM120 constraints, demonstrating the improved sensitivity achieved in this analysis through the leveraging of Cherenkov radiation to increase background suppression.

\section{Future Prospects}
This ALP analysis represents the first demonstration of a physics search using a hybrid Cherenkov radiation and scintillation optical detector in a beam dump environment. By combining four complementary observables, sensitive to differences in timing, wavelength, and directionality between scintillation and Cherenkov light, this analysis achieves robust identification of electromagnetic-like events while maintaining strong background rejection.

The present work builds directly on the $^{22}$Na calibration studies presented in this thesis and is therefore most sensitive to lower-energy signals. Ongoing efforts within the collaboration aim to extend this calibration program using Michel electrons, which provide a well-understood higher-energy standard candle. Incorporating these datasets will enable future analyses to expand the accessible energy range and improve sensitivity to higher-mass ALP parameter space.

Looking ahead, the demonstrated ability to disentangle scintillation and Cherenkov components opens a broader path for precision event reconstruction in liquid argon detectors. With improved calibrations and increased statistics, this approach can be extended beyond ALP searches to a wider class of rare-event and neutrino measurements, enhancing both signal identification and background discrimination.

\chapter{Phenomenological Study of Supernovae Neutrinos at DUNE}\label{chap:pheno}
In addition to work on the CCM experiment, this thesis engages with the broader liquid argon neutrino physics program through a phenomenological study complementary to the Deep Underground Neutrino Experiment (DUNE)~\cite{DUNE:2020ypp}.

The use of liquid argon in the CCM experiment was driven by specific physics interests discussed in Chapter~\ref{chap:intro}. It was also driven by overlap with the larger US Liquid Argon Program that makes use of liquid argon time projection chamber (LArTPC) detectors. LArTPC detectors instrument liquid argon with both charge collection planes and photo-sensors. By applying a strong electric field across the active volume, ionization electrons created when charged particles pass through the argon are drifted to the charge collection planes. Typically, these planes are based on wire readouts that provide charge collection in two dimensions. Additionally, LArTPCs include a photo-detection system which provides precise event timing through the very prompt scintillation light. Together, both readouts allow for full reconstruction of event topology in the three spatial dimensions with $\mathcal{O}(\text{mm})$ resolution. 

The US LArTPC program is a sequential plan working toward the establishment of the DUNE experimental program. DUNE combines an intense proton beam at Fermi National Accelerator Laboratory (Fermilab) in Batavia, Illinois with a complex of both near and far detectors. This beamline will create a very intense source of neutrinos, which are constrained using measurements at a near detector complex at Fermilab. The beam of neutrinos is directed approximately 1,300~km across the earth and 1,500~m underground towards Sanford Underground Research Laboratory (SURF) in Lead, South Dakota. The far detector site at SURF will be instrumented with a complex of four ultra large LArTPC detectors, with approximately 40~kton of total fiducial mass. 

The primary goal of DUNE is accelerator-based measurements of long-baseline neutrino oscillations parameters, including the atmospheric mass splitting $\Delta m^2_{31}$, neutrino mass ordering, and $\delta_{CP}$. However, the beam is delivered in very short bursts during the run-periods, which will last less than one full year. The result is that there is substantial time in which this detector can be used for other physics. A priority for the US community is to develop the phenomenology that can lead to new measurements not dependent on the beamline. To this end, this work investigates DUNE’s expected sensitivity to the total flux of neutrinos produced from the core-collapse of supernovae through the neutral current channel.

While the neutral current interaction channel has a cross section roughly two orders of magnitude smaller than that of charged current interactions, it provides the only flavor-agnostic probe in liquid argon. As such, it serves as a critical complement to charged current measurements on both argon and water, which primarily constrain $\nu_e$ and $\overline{\nu}_e$ flavors of supernovae neutrinos.

The results presented here, published in Ref.~\cite{Newmark:2023vup} and reprinted below, build on calculations of the incoherent neutral current neutrino cross section on $^{40}$Ar~\cite{Tornow:2022kmo}. In this process, neutrinos of any flavor can excite argon nuclei, producing de-excitation gamma rays. This study evaluates the ability of the DUNE far detector photon detection system to resolve these MeV-scale signals, thereby opening a new channel for supernova neutrino detection. Combined with charged current constraints on $\nu_e$ and $\overline{\nu}_e$, the neutral current channel enables an estimated $\sim$30\% resolution on the $\nu_x$ temperature.



# Sensitivity to supernovae average $\nu_x$ temperature with neutral current interactions in DUNE

Darcy A. Newmark[1] and Austin Schneider[2,1]

[1]*Department of Physics, Massachusetts Institute of Technology, Cambridge, Massachusetts 02139, USA*
[2]*Los Alamos Laboratory, Los Alamos, New Mexico 87545, USA*



We explore a novel method for measuring the time averaged temperature of the $\nu_x$ component in type-II core-collapse supernovae. By measuring neutral current incoherent neutrino-argon interactions in DUNE we can obtain spectral information for the combination of all active neutrino species. Combining this all-neutrino spectral information with detailed charged current measurements of the electron neutrino and electron antineutrino fluxes from DUNE and Hyper-Kamiokande, we can infer the time averaged temperature for the remaining neutrino species in the $\nu_x$ component to within a factor 2 for most cases and to 30% for a small range of time averaged $\nu_x$ temperatures. Because of the limited energy range of the emitted photons from incoherent neutral current interactions on argon, the $\nu_x$ temperature reconstruction demonstrates a degeneracy in the one and two sigma credible regions. Furthermore, while large uncertainties on the neutral current (NC) cross section penalize this measurement, we examined the efficacy of constraining NC cross section uncertainties on improving $\nu_x$ measurements. We found that if additional measurements of $B(M1\uparrow)\ 1^+$ excited state transitions in argon are able to reduce correlated cross section uncertainties from 15% to 7%, the size of the $1\sigma$ allowed regions for $T_{\nu_x}$ becomes sample size limited, and approaches the case where there are no uncertainties on the cross section.



## I. INTRODUCTION

Type-II core-collapse supernova events release approximately $10^{53}$ ergs of energy, carried away almost exclusively by neutrinos and antineutrinos of all flavors [1]. These neutrinos travel nearly unimpeded to Earth, where they can be detected. In 1987, three neutrino detectors on Earth measured 20 events within 13 seconds, which coincided with the release of energy from the collapse of SN1987A [2–4]. If a core-collapse supernova event were to occur within our Galaxy today, the many operating neutrino detectors would provide precise measurements of neutrino and antineutrino energy spectra. Such measurements are crucial for reconstructing the initial conditions of supernovae just before collapse, understanding their evolution, and modeling their collapse.

Large detectors such as the Deep Underground Neutrino Experiment (DUNE) and Hyper-Kamiokande (Hyper-K) will provide large sample measurements of $\nu_e$ and $\bar{\nu}_e$ charged current (CC) interactions, respectively, with excellent energy resolution [5,6] However, the remaining neutrino species, summarized as $\nu_x$, are below threshold for CC interactions and can only be measured through neutral current (NC) interactions. While there are many theoretical models of supernova collapse, they vary greatly in the prediction and treatment of $\nu_x$ [7–9]. With little existing experimental information and large uncertainties in the model space, measurements of NC neutrino interactions will be crucial for understanding the $\nu_x$ component of core-collapse supernovae.

In this paper, we explore how DUNE's observation of gamma rays generated by NC interactions of supernova neutrinos can be used in conjunction with CC observations to infer the time averaged temperature of the supernova $\nu_x$ component. NC interactions of all supernova neutrino species in DUNE will excite argon nuclei, which subsequently emit a gamma ray at one of several known energies, providing a weak correlation between incident neutrino energy and the emitted gamma-ray energies. We then combine the flavor-agnostic measurement of this NC gamma-ray spectrum with CC spectral measurements of the $\nu_e$ and $\bar{\nu}_e$ components to infer the temperature of the $\nu_x$ component. We also explore how uncertainties in the NC cross section impact sensitivity to the $\nu_x$ temperature, using recently updated calculations of the NC neutrino-nucleus cross section and its associated uncertainties [10].

## II. NEUTRAL CURRENT MEASUREMENTS

In order to examine DUNE's sensitivity to supernova $\nu_x$ temperature reconstruction, we must first consider the signatures of the NC interaction channel.

In the incoherent NC interaction, argon nuclei are excited via interaction with a neutrino. As the nucleus deexcites, $\gamma$ rays are emitted at specific energy levels. The magnetic dipole strengths of these $1^+$ excited state transitions are displayed in Fig. 1. These measurements, combined with shell model calculations, guide neutral current cross section predictions and dictate experimental reconstruction of incident neutrino spectra.

NC neutrino-argon interactions are difficult to measure due to the small cross section of this channel and low energy at which the interactions occur [10]. For supernova neutrino energies, the NC interaction cross section is at least 2 orders of magnitude smaller than the corresponding CC cross section, given by [11]. For this reason, projected measurements of supernova neutrinos tend to focus on CC channels.

While NC measurements are difficult to make compared to the CC channel, they provide the only source of information about the neutrino species of $\nu_x$. Flavor nonspecific information from supernova neutrinos is crucial to understanding the neutrino flux emitted without uncertainties related to neutrino oscillations, matter effects, or neutrino self-interactions that occur within the supernovae. Combining flavor nonspecific NC measurements with $\nu_e$ and $\overline{\nu_e}$ CC measurements, we can indirectly probe the properties of the supernova $\nu_x$ flux.

Recent developments in detector technology will allow us to overcome this challenge with ultralarge detectors that are sensitive to these low energy NC interactions. With 40 kt of combined fiducial mass and energy resolution down to the MeV scale, the DUNE far detector (FD) modules will be an ideal candidate to study the $\nu_x$ component, providing the largest sample of NC neutrino-argon interactions [12]. DUNE will be able to separate CC from NC interactions using the distribution of deposited energy as a function of interaction length within the detector. The DUNE FD modules are planned to have a high efficiency trigger system down to the few MeV threshold, with good energy resolution, and moderate spatial resolution [12,13]. With these anticipated capabilities, DUNE will be able to resolve photon emissions from NC interactions in the energy range of interest for this analysis.

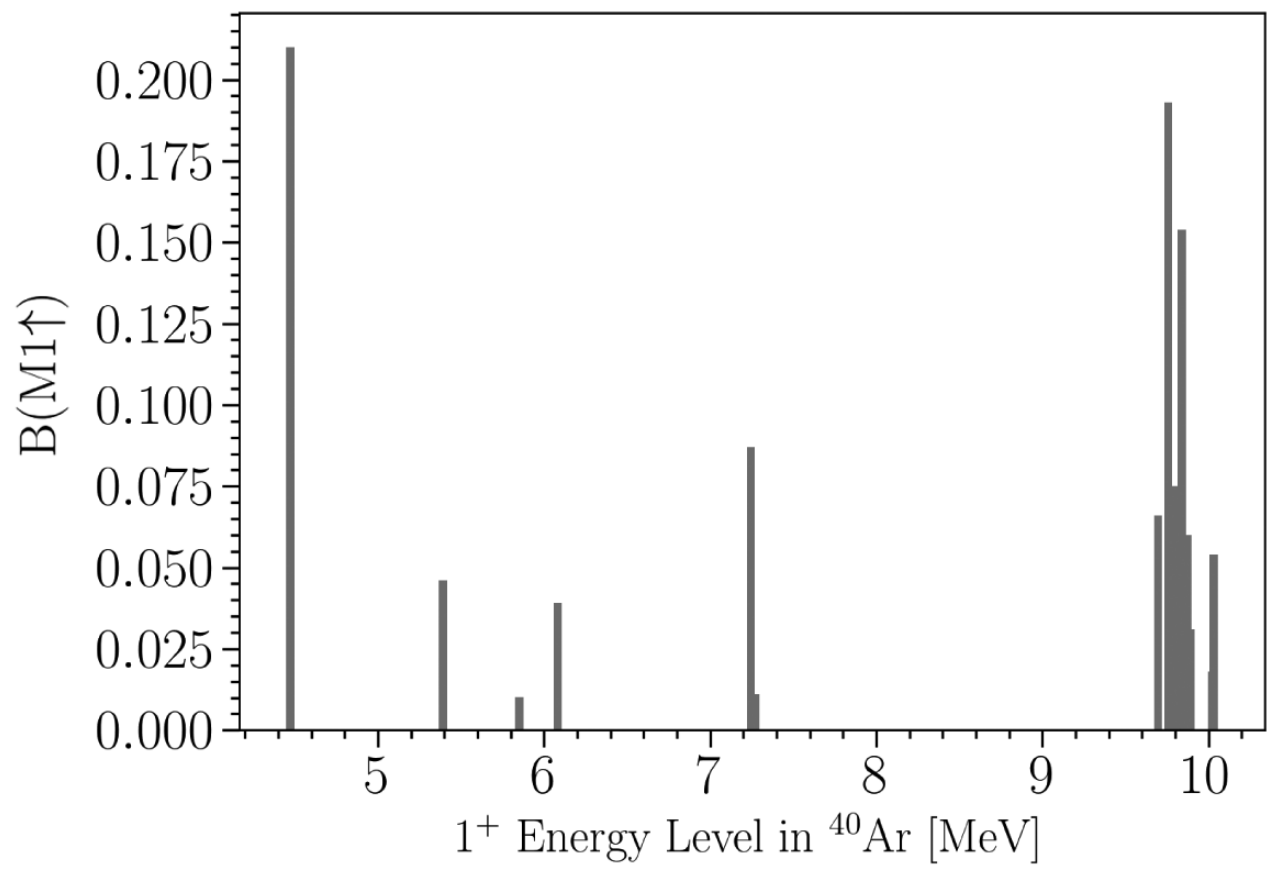


FIG. 1. B(M1↑) neutral current $1^+$ transition strengths for $^{40}$Ar. The B(M1↑) value for each $1^+$ is shown as a function of the transition energy. Fourteen of these 1+ transitions exist in total.

While this work is directly relevant to DUNE, SBND, ICARUS and other liquid argon (LAr) detectors, supernova neutrino interaction measurements on other nuclei will provide further insights into the $\nu_x$ spectrum. The Jiangmen Underground Neutrino Observatory (JUNO) experiment will be able to observe incoherent neutral current neutrino interactions on carbon nuclei, with a predominant excitation state of 15.1 MeV [14]. With a 20 kt fiducial mass of liquid scintillator and excellent energy resolution of $3\%/\sqrt{E(\mathrm{MeV})}$, JUNO is another large neutrino detector that will be sensitive to the NC channel from supernova neutrinos [14]. Given the different photon excitation energies of NC neutrino-nucleus interactions on argon and carbon, a combined analysis of NC neutrino interactions on $^{40}$Ar and $^{12}$C would be sensitive to greater parameter space than either individual analyses. This combined analysis of data from DUNE and JUNO could provide a more robust and more sensitive measurement of $\nu_x$ supernova neutrinos than either individual analysis.

In addition to NC measurements, DUNE and Hyper-K will provide leading constraints on the $\nu_e$ and $\overline{\nu_e}$ spectrum through CC measurements on liquid argon and water, respectively. By incorporating DUNE CC, DUNE NC, and Hyper-K CC measurements in a combined analysis, we can make definitive statements about the $\nu_x$ component of the neutrino flux. Although sample sizes for NC events are much smaller, and energy resolution is poorer, the information we obtain from NC interaction on argon nuclei in DUNE provides enough discriminating power to reconstruct the time averaged $\nu_x$ temperature.

## III. METHODOLOGY

Modeling the expected neutral current event rate requires (1) the supernova neutrino spectrum, (2) the neutrino-nucleon NC interaction cross section, (3) the detector response of DUNE's far detector modules, and (4) an event selection.

For the supernova neutrino spectrum, we assume a representative supernova neutrino flux with a progenitor mass of ten solar masses, a distance of 10 kpc, and a Fermi-Dirac energy spectrum [15] given by

$$\frac{dF_\nu}{d\varepsilon_\nu} = \frac{L_\nu}{4\pi D^2 T_\nu^4 F_3(\eta)} \frac{\varepsilon_\nu^2}{e^{\beta(\varepsilon_\nu - \mu)} + 1}, \tag{1}$$

where $\varepsilon_\nu$ is incident neutrino energy, $L_\nu$ is the luminosity for each neutrino species of SN1987A integrated over total time of collapse for a total energy of approximately $3\times10^{53}$ ergs, $D$ is the distance from Earth to the supernova, $T_\nu$ is the time averaged temperature of the neutrino species, $\mu$ is the chemical potential, $\beta = 1/T_\nu$, and $\eta = \mu/T_\nu$ (the Boltzmann constant is set to unity). $F_3(\eta)$ is defined by

$$F_3(\eta) \equiv \int_0^\infty \frac{x^3}{e^{x-\eta}+1}dx, \tag{2}$$

and in this work $\eta$ is set to 0. Figure 2 provides an example of this calculation, setting $T_{\nu_e} = 3.3$ MeV, $T_{\overline{\nu_e}} = 4.6$ MeV, and $T_{\nu_x} = 6.4$ MeV. Because the NC interaction is flavor independent, we compute the fluence assuming a no neutrino oscillation scenario. In this work we neglect further modification of the spectrum from a pinching parameter [1,7–9,16] because such features in the energy spectrum will not be resolvable in the $\nu_x$ component through NC measurements.

For the neutrino-nucleon NC interaction cross section, we use the most recently published NC cross sections from W. Tornow *et al.* [10]. Figure 3 shows this cross section for neutrinos and antineutrinos with 43% theoretical uncertainty. We model the NC cross section uncertainties as a 15% correlated uncertainty within each excitation mode, corresponding to the uncertainty on B(M1↑) measurements, and an additional 40% uncorrelated uncertainty across the entire energy spectrum to account for the theoretical uncertainties between different choices of shell-model calculations. Figure 4 shows an example fractional covariance matrix used in this calculation.

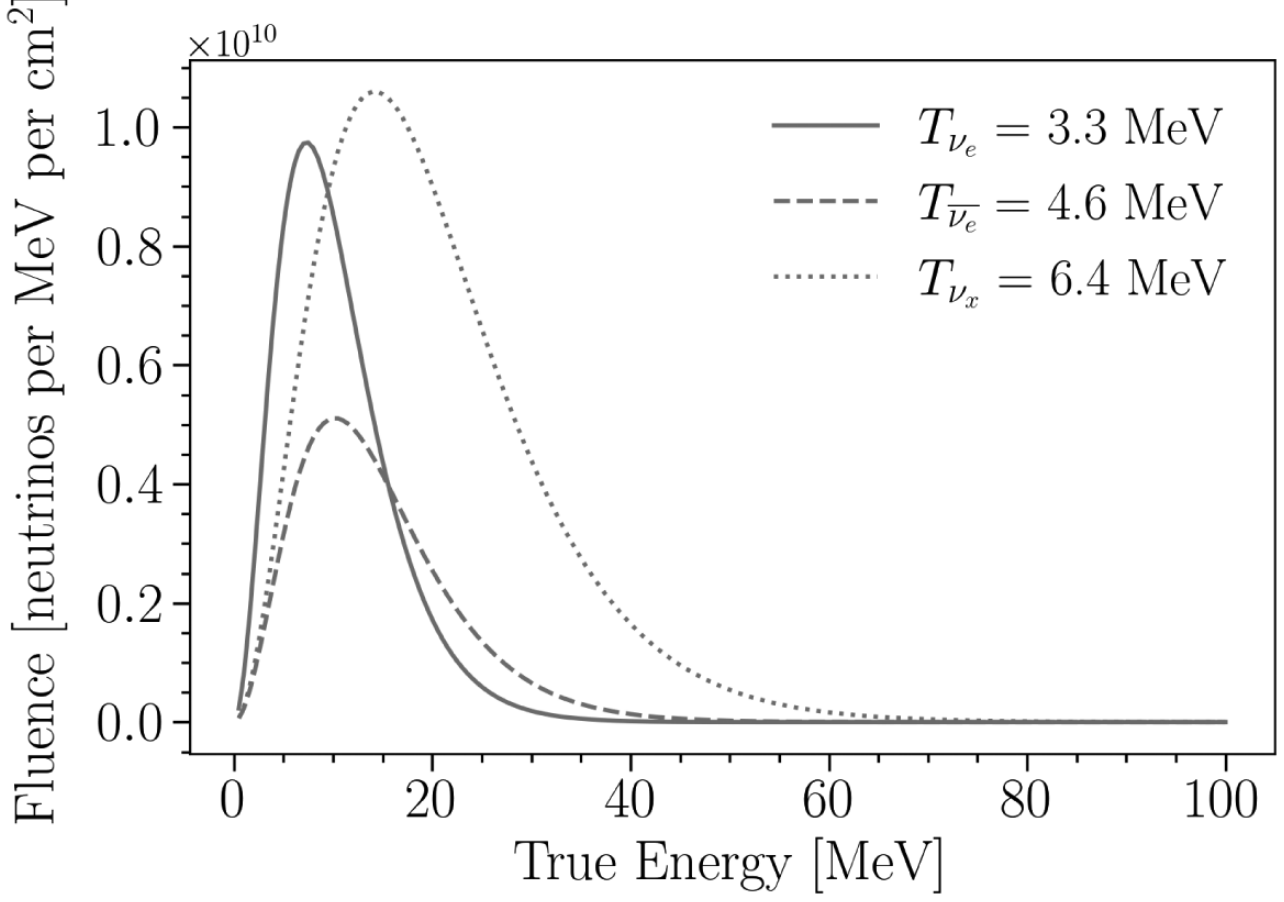


FIG. 2. Supernova neutrino differential fluence for a Fermi-Dirac spectrum. The differential fluence is shown as a function of neutrino energy for each of the three neutrino flux components, assuming a Fermi-Dirac energy distribution, and the baseline supernova scenario described in Sec. III.

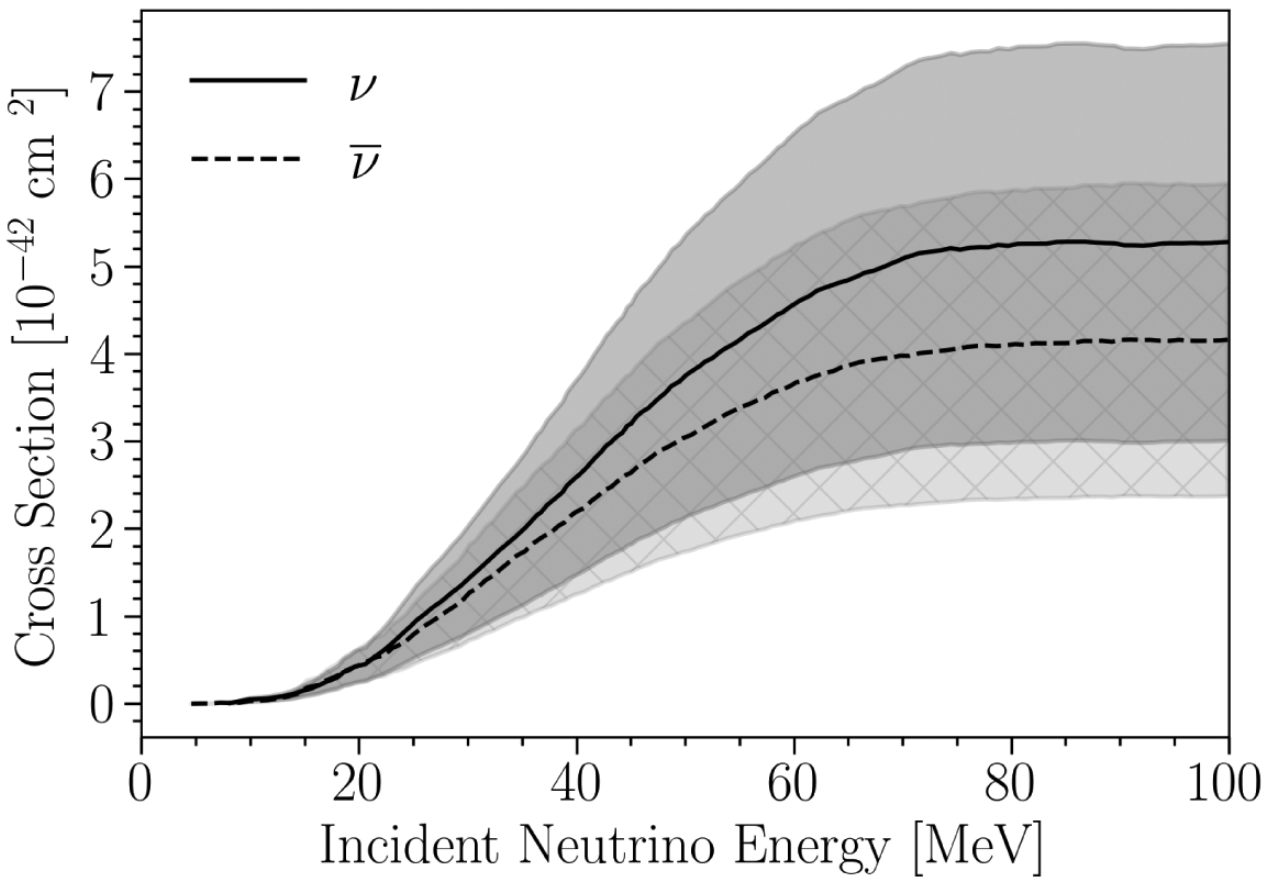


FIG. 3. Incoherent NC neutrino-argon cross sections and uncertainties. The NC neutrino-argon cross sections derived in [10] are shown for $\nu$ and $\bar{\nu}$ as a function of incident neutrino energy with bands showing their associated 43% theoretical uncertainties. The solid shading shows $\nu$ cross section uncertainties, and hatched shading shows the same for $\bar{\nu}$.

For the detector response of DUNE's far detector modules, we assume the 15% photon energy resolution expected in DUNE's Time Projection Chambers (TPCs) at the energies of relevance for supernova neutrino measurements [12]. We also note that variation in the detectors response to NC neutrino interactions of different incident neutrino energies predominantly arises from the threshold of different nuclear excitations and changes in the branching ratio of these same excitations. Using this predicted energy resolution and weighting it by the cross sections of the individual photon excitation modes, Fig. 5 shows the response matrix for the $\nu$ component of the NC cross section.

Finally, for the event selection, we consider the relevant background processes for NC events and choose

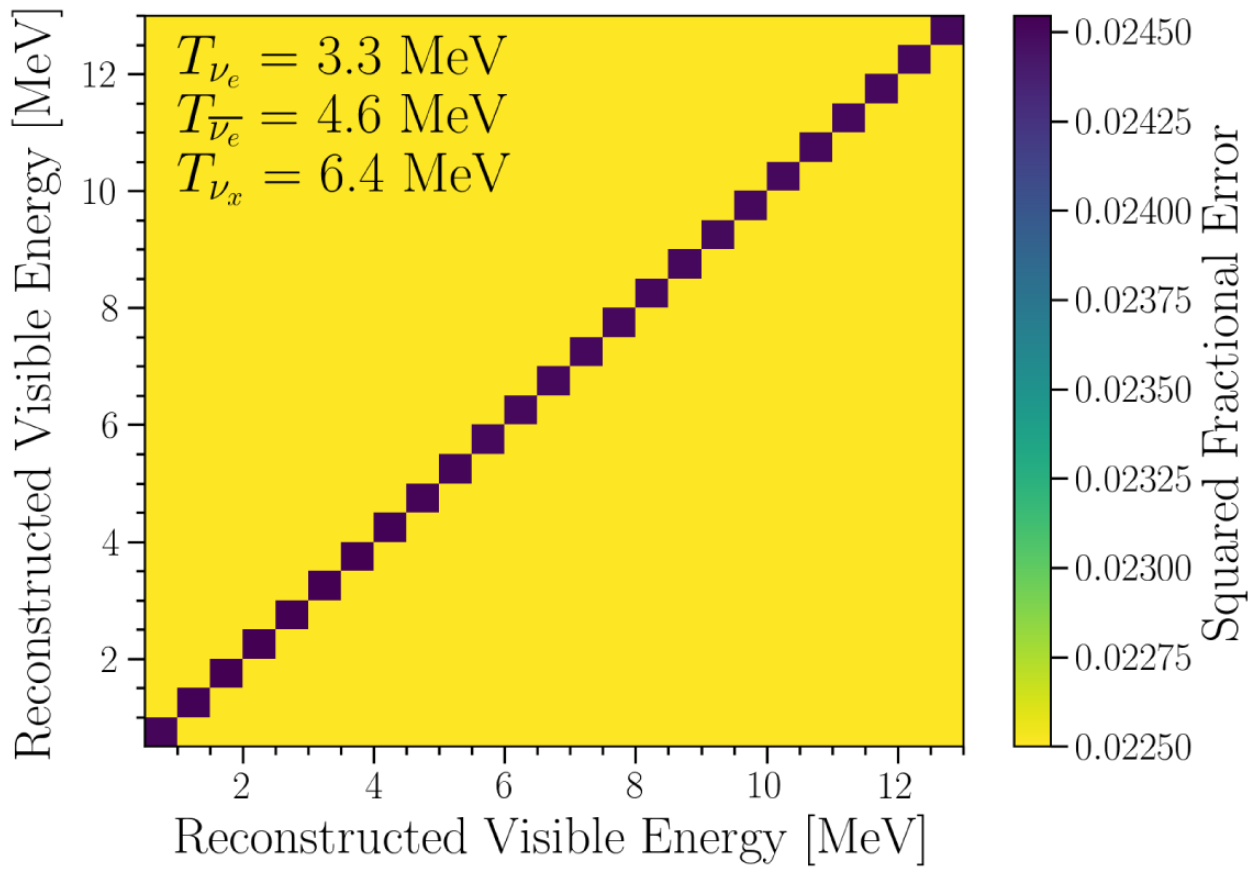


FIG. 4. DUNE FD NC fractional covariance matrix. The uncertainties on the cross section are composed of 15% correlated uncertainties and 40% uncorrelated uncertainties, as described in [10].

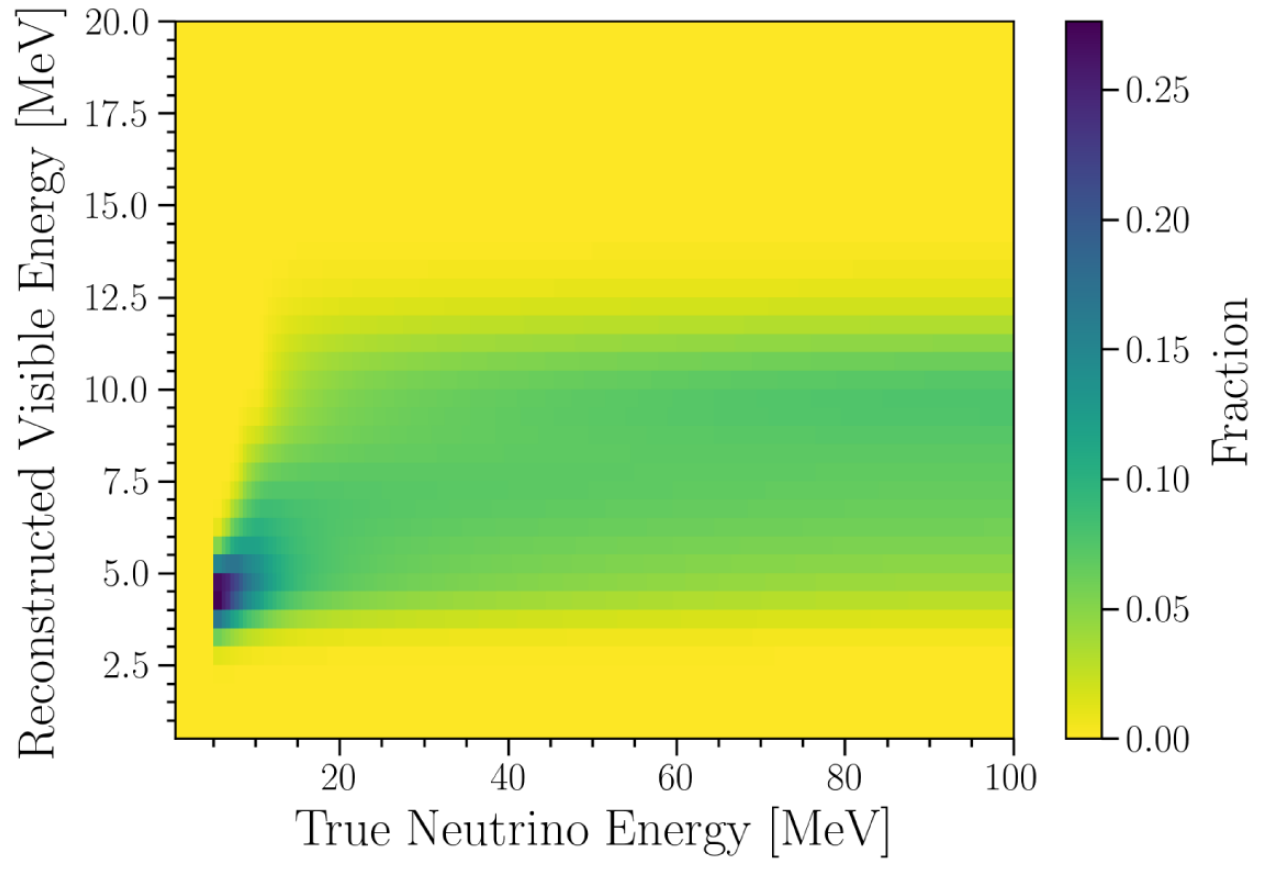


FIG. 5. DUNE FD NC response matrix. Response matrix for the $\nu$ component of the incoherent NC neutrino interaction in LAr assuming 15% energy resolution for the DUNE FD TPC as described in [12].

appropriate cuts to model the signal and background efficiencies. In the DUNE far detector modules, the primary background for NC interactions from supernova neutrinos is CC interactions from the $\nu_e$ neutrino flux. These CC interactions mostly occur at energies above the energy of the photons from NC interactions, and so we restrict our discussion of the CC background to reconstructed visible energies between 0.5 and 13 MeV. The electrons produced in CC interactions can be further differentiated from the photons produced in NC interactions using the fine grained information available from the LAr TPC. For simplicity we assume that most of this selection power is captured in the dE/dx distribution. Figure 6 shows the simulated dE/dx distributions, described in [12], scaled to the expected event rates of the NC and CC channels within the reconstructed energy region of interest between 0.5 and 13 MeV for $T_{\nu_e} = 3.3$ MeV, $T_{\overline{\nu_e}} = 4.6$ MeV, and $T_{\nu_x} = 6.4$ MeV. We found that a single cut at dE/dx $= 3.1$ MeV/cm includes 95.26% of NC signal events, while rejecting all but 15.83% of background CC interactions from the $\nu_e$ supernova neutrino flux.

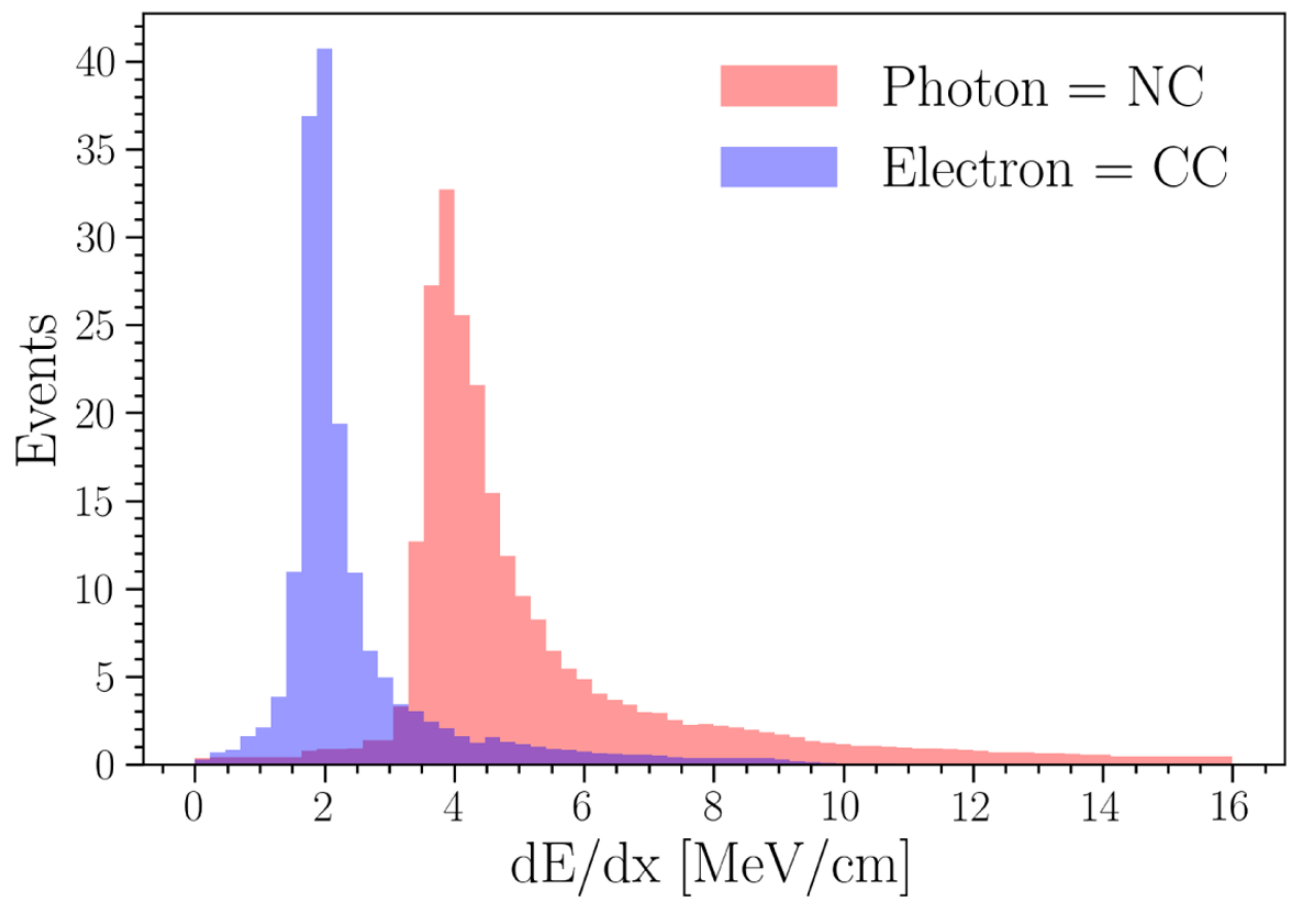


FIG. 6. DUNE dE/dx distribution for electron and photonlike final states. Distribution is normalized to the expected NC and CC event rates for a representative sample where $T_{\nu_x} = 3.3$ MeV, $T_{\overline{\nu_e}} = 4.6$ MeV, and $T_{\nu_x} = 6.4$ MeV.

Figure 7 shows an example of the expected reconstructed event rate in DUNE FD, after combining the fluence for the specified neutrino temperatures, NC cross section, response matrix, and efficiency cut. This plots shows the expected events for $T_{\nu_e} = 3.3$ MeV and $T_{\overline{\nu_e}} = 4.6$ MeV while $T_{\nu_x}$ varies between 1 and 30 MeV. The dashed black line represents the CC events due to the $\nu_e$ component of the supernova neutrino flux in DUNE. Although the NC channel is suppressed compared to the CC channel, DUNE's large detector size combined with the high fluence of expected neutrinos for a supernova collapse provides us with a large enough sample to perform spectral analysis.

With the modeling of DUNE's NC measurements complete, we must now combine this flavor-agnostic information with the CC measurements of the $\nu_e$ and $\overline{\nu}_e$ temperatures from DUNE and Hyper-K, respectively. This combination of three measurements will allow us to infer the temperature of the $\nu_x$ component. DUNE TPCs are predicted to resolve energies in the $\nu_e$ CC region of interest to around 15% [12]. Using this energy resolution to calculate the CC response matrix, $\nu_e$ fluence described in Eq. (1), and CC cross section [11], we calculate the predicted CC event rate as a function of $\nu_e$ temperature in DUNE and are able to reconstruct injected temperatures to $1\sigma$ credible region with 0.5% fractional error. We expect that Hyper-K will be able to reconstruct the $\overline{\nu_e}$ temperature to similar precision, if not better, given their similar energy resolution [17] as the DUNE TPC, greater fiducial mass [18] than the DUNE FD modules, and greater cross section for inverse beta decay events below ~15 MeV than $\nu_e$ CC events on argon [19].

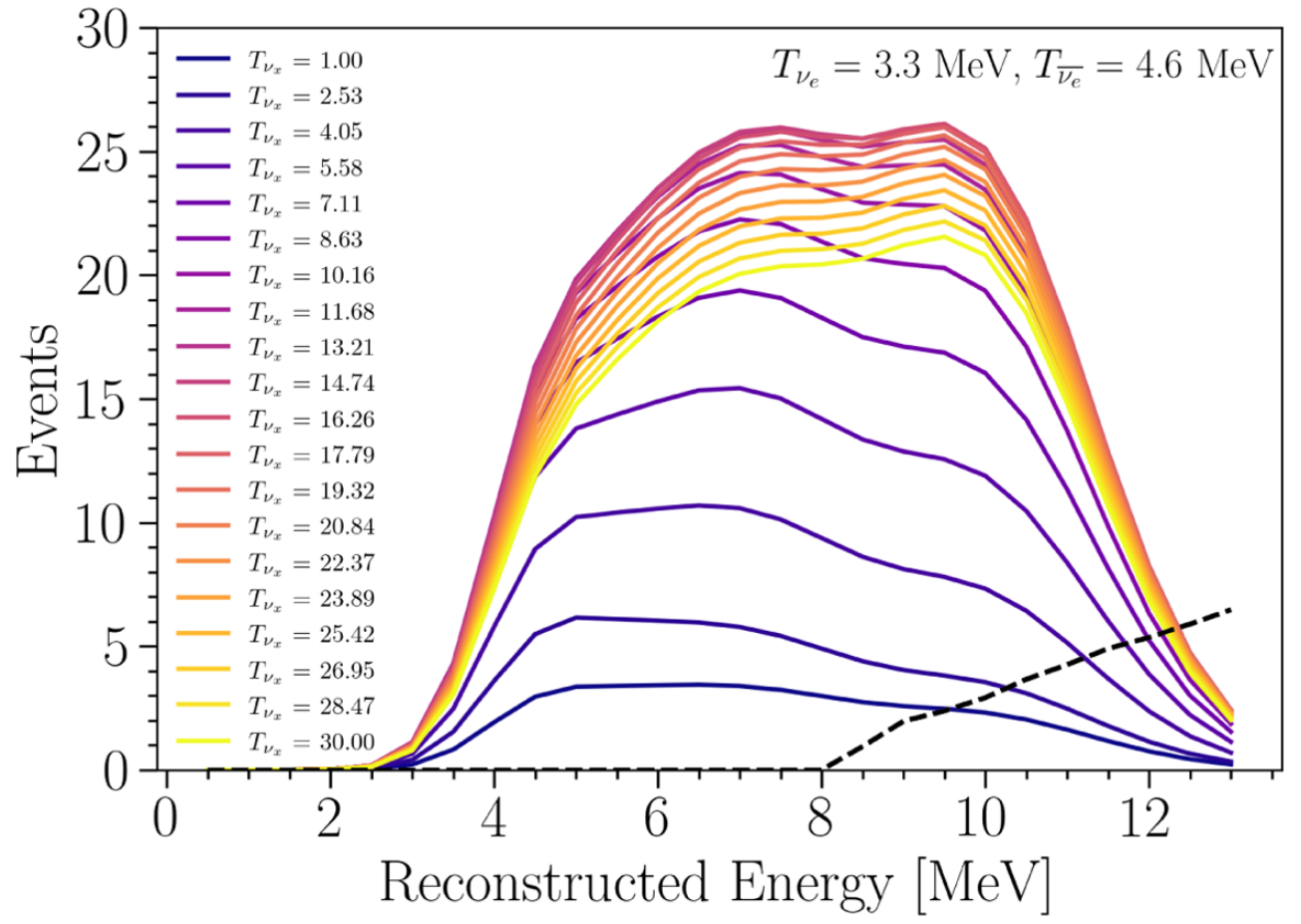


FIG. 7. Expected NC event rate in DUNE FD as a function of $\nu_x$ temperature. $\nu_e$ and $\overline{\nu_e}$ temperatures are fixed to 3.3 and 4.6 MeV, respectively. The black dashed line represents events due to CC interactions in the DUNE FD from the $\nu_e$ component of the fluence.

## IV. STATISTICAL METHODS

With the methods described above, we can now predict the NC event rates in DUNE for any combination of temperatures of the three neutrino flux components. The NC data from DUNE provides information about the total neutrino flux, however, to obtain sensitivity to the $\nu_x$ temperature we must simultaneously constrain the temperature of the $\nu_e$ and $\overline{\nu_e}$ components. We construct the binned-log-likelihood comparing expected NC DUNE data to our predictions, and combine this with external statistical constraints on $\nu_e$ and $\overline{\nu_e}$. This combination produces a constrained log likelihood given by

$$\mathcal{L}(\vec{T}) = \left[\prod_i \frac{\lambda_i(\vec{T},\alpha_i)^{k_i} e^{-\lambda_i(\vec{T},\alpha_i)}}{k_i!}\right]\mathcal{P}(T_{\nu_e})\mathcal{P}(T_{\overline{\nu_e}})\mathcal{P}(\vec{\alpha}), \quad (3)$$

where $\lambda_i(\vec{T},\alpha_i)$ is the expected number of events in bin $i$ for the combination of temperatures $\vec{T}$, and $k_i$ is the number of observed data events in bin $i$. For the two external constraints, $\mathcal{P}(T_{\nu_e})$ and $\mathcal{P}(T_{\overline{\nu_e}})$, we assume normally distributed constraints centered on the true values of the temperatures with widths that correspond to the uncertainties on $\nu_e$ and $\overline{\nu_e}$ temperatures that we expect to obtain from the CC measurements of DUNE and Hyper-K, conservatively assumed to be 0.5%. To model the cross section uncertainties, we introduce the parameters $\vec{\alpha}$ which fractionally modify the expected number of events in each bin, such that $\lambda_i(\vec{T},\alpha_i) = (1+\alpha_i)\cdot\lambda_i(\vec{T})$. The parameters $\vec{\alpha}$ are constrained by the multivariate normal distribution denoted by $\mathcal{P}(\vec{\alpha})$, which has a fractional covariance matrix derived from the cross section uncertainties.

To explore the sensitivity of this combined analysis to the $\nu_x$ timed averaged temperature, we examine a set of representative scenarios where $T_{\nu_e} = 3.3$ MeV, $T_{\overline{\nu_e}} = 4.6$ MeV, and $T_{\nu_x}$ has a value between 0.5 and 30 MeV. For each value of $T_{\nu_x}$ we perform an Asimov test, where the nominal expected event distribution is injected as data. We use a Markov chain Monte Carlo with adaptive parallel tempering [20,21] to explore the parameter space, and derive Bayesian credible regions (CRs) for $T_{\nu_x}$ that are marginalized over $T_{\nu_e}$, $T_{\overline{\nu_e}}$, and the nuisance parameters $\vec{\alpha}$. The two electron flavor temperatures have normal priors modeled after the expected constraints from CC data, and the nuisance parameters $\vec{\alpha}$ have a multivariate-normal prior, as described in Eq. (3). We use a uniform prior for $T_{\nu_x}$ across most of the parameter space, but introduce a hyperbolic tangent cutoff at 60 MeV, with a characteristic transition width of 3 MeV. For injected $T_{\nu_x}$ less than 10 MeV a degenerate region of the parameter space becomes apparent, which extends to reconstructed temperatures above 100 MeV for the lowest injected temperatures. We focus the results presented here on temperatures below 60 MeV, because such large temperatures are not within the expected range, and will be ruled out by other observations [22–27]. This is accomplished with the hyperbolic tangent cutoff, the primary effect of which is a reduction in the size of the derived CRs for injected $T_{\nu_x}$ less than 5 MeV, because there is no longer appreciable posterior mass above 60 MeV. Additional discussion of this prior, and CRs derived with a uniform $T_{\nu_x}$ prior from 0.1 to 300 MeV are given in the Appendix.

## V. RESULTS

The time averaged $\nu_e$ and $\overline{\nu_e}$ temperatures will be constrained very well by CC measurements from DUNE and Hyper-K. As a result, the sensitivity to $T_{\nu_x}$ does not significantly depend on the choice of injected $T_{\nu_e}$ and $T_{\overline{\nu_e}}$. We can therefore examine the sensitivity to reconstructed $\nu_x$ temperature as a function of true injected $\nu_x$ temperature without significant bias from the choice of injected $\nu_e$ and $\overline{\nu_e}$ temperatures. We present sensitivities to $T_{\nu_x}$, using $T_{\nu_e} = 3.3$ MeV and $T_{\nu_e} = 4.6$ MeV as a representative set of temperatures.

We explore sensitivities for three scenarios with different cross section uncertainties: a case without cross section uncertainties, a case with full cross section uncertainties, and a case with reduced cross section uncertainties. The case with full cross section uncertainties assumes 40% fully uncorrelated uncertainty in the cross section and a 15% correlated uncertainty in the cross section as described in Sec. III. The case with reduced cross section uncertainties retains the 40% fully uncorrelated uncertainty, but reduces the correlated uncertainty to 7% to model the effect of improved B(M1↑) measurements. Figures 8–10 show the expected sensitivity to $T_{\nu_x}$ as Asimov credible regions for

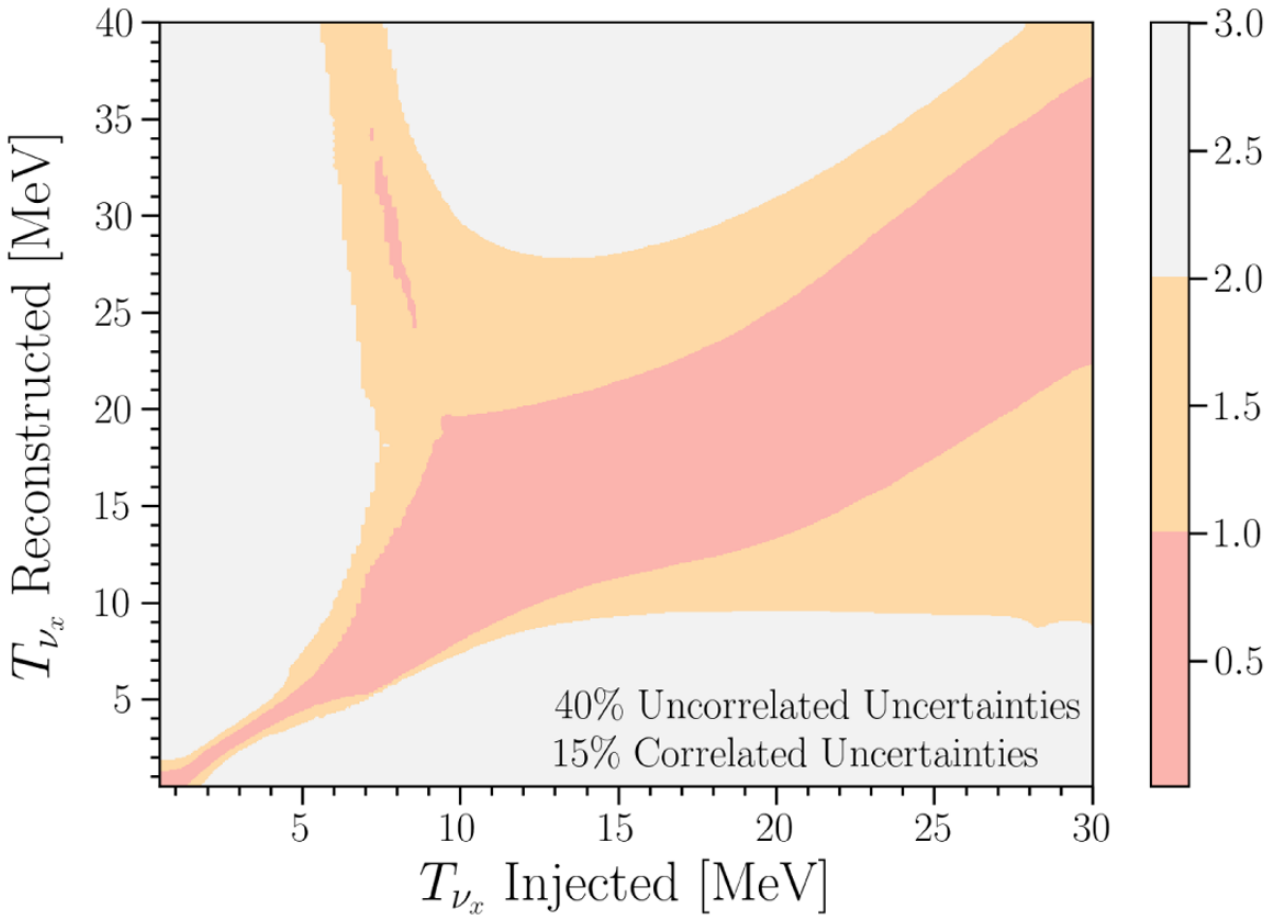


FIG. 8. Expected $T_{\nu_x}$ credible regions with full cross section uncertainties. The Asimov credible regions for $T_{\nu_x}$ are shown as a function of injected $T_{\nu_x}$ by the colored regions. Average neutrino temperatures of $T_{\nu_e} = 3.3$ MeV and $T_{\overline{\nu_e}} = 4.6$ MeV are assumed, along with 40% uncorrelated uncertainties and 15% correlated uncertainties on the NC cross section.

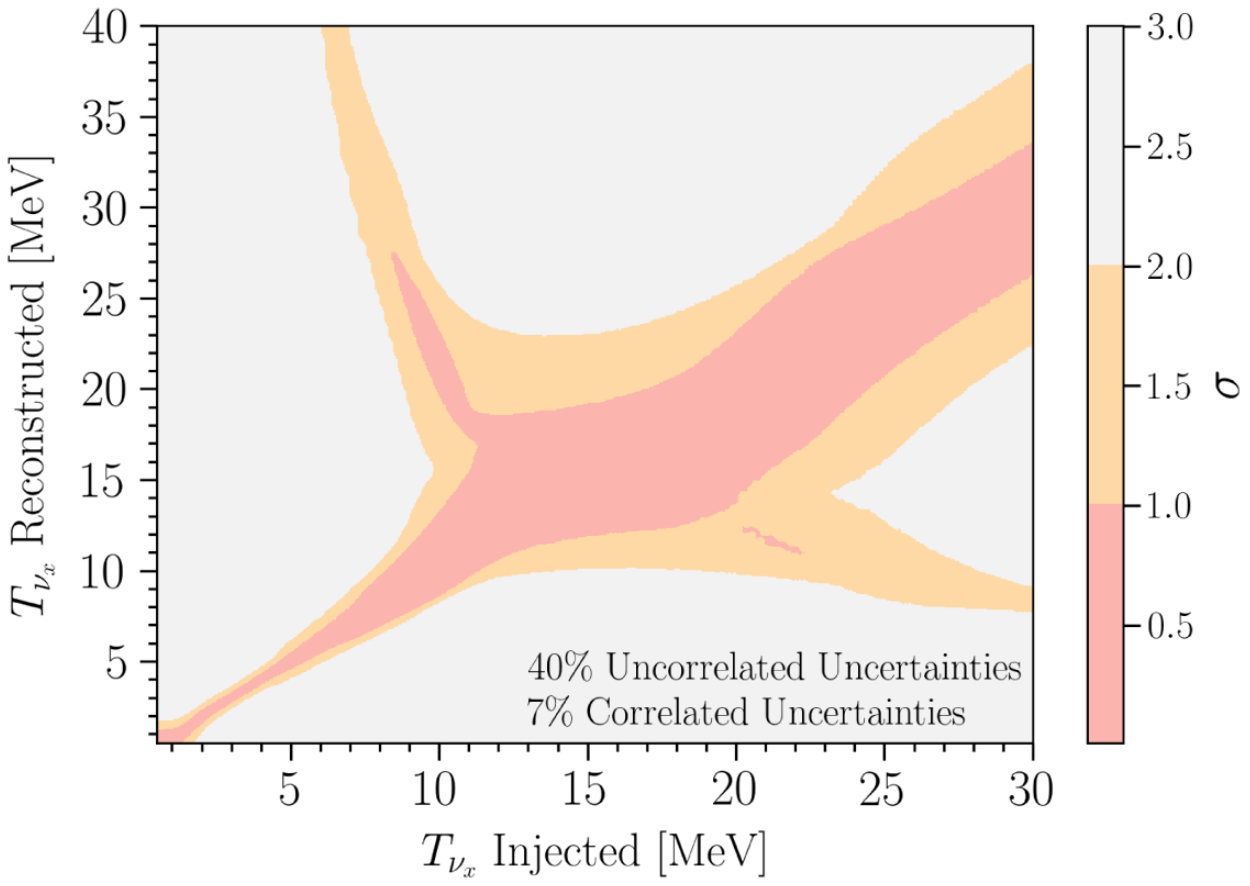


FIG. 9. Expected $T_{\nu_x}$ credible regions with reduced cross section uncertainties. The Asimov credible regions for $T_{\nu_x}$ are shown as a function of injected $T_{\nu_x}$ by the colored regions. Average neutrino temperatures of $T_{\nu_e} = 3.3$ MeV and $T_{\overline{\nu_e}} = 4.6$ MeV are assumed, along with 40% uncorrelated uncertainties and 7% correlated uncertainties on the NC cross section.

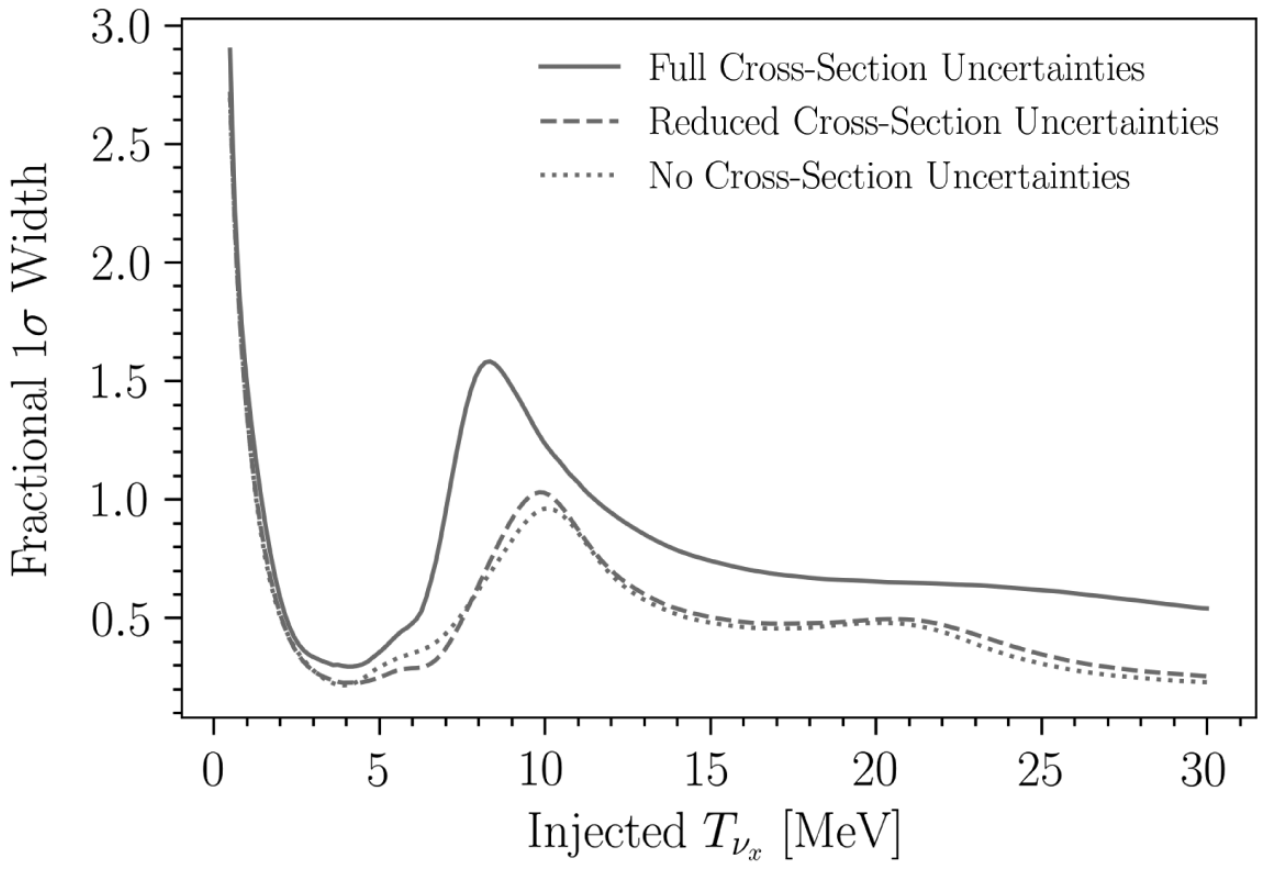


FIG. 11. Fractional width of the $1\sigma$ credible region. The solid line represents the scenario with 40% uncorrelated uncertainties and 15% correlated uncertainties on the NC cross section. The dashed line represents the scenario with 40% uncorrelated uncertainties and 7% correlated uncertainties on the NC cross section. The dotted line represents the scenario with no uncertainties on the NC cross section.

the no uncertainty, full uncertainty, and reduced uncertainty cases, respectively.

There are two distinct degenerate regions that arise in the measurement space from multiple factors. The primary cause is the discrete $\gamma$-ray energies of the NC cross section that results in a decoupling between neutrino energy and observed energy. In addition to that, for $T_{\nu_x}$ above approximately 8 MeV the total fluence becomes much more uniform across the energy range of the NC excitations. These characteristics of the NC cross section combined with the Fermi-Dirac spectrum at higher temperatures leads to a degeneracy in the $\nu_x$ temperature measurement.

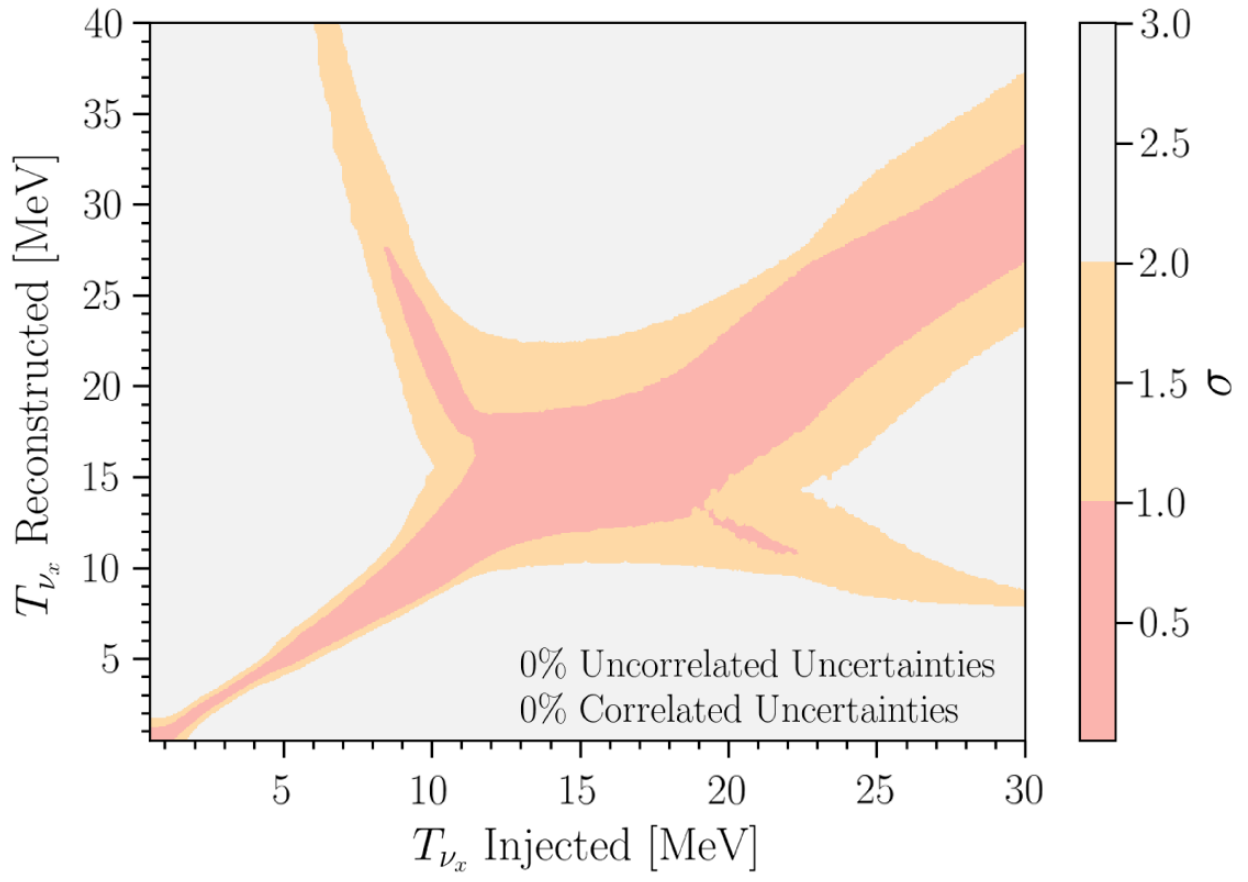


FIG. 10. Expected $T_{\nu_x}$ credible regions with no cross section uncertainties. The Asimov credible regions for $T_{\nu_x}$ are shown as a function of injected $T_{\nu_x}$ by the colored regions. Average neutrino temperatures of $T_{\nu_e} = 3.3$ MeV and $T_{\overline{\nu_e}} = 4.6$ MeV are assumed, along no uncertainties on the NC cross section.

Figure 11 shows the fractional width of the $1\sigma$ credible region for these measurements. The peak in the fractional width of the $1\sigma$ credible region around 10 MeV true $\nu_x$ temperature is due to the degeneracy in the measurement. Reducing the correlated uncertainties from 15% to 7%, without changes to the 40% uncorrelated uncertainties, reduces the fractional width of the $1\sigma$ credible region almost to that of the measurement without any uncertainties on the cross section. This demonstrates that even modest improvements to the argon $B(M1\uparrow)$ measurements provide substantial improvements to the $T_{\nu_x}$ sensitivity.

## VI. CONCLUSIONS

While NC channel measurements of supernova neutrinos are often overlooked due to the technical challenges and small sample sizes, this analysis has shown the power of NC measurements in measuring $\nu_x$ spectral information from type-II core-collapse supernova events. Because of the discrete excitation nature of this interaction channel, the energy spectrum from supernova neutrinos cannot be resolved, but general parameters of the fluence can be measured to the $1\sigma$ and $2\sigma$ level. For this analysis we focused on measuring the time averaged $\nu_x$ temperature from the incoherent NC neutrino-argon interaction in the DUNE FD modules. To make this measurement we combined the expected incoherent NC sample from the DUNE FD with expected constraints from the CC channel measurements of the DUNE and Hyper-K experiments on $T_{\nu_e}$ and $T_{\overline{\nu}_e}$ of the supernova flux. We find that such a combined analysis can measure the $\nu_x$ temperature to within a factor of 2 in most cases, and to within 30% in the best case.

In addition to exploring DUNE's sensitivity to time averaged $\nu_x$ temperature using the NC channel, we explore

the effects of reducing the uncertainty in NC cross section predictions. By reducing correlated uncertainties from 15% to 7%, the $T_{\nu_x}$ measurement becomes sample size limited, and approaches the case where there are no uncertainties on the cross section. More precise measurements of $B(M1\uparrow)$ transition strengths will provide even greater spectral information for the $\nu_x$ component of supernova flux. While this measurement will be limited by the small energy range of argon-nucleus gamma emissions from NC interactions, combining NC measurements of neutrinos on argon with NC measurements of neutrinos on carbon from JUNO could provide stronger constraints on time averaged $\nu_x$ temperature.

## ACKNOWLEDGMENTS

We want to thank Janet M. Conrad for insights on liquid argon TPCs. We would also like to thank Anna C. Hayes for her dedicated work on modeling neutral current cross sections. D. A. N. is supported by the NSF Graduate Research Fellowship under Grant No. 2141064. A. S. is supported by the U.S. Department of Energy through the Los Alamos National Laboratory. Los Alamos National Laboratory is operated by Triad National Security, LLC, for the National Nuclear Security Administration of U.S. Department of Energy (Contract No. 89233218CNA000001).

## APPENDIX: $T_{\nu_x}$ PRIOR

In this work we demonstrate the potential sensitivity of the DUNE FD to the time averaged temperature of the supernova $\nu_x$ component with credible regions derived from the highest posterior density regions (HPDs). A side effect of this methodology is a dependence of the local HPD width on the global distribution of posterior mass. For these NC derived constraints on $T_{\nu_x}$, a degenerate region of allowed $\nu_x$ temperatures is present for the entire energy range. Below ~10 MeV and above ~20 MeV, this degeneracy manifests as two distinct allowed regions at both the $1\sigma$ and $2\sigma$ level. For some values of injected $T_{\nu_x}$ the allowed region not centered on the injected temperature, referred to as the degenerate region, lies within the range of expected supernova neutrino temperatures. However, for injected $T_{\nu_x}$ less than ~5 MeV this degenerate region exists at temperatures far above the expected range of supernova neutrino

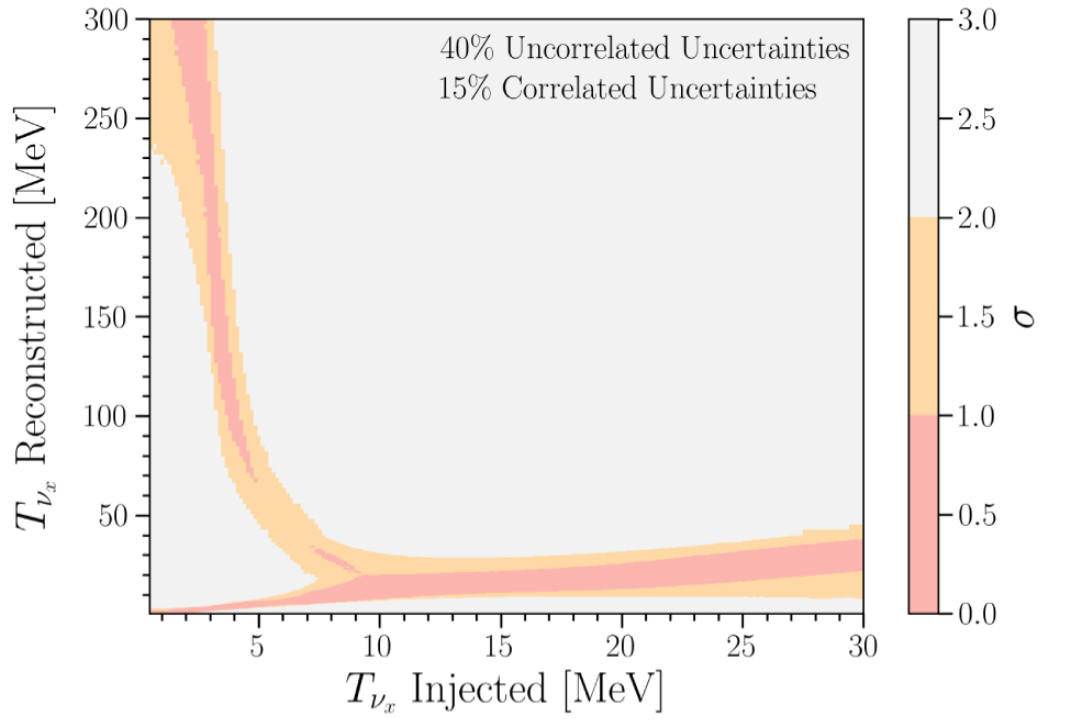


(a) 40% uncorrelated uncertainties and 15% correlated uncertainties on the NC cross-section

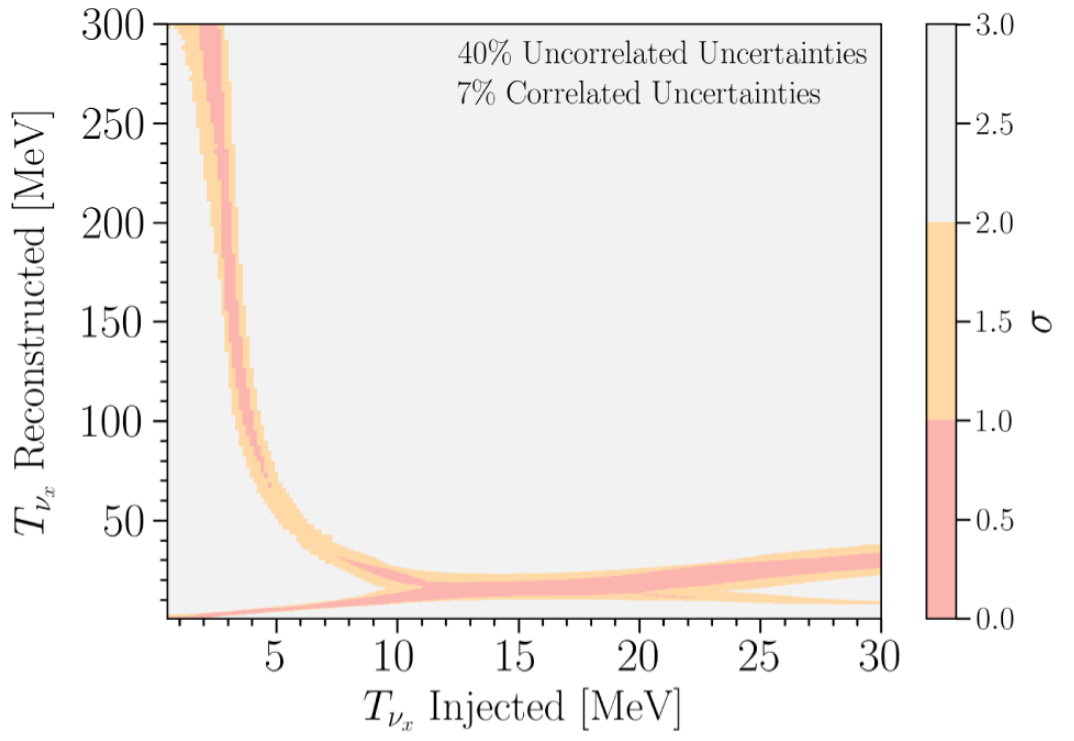


(b) 40% uncorrelated uncertainties and 7% correlated uncertainties on the NC cross-section

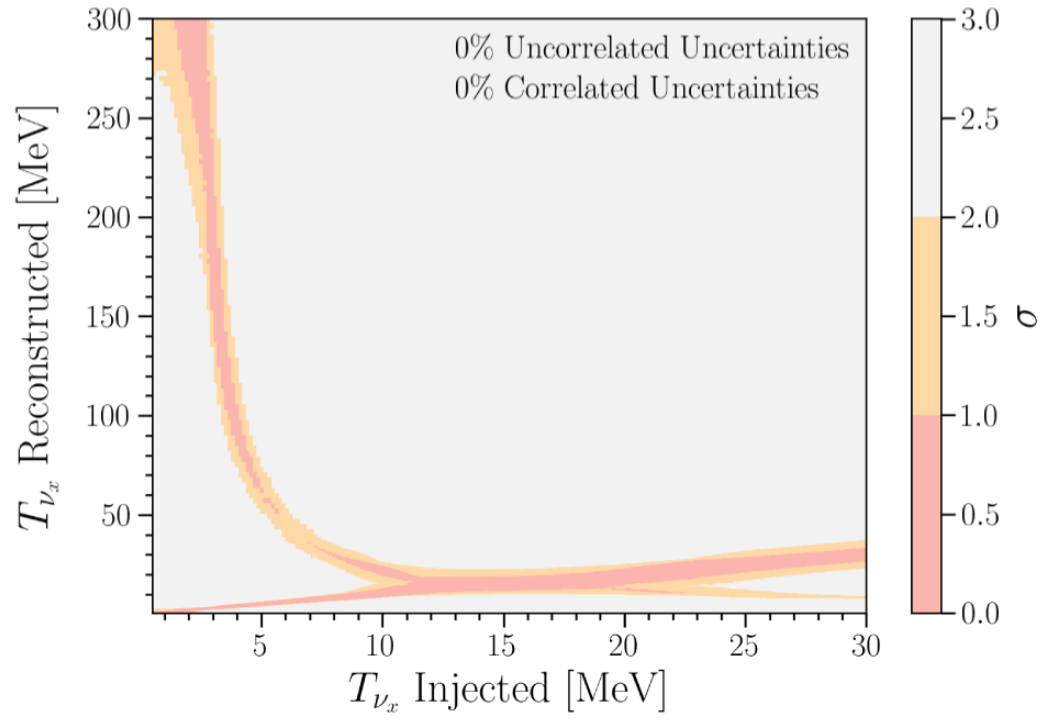


(c) No uncertainties on the NC cross-section

FIG. 12. Expected $T_{\nu_x}$ credible regions with a uniform prior from 0.1 to 300 MeV. The Asimov credible regions for $T_{\nu_x}$ are shown as a function of injected $T_{\nu_x}$ by the colored regions. Average neutrino temperatures of $T_{\nu_e} = 3.3$ MeV and $T_{\overline{\nu_e}} = 4.6$ MeV are assumed. The credible regions in this figure are computed with a uniform $T_{\nu_x}$ prior that extends up 300 MeV; this is in contrast to the credible regions shown in Figs. 8–10 which limited values of $T_{\nu_x}$ to be below 60 MeV.

temperatures. In the main text of this paper, we choose to apply a prior that allows only values of $T_{\nu_x}$ below 60 MeV. This is motivated by the low temperatures observed from SN1987A [23], predicted by supernova simulation studies [24], required by constraints from neutrino induced nucleosynthesis [25,26], and disfavored by diffuse supernova background searches [27]. This results in expected sensitivities that are not biased by a degeneracy that is both far outside of the range of expected supernova behavior and which would be ruled out by other measurements.

In this Appendix we show the expected sensitivities under a different prior assumption, namely a uniform prior on $T_{\nu_x}$ between 0.1 and 300 MeV. Because the degenerate region extends above 60 MeV at low $\nu_x$ temperatures and contains appreciable posterior mass, this has the effect of widening the allowed regions of $T_{\nu_x}$ that are centered on the injected $T_{\nu_x}$. Figure 12 show the expected sensitivity to $T_{\nu_x}$ with this wider range of allowed $T_{\nu_x}$ values for the three different cross section uncertainty scenarios explored in the main text. The degenerate region below ~5 MeV extends up into the 100s of MeV, and below ~3 MeV the degenerate region begins to intersect with the prior boundary at 300 MeV.

To examine the effect of this prior we compare the width of the 1$\sigma$ credible region across three scenarios: the full width derived from a uniform $T_{\nu_x}$ prior between 0.1 and 300 MeV, the width of the allowed region below 60 MeV using the same uniform $T_{\nu_x}$ prior between 0.1 and 300 MeV, and finally the width derived using the prior from the main text (a hyperbolic-tangent cutoff prior that penalizes $T_{\nu_x}$ above 60 MeV and has a 3 MeV characteristic width). Figure 13 shows the fractional width of these 1$\sigma$ credible regions for the three scenarios. The solid lines

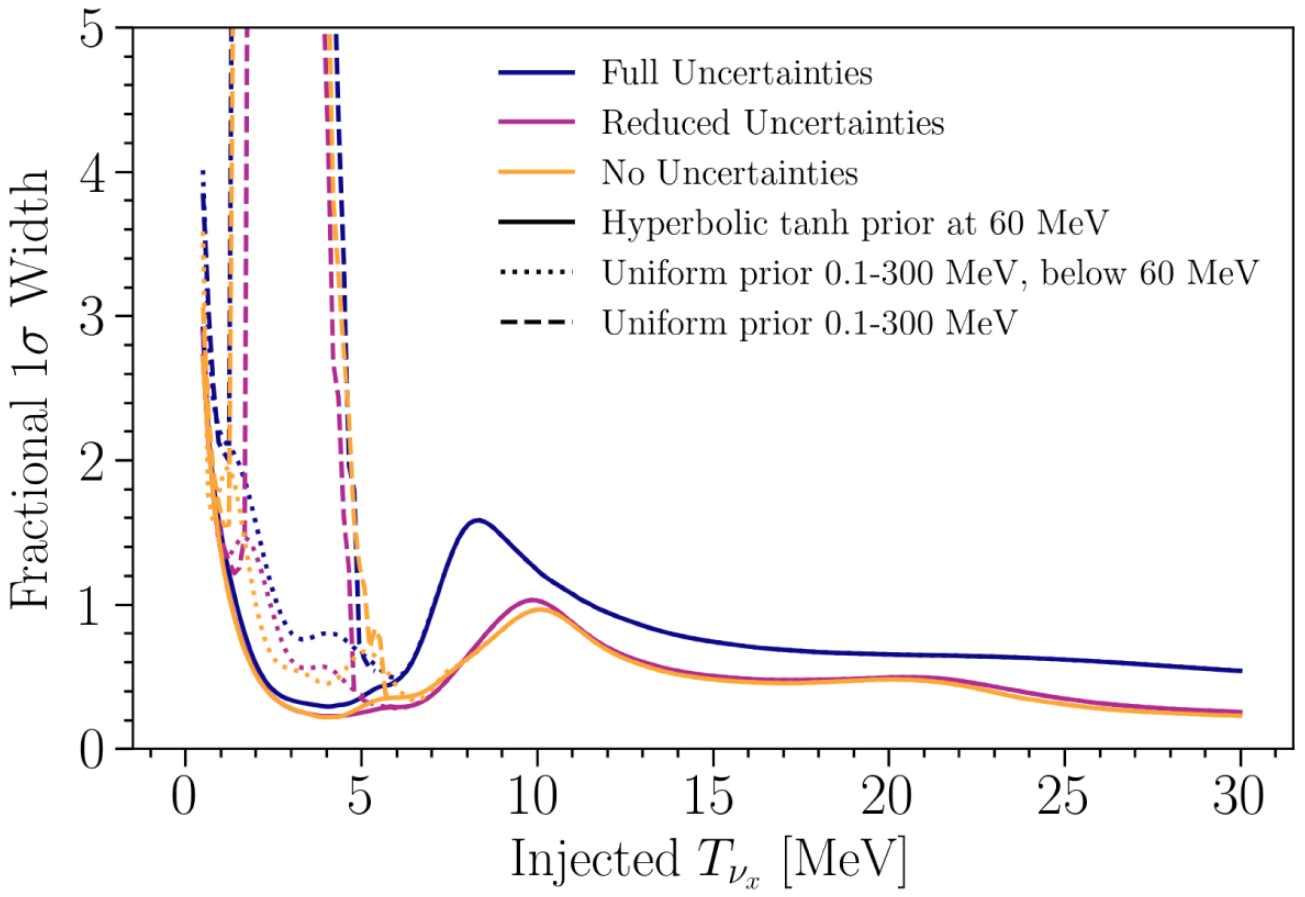


FIG. 13. Fractional width of 1$\sigma$ credible regions. The solid lines correspond to the width from the hyperbolic tangent prior, the dotted lines correspond to the width below 60 MeV from the 0.1–300 MeV uniform prior, and the dashed lines correspond to the full width the from the 0.1–300 MeV uniform prior (which extends up to 35). The three colors correspond to the three cases of cross section uncertainties.

denote the case where $T_{\nu_x}$ prior is in the form of a hyperbolic tangent cutoff at 60 MeV. Relaxing the prior assumptions on $T_{\nu_x}$ significantly widens the allowed regions below 60 MeV for injected $T_{\nu_x}$ between 1 and 5 MeV, but does not significantly affect the expected allowed regions outside of this region. The width of the allowed regions considering the full $T_{\nu_x}$ parameter space up to 300 remains largely unchanged outside of this 1 to 5 MeV region of injected $T_{\nu_x}$, but between 1 and 5 MeV is dominated by the large width of the high-temperature degenerate region.




\chapter{Summary and Future Prospects}\label{chap:conclusion}
This thesis has presented the first demonstration of a hybrid Cherenkov and scintillation optical detector operating at a beam dump, applied to a proof-of-principle search for axion-like particles (ALPs). This achievement was enabled by fast 2~ns time resolution and the combination of photo-sensors coated in wavelength shifter and uncoated. Detector calibration, including measurement of liquid argon light production and propagation parameters, and identification of Cherenkov light from sub-MeV electrons on an event-by-event basis was performed on data from a $^{22}$Na radioactive calibration source. This work resulted in Ref.~\cite{CCM:2025dbq} and Ref.~\cite{CCM:2025kal}, for which the thesis author was the principal author.

Additionally, this thesis advanced event reconstruction techniques for the CCM experiment. A machine learning–based position reconstruction algorithm achieved $\mathcal{O}(5~\text{cm})$ resolution in each spatial dimension. Energy reconstruction was performed using a position-dependent scaling of the collected charge within an $\sim$100~ns time window, providing a consistent framework for low-energy event characterization, achieving $\mathcal{O}(10\%)$ energy resolution.

Building on these developments, a complete ALP analysis pipeline was constructed. A detailed simulation of the tungsten target was employed to model the gamma-ray flux necessary for ALP production via the Primakoff process. The subsequent ALP production, propagation, and detection was simulated with a realistic implementation of the experimental geometry in the \texttt{SIREN} injection framework. Two detection channels were considered: inverse Primakoff scattering on argon, yielding a single gamma-ray final state, and diphoton decay, which dominates at higher ALP masses, resulting in two gamma-ray final states. The signal was evaluated over a dense grid of $10^4$ points in the mass–coupling parameter space.

To fully exploit the hybrid detection concept, four discriminating observables were developed to isolate electromagnetic final state particles. These include sensitivity to prompt Cherenkov light in uncoated PMTs, directional information of early photons, pulse shape characteristics, and the spatial extent of charge deposition. Combined into a likelihood ratio test statistic, these variables provided strong suppression of steady-state and neutron-induced backgrounds. 

Within a frequentist framework, no significant excess consistent with ALPs was observed. The improved background rejection enabled this analysis to extend 90\% confidence level excluded regions of parameter space beyond previous CCM120 results despite a smaller exposure of protons on target. This work establishes analysis techniques applicable to hybrid Cherenkov and scintillation detectors in high background environments like beam dump facilities.  

In addition to the ALP search, this thesis explored the use of neutral current interactions of neutrinos on argon to constrain the total neutrino flux from core-collapse supernovae in the DUNE far detector modules, highlighting complementary physics opportunities for liquid argon detectors.

Ongoing work within the collaboration builds on these developments. The instrumentation techniques presented here are being utilized to study higher-energy calibration samples, including cosmic-ray muons and Michel electrons. This will further extend the sensitivity and validity of future CCM200 analyses above 10~MeV.

Overall, this thesis establishes hybrid Cherenkov and scintillation detection as a powerful technique for rare-event searches at beam dump facilities, and lays the groundwork for its application in next-generation neutrino and dark sector experiments.

\appendix

\chapter{Developments of an Ultra Large Liquid Argon Hybrid Detector}\label{chap:appendix}

Building on the motivations behind the work presented in Chapter~\ref{chap:pheno}, this study explores the physics reach of an optical-only detector as part of the DUNE far detector complex. Leveraging the demonstrated separation of Cherenkov and scintillation light in liquid argon with the CCM200 detector, this work evaluates the performance of such a detector concept. In particular, the ongoing investigations probe whether an optical-only approach could serve as a viable alternative to a traditional LArTPC for one of the far detector modules to extend the potential physics reach.

\section{Ultra Large Liquid Argon Hybrid Detector}
A large experimental program such as DUNE naturally enables a broad range of complementary physics opportunities beyond its core long-baseline oscillation measurement mission. One proposed concept is Theia~\cite{Theia:2019non}, an ultra-large optical detector designed to extend sensitivity to solar neutrinos and potentially neutrinoless double beta decay while remaining competitive for oscillation physics. Named after the Greek goddess of light, Theia is an optical detector that would exploit both Cherenkov and scintillation signals to support a wide-ranging physics program.

The proposed Theia-25 design consists of approximately 25~kton of water-based liquid scintillator (WbLS) instrumented within a detector volume comparable to the DUNE far detector caverns. Combined with fast photo-detectors and potential spectral sorting~\cite{Lyashenko:2019tdj, Kaptanoglu:2018sus}, this approach aims to fully leverage the timing and wavelength differences between Cherenkov and scintillation light to create the optimal hybrid detector. The resulting physics reach includes sensitivity to long-baseline oscillations, burst of neutrinos from core-collapse supernovae, the diffuse supernova neutrino background, solar neutrinos, reactor and geoneutrinos, neutrinoless double beta decay, and proton decay.

To explore how to optimize such a hybrid optical detector, this phenomenological study investigates liquid argon as an alternative target medium. Although liquid argon is generally more expensive than WbLS and requires cryogenic operation, it offers several compelling advantages. It aligns naturally with the existing liquid argon infrastructure planned for DUNE and benefits from extensive prior research and development, whereas WbLS remains under active development and testing~\cite{Anderson:2022lbb, Yeh:2011zz, Bignell:2015oqa, Caravaca:2016fjg, Caravaca:2016ryf}.

Moreover, using liquid argon in both near and far detectors could reduce the impact of systematic uncertainties in flux measurements, particularly those associated with hadronic interaction models, by providing a consistent nuclear target across the experimental program. Additionally, the work presented in this thesis explores some of the benefits of liquid argon as a hybrid detector medium over liquid scintillators, highlighted in Chapter~\ref{chap:cherenkov}. This study aims to assess the capabilities of liquid argon for an ultra-large hybrid optical detector. 

\subsection{Detector Concept}
\begin{figure}[h]
  \centering
  \includegraphics[width=0.7\linewidth]{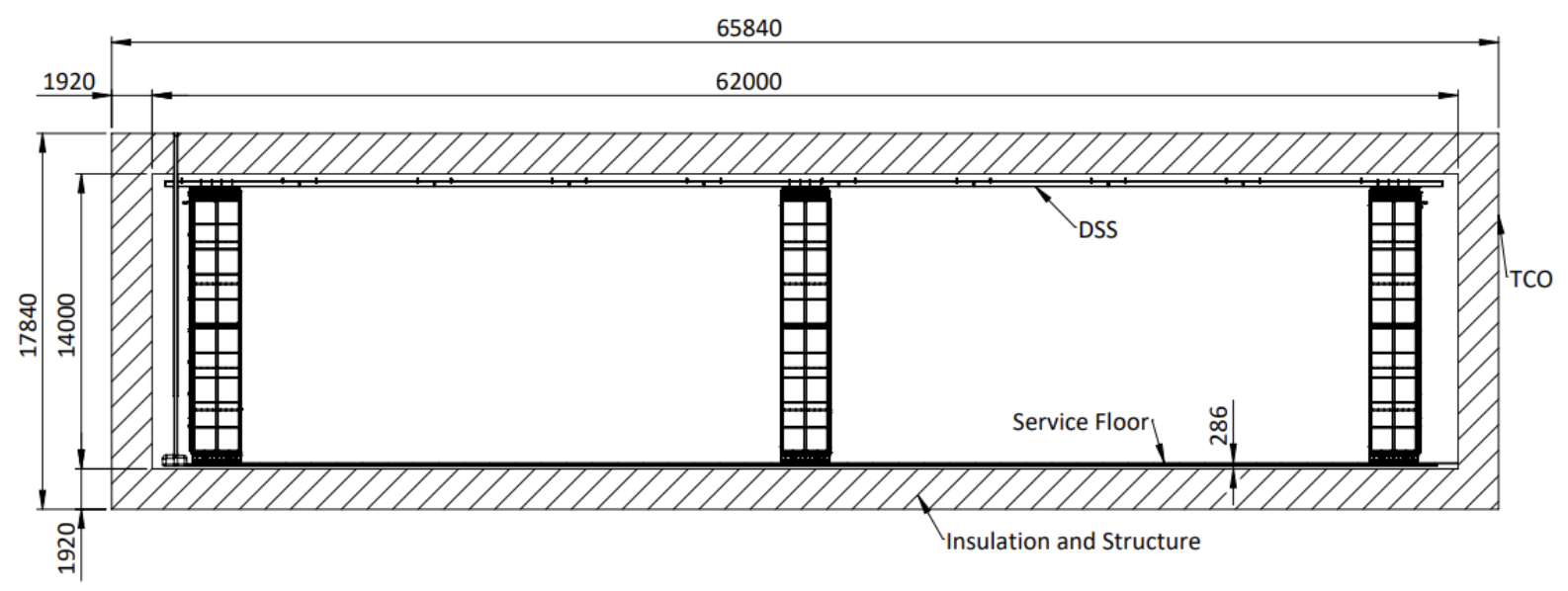}
  \caption{Illustration of the DUNE far detector module cryostat. For this work, we assume a rectangular prism detector with inner dimensions of $14~\text{m} \times 15.1~\text{m} \times 62~\text{m}$ and a 13.2~mm thick steel structure. Figure from Ref.~\cite{DUNE:2020mra}.}
  \label{fig:moo_geo}
\end{figure}

This work presents a \texttt{GEANT4}-based simulation study of an ultra-large liquid argon optical detector. We model a single DUNE far detector module cryostat instrumented with photomultiplier tubes (PMTs) along its inner surfaces to detect optical photons. The detector geometry, shown in Fig.~\ref{fig:moo_geo}, is a rectangular prism with inner dimensions of $14~\text{m} \times 15.1~\text{m} \times 62~\text{m}$~\cite{DUNE:2020mra}.

The inner surfaces of the cryostat are instrumented with $83{,}670$ 8-inch cryogenic PMTs, uniformly distributed across all six faces, yielding an effective photocathode coverage of approximately $70\%$. The remaining surface area is coated with a 2.8~$\mu$m thick layer of the wavelength shifter tetraphenyl butadiene (TPB), implemented using optical properties from Ref.~\cite{Benson:2017vbw}. We use the Hamamatsu Photonics reference quantum efficiency for 8-inch cryogenic PMTs with a maximum efficiency of approximately 18\% at 425~nm~\cite{hamamatsu_pmt_handbook}. For this study, 2~ns time resolution is assumed.

For the liquid argon model, we assume ultra-high purity conditions consistent with requirements for TPC operation. The optical properties include the wavelength-dependent absorption length from the \texttt{LArSoft} framework~\cite{Church:2013hea}, scintillation parameters measured by the DEAP experiment~\cite{DEAP:2020hms}, and the scintillation emission spectrum from Ref.~\cite{Heindl:2010zz}. The index of refraction is taken from a fit using a damped harmonic oscillator model~\cite{Rahman:2024zhp}, and Rayleigh scattering is modeled with a reference length of 100~cm at 128~nm, including its wavelength and refractive index dependence~\cite{landau1984electrodynamics}.

\subsection{Preliminary Results}
\begin{figure}[h]
  \centering
  \includegraphics[width=\linewidth]{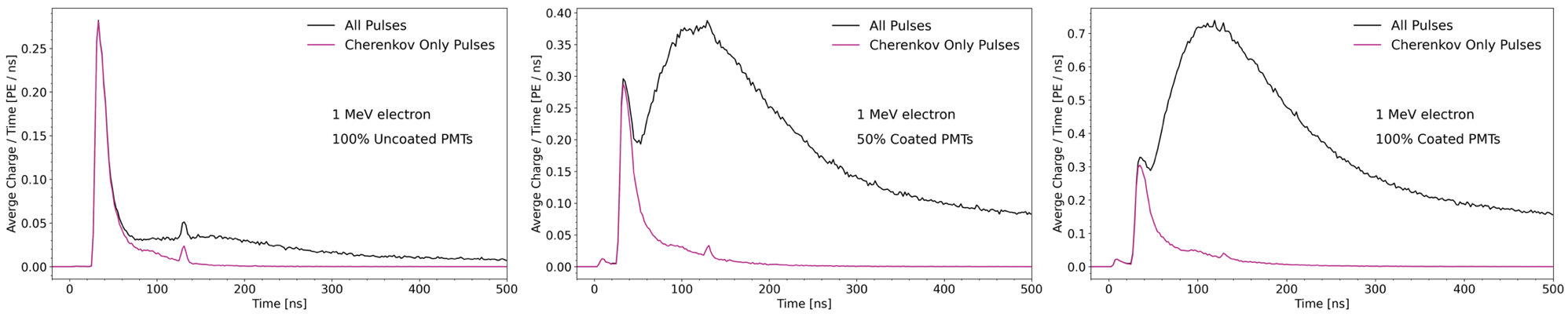}
  \caption{Average photoelectron yields from simulation of 1~MeV electrons with different detector configurations. From left to right, none of the PMTs are coated in TPB, 50\% of the PMTs are coated in TPB, and 100\% of the PMTs are coated in TPB. The black distribution represents total scintillation and Cherenkov
  hits while the magenta distribution is only from Cherenkov radiation.}
  \label{fig:moo_sim_1mev_config}
\end{figure}

\begin{figure}[h]
  \centering
  \includegraphics[width=\linewidth]{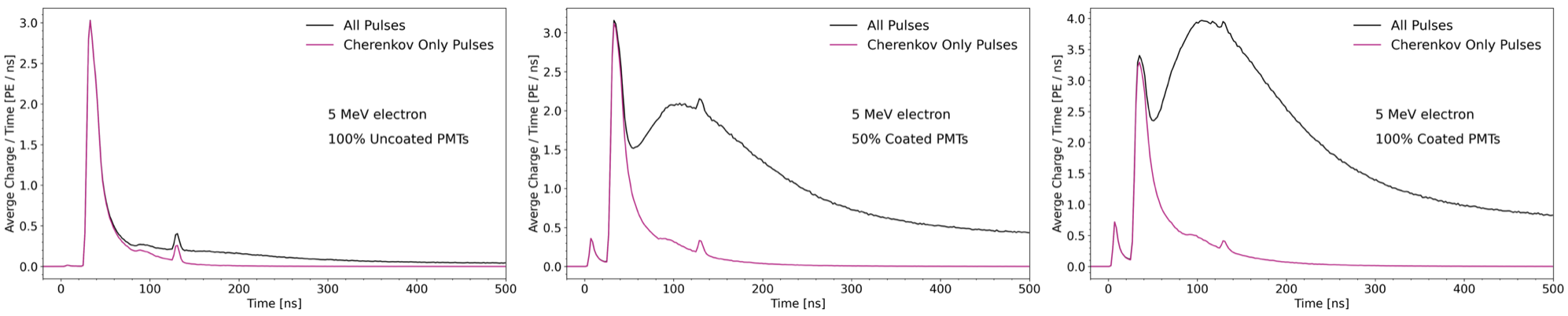}
  \caption{Average photoelectron yields from simulation of 5~MeV electrons with the same detector configurations as described in Fig~\ref{fig:moo_sim_1mev_config}.}
  \label{fig:moo_sim_5mev_config}
\end{figure}

Motivated by studies of Cherenkov and scintillation separation in the CCM200 detector, where uncoated PMTs provide enhanced sensitivity to the prompt, visible component of Cherenkov radiation, this study explores how varying levels of TPB coverage impact hybrid detector performance.

We simulate 1~MeV and 5~MeV electrons to benchmark performance across different energy regimes. Three detector configurations are considered: (i) all PMTs left uncoated, with wavelength shifting provided only by TPB-coated detector walls; (ii) 50\% of PMTs additionally coated with a 2.0~$\mu$m TPB layer; and (iii) full TPB coverage of all PMTs, again in addition to the TPB-coated walls.

Figures~\ref{fig:moo_sim_1mev_config} and~\ref{fig:moo_sim_5mev_config} show the resulting average photoelectron (PE) time profiles for 15,000 simulated events at 1~MeV and 5~MeV, respectively. In each case, electrons are generated at the detector center with isotropically distributed initial directions. The vertical axis corresponds to the average PE yield per 2~ns time bin. The total detected signal (scintillation light plus Cherenkov radiation) is shown in black, while the Cherenkov-only contribution is shown in magenta. A recurring feature around $\sim 130$~ns across all configurations is attributed to optical reflections from the detector boundaries.

In the fully uncoated configuration, only a fraction of the detector surface is wavelength-shifting, resulting in reduced overall scintillation light collection. As TPB coverage of the PMTs increases, scintillation light yield correspondingly increases, while the relative visibility of the prompt Cherenkov component becomes more challenging to isolate.

In the 50\% TPB-coated configuration, a distinct prompt Cherenkov-dominated component remains visible in the early time region, even for 1~MeV electrons. In contrast, full TPB coverage significantly enhances scintillation light collection but reduces time-based separation between Cherenkov and scintillation signals, particularly at lower energies. Nevertheless, separation remains promising for the 5~MeV case, where the Cherenkov component is more pronounced.

Overall, these preliminary results suggest that Cherenkov radiation and scintillation separation remains feasible across a range of TPB coverage scenarios, though with a clear trade-off between total light yield and prompt-component visibility. Further work is ongoing to quantify the impact of these configurations on reconstructed event-level observables and particle identification performance.

\subsection{Ongoing Collaboration}
This work has been undertaken in collaboration with Professor Gabriel D. Orebi Gann at the University of California, Berkeley and Lawrence Berkeley National Laboratory, spokesperson of the Theia Consortium. 

Within this collaboration, this thesis contributes to simulation studies of various detector configurations. Ongoing work is comparing reconstruction techniques, using both likelihood based methods and machine learning techniques, to assess the physics capabilities of an ultra large liquid argon hybrid optical detector. 

This effort is expected to result in one to two publications detailing detector concepts and associated physics sensitivity studies.


\defbibheading{bibintoc}{\chapter*{#1}\addcontentsline{toc}{backmatter}{\refname}} 

\printbibliography[heading=bibintoc]


\end{document}